\documentclass[letterpaper,12pt]{report}	

\usepackage[authoryear, round]{natbib}
\usepackage{styles/utformat}  		

\usepackage{pdflscape}
\usepackage{graphicx}
\usepackage[linesnumbered,lined,ruled,commentsnumbered]{algorithm2e}
\usepackage{amsmath,amsthm,amsfonts,amscd}
\usepackage{eucal} 	 	
\usepackage{verbatim}      	
\usepackage{styles/citesort}         	%
\usepackage{float}
\usepackage{bibentry}
\usepackage{xcolor}
\usepackage{markdown}
\usepackage{booktabs}
\usepackage{url}
\usepackage{breakurl}
\usepackage{comment}
\usepackage{subfig}
\usepackage{multirow}
\usepackage{makecell}
\usepackage{tabularx}
\usepackage{bbm}
\usepackage{mathtools}
\usepackage{hyperref}
\usepackage{cleveref}
\hypersetup{colorlinks=false}
\usepackage{svg}
\usepackage{colortbl}
\usepackage[table]{xcolor}
\usepackage{listings}
\usepackage{tablefootnote}
\usepackage[flushleft]{threeparttable}
\usepackage[symbol]{footmisc}
\usepackage[colorinlistoftodos]{todonotes}

\newcommand{\arraylength}[1][]{\textit{TB\textsubscript{len}}#1}
\newcommand{\Mexpansion}[1][]{\textit{M\textsubscript{p}}#1}
\newcommand{\Nexpansion}[1][]{\textit{N\textsubscript{p}}#1}
\newcommand{\Kexpansion}[1][]{\textit{K\textsubscript{p}}#1}

\definecolor{LightBlue}{rgb}{0.83, 0.91, 1}

\let\oldnl\nl
\newcommand{\nonl}{\renewcommand{\nl}{\let\nl\oldnl}}
\definecolor{codegreen}{rgb}{0,0.6,0}

\oneandonehalfspacing

\theoremstyle{definition}

\theoremstyle{remark}

\newcommand{\latexe}{{\LaTeX\kern.125em2%
                      \lower.5ex\hbox{$\varepsilon$}}}

\chardef\bslash=`\\	

\makeatletter		
\def\square{\RIfM@\bgroup\else$\bgroup\aftergroup$\fi
  \vcenter{\hrule\hbox{\vrule\@height.6em\kern.6em\vrule}%
                                              \hrule}\egroup}

\makeatother		

\nobibliography*

\author{Endri Taka}  	

\address{endri.taka@utexas.edu}  

\title{Efficient Reconfigurable Architectures for Deep Learning} 

\supervisor
	{Diana Marculescu}

\committeemembers
	[Lizy Kurian John]
	[Andreas Gerstlauer]
    [Michael Schulte]
	{Aman Arora}

\degree{Doctor of Philosophy}
\degreeabbr{PhD} 

\dissertation  

\begin{document}


\copyrightpage  

\makeatletter
\def\certif@words{
  \centerline{The Dissertation Committee for \au@thor}
  \centerline{certifies that this is the approved version of the following dissertation:}
}
\makeatother

\commcertpage   

\titlepage      

\utabstract
The rapid growth in computational demands of contemporary deep learning (DL) workloads has driven the development of diverse AI-optimized hardware architectures, including GPUs, ASICs, and FPGAs. Among these, FPGAs are particularly appealing for the fast-evolving DL landscape due to their high performance, low power consumption, and reconfigurability. Recognizing this potential, vendors have introduced AI-optimized reconfigurable devices that integrate dedicated compute engines, such as the AI Engines (AIEs) in the AMD Versal ACAP, the Tensor Blocks (TBs) in the Intel Stratix 10 NX FPGA, and the AIE-ML tiles in the AMD Ryzen AI NPUs.

However, these devices introduce new design challenges. Their heterogeneous compute and memory resources and novel dataflow models differ fundamentally from conventional reconfigurable devices, thus established optimization techniques are not directly applicable. As a result, exploiting these architectures efficiently is nontrivial, and the optimization of fundamental DL kernels on them is not well understood. Furthermore, unlike GPUs, which provide dedicated hardware for structured sparsity, FPGAs lack native support for it, leaving a major source of efficiency unexploited. This dissertation addresses these gaps by optimizing dense general matrix multiplication (GEMM), the dominant computational kernel in most DL workloads, on state-of-the-art AI-enabled FPGAs. It then proposes architectural enhancements to FPGAs for structured sparsity and extends these optimization techniques to next-generation AMD Ryzen AI NPU architectures.

First, this dissertation presents MaxEVA, a framework that optimizes GEMM performance on the Versal AIE devices by efficiently exploiting the AIE’s architectural features and resolving performance bottlenecks at multiple levels. MaxEVA achieves up to 2.19$\times$ higher throughput and up to 20.4\% higher energy efficiency compared to the state-of-the-art approach, providing insights that help close the knowledge gap in optimizing GEMM workloads on these devices.

The AMD Versal and Intel Stratix 10 NX take contrasting approaches to AI acceleration. The Versal pairs out-of-fabric programmable vector processors (the AIEs) with traditional reconfigurable logic, whereas the Stratix 10 NX incorporates in-fabric hard blocks (the TBs), each comprising multiple dot-product engines. To this end, the MaxEVA framework is extended to support arbitrary matrix sizes by incorporating an additional level of memory hierarchy utilizing  the Versal’s FPGA memory resources, and proposes a systematic framework for optimizing GEMM workloads by efficiently exploiting the Stratix 10 NX’s TBs. Both frameworks maximize performance and energy efficiency by leveraging each device’s architectural characteristics, achieving state-of-the-art throughput for 8-bit integer (int8) precision of up to 77 and 68 TOPS and energy efficiency of up to 0.94 and 1.35 TOPS/W on the Versal and Stratix, respectively. This study provides guidelines for optimizing GEMM-based applications on both platforms while highlighting their respective programmability trade-offs and challenges.

To address the sparsity gap, Systolic Sparse Tensor slices (SSTs) are proposed as in-fabric hard blocks that bring structured sparsity support directly into FPGA architectures. SSTs efficiently accelerate a wide variety of DL workloads across multiple degrees of sparsity, supporting dense operation, 2:4 (50\%) and 1:4 (75\%) sparsity, and a new 1:3 (66.7\%) level for additional flexibility. On GEMM accelerators, SST-based designs achieve up to 5$\times$ higher frequency and up to 10.9$\times$ lower area than conventional FPGA logic. On state-of-the-art sparse vision transformer (ViT) and convolutional neural network (CNN) models, SSTs deliver up to 3.52$\times$ speedup over dense in-fabric acceleration with a moderate area increase of up to 13.3\%.

Finally, AMD Ryzen AI processors integrate a reconfigurable NPU that represents an evolution of the Versal AIE architecture. A common methodology is proposed to optimize GEMM workloads across both NPU generations, XDNA and XDNA2, exploiting their architectural features and addressing performance bottlenecks at the system level. End-to-end evaluation across various GEMM sizes demonstrates state-of-the-art throughput of up to 6.76 TOPS (XDNA) and 38.05 TOPS (XDNA2) for int8 precision, and up to 3.14 TOPS (XDNA) and 14.71 TOPS (XDNA2) for brain floating point (bf16) precision. Collectively, this dissertation advances GEMM acceleration across reconfigurable and AI-specialized hardware, spanning dense and sparse acceleration on FPGAs and dense acceleration on NPUs.    

\tableofcontents   

\listoftables      
\listoffigures     

%
%

\chapter{Introduction}
\label{ch:Introduction}

The rapid growth of deep learning (DL) workloads \cite{TPU_paper2017, LLM_explosion} has driven the development of AI-optimized hardware, including GPUs \cite{NVIDIA_A100, NVIDIA_H100, AMD_CDNA_3, AMD_CDNA_4} and ASICs \cite{TPU_paper2017, TPUv42021, MTIA_Meta_2023, Intel_Goya_inference, Amazon_Inferentia2}. Reconfigurable platforms have followed the same trend, with AI-optimized architectures proposed by both industry \cite{langhammer2021stratix, Versal_AI_Engines_FPGA_paper, TensorTile, tensor_ram, Achronix_Speedster} and academia \cite{aman_tensor_slices_FPGA, aman_tensor_slices_TRETS, comefa_trets, math_no_hard, logic_block_mohd, embrace_div_dsp}.

Within this landscape, reconfigurable architectures are particularly attractive for accelerating DL workloads. Unlike GPUs and ASICs, whose datapaths, supported numerical precisions, and dataflows are fixed at design time, and whose long design cycles hinder adaptation to rapidly evolving DL models, FPGAs allow these aspects to be tailored to the workload, while offering low-latency streaming and high energy efficiency \cite{FPGA_for_DL_2024, Eriko_FPGAvsGPU_FPGA17, brainwave_NPU}. The recent integration of AI-specific hardened blocks, such as the AMD Versal AI Engines and Altera/Intel Stratix 10 NX tensor blocks, combines this flexibility with substantially higher compute density. Within these platforms, general matrix multiplication (GEMM) is the natural optimization target, as it is the computational core of DL workloads: fully connected layers, attention projections in transformers, and convolutions lowered to matrix form all reduce to GEMM, which accounts for up to 90\% of execution time \cite{fathom_harvard_16, wang_gemm_breakdown}.

Improving GEMM efficiency therefore translates directly into end-to-end gains in performance and energy. However, realizing the theoretical peak throughput of these platforms is challenging, as it requires careful management of data movement, on-chip memory, as well as placement and routing. Moreover, they lack native support for features such as structured sparsity, which GPUs already exploit. This dissertation addresses these challenges by analyzing the Versal and Stratix architectures in depth, developing frameworks that bridge the gap between peak and achieved GEMM performance, and proposing hardware and mapping extensions that overcome their current limitations.

This chapter introduces the three main directions explored in this dissertation.
First, we present the leading AI-optimized FPGA architectures targeted in this work — namely the AMD Versal AI Engine and the Altera/Intel Stratix 10 NX — and the frameworks developed to optimize dense GEMM performance on each.
Second, we discuss the lack of native sparsity support in these FPGA architectures, introducing the systolic sparse tensor slices (SSTs), in-fabric hardware blocks designed to accelerate structured sparsity in DL workloads.
Finally, we extend our GEMM framework to include reconfigurable NPUs, presenting a methodology for optimizing GEMM workloads across AMD's Ryzen AI  NPU generations.

\subsection{AI-Optimized FPGA Architectures}
 
The two major FPGA vendors have taken different approaches to optimizing their devices for DL. 
AMD introduced the Versal Adaptive Compute Acceleration Platform (ACAP) \cite{VC1902_hot_chips, Alok2020apocalypse}, which combines the AI Engine (AIE) with programmable logic and scalar processors. 
In contrast, Altera/Intel released the Stratix 10 NX \cite{langhammer2021stratix}, which retains the conventional FPGA architecture but replaces DSP blocks with new AI Tensor Blocks (TBs). 
Consequently, the Versal AIE is an out-of-fabric solution consisting of programmable vector processors operating at high frequency, whereas TBs are in-fabric blocks comprising multiple dot-product engines that run at lower FPGA fabric frequency.

Given these differences, each platform requires its own design approach. We therefore present systematic methodologies and optimization techniques for mapping GEMM onto each device, and propose frameworks that maximize GEMM throughput and energy efficiency on Versal and Stratix while achieving high resource utilization. As discussed earlier, GEMM occupies up to 90\% of DL execution time, which makes it a representative workload for evaluating these architectures.

\subsection{Sparsity Support in Modern DL Architectures}

Contemporary DL workloads exhibit varying degrees of sparsity, \emph{i.e.}, the presence of zero values in weight or activation tensors \cite{Sparsity_Hoefler_2021}. Sparsity arises naturally in activations (as a by-product of non-linear activation functions such as ReLU) and is induced in weights through pruning, which removes redundant parameters from over-parameterized models. Since zero-valued operands do not need to be stored or computed, exploiting sparsity reduces computational and memory requirements, and has become a common strategy in efficient DL accelerators \cite{S2TA_HPCA_2022}.

Typically, zeros in modern deep neural networks
(DNNs) are distributed randomly within the data tensors \cite{Sparsity_Hoefler_2021, S2TA_HPCA_2022}. 
This random sparsity is commonly referred to as \textit{unstructured} sparsity.
Leveraging unstructured sparsity has been the main focus of numerous sparse hardware accelerators over the past years, targeting both ASICs \cite{EIE_ISCA_2016, Cambricon_X_MICRO_2016, OuterSpace_HPCA_2018, SMT_SA_2019, ExTensor_2019, SparTen_2019, Eyeriss_v2_2019, SpArch_2020} as well as FPGAs \cite{unstructured_sparsity_SpWA_FPGA_DAC_2018,
unstructured_sparsity_FPGA_Abhishek_2021,
unstructured_sparsity_CNN_TCS_2021, unstructured_FPGA_TCAD_2022}.
However, unstructured sparse accelerators require complex hardware structures, which result in high area overheads \cite{HighLight_MIT_2023}. 
The random location of zeros in unstructured sparsity causes unpredictable and low hardware utilization, rendering inference speedup inefficient \cite{STA_arxiv_2020, STA_arch_letters_2020, EIE_ISCA_2016}.

To address these challenges in sparsity acceleration, 
\textit{structured} sparsity has recently been proposed  \cite{Nvidia_accelerate_sparse_2021, S2TA_HPCA_2022, Vegeta_HPCA_2023}.
Structured sparsity imposes constraints on the sparsity patterns in data tensors, enabling low area overheads and highly energy-efficient hardware enhancements for sparsity exploitation.
To this end, Nvidia A100 GPUs introduced support for the 2:4 structured sparsity, 
where two out of every four \textit{consecutive} values must be non-zero.
Furthermore, structured sparse ASIC accelerators have also been proposed, demonstrating up to 3.1$\times$ less energy consumption per inference compared to state-of-the-art unstructured sparse accelerators \cite{S2TA_HPCA_2022}.

Although industrial FPGA architectures \cite{Stratix_10_NX_FPGA_2021, Sergey_Intel_TB_Agilex_5_FCCM_2024} and academic proposals for in-fabric tensor blocks \cite{aman_tensor_slices_FPGA, aman_tensor_slices_TRETS} have improved support for the computational demands of DL, they primarily target dense computation.
This results in a significant limitation of their efficiency and applicability to most real-world DL models, which inherently exhibit sparsity.
To address this challenge, this dissertation proposes incorporating flexible in-fabric slices into the FPGA architecture to efficiently support both \textit{sparse} and \textit{dense} DL workloads.
We employ 2D systolic slices similar to \cite{TS_Aman_FPGA_2021, Aman_TS_TRETS_2022}, which are augmented with structured sparsity features and FPGA-specific architecture optimizations.

\subsection{AMD Ryzen AI Reconfigurable NPU Architecture}

AMD released the Ryzen AI processors \cite{NPU_arch_MICRO_2024}, featuring multi-core CPUs, an integrated GPU, and a new neural processing unit (NPU).
The NPU features the XDNA architecture and serves as a dedicated DL accelerator integrated with x86 processors.
The NPU architecture leverages the characteristics of modern DL workloads, where compilers can determine most of the control flow statically.
In particular, AMD NPUs provide an explicit data movement architecture, which both reduces hardware complexity and enables high performance, while offering substantially higher energy-efficiency compared to dynamically scheduled architectures, such as CPUs and GPUs \cite{NPU_arch_MICRO_2024}.
This lower power and energy footprint has driven the integration of NPUs into edge platforms, enabling DL applications to run efficiently under tight power and thermal budgets.

The Ryzen AI NPU is an evolution of the Versal AIE architecture \cite{NPU_arch_MICRO_2024}, retaining its spatial array of programmable vector processors with local memory while introducing architectural changes that significantly affect GEMM mapping, such as data reuse, tiling, and data movement. In contrast to the NPU, Versal devices integrate an FPGA fabric, which acts as an additional level of memory hierarchy and enables efficient exploitation of data reuse in GEMM. Furthermore, the flexibility of the fabric allows customized tiling schemes \cite{Versal_vs_Stratix_FCCM_2024, CHARM_FPGA_2023} and tailored data layout transformations \cite{RSN_ISCA_2025}, which the NPU must instead handle within its fixed memory hierarchy and DMA capabilities.

In this dissertation, we conduct a comprehensive study to optimize GEMM workloads on AMD Ryzen AI NPUs.
The architectural differences discussed above introduce new design challenges, highlighting the necessity for a separate methodology rather than a direct port of Versal approaches.
Having identified this important research gap, we propose a \textit{unified} systematic framework to maximize GEMM performance across the two current generations, namely XDNA and XDNA2.
The framework targets end-to-end GEMM performance and introduces a novel procedure for system-level optimization.


\section{Research Objectives}
\label{sec:Intro:ResearchObj}

First, we focus on the Versal AIE, a new class of reconfigurable architecture. To address the design challenges this architecture introduces, we propose the MaxEVA framework \cite{Taka2023MaxEVA}, a systematic methodology that maximizes the performance and energy efficiency of GEMM workloads on the AIE array. MaxEVA exploits key architectural features of the AIE, namely local memory sharing, static circuit switching, and broadcasting, while overcoming challenges that lead to sub-optimal performance: limited I/O and switch bandwidth, reduced number of AIE-FPGA interface tiles, and routing congestion.

Second, we build on MaxEVA \cite{Taka2023MaxEVA} and extend it with an additional memory hierarchy level that uses the Versal FPGA’s on-chip resources. 
Since Versal and Stratix 10 NX take fundamentally different approaches to AI acceleration (out-of-fabric vector processors versus in-fabric tensor blocks), comparing them fairly requires a GEMM design that is equally well optimized for each. 
We therefore develop a framework to design, map, and optimize a configurable GEMM accelerator on Stratix 10 NX that exploits its in-fabric TBs, enabling a direct comparison of the two architectures \cite{Versal_vs_Stratix_FCCM_2024}.
Both frameworks \cite{Versal_vs_Stratix_FCCM_2024} involve extensive design space exploration (DSE) and analytical modeling to maximize GEMM performance, data reuse, and memory utilization, and thus the overall energy efficiency.

Third, we address the lack of sparsity support in existing in-fabric blocks by proposing the SST slices \cite{SST_slices_FPGA_2025}, which accelerate a wide range of DL workloads. To the best of our knowledge, SSTs are the first in-fabric FPGA blocks to natively exploit structured sparsity. Specifically, they support 2:4 (50\% sparsity), 1:3 (66.7\%), and 1:4 (75\%) patterns, in addition to dense operation.
These levels align with the most \textit{common} sparsity levels in DNNs, since sparsity higher than 75\% typically leads to significant accuracy degradation \cite{Sparse_tensor_GPUs_2019, STA_arxiv_2020, N_M_sparse_transformers_FPGA_VLSI_2022, Learning_N_M_ICLR_2021, Learning_best_N_M_NeurIPS_2022}.

The 1:3 sparsity bridges the gap between 50\% and 75\% sparsity, increasing flexibility and providing both acceleration and storage benefits, while can be supported in SSTs 
\textit{without incurring additional area overheads} (by reusing the area allocated for 2:4 and 1:4).
This sparsity flexibility enables the development of tailored solutions for DL models, since each model exhibits distinct levels of sparsity.
Moreover, we introduce \textit{dedicated} interconnects between the SST slices to facilitate their efficient integration within the FPGA architecture.
These dedicated interconnects demonstrate substantial routing savings in FPGAs compared to prior work on 2D systolic blocks \cite{TS_Aman_FPGA_2021, Aman_TS_TRETS_2022}, where such interconnects are not employed.

Finally, we propose a new systematic methodology to optimize GEMM performance on Ryzen AI NPUs \cite{GEMM_NPU_Taka_FPGA_Poster, GEMM_NPU_Taka_arxiv}.
A key requirement of our methodology is that matrices remain in standard layout format (\textit{row-} and \textit{column-major}) in main memory (DRAM), rather than requiring custom or pre-transformed data layouts.
This enables seamless integration with DL tensor libraries, such as GGML \cite{GGML_github}, and supports the development of high-performance GEMM libraries, similar to those available for GPUs \cite{cublas_doc, cutlass_doc, hipBLAS_doc, rocBLAS_doc}.
Furthermore, our GEMM designs efficiently exploit several architectural features of the NPU, \emph{e.g.,} on-the-fly tensor transformations, to address critical performance bottlenecks, particularly through maximization of DRAM bandwidth and sophisticated data movement design between the NPU and DRAM.

\section{Dissertation Statement}
\label{sec:Intro:Statement}

Currently, the dominant hardware platforms for DL workloads are GPUs \cite{NVIDIA_H100, AMD_CDNA_3} and specialized ASIC accelerators \cite{TPUv42021, MTIA_Meta_2023, Amazon_Inferentia2}.
However, the rapid evolution of DL workloads creates a strong demand for hardware platforms that combine high efficiency with architectural flexibility.
Reconfigurable architectures offer a promising alternative due to their ability to adapt to emerging workloads while maintaining competitive performance and efficiency. 
Although AI-optimized reconfigurable platforms have been proposed both in industry \cite{langhammer2021stratix, Versal_AI_Engines_FPGA_paper, tensor_ram} and academia \cite{aman_tensor_slices_FPGA,comefa_trets,embrace_div_dsp}, their large-scale adoption for DL acceleration has yet to be fully realized.

This work presents a comprehensive evaluation of prominent AI-optimized reconfigurable architectures, \emph{i.e.,} AMD Versal ACAP, Altera/Intel Stratix 10 NX and AMD Ryzen AI NPUs, using GEMM as the primary workload, and provides insights into their performance and efficiency characteristics. 
Building on these insights, we propose architectural enhancements to reconfigurable platforms that enable efficient acceleration of a wide range of dense and sparse DL workloads, including in-fabric support for structured sparsity, a capability absent from existing FPGA architectures.
Ultimately, this dissertation aims to facilitate the broader adoption of reconfigurable architectures for DL acceleration through a detailed analysis of their efficiency, programmability trade-offs, and associated architectural challenges.


\section{Contributions}
\label{sec:Intro:Contributions}

\noindent This dissertation makes the following key contributions:

\vspace{-7.5pt}

\begin{itemize}
\item \textit{MaxEVA Framework} \cite{Taka2023MaxEVA}: We propose an optimization methodology based on analytical modeling to maximize the performance of GEMM on the Versal AIE. 
Our methodology is generalizable to any Versal AIE device and addresses various performance bottlenecks, leading to maximized utilization of the AIE resources.  Furthermore, we propose a sophisticated AIE kernel placement strategy to effectively leverage the most efficient data movement mechanisms of the Versal AIE architecture.
Demonstration on the AMD VC1902 device shows up to 5.44 and 77.01 TOPS throughput for single-precision floating point (fp32) and 8-bit integer (int8) precisions, respectively, outperforming the state-of-the-art approach by showcasing up to 2.19$\times$ higher performance and 20.4\% higher energy efficiency.
 
\item \textit{Efficient GEMM Frameworks for Leading AI-Optimized FPGAs} \cite{Versal_vs_Stratix_FCCM_2024}: We propose novel systematic frameworks to optimize GEMM performance on leading commercial AI-optimized FPGAs.
For Versal, we leverage the MaxEVA framework and extend it
to incorporate an additional memory hierarchy level utilizing the Versal FPGA's on-chip resources. 
We maximize performance via DSE and analytical modeling, and we propose a novel RAM optimization scheme to overcome severe limitations of Vitis high-level synthesis (HLS).
For Stratix, we develop a novel framework to design, map and optimize a configurable GEMM accelerator by exploiting the device's in-fabric TBs.
Our framework involves extensive DSE and analytical modeling to maximize GEMM performance.
Demonstration of our frameworks on various GEMM workloads for int8 precision, exhibits up to 77 and 68 TOPS throughput with 100\% AIE and 91\% TB utilization for Versal and Stratix, respectively. Moreover, our designs attain up to 0.94 and 1.35 TOPS/W energy efficiency, with 88\% and 94\% on-chip memory for Versal and Stratix, respectively.

\item \textit{Systolic Sparse Tensor Slices} \cite{SST_slices_FPGA_2025}: We propose incorporating structured sparsity support into FPGA architectures to address the challenge of efficiently accelerating sparse DL workloads. 
We architect 2D systolic in-fabric blocks,
named SST slices, that support multiple
degrees of sparsity to efficiently accelerate a wide variety of DL workloads.
Our sparse SST-based designs attain up to 5× higher FPGA frequency and 10.9× lower area, compared to traditional FPGAs. 
Moreover,
evaluation of the proposed SSTs on state-of-the-art sparse vision transformers (ViTs)
and convolutional neural networks (CNNs) exhibits up to 3.52$\times$ speedup with minimal area
increase of up to 13.3\%, compared to dense in-fabric acceleration.

\item \textit{GEMM Optimization on AMD Ryzen AI NPUs} \cite{GEMM_NPU_Taka_FPGA_Poster, GEMM_NPU_Taka_arxiv}: We propose a systematic methodology to optimize GEMM workloads on Ryzen AI NPUs, combining analytical modeling with hardware profiling. Our GEMM implementation generates sufficiently long contiguous DRAM accesses to maximize main-memory bandwidth utilization, which in turn maximizes GEMM performance across the NPU array. We leverage novel tensor transformations on the NPU’s DMA engines to optimize DRAM bandwidth while keeping matrices in standard row- or column-major layout in DRAM, without explicit pre-tiling.
Evaluated on two mini PCs with NPUs from different generations, our approach achieves state-of-the-art performance of up to 6.76 TOPS (XDNA) and 38.05 TOPS (XDNA2) for int8, and 3.14 TOPS (XDNA) and 14.71 TOPS (XDNA2) for brain floating point (bf16) precision.
    
\end{itemize}

\section{Dissertation Organization}
\label{sec:Intro:Organization}

The remainder of this dissertation is organized as follows. ~\Cref{ch:Related Work} reviews related work in AI-optimized FPGA architectures, structured sparsity support for DL accelerators, and prior applications of the AMD Ryzen AI NPUs. 
~\Cref{ch:Background} provides background on the three reconfigurable architectures central to this dissertation: the AMD Versal ACAP, the Intel Stratix 10 NX, and the AMD Ryzen AI NPUs. 
~\Cref{ch:MAXEVA} presents the MaxEVA framework, which addresses the design, mapping, and optimization of GEMM on the Versal AI Engine. 
~\Cref{ch:Versal_vs_Stratix} extends MaxEVA to support arbitrary matrix sizes and introduces a parallel framework for the Altera/Intel Stratix 10 NX. 
~\Cref{ch:SST_Slices} presents the architectural design of the SST slices and their application to sparse GEMM and DL acceleration. 
~\Cref{ch:RyzenAI_NPUs} presents a methodology for optimizing GEMM workloads across the XDNA and XDNA2 generations of AMD Ryzen AI NPUs. 
Finally, ~\Cref{ch:Summary} concludes the dissertation and discusses directions for future work.

\chapter{Related Work}
\label{ch:Related Work}

This chapter reviews prior work in three areas that correspond to the contributions of this dissertation. 
First, we discuss work targeting the two leading AI-optimized FPGA architectures from industry: AMD Versal and Intel Stratix 10 NX. 
Second, we survey prior work on structured sparsity support in modern DL accelerators. 
Finally, we review prior work on the AMD Ryzen AI NPU across across a range of workloads.

\section{Leading AI-Optimized FPGAs}
\label{sec:RelatedWork:AI_Optimized_FPGAs}

\subsection{AMD Versal ACAP}
\label{sec:RelatedWork:AMD_Versal_ACAP}

Several prior works leverage the Versal ACAP architecture across multiple application domains.
In particular, CHARM \cite{charm2023fpga, charm_DAC23} proposes a framework to automate the process of GEMM acceleration on Versal ACAP.
Their experimental results on the VC1902 exhibit higher energy efficiency of up to 7.2$\times$ and 1.7$\times$ compared to traditional FPGAs (AMD/Xilinx U250) and GPUs (NVIDIA A100), respectively.
Furthermore, additional research on Versal focuses on accelerating specific DL workloads, such as CNNs \cite{XVDPU_AIE_FPL22}, \cite{CNN_acc_Versal_FPL22}, and graph neural networks (GNNs) \cite{H_GCN_FPL2022, chen2023exploiting}.

\cite{Perf_analysis_Kaustubh} presents a thorough analysis of GEMM workloads of varying shapes and sizes on AMD Versal, aiming to provide insights for optimal performance.
Moreover, GAMA~\cite{GAMA_FPL_2025} proposes a design that maximizes memory utilization on the second generation of Versal AIE (AIE-ML), attaining state-of-the-art performance of up to 165 TOPS.
Other works include AIE compilers \cite{AIE_compiler_HPEC20}, arbitrary precision integer multiplication \cite{yang2023aim}, as well as acceleration of atmospheric simulations \cite{sparta_ICS23, Stensil_AIE_FPGA23}.

Considering all prior work, our  MaxEVA framework~\cite{Taka2023MaxEVA} detailed in this dissertation, represents the state-of-the-art GEMM implementation on Versal AIE, achieving notable performance gains of 2.19$\times$ and 20.8\% for int8 and fp32, respectively, as well as up to 20.4\% higher energy efficiency, over prior works. 
Nevertheless, MaxEVA only targets small matrix sizes that fit within the Versal AIE. 
We address this limitation in our subsequent work~\cite{Versal_vs_Stratix_FCCM_2024}, extending MaxEVA to support arbitrary GEMM sizes by implementing an additional level of memory hierarchy exploiting Versal's FPGA resources.

\subsection{Intel Stratix 10 NX}
\label{sec:RelatedWork:Intel_Stratix_10_NX}

The Intel Stratix 10 NX architecture was introduced in~\cite{langhammer2021stratix}, including a discussion of the TB operating modes and design trade-offs. 
This work also presents a TB design used for feneral matrix-vector (GEMV) as well as GEMM operations.
Multiple other works have targeted the Stratix 10 NX for DL workloads.
In \cite{boutros2020beyond}, a GEMV accelerator is mapped on AI TBs and incorporated into an enhanced Brainwave NPU overlay~\cite{brainwave_NPU}. 
They show a speedup of up to 3.5$\times$ compared to all prior works on FPGA-based acceleration of recurrent neural networks (RNN)~\cite{Guan_LSTM_FPGA,Vladimir_FINNL,Que_RNN_FPGA}, as well as the baseline NPU with legacy DSPs~\cite{Stratix10GX_NPU}. 
Additionally, in~\cite{hpipex_NX}, the Stratix 10 NX is utilized to enhance the CNN HPIPE accelerator~\cite{HPIPE_baseline}.
They demonstrate a 4$\times$ speedup over prior FPGA accelerators for CNN.
In~\cite{denseint16_NX}, a method to assemble higher than int8 precision multipliers is presented.
Finally, in~\cite{WaveNet_NX}, 
a speech-generation model is implemented on Stratix 10 NX, substantially outperforming a V100 GPU implementation.

In this dissertation, we develop a framework, which includes a detailed, systematic approach for automatically generating a configurable GEMM accelerator on Stratix 10 NX \cite{Versal_vs_Stratix_FCCM_2024}.
Furthermore, we perform an extensive DSE and we explore various trade-offs in GEMM design, which are not thoroughly examined in prior work targeting the Intel Stratix 10 NX device.

\section{Architectural Support for Sparsity in AI Accelerators}
\label{sec:RelatedWork:Sparsity_architectural_support}

Structured sparsity is currently supported by various commercial architectures.
In particular, the 2:4 sparsity is supported by state-of-the-art GPUs, \emph{i.e.,} Nvidia A100 \cite{NVIDIA_A100}, Nvidia H100 \cite{NVIDIA_H100} and AMD MI300 \cite{AMD_CDNA_3}, as well as the new AMD Versal AIE-ML 
processors \cite{AMD_AIE_ML_architecture_manual, AMD_AIE_ML_kernel_guide}.
In academic research, hardware support for structured sparsity has also been investigated in ASIC DNN accelerators \cite{S2TA_HPCA_2022, STA_arch_letters_2020, HighLight_MIT_2023}.
Moreover, prior academic research \cite{Sparse_tensor_GPUs_2019} introduces structured sparsity features in the tensor cores of modern GPUs.
Finally, structured sparsity support has also been proposed inside the matrix engines \cite{Vegeta_HPCA_2023} and vector units \cite{RISC_V_CPU_structured_sparsity_DATE_2024} of CPUs.
Considering all prior work, we are the first to incorporate structured sparsity support into the FPGA architecture.
Moreover, our proposed in-fabric blocks support multiple levels of structured sparsity (beyond merely 2:4), to enable efficient acceleration for most contemporary DNN models.

In-fabric hard blocks have been commercially incorporated into several AI-optimized FPGAs.
More specifically, the Intel Stratix 10 NX \cite{Stratix_10_NX_FPGA_2021} replaces the traditional DSP slices with tensor blocks.
These tensor blocks comprise multiple dot-product engines to more efficiently support AI applications.
More recently, Intel announced the Agilex-5 FPGAs \cite{Sergey_Intel_TB_Agilex_5_FCCM_2024, Intel_Agilex_5_tensor_blocks}, which introduce new AI-enhanced DSP blocks.
These new DSPs include multiple dot-product operations, while also maintaining various features of the traditional Agilex DSP blocks.
Finally, the Achronix Speedster7t FPGAs incorporate machine learning processor (MLP) blocks \cite{Achronix_Speedster, Achronix_Speedster_2024}.
These MLP hard blocks feature multiplier arrays, adder trees, accumulators and tightly coupled memories to the computational blocks.

In-fabric enhancements have also been proposed in academia.
In \cite{TS_Aman_FPGA_2021, Aman_TS_TRETS_2022}, the authors enrich the FPGA architecture with 2D systolic tensor slices, showing substantial efficiency benefits in DNN acceleration over traditional FPGAs.
Moreover, in \cite{PIR_DSP_FCCM_2019}, a special pattern of dedicated wires between DSP blocks is proposed, which enables more efficient mapping of systolic arrays on FPGAs.
However, all aforementioned commercial and academic in-fabric blocks target primarily dense computation.
In this dissertation, we propose incorporating the SST slices \cite{SST_slices_FPGA_2025}, which are 
architected to enable efficient computation for both sparse and dense AI workloads.
The SST slices employ a 2D systolic dataflow similar to \cite{TS_Aman_FPGA_2021, Aman_TS_TRETS_2022}, but are augmented with sparsity features.
Moreover, we introduce vertical dedicated wires between the SSTs to more efficiently map GEMM accelerators into FPGA architectures.

Finally, prior research \cite{LSTM_fine_grained_sparsity_FPGA_2019, Fine_grained_structured_sparsity_FPL_2021, N_M_sparse_transformers_FPGA_VLSI_2022, Fine_grained_Neural_ODE_FPGA_2023, LAMPS_FCCM_2024} implement DNN accelerators that support multi-level structured sparsity on traditional (non AI-optimized) FPGAs. 
In contrast, we also introduce the novel 1:3 sparsity pattern \cite{SST_slices_FPGA_2025}, which is absent from any prior work and offers increased sparsity flexibility.
In this dissertation, we demonstrate the importance of this flexibility for various state-of-the-art DNNs.
Furthermore, our SST-based GEMM designs show remarkable performance and area efficiency gains compared to traditional FPGAs.

\section{Application Mapping on Ryzen AI NPUs}
\label{sec:RelatedWork:Ryzen_AI_NPUs}

Several workload mapping on Ryzen AI NPUs are covered in prior publications.
The authors in \cite{ARIES_FPGA_2025} map a ResNet layer on XDNA, utilizing only 20\% of the XDNA cores. 
In \cite{Zen_attention_AMD_paper}, a compiler framework for dynamic attention folding on AMD NPUs is presented.
Moreover, in \cite{karami2025exploringdynamicschedulingspace} the authors perform characterization of generative AI workloads on Ryzen AI processors.
Other works leverage the NPUs to perform stencil applications \cite{Stensil_NPU_FPGA2025}, accelerate Fortran intrinsics \cite{Fortran_instrinsics_FPGA2025}, and implement a variant of the fast Fourier transform (FFT) \cite{Variant_FFT_NPU_2024}.

Moreover, some prior works have also explored mapping GEMM on AMD NPUs. In \cite{roesti2025unlocking}, the authors explore fine-tuning GPT-2 on the Ryzen AI processors.
Specifically, they offload the time-intensive GEMM computation on the first-generation XDNA and attain 2.8$\times$ speedup compared to a CPU-only implementation.
Furthermore, they utilize a GEMM implementation similar to the non-optimized programming example in \cite{GEMM_example_mlir_aie}.
In \cite{fang2025dato_paper}, a task-based programming model for dataflow accelerators is presented, demonstrating a GEMM implementation on XDNA. 
In particular, they attain up to 5.04 TOPS and 1.95 TOPS for int8 and bf16 data types, respectively, which closely matches the performance in \cite{GEMM_example_mlir_aie}.

Furthermore, in \cite{Erwei_MLIR_AIR} the authors introduce MLIR-AIR, which is a novel open-source compiler stack, operating at a higher level compared to the lower-level MLIR-AIE framework \cite{mlir_aie_github_repo}.
They demonstrate up to 79\% compute efficiency compared to MLIR-AIE.
More recently, \cite{wang2025asymmetrictilebufferingbeneficial} introduced asymmetric tile buffering, which decouples the buffered tile dimensions of the input and output operands. They demonstrate up to 28.5 TOPS GEMM performance on XDNA2 using block floating-point (bfp16), a datatype in which a block of values shares one common exponent.
Finally, \cite{pang2026tilefusefusedmixedprecisionkernel} presents the TileFuse library, targeting mixed-precision quantization formats such as W4A16, and attaining up to 12 TOPS GEMM performance on XDNA2.

In this dissertation, we present a systematic framework~\cite{GEMM_NPU_Taka_FPGA_Poster, GEMM_NPU_Taka_arxiv} to optimize GEMM performance across both NPU generations (XDNA and XDNA2), while retaining matrices in standard layout in DRAM, \textit{i.e.}, \textit{row-} and \textit{column-}major order, without explicit pre-tiling. Our designs leverage the multi-dimensional tensor addressing feature across the entire NPU hierarchy to transform matrices into the tiled layout expected by NPU cores. We show that our optimized GEMM implementation on the first-generation XDNA achieves superior performance over all prior work, reaching up to 6.76 TOPS (34\% higher) for int8 and 3.14 TOPS (62\% higher) for bf16. For the second-generation XDNA2, we attain state-of-the-art performance of up to 38 TOPS for int8 and 14.7 TOPS for bf16.

We note that \cite{wang2025asymmetrictilebufferingbeneficial} assumes explicit matrix pre-tiling, which provides optimal DRAM bandwidth, and focuses on bfp16 precision. However, retaining matrices in standard layout in DRAM and exploiting the multi-dimensional DMA addressing features of the NPUs for bfp16 remains an open problem, due to significant challenges in data storage and limitations of NPU DMA transformations. A direct comparison is therefore not directly applicable, since our work in this dissertation focuses on int8 and bf16 formats, for which we implement sophisticated DMA transformations. Similarly, \cite{pang2026tilefusefusedmixedprecisionkernel} targets mixed-precision formats and achieves only up to 12 TOPS on XDNA2 (we attain up to 38 TOPS); however, since we focus on single-precision formats in this work, a direct comparison is not applicable here either.

\chapter{Background \& Architecture Overview}
\label{ch:Background}

This chapter provides background on the three reconfigurable architectures central to this dissertation: the AMD Versal ACAP, the Intel Stratix 10 NX, and the AMD Ryzen AI NPUs. For each architecture, we describe the main architectural features, the primary data movement and communication mechanisms, as well as their associated programming models and tools.

\section{AMD Versal ACAP}
\label{sec:Background:Versal_ACAP}

\subsection{Architecture Overview}
\label{subsec:Versal_architecture}

\begin{figure}[tbp]
\centering
\includegraphics[width=0.72\linewidth]{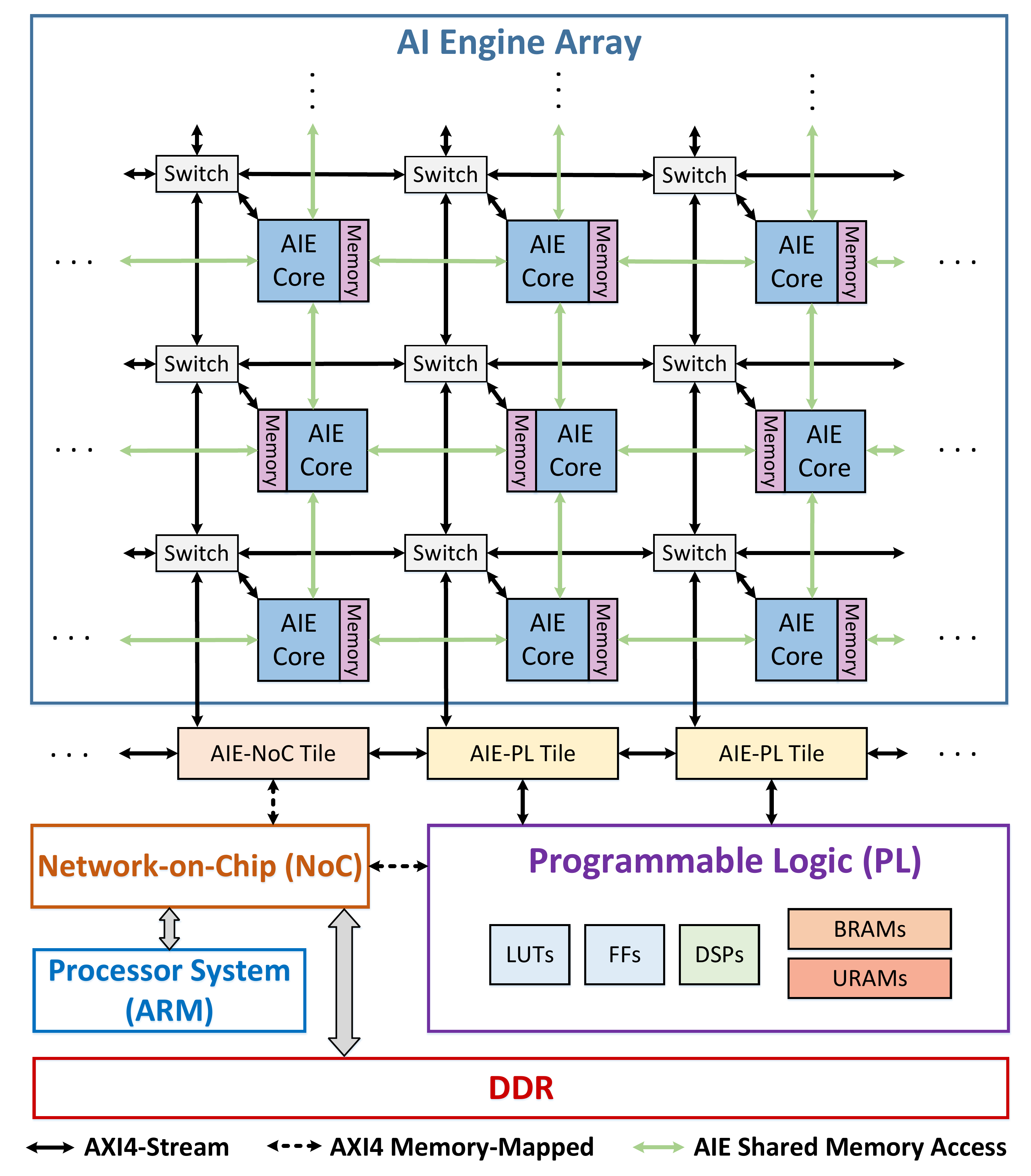}
\caption{Versal ACAP architecture.}
\label{fig:Versal_ACAP_Architecture}
\end{figure}

The architecture of the Versal ACAP is depicted in Fig. \ref{fig:Versal_ACAP_Architecture}.
The Versal ACAP comprises the processor system (PS), the programmable logic (PL), as well as the novel AIE array \cite{Versal_ACAP_design_guide}.
The PS consists of scalar ARM processors, while
the PL includes the traditional FPGA resources, \emph{e.g.}, LUTs, FFs, DSP slices, and on-chip memory resources (BRAMs/URAMs).

The Versal AIE is a 2D array consisting of identical AIE tiles.
Each tile includes an AIE core, a memory unit and an interconnect (switch)~\cite{Versal_acap_architecture_manual}.
The AIE cores are architected as  very-long instruction word (VLIW) programmable processors featuring single-instruction multiple-data (SIMD) vector units.
The AIE array provides three levels of parallelism.
First, \emph{instruction-level} parallelism is realized by executing up to 7 instructions every clock cycle (7-way VLIW).
Second, \emph{data-level} parallelism is achieved through vector operations, where multiple data can be processed each clock cycle (SIMD).
Third, \emph{spatial-level} parallelism is attained via the concurrent execution of multiple AIE cores (up to 400).
Communication between different AIE cores is achieved by local memory sharing access for neighboring cores, or by programmable switches for distant cores (Fig. \ref{fig:Versal_ACAP_Architecture}).

The AIE array communicates efficiently with the PL via the dedicated AIE-PL tiles, located on the last row of the AIE, as shown in Fig. \ref{fig:Versal_ACAP_Architecture}.
These tiles provide AXI4-Streaming interface with the PL, while also supporting clock domain crossing between the AIE and the PL.
The Versal ACAP additionally includes a network-on-chip (NoC) to enable flexible communication with the off-chip memory (DDR).

\subsection{AI Engine Array Data Movement}
\label{subsec:AIE_data_movement}

\begin{figure}[tbp]
\centering
\includegraphics[width=0.73\linewidth]{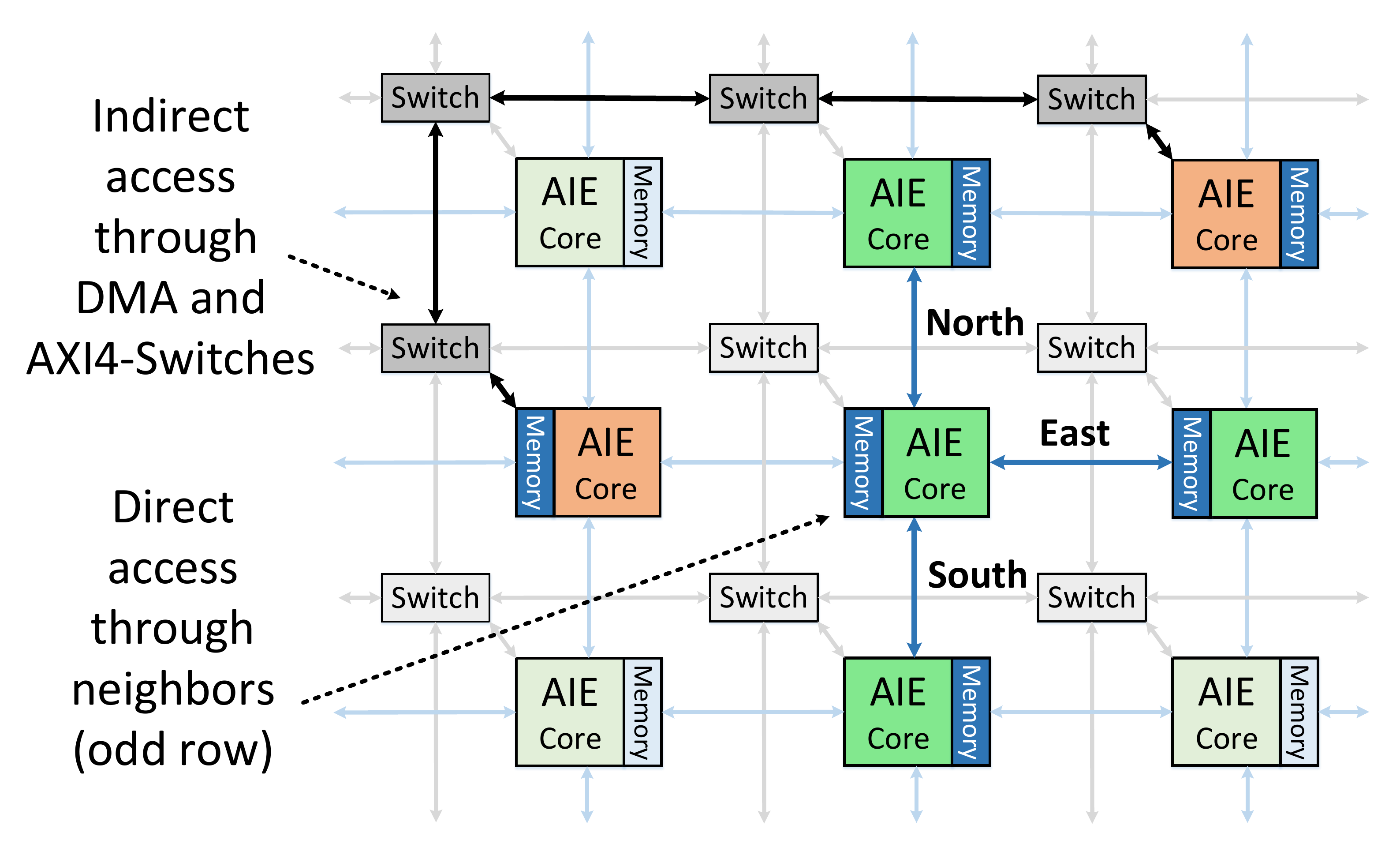}
\caption{Data movement mechanisms in AI Engine array.}
\label{fig:Data_movement_mechanims}
\end{figure}

Each AIE tile has 16KB of program memory to store VLIW instructions, as well as 32KB of data memory divided into 8 banks of 4KB.
For higher memory requirements, AIEs can access data memory directly from their neighbors, for a total of 128KB.
Fig. \ref{fig:Data_movement_mechanims} shows this direct access (AIEs highlighted in green), which constitutes the main data movement mechanism of the AIE array \cite{Versal_acap_architecture_manual}.
However, notice that while each AIE is able to directly access memory from its north and south directions, the east and west access depend on the row location of the AIE.
In particular, the AIE array is arranged on alternate even and odd rows, where the cores in even rows can access memory on the west direction, while in odd rows, the east module is accessed.
Finally, we note 
that AIEs on the edges of the array have fewer memory accesses on both north/south and east/west directions following the pattern described above.


For non-neighboring AIEs, the communication is realized through the direct memory access (DMA) mechanism using the programmable switches, as shown in Fig. \ref{fig:Data_movement_mechanims} (AIEs highlighted in orange).
Compared to direct access, non-neighboring communication through DMA has increased communication latency and requires more memory resources.
The AXI4-Stream switches can be configured for either circuit-switching or packet-switching.
Circuit-switching provides dedicated connections which are  statically configured at compilation time.
In contrast, packet-switching allows routing to different destinations by dynamically setting a destination header at the beginning of each packet.
Due to static configuration, circuit-switching exhibits deterministic latency between connections, while also supporting broadcast to multiple output channels.
Conversely, packet-switching can cause resource contention, leading to non-deterministic latency \cite{Versal_acap_architecture_manual}.
In this work, we only exploit the most efficient circuit-switching mechanism, without the need of explicit packet-switching proposed in \cite{charm2023fpga, dac23automm}.

\subsection{Programming Model \& Tools}
\label{subsec:Programming_tools_Versal}

An AIE kernel running on a single AIE core can be programmed in high-level C/C++~\cite{AIE_API_user_guide}, or low-level SIMD intrinsics \cite{AI_Engine_programming_guide}.
The mapping of multiple kernels on the AIE array is realized through the adaptive data flow (ADF) graph modeling.
The nodes in the ADF correspond to AIE kernels while the edges represent connections between them \cite{AI_Engine_programming_environment}.
The PL can be programmed in C/C++ using Vitis 
HLS \cite{Vitis_HLS_guide} or low-level RTL.
Finally, AMD/Xilinx provides the Vitis V++ tool~\cite{Vitis_software_guide} to integrate the AIE graph system and the PL kernels.




\section{Intel Stratix 10 NX}
\label{sec:Background:Stratix_10_NX}

\begin{figure}[t]
\centering
\includegraphics[width=0.85\columnwidth]{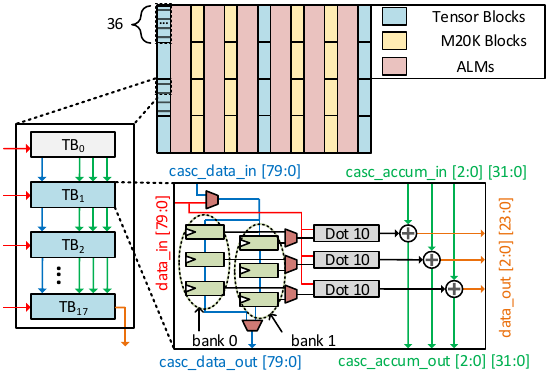}
\caption{Architecture of Stratix 10 NX Tensor Blocks.}
\label{fig_tensor_block_diagram}
\end{figure}

\subsection{Architecture Overview}
\label{subsec:Stratix_10_Nx_architecture}

The Intel Stratix 10 NX device's PL comprises the traditional FPGA resources, \textit{i.e.,} ALMs (adaptive logic modules), M20K blocks (Intel's BRAMs, 20Kbit in size~\cite{Intel_M20K_guide}), as well as the new AI-optimized TBs. 
TBs replace the traditional variable-precision DSP blocks (Fig.~\ref{fig_tensor_block_diagram}), by dropping many of their legacy, high-precision operating modes and replacing them with scalar, vector, and tensor operating modes for several DL-optimized data types. 
The TB maintains the interface of the legacy DSP. 
In this work, we focus on the Tensor int8 mode, which supports three 10-element 8$\times$8 signed fixed-point 
dot-product operations and three 32-bit fixed-point 
additions for accumulation on partial products.
TBs in an FPGA column are \textit{physically} grouped in chains, each consisting of 36 TBs.
The TBs within a chain are cascaded using dedicated wires to propagate operands and accumulated products. Neighboring chains are not cascaded. Within a chain, TBs can be \textit{logically} grouped in independent \textit{arrays} of configurable length.

Fig.~\ref{fig_tensor_block_diagram} illustrates an example of an \textit{array} of 18 TBs that lies within a TB chain, as well as a simplified block diagram of the 
TB operating in the int8 mode. Each TB contains two banks of three 
ping-pong registers for storing operands ($bank$ $0, 1$), where each register holds ten 8-bit values.
Moreover, it includes three 10-element dot-product engines ($Dot$ $10$). 
The first input to each dot-product engine comes from its corresponding register, while the second input is broadcast from the 80-bit wide $data\_in$ port to all three engines. Finally, three independent 32-bit fixed-point adders are responsible for adding the generated dot-products to the cascade input from the previous TB. As shown in Fig.~\ref{fig_tensor_block_diagram}, accumulation results are propagated along the TB \textit{array} through cascade connections between a block's $casc\_accum\_out$ port and the following block's $casc\_accum\_in$ port. The latency of the dot-product calculation plus the cascade accumulation is equal to two cycles. The \textit{array's} final TB outputs are exposed through three 24-bit wide $data\_out$ ports.

There are three methods for loading operands in an \textit{array's} TB registers: (i) parallel load mode, (ii) side load mode and (iii) cascade mode. In this work, we focus on the cascade mode, since it leads to less routing congestion, as mentioned in~\cite{langhammer2021stratix}. We refer the reader to~\cite{langhammer2021stratix} for a detailed description of the other modes and their trade-off analysis. In cascade mode, $TB_0$ (Fig.~\ref{fig_tensor_block_diagram}) performs no computation and acts only as a loading port, where operands enter the array through its 80-bit $data\_in$ port. These operands are then propagated to subsequent TBs ($TB_1$--$TB_{17}$ in Fig.~\ref{fig_tensor_block_diagram}) from the $casc\_data\_out$ port of one TB to the next's $casc\_data\_in$ port and eventually stored in the TB registers. This requires three cycles per TB, leading to longer loading latencies as the array grows in size. However, computation can occur concurrently due to the ping-pong registers, allowing to hide the loading latency.


\subsection{Programming \& Tools}
\label{subsec:Programming_tools_Stratix}

While the TBs support many modes of operation, there is currently no support for programming them using Intel HLS~\cite{Intel_HLS_guide} or other high-level synthesis tools. 
Consequently, the developer is responsible for implementing a TB-based design exclusively at the register transfer level (RTL), using languages such as Verilog or VHDL, within Intel's Quartus Prime software~\cite{Quartus_Prime_tool}.

\section{AMD Ryzen AI NPUs}
\label{sec:Background:Ryzen_NPU}

\subsection{Architecture Overview}
\label{subsec:NPU_architecture}

\begin{figure}[tbp]
\centering
\includegraphics[width=0.65\linewidth]{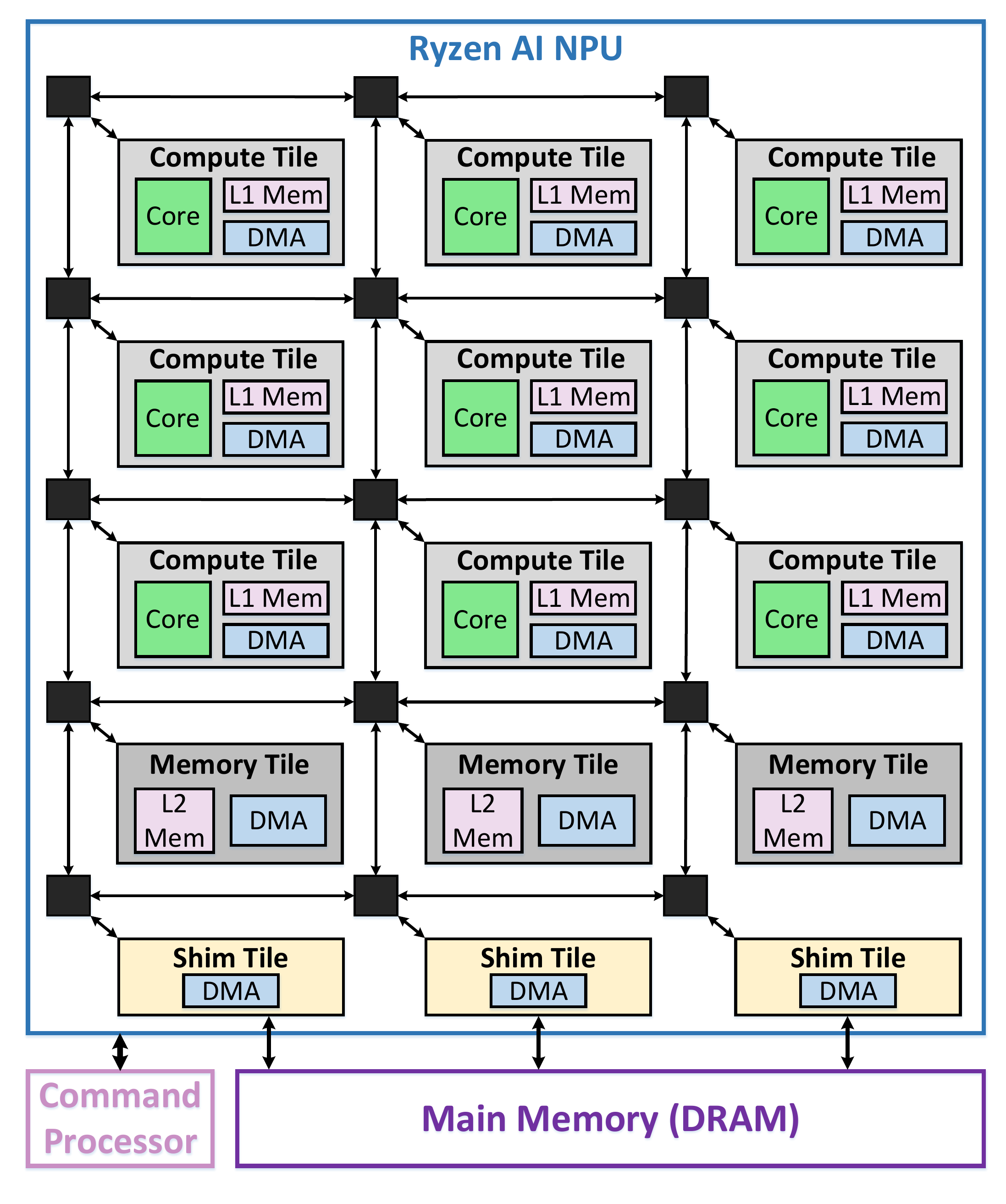}
\caption{Architecture of Ryzen AI NPUs.}
\label{fig:NPU_architecture}
\end{figure}

The architecture of both XDNA and XDNA2 is illustrated in  Fig. \ref{fig:NPU_architecture}.
The NPU is a modular and scalable 
architecture, comprising a 2D array of identical compute tiles (CompTiles) \cite{NPU_arch_MICRO_2024, AMD_AIE_ML_architecture_manual}. 
CompTiles include the processing cores which operate out of a local memory (L1).
The NPU core is a VLIW processor with SIMD datapath, supporting both fixed-point and floating-point operations.
NPUs incorporate a second level of on-chip memory (L2) via the memory tiles (MemTiles).
These are arranged in a single row, located below the array of CompTiles (Fig. \ref{fig:NPU_architecture}).
Finally, NPUs include a last row of interface tiles (ShimTiles) to provide communication with the main off-chip memory (DRAM).

Data movement between the levels of memory hierarchy is facilitated by DMA engines, which are integrated across all NPU tiles.
DMAs move data between the NPU tiles by utilizing the configurable interconnects (switches), shown as black squares in Fig. \ref{fig:NPU_architecture}.
The DMA engines of ShimTiles read/write data to DRAM via the NPU NoC and SoC-level fabric of the Ryzen AI chips \cite{NPU_arch_MICRO_2024, NPU_HOT_CHIPS_2023}.
Task scheduling and data movement are orchestrated by an on-chip command processor.
This processor controls the data movement between ShimTiles and DRAM at runtime, allowing the NPU to focus exclusively on the main computation.
The command processor is also responsible for (re-)configuring the NPU compute kernels, switches, and DMA transfers.


The XDNA has 20 compute cores, which are organized as a 4$\times$5 array (rows $\times$ columns) of CompTiles \cite{NPU_arch_MICRO_2024}, while the XDNA2 contains 32 cores as a 4$\times$8 array \cite{IRON_MICRO_2024}.
Both NPU generations have 64 KB of memory per L1 tile and 512 KB of memory per L2 tile \cite{NPU_arch_MICRO_2024, AMD_AIE_ML_architecture_manual, Zen_attention_AMD_paper}.
XDNA and XDNA2 natively support int8, int16 and bf16 precisions.
Additionally, XDNA2 has hardware support for bfp16 datatype \cite{AMD_AIE_API_user_guide_2025.1}, where a block of eight numbers shares one common exponent \cite{AMD_quark_bfp16, Microsoft_block_floating_point_NEURIPS_2020}.
XDNA2 offers increased theoretical peak compute capabilities, delivering up to 50 TOPS \cite{AMD_CES, IRON_MICRO_2024}, compared to the 10 TOPS of XDNA \cite{NPU_arch_MICRO_2024}.




\subsection{Data Movement Architecture}
\label{subsec:data_movement_architecture}

Ryzen AI NPUs incorporate a dedicated data movement architecture that allows programmers to explicitly configure data transfers between all levels of the memory hierarchy. 
Data moves from a source DMA channel to a destination DMA channel through a circuit- or packet-switched stream, using one or more configurable switches between them.
The source channel is a memory-map to stream (MM2S) channel which reads data from memory and pushes it to the stream switches.
Conversely, the destination channel receives data from the stream switches and writes it to the memory-mapped space (S2MM).
Each CompTile and ShimTile includes two MM2S and two S2MM DMA channels, while MemTiles incorporate six MM2S and six S2MM channels \cite{AMD_AIE_ML_architecture_manual}.

\begin{figure}[t]
\centering
\subfloat[]
{\includegraphics[width=0.35\linewidth]{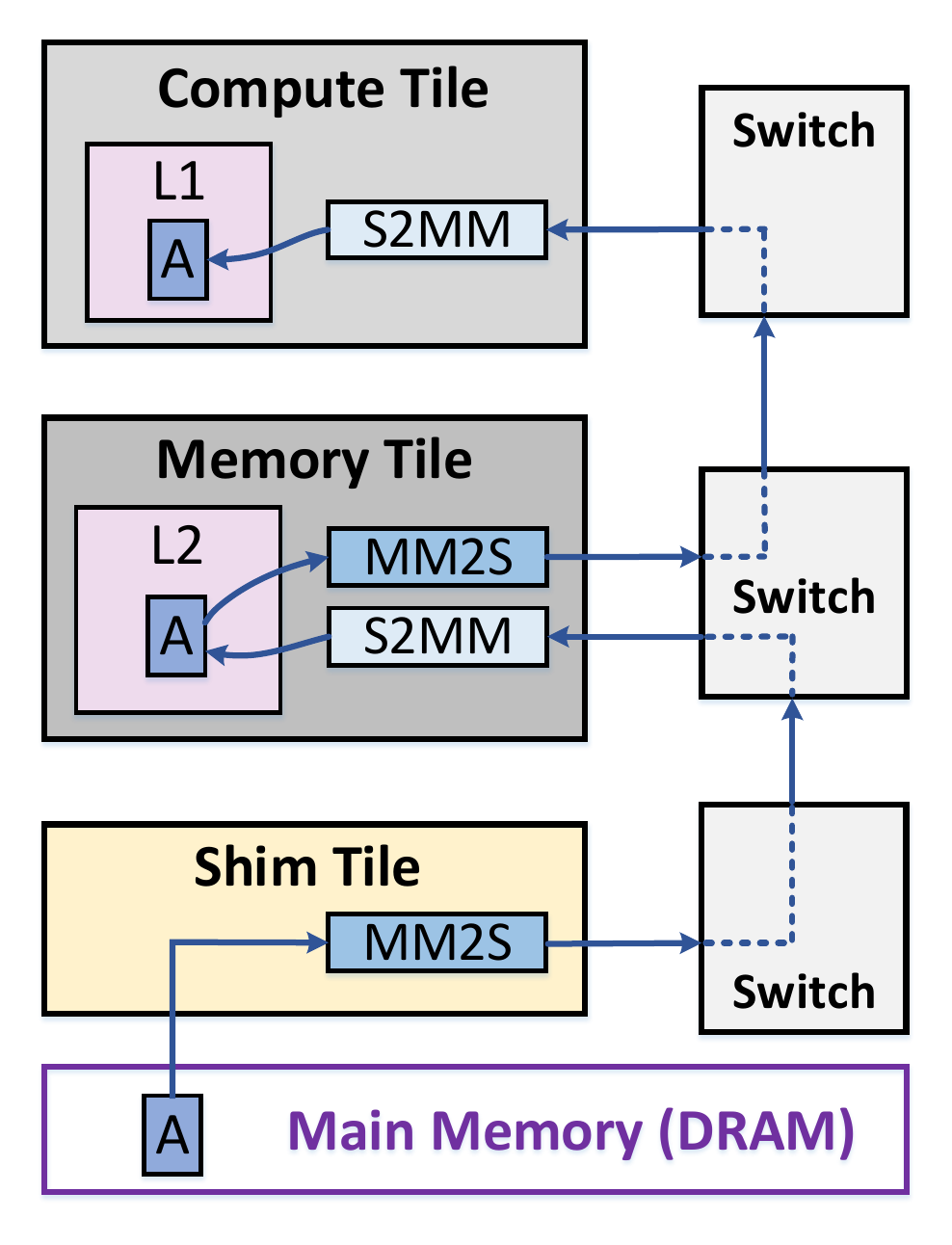}
\label{fig:Data_movement_matrix_a}}
\hspace{40pt}
\subfloat[]{\includegraphics[width=0.35\linewidth]{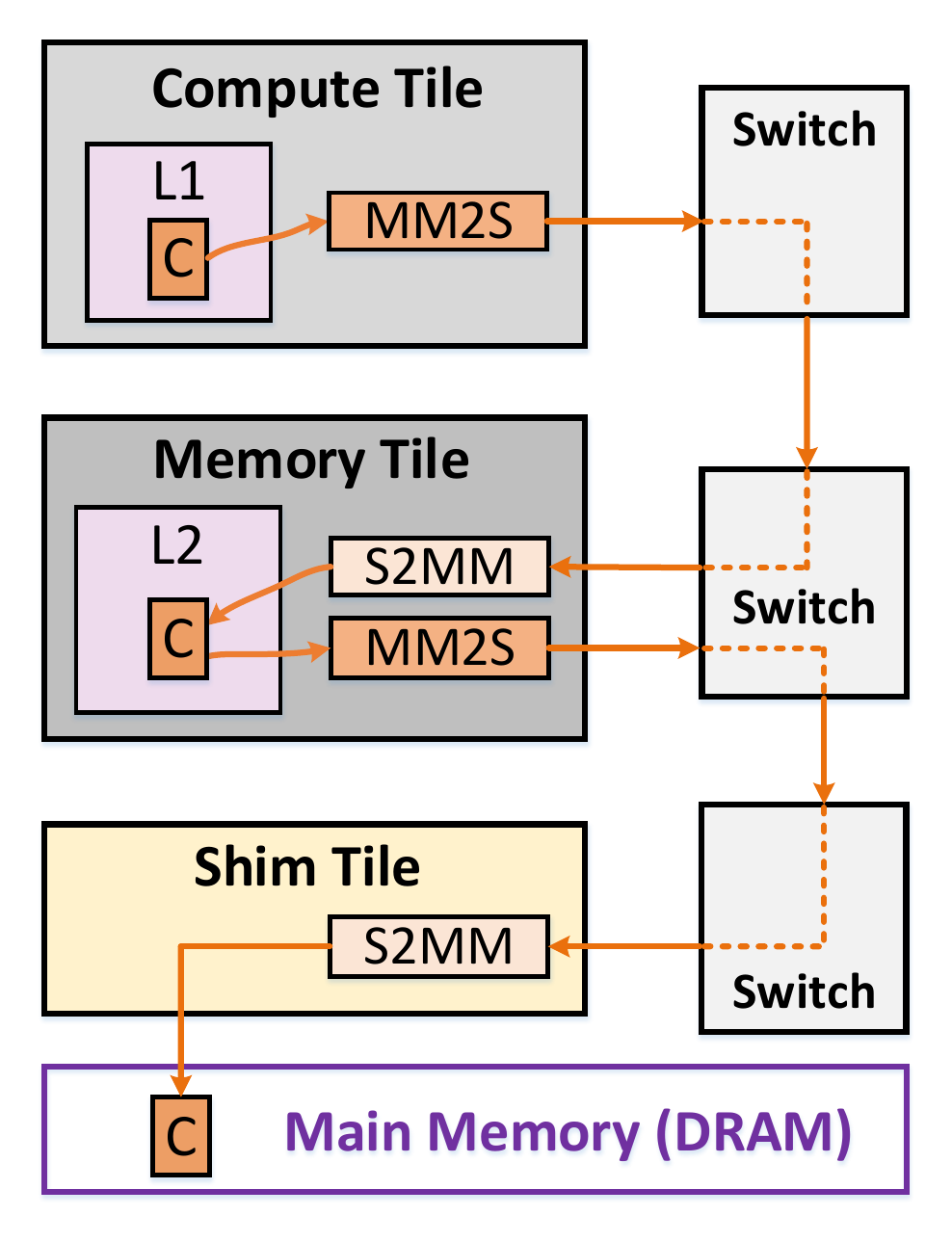}
\label{fig:Data_movement_matrix_c}} 
\caption{Data movement across the memory hierarchy in Ryzen AI NPUs: input buffer A (a) and output buffer C (b).} 
\label{fig:data_movement_architecture}
\end{figure}

Fig. \ref{fig:Data_movement_matrix_a} shows an example of moving an input buffer $A$ from DRAM to an L2 MemTile, and eventually to L1, where it can be utilized by the corresponding core.
In a similar fashion, Fig. \ref{fig:Data_movement_matrix_c} depicts the data movement of an output buffer from L1 to L2, and eventually to DRAM. 
The synchronization of data buffers between the DMAs and the corresponding module, \emph{e.g.,} the NPU core or DRAM, is managed by hardware lock units \cite{AMD_AIE_ML_architecture_manual, NPU_arch_MICRO_2024}.

DMAs run independently and in parallel with computation on cores.
They are programmed by configuring a sequence of  buffer descriptors (BDs).
A BD contains all the required information associated with a specific DMA transfer, such as the amount of data to read/write, the memory addresses involved, and the locks to acquire/release before and after each transfer \cite{AMD_AIE_ML_architecture_manual, AMD_AIE_ML_kernel_guide}.
BDs support both linear memory addressing and multi-dimensional address generation, enabling on-the-fly data layout transformations required for DL tensors.
CompTiles and ShimTiles support each 3D tensor addressing, while MemTiles incorporate 4D addressing.
This important DMA addressing feature is extensively exploited by our GEMM implementation across all tiles in the NPU architecture, allowing tensors to be stored in regular order in DRAM.

\subsection{Programming \& Tools}
\label{subsec:Programming_tools_Ryzen}

In this work, we use IRON, an open-source close-to-metal toolchain for developing programs on Ryzen AI NPUs \cite{mlir_aie_github_repo}.
As a low-level toolkit, IRON enables fine-grained control of the NPU architectural attributes, such as explicit data movement and complex access patterns supported in DMAs, while also providing convenient programming abstractions \cite{Erika_IRON_FCCM_2025}.
These characteristics allow explicit control of the NPU architectural features, which is critical for achieving high performance.

IRON is based on a multi-level intermediate representation (MLIR) \cite{mlir_paper} dialect, named ``\textit{AIE}'', and offers a user-friendly interface to this dialect through Python bindings \cite{mlir_python_bindings}.
Hence, Python scripts can be used to generate code that describes both the data movement across the NPU hierarchy and the compute kernels that run on the cores.
The single-core compute kernels can be written in high-level C++ \cite{AMD_AIE_API_user_guide_2025.1}, or low-level SIMD intrinsics \cite{AIE_intrinsics}.
Finally, two options are available for single-core kernel compilation: the proprietary \emph{xchesscc} compiler \cite{AMD_AIE_ML_kernel_guide, mlir_aie_github_repo} and the open-source \emph{Peano} tool \cite{peano_github_repo}.

\chapter{MaxEVA: Maximizing the Efficiency of Matrix
Multiplication on Versal AI Engine}
\label{ch:MAXEVA}

In this chapter, we discuss the details of the MaxEVA framework \cite{Taka2023MaxEVA, maxeva_github}. The MaxEVA framework addresses the design, mapping, and optimization of GEMM on the AIE array. MaxEVA assumes that input/output data buffers are placed in PL BRAMs, as repeatedly used in practice to efficiently exploit data reuse in large matrices \cite{charm2023fpga, AutoSA_FPGA21, dac23automm, Caffeine_CNN_FPGA19}. This assumption restricts the matrix sizes that can be processed entirely on-chip; we revisit and resolve this limitation in Chapter~\ref{ch:Versal_vs_Stratix}, where we extend MaxEVA with an additional level of memory hierarchy to support arbitrary matrix sizes.


In this work, we focus on VC1902 device of the VCK190 board \cite{VCK_190_manual}, which has a total of 400 AIEs organized as an 8 rows $\times$ 50 columns array.
The AIE/PL communication is established through the AIE-PL interface tiles.
However, not all existing columns in the AIE array can interface with PL.
For instance, in VC1902 there are only 
39 AIE-PL tiles \cite{Versal_AI_DC_AC_switching}.
The small AIE-PL bandwidth is one of the main challenges when designing GEMM workloads on AIEs, which MaxEVA successfully overcomes.
Finally, we note that although we show our method 
on the VC1902, our work can be 
generalized in straightforward fashion to any Versal device.

\section{MaxEVA Framework}
\label{sec:MaxEVA:Framework}

\subsection{Matrix Multiplication Tiling Scheme}
\label{subsec:tiling_scheme}

\begin{figure}[tbp]
\centering
\includegraphics[width=0.60\linewidth]{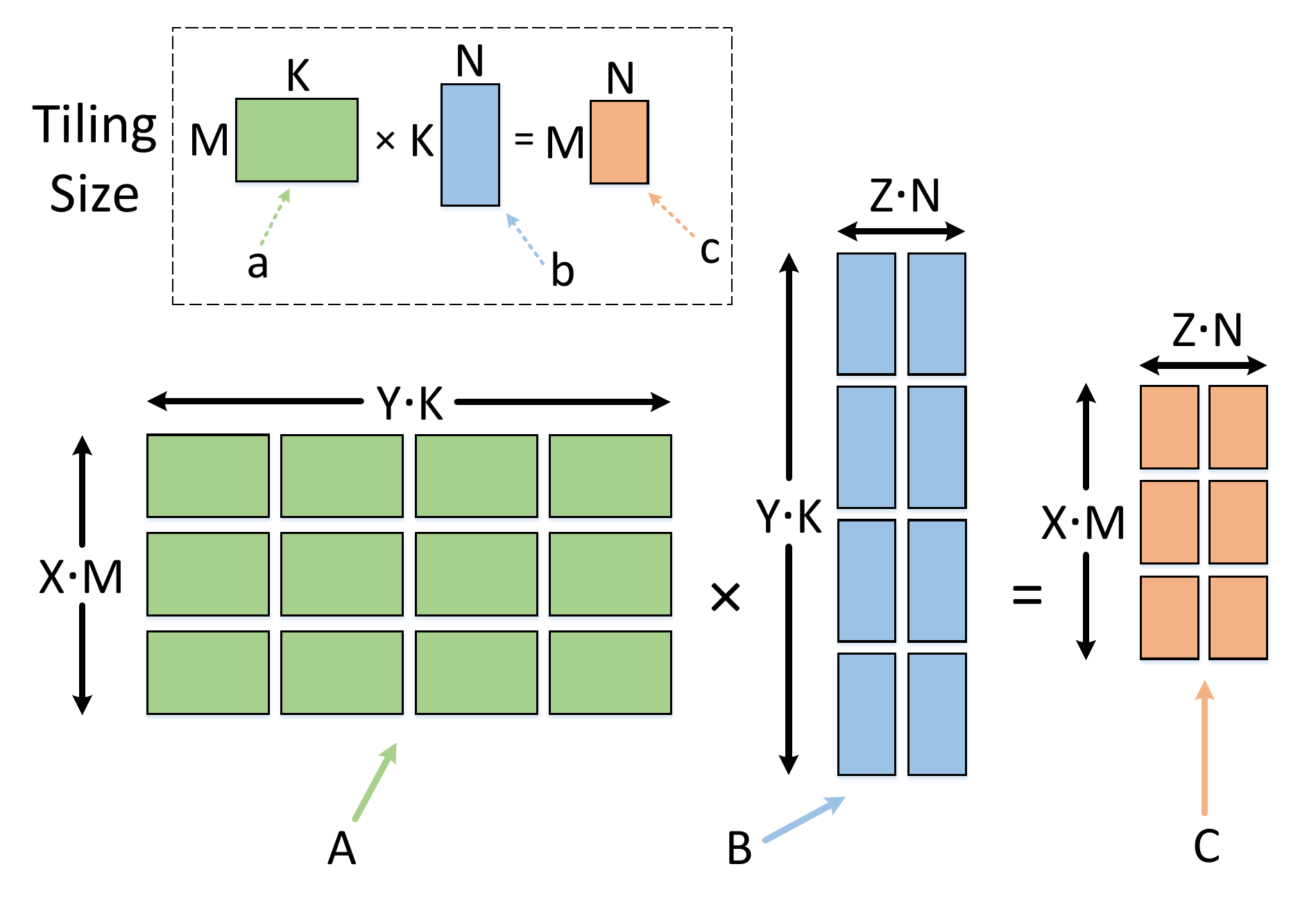}
\caption{Simplified view of tiling scheme for X=3, Y=4, Z=2.}
\label{fig:tiling_scheme}
\end{figure}

Fig. \ref{fig:tiling_scheme} depicts a simplified example of our proposed tiling scheme.
In this design, the single-core tiling size ($M$$\times$$K$$\times$$N$) 
is determined by the single GEMM kernel, which is mapped to exactly one AIE core.
Since the Versal AIE comprises multiple cores, we map multiple GEMM kernels on the AIE array
(described by the parameters $X, Y, Z$ as explained below).
With this scheme, the final GEMM size running on the entire AIE array is 
$(X \cdot M) \times (Y \cdot K) \times (Z \cdot N)$. 
To this end, $X, Y, Z, M, K, N$ constitute the configurable integer parameters which are optimized by the MaxEVA framework.

\subsection{Matrix Multiplication Mapping on AI Engine Array}
\label{subsec:GEMM_mapping_on_AIE}

\begin{figure}[htbp]
\centering
\includegraphics[width=0.75\linewidth]{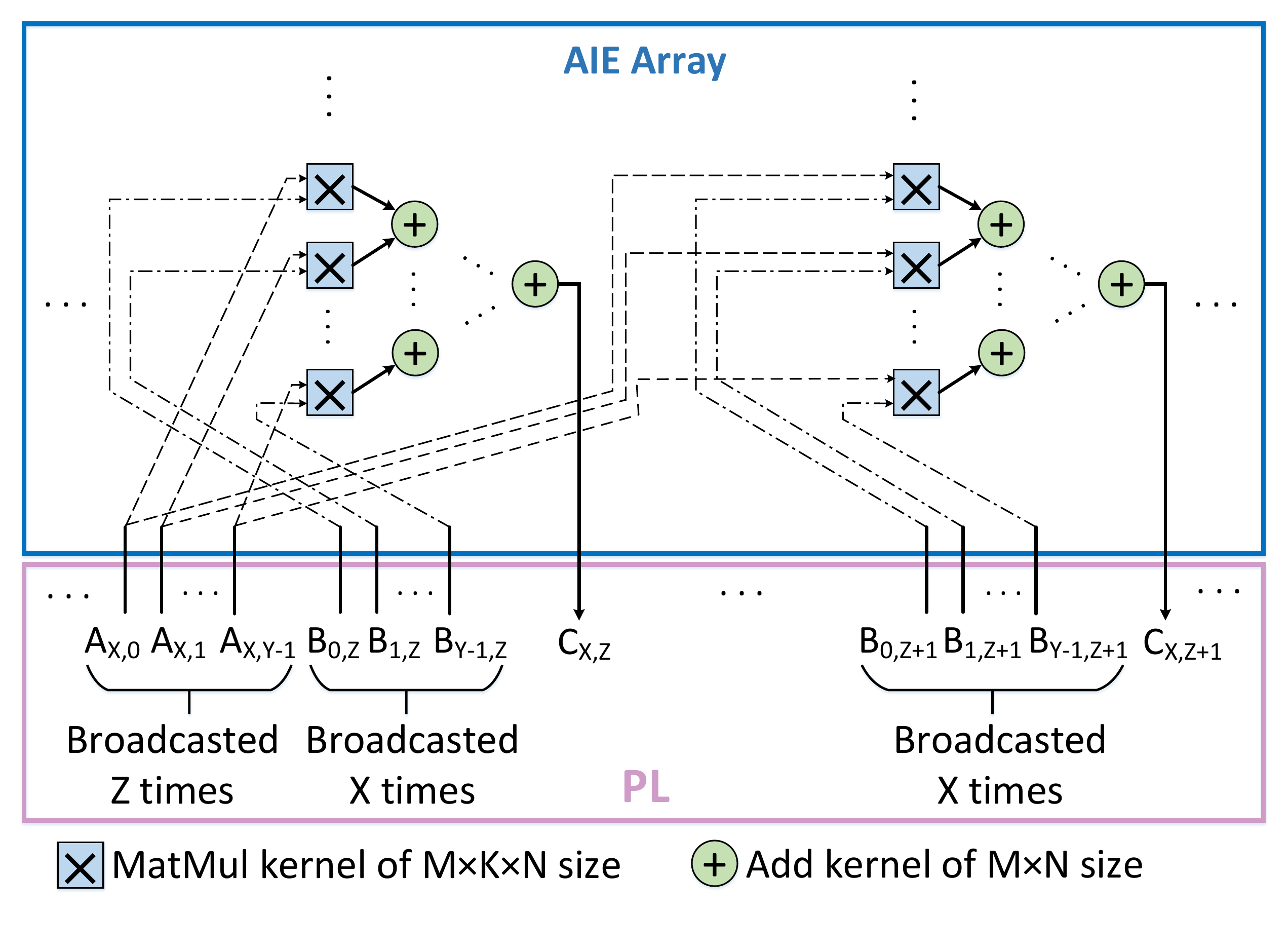}
\caption{GEMM mapping on AIE array by leveraging input broadcasting and output adder tree reduction.}
\label{fig:GEMM_mapping_on_AIE_array}
\end{figure}

To overcome the reduced number of AIE-PL interface tiles, and thus avoid the PL Input/Output (PLIO) bottleneck, we leverage  two principal properties of the GEMM algorithm.
First, we exploit the inherent data reuse in GEMM to reduce the number of incoming PLIOs (inputs to AIE array), by broadcasting inputs $A$ and $B$ (Fig. \ref{fig:tiling_scheme}) to multiple AIEs.
Second, we utilize the native reduction of the $Y \cdot K$ dimension in GEMM to decrease the number of outgoing PLIOs (outputs of AIE array) by performing reduction on the AIE itself, instead of the PL.
With this method, we are able to efficiently map $X \cdot Y\cdot Z$ GEMM kernels 
(each performing a GEMM computation of $M$$\times$$K$$\times$$N$ size), and $X \cdot (Y-1)\cdot Z$ Add kernels (each reducing partial results of $M$$\times$$N$ size).

Fig. \ref{fig:GEMM_mapping_on_AIE_array} shows a high-level mapping diagram of GEMM and Add kernels on the AIE array.
In our design, there exist groups of $Y$ GEMM kernels along with their corresponding adder trees ($Y-1$ adders on each group).
More specifically, there are $X \cdot Z$ of such groups in total, all executing in a parallel fashion.
Each $A_{x,y}$ and $B_{y,z}$ PLIO input data are broadcasted to their corresponding GEMM kernels, $Z$ and $X$ times, respectively, as governed by the GEMM algorithm.
In this design, there are in total $X \cdot Y + Y \cdot Z$ PLIO inputs and $X \cdot Z$ PLIO outputs.
We note here that broadcasts are realized through the programmable switches of the AIE array and are statically configured during compilation (via circuit-switching).

\begin{figure}[tbp]
\centering
\includegraphics[width=0.60\linewidth]{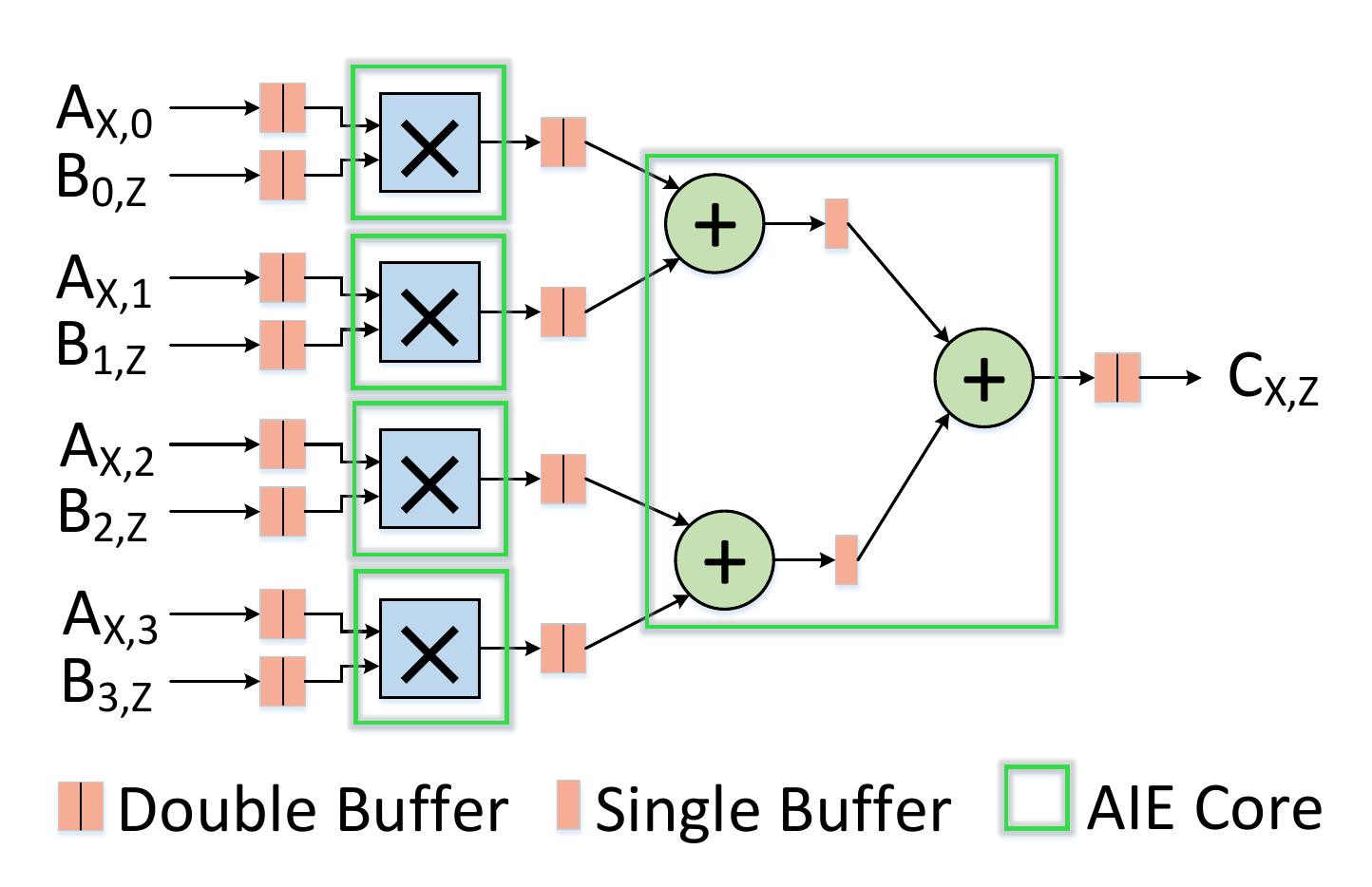}
\caption{Group of GEMM kernels (each mapped to one AIE core) and its adder tree (whole tree mapped to a single AIE core).}
\label{fig:adder_tree_mults_double_single_buffers}
\end{figure}

Fig. \ref{fig:adder_tree_mults_double_single_buffers} illustrates a group comprising 4 GEMM kernels along with its adder tree (3 Add kernels).
As mentioned above, each GEMM kernel executes on a separate AIE core.
However, we map the whole adder tree to a single AIE core, where all Add kernels are executed in a sequential fashion.
We make such design choice for various reasons.
First, we note that only GEMM kernels contribute to throughput, while Add kernels are only used to reduce the output PLIOs.
Thus, we maximize the number of GEMM kernels, by minimizing the AIE cores used to run the Add kernels.
Second, Add kernel's execution time (latency) is much lower than GEMM kernel's latency.
Therefore, we are able to map multiple Add kernels to a single AIE core, without any performance degradation (as we show in Section \ref{subsec:Single_kernel_results}).
Third, when mapping multiple kernels to a single AIE core, memory resources are reduced compared to mapping to several AIE cores.
As depicted in Fig. \ref{fig:adder_tree_mults_double_single_buffers}, double buffers are inserted between separate AIE cores to effectively overlap computation with communication (and thus increasing the compute utilization of AIEs).
However, between the Add kernels only single buffers are used, since all Add kernels are executed sequentially.
This results in twofold memory buffer savings compared to if they were executed on separate AIEs.
Overall, the throughput of the entire design is determined by the computationally heavy GEMM kernels, since the whole adder tree latency is lower than GEMM latency (Section \ref{subsec:Single_kernel_results}).

\subsection{AI Engine Kernels Modeling \& Optimization}
\label{subsec:AIE_kernels_optimization}

Because the design space of the configurable integer parameters $X, Y, Z, M, K, N$ is large, we propose an analytical model to maximize the throughput of GEMM on the AIE array.
Our model takes as input device-specific parameters and constraints, \textit{e.g.}, I/O bandwidth, AIE peak throughput, number of PLIOs, and finds the optimal solution based on our mapping methodology.
In this section, first, we describe the optimization of the single GEMM kernel, and then we discuss mapping the entire GEMM to the AIE array.


\textbf{1) \textit{Single AI Engine Kernel Optimization: M, K, N parameters}}.
Our model takes as input a lower bound of the  efficiency of the single GEMM kernel, $eff\_lb$,
to ensure that the achieved throughput is very close to the theoretical AIE throughput.
Here, we define efficiency ($eff$) as the fraction of the achieved throughput to the peak throughput of the vector processor inside the AIE core.
In particular, each AIE vector processor is able to achieve up to 128 MACs/cyc (multiply-accumulate operations per clock cycle) for int8 precision, while for fp32 the peak throughput is 8 MACs/cyc \cite{Versal_acap_architecture_manual}. 
Defining $peak\_MACs$ as the AIE core peak throughput and $kernel\_cyc$ as the latency (in clock cycles) of the GEMM kernel of $M$$\times$$K$$\times$$N$ size, we obtain the following constraint: 
\begin{gather} 
eff \geq eff\_lb \Rightarrow \nonumber \\
\left(\frac{M \cdot K \cdot N}{kernel\_cyc}\right)/peak\_MACs \geq eff\_lb \Rightarrow \nonumber \\
kernel\_cyc \leq M \cdot K \cdot N / \left(eff\_lb \cdot peak\_MACs\right) \label{eq:kernel_throughput}
\end{gather}
Next, we optimize our design by ensuring that I/O bandwidth does not become a performance bottleneck.
There are two I/O bandwidth considerations: the PLIO bandwidth for both inputs and outputs, as well as the bandwidth of the AXI4-Stream switches. 
According to \cite{Versal_acap_architecture_manual}, the bandwidth for both I/Os is $BW\_IO = 4Bytes/cyc$. 
To assure that our design is not I/O limited, we require the latency of input ($a\_cyc$, $b\_cyc$) and output ($c\_cyc$) data transmission to be lower than the GEMM kernel latency. Therefore, we get:
\begin{align}
\{a\_cyc, \ b\_cyc, \ c\_cyc\} \le & kernel\_cyc \Rightarrow \nonumber \\
\{ M \cdot K \cdot sizeof(a)/BW\_IO &\le kernel\_cyc ,\nonumber\\
K \cdot N \cdot sizeof(b)/BW\_IO &\le kernel\_cyc , \nonumber\\
M \cdot N \cdot sizeof(c)/BW\_IO &\le kernel\_cyc \} 
\label{eq:plio}
\end{align}
By combining equations \ref{eq:kernel_throughput} and \ref{eq:plio}, we have the following:
\begin{align}
N \geq eff\_lb \cdot peak\_MACs &\cdot sizeof(a) / BW\_IO \label{eq:a_BW_constr}\\
M \geq eff\_lb \cdot peak\_MACs &\cdot sizeof(b) / BW\_IO \label{eq:b_BW_constr}\\
K \geq eff\_lb \cdot peak\_MACs &\cdot sizeof(c) / BW\_IO \label{eq:c_BW_constr}
\end{align}
Notice that $a$, $b$ are the inputs of the GEMM kernel, while $c$ is the output of either the GEMM or the Add kernel (both have same output size of $M$$\times$$N$).
The $sizeof()$ function calculates the size (in Bytes) of input/output \textit{data types}.
This is particularly important for int8 computation, since we retain all accumulations in 32-bits (int32).
In this case, the size of inputs $a$, $b$ is 1 Byte, while the size of output $c$ is 4 Bytes.

Finally, we define a constraint that all input/output buffers of the single GEMM kernel should fit within the local AIE memory.
By not exceeding the local AIE memory, we are able to maximize the number of GEMM kernels that execute in parallel on the AIE array.
Each AIE core needs some system memory for its operation, \textit{e.g.}, stack. The AIE data memory is partitioned in 4KB banks; we leave one bank for system memory. This implies an available space of 28KB for our/user buffers.
Because both input and output buffers of GEMM kernels are double buffered (see Fig. \ref{fig:adder_tree_mults_double_single_buffers}), we get:
\begin{equation}
\{M \cdot K \cdot sizeof(a) + K \cdot N \cdot sizeof(b) 
+ M \cdot N \cdot sizeof(c)\} \le 14KB \label{eq:memory_constr}
\end{equation}

The solution of $M, K, N$ can be formulated as an integer programming (IP) optimization problem, where we maximize the number of MACs of the single GEMM kernel by having eq. \ref{eq:a_BW_constr}--\ref{eq:memory_constr} as constraints.
Increasing the number of MACs will lead to more data reuse in the vector registers of the AIE core, resulting in higher kernel efficiency.
The lower bound of the efficiency ($eff\_lb$) can be assigned based on the performance constraints of the application and the achievable throughput.
In this work, we are able to achieve 95\% of GEMM kernel efficiency (Section \ref{subsec:Single_kernel_results}), which we set it as our lower bound.
The configurable parameters are evaluated through exhaustive search and top-ranked design points are reported.
We note that solving the IP exhaustively is a suitable approach, since the search space is significantly reduced by considering only powers of two for $M, K, N$ (as discussed in Section \ref{subsec:Single_kernel_results}).


\textbf{2) \textit{Multiple AI Engine Kernels Optimization: X, Y, Z parameters}}.
To obtain an optimal mapping onto the AIE array, we require our entire design to fit in the total number of AIE cores ($AIE\_cores$).
We also require the number of utilized input/output PLIOs to not exceed the available PLIOs of the device.
In particular, for VC1902, $PLIO\_in = 78$ and $PLIO\_out = 117$ \cite{Versal_acap_architecture_manual, Versal_AI_DC_AC_switching}.
Based on the discussion at Section \ref{subsec:GEMM_mapping_on_AIE} and the above, the following can be expressed:
\begin{gather}
X \cdot Y \cdot Z + X \cdot Z \le AIE\_cores \label{eq:total_cores_constr} \\
X \cdot Y + Y \cdot Z \le PLIO\_in  \label{eq:PLIO_in_constr}\\
X \cdot Z \le PLIO\_out \label{eq:PLIO_out_constr}
\end{gather}

The optimization of $X, Y, Z$ is evaluated through exhaustive search by maximizing the total number of GEMM kernels ($X \cdot Y \cdot Z$), which leads to maximized throughput in GEMM.
Again, exhaustive search is sufficient because all constants in eq. \ref{eq:total_cores_constr}--\ref{eq:PLIO_out_constr} are in the order of hundreds (reasonably small) \cite{Versal_acap_architecture_manual}.
Multiple top-ranked design points are reported, from which we explore various options (refer to Section \ref{subsec:Multiple_kernels_results} for details).

\subsection{AI Engine Kernel Placement}
\label{subsec:Placement_strategy}

\begin{figure}[bp]
\centering
\includegraphics[width=0.50\linewidth]{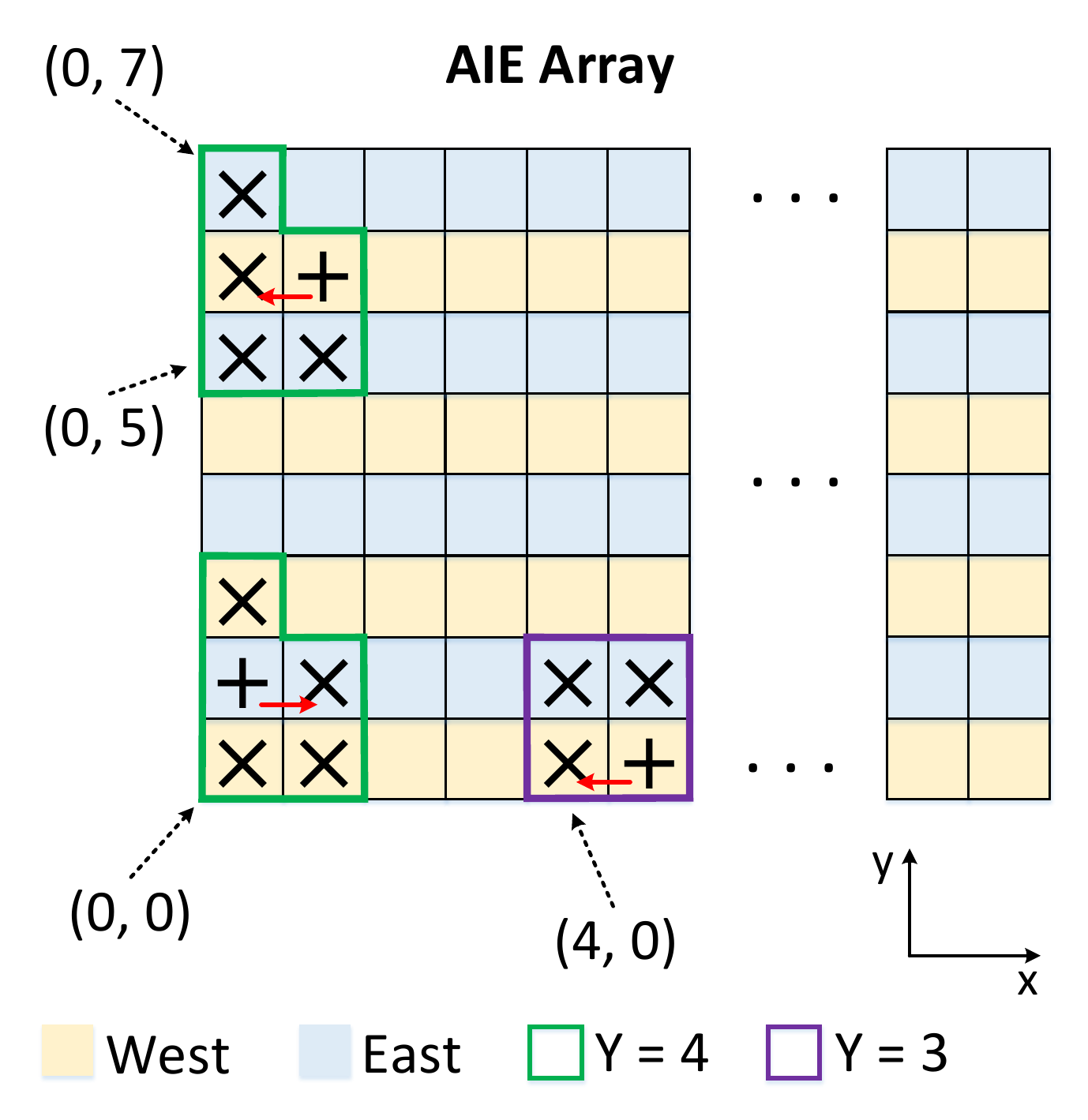}
\caption{Examples of proposed placement strategy for $Y=3, 4$.}
\label{fig:placement_example}
\end{figure}

To leverage the most efficient neighboring data sharing mechanism of the AIE, we propose a sophisticated kernel placement strategy.
Fig. \ref{fig:placement_example} illustrates an example of the placement procedure, where each \textit{multiplication} symbol denotes a GEMM kernel, while the \textit{addition} symbol indicates the adder tree mapped to a single AIE core.
We place each group of $Y$ GEMM kernels along with its adder tree, in a manner to avoid any DMA usage in the buffer connections \textit{between} GEMM and Add kernels (Fig. \ref{fig:adder_tree_mults_double_single_buffers}).
For instance, when considering the group starting at (0, 0) location in Fig. \ref{fig:placement_example}, the adder tree is able to access the memory of 3 (out of 4) GEMM kernels directly (along the north, south and east direction).
Notice that in this case, the GEMM kernel located at (1, 0) does not directly communicate with the adder tree.
However, the output buffer of this GEMM kernel can be placed to its north location (1, 1); this is possible because of direct memory sharing between neighboring AIEs as shown in Fig \ref{fig:Data_movement_mechanims}. From here the adder tree AIE can access it directly, thus ensuring no DMA usage.

\begin{figure}[htbp]
\centering
\includegraphics[width=0.85\linewidth]{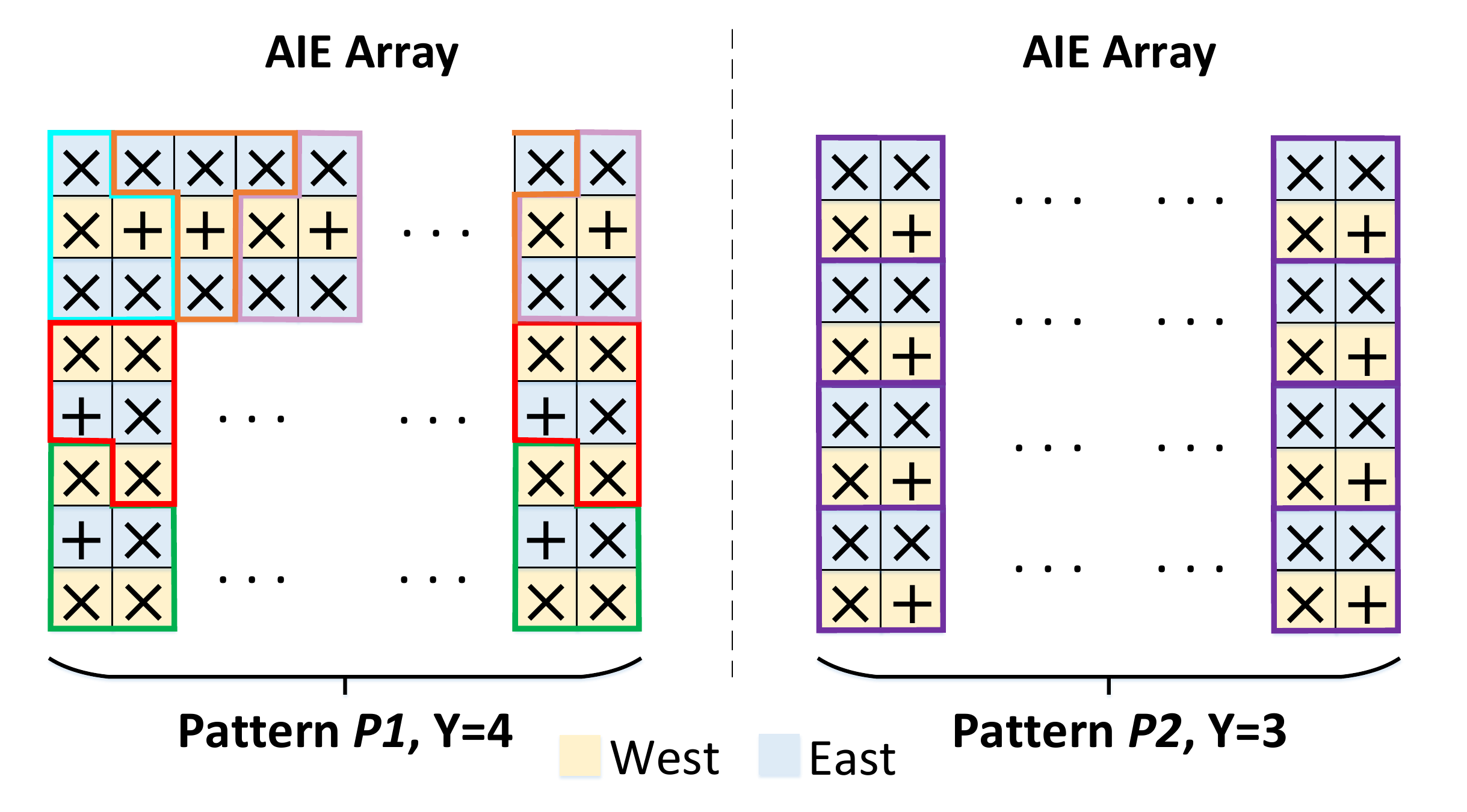}
\caption{Proposed placement patterns.}
\label{fig:placement_patterns}
\end{figure}

Another placement example is the group starting at (0, 5).
Although, this placement is similar to the (0, 0) case described above, notice that the adder tree is located on the opposite side.
This is because the local data sharing is realized only on the west direction in even rows.
We observe that in this case, the adder tree can only access 2 out of 4 GEMM kernels directly (located at (0, 6) and (1, 5)). 
However, the output buffers of all GEMM kernels can be placed such that no DMA is used.
For instance, the output buffers of (0, 5) and (0, 7) kernels can be placed at (1, 5) and (1, 7) locations (east access), respectively, which can be both directly accessed by the adder tree.



An example for a group of $Y=3$ kernels is also shown in Fig. \ref{fig:placement_example}, where the explanation of its placement is similar to the examples discussed above.
Finally, we note that the input data buffers of the GEMM kernels, as well as the output data buffer of the adder tree (refer to Fig. \ref{fig:adder_tree_mults_double_single_buffers}) can be placed on any free memory space, again only accessing memory directly.
We let the AMD/Xilinx AIE PnR (place and route) tool to make such decisions and optimize the whole AIE array mapping.


We exploit the aforementioned placement strategy to fill the entire AIE array, thus mapping multiple GEMM/Add kernels.
Fig. \ref{fig:placement_patterns} shows our two proposed placement patterns, $P1$ and $P2$. 
Notice that in pattern $P1$, we use a new \emph{``T"--like} shape outlined in orange color, while all other shapes of both patterns are similar to the examples in Fig. \ref{fig:placement_example}.
This is required to fill the whole AIE array and ensure that no AIEs remain unutilized.
However, each \emph{``T"--like} shape will lead to a small DMA usage (one GEMM output buffer), as we show in the next section.
Finally, we only propose patterns for $Y=3, 4$, because these constitute the top-ranked solutions based on our optimization approach, as we also present in the next section.

\section{Evaluation}
\label{sec:MaxEVA:Evaluation}
In this section, we first report single AIE kernel experimental results.
Second, we present results and evaluation of full  GEMM mapping on the AIE array.
Finally, we perform a comprehensive comparison with the state-of-the-art approach, which exhibits the superiority of the MaxEVA framework.

We compile and simulate our designs by using the AMD/Xilinx Vitis 2022.1 version.
Across all experiments, the AIE frequency is set to the maximum frequency of the VC1902 on VCK190, \textit{i.e.}, 1.25 GHz, while the PL operates at 312.5 MHz, as recommended in \cite{AI_Engine_programming_guide, AI_Engine_programming_environment}.
To ensure rate matching without performance reduction between AIE and PL, we set the PLIO width to 128-bits \cite{AI_Engine_programming_guide}.
Moreover, we use the AIE simulator \cite{AI_Engine_programming_environment}   
to calculate the throughput of our designs, which we report as an average of 10 runs.
Finally, power consumption is estimated through the AIE XPE tool \cite{AIE_XPE}.

Since our focus is to achieve maximum GEMM throughput from the AIE array, 
isolating the AIE implementation from any source of performance degradation is crucial to obtain accurate evaluation.
Such sources of performance reduction may arise from any design mapped to PL causing stalls, as well as the limited DRAM bandwidth.
Therefore, we simulate our AIE designs without utilizing the PL and DRAM.
We simulate the state-of-the-art CHARM framework as well by leveraging their open-source code \cite{charm2023fpga, dac23automm}, in the exactly same manner, with the same assumptions.
This ensures a thorough and fair comparison between MaxEVA and CHARM frameworks, thus enabling us to report accurate throughput, power, and AIE resource utilization.
However, we note that the CHARM code includes only fp32 implementation.
In \cite{dac23automm} the authors report results for int8 implementation as well, but their code is not publicly available.
Therefore, for int8, we show a qualitative comparison based on their reported results.

\begin{table}[t]
 \centering
\caption{Single AI Engine kernel results.}
\resizebox{0.94\linewidth}{!}{
\begin{tabular}{c|c|c|c|c}
\Xhline{2.5\arrayrulewidth}
\textbf{Kernel}  & \textbf{Kernel size}  & \textbf{Latency}  & \textbf{Throughput}  & \textbf{Effic-} \\

\textbf{type} & \textbf{M$\times$K$\times$N} & \textbf{(cyc)} & \textbf{(MACs/cyc)} & \textbf{iency} \\
\hline
\hline

GEMM int8 & 32$\times$128$\times$32 & 1075 (1$\times$) & 121.93 & 95.26\% \\
Add int32 & 32$\times$32 & 164 (0.15$\times$) & 6.24 & 78.05\% \\
\hline
GEMM fp32 \cite{charm2023fpga, dac23automm} & 32$\times$32$\times$32 & 4329 (1$\times$) & 7.57 & 94.70\% \\
Add fp32 & 32$\times$32 & 167 (0.04$\times$) & 6.13 & 76.65\% \\

\Xhline{2.5\arrayrulewidth}

\end{tabular}

}
\label{tb:single_AI_engine_results}
\end{table}

\subsection{Performance of Single AI Engine Kernels}
\label{subsec:Single_kernel_results}

For maximum efficiency, GEMM and Add kernels have been designed to leverage the vector processors of the AIEs.
Our kernels are written in C/C++ utilizing the AIE APIs and several optimization compiler directives (pragmas) that perform software pipelining, loop unrolling/flattening and explicit independence between data \cite{AI_Engine_programming_guide}.
Moreover, during our kernel design experimentation, we found that GEMM kernels with powers of two dimensions produce higher efficiency.
Hence, in this work, we use powers of two during the optimization of the $M, K, N$ parameters, as also proposed in \cite{charm2023fpga, H_GCN_FPL2022}.

Table \ref{tb:single_AI_engine_results} presents the single AIE kernel results.
For int8, the 32$\times$128$\times$32 GEMM kernel was the only solution that satisfied all constraints in our optimization procedure (Section \ref{subsec:AIE_kernels_optimization}).
All other values for $M, K, N$ are either I/O bandwidth limited or exceed the memory constraint of 14KB (eq. \ref{eq:memory_constr}).
In contrast, for fp32, there are many top-ranked solutions that maximize the number of MACs, \textit{e.g.}, 16$\times$64$\times$32, 64$\times$16$\times$32, 32$\times$32$\times$32, \textit{etc.}
However, we notice that all best solutions exhibit the same number of MACs (equal to 32768).
Since 32$\times$32$\times$32 GEMM kernel is one of our optimized solutions and also used in state-of-the-art CHARM \cite{charm2023fpga, dac23automm}, we obtain its code from their open-source code base.
This ensures a fair comparison between their approach and ours. 

From Table \ref{tb:single_AI_engine_results}, we observe that the int8 GEMM kernel utilizes the vector processor of the AIE efficiently - a very high efficiency of 95.26\% is achieved.
The fp32 GEMM kernel obtained from \cite{charm2023fpga, dac23automm} also presents very high efficiency (94.70\%), and is designed by using AIE intrinsics.
Moreover, we observe that int8 and fp32  Add kernels have very similar latencies, which are both significantly lower than GEMM latencies (164 \textit{vs.} 1075 cycles for int8, and 167 \textit{vs.} 4329 cycles for fp32). 
This validates that multiple Add kernels are able to run sequentially into a single AIE, without causing any performance degradation.
We also observe that the relative latency ratio of Add kernel to GEMM kernel is notably lower for fp32 (0.04$\times$) compared to int8 (0.15$\times$).
These relative ratios indicate that the AIE core running even multiple Add kernels remains idle for substantially longer for the fp32 case, affecting its power consumption accordingly (Section \ref{subsec:Multiple_kernels_results}).
Finally, we note that Add kernels also exhibit high efficiency (78.05\% and 76.65\% for int8 and fp32, respectively), though not as high as the efficiency of GEMM kernels.
This performance difference is due to less data reuse on the AIE vector registers by Add kernels compared to GEMM kernels. 



\subsection{Performance of Matrix Multiplication on AI Engine Array}
\label{subsec:Multiple_kernels_results}

We utilize the MaxEVA framework to optimize the performance of the entire GEMM application.
To map multiple kernels to the AIE array we wrote a parameterized C++ code for any values of $X, Y, Z$ by exploiting the ADF graph model.



\subsubsection{MaxEVA \textit{vs.} state-of-the-art CHARM for fp32 precision}
From our multiple AIEs optimization of $X$$\times$$Y$$\times$$Z$ parameters, we found that the 10$\times$4$\times$8 solution maximizes the number of GEMM kernels. 
In this case, there are 320 GEMM kernels and 80 cores which run Add kernels, hence, all 400 AIE cores are utilized.
However, this solution was not feasible because the AIE PnR tool failed due to routing congestion.
This is due to the extra routing needed because of DMA usage (pattern $P1$), as well as the 100\% utilization of the AIE cores, leaving no free space for successful routing.
Our second top-ranked solution, \textit{i.e.}, 13$\times$4$\times$6, does not present any routing issues and is successfully mapped to the AIE array.
In Table \ref{tb:MaxEVA_results_fp32} we show this solution (row 1), which achieves a very high throughput of 5442.11 GOPS, outperforming the state-of-the-art approach by 20.8\% 
(CHARM presents 4504.46 GOPS).

\begin{table*}[t]
 \centering
 \renewcommand{\arraystretch}{1.2}
\caption{Evaluation of several MaxEVA configurations for fp32 and comparison with state-of-the-art approach.}
\resizebox{1.00\textwidth}{!}{
\begin{tabular}{c|c|c|c|c|c|c|c|c}
\Xhline{2.5\arrayrulewidth}
\textbf{MaxEVA Cfg.}  & \textbf{GEMM}  & \textbf{Total}  & \textbf{Memory} & \textbf{DMA}  & \textbf{PLIOs}  & \textbf{Throughput} & \textbf{Power} & \textbf{Energy Eff.}\\
\textbf{X$\times$Y$\times$Z (pat.)} & \textbf{kernels} & \textbf{AIE cores} & \textbf{banks} & \textbf{banks} & \textbf{} & \textbf{(GOPS)} & \textbf{(W)} & \textbf{(GOPS/W)}\\
\hline
\hline
1. \hspace{1.5mm} 13$\times$4$\times$6 (P1) & 312 & 390 (97.5\%) & 3138 (98.1\%) & 18 & 154 (79.0\%) & \textbf{5442.11 (+20.8\%)} &  43.83 & \textbf{124.16 (+20.4\%)} \\
2. \hspace{0.2mm} 10$\times$3$\times$10 (P2) & 300 & 400 (100\%) & 3190 (99.7\%) & 0 & 160 (82.1\%) & 5405.33 (+20.0\%) & 44.66 & 121.03 (+17.4\%) \\
3. \hspace{1.5mm} 11$\times$4$\times$7 (P1) & 308 & 385 (96.3\%) & 3106 (97.1\%) & 18 & 149 (76.4\%) & 5414.39 (+20.2\%) & 44.01 & 123.03 (+19.3\%) \\
4. \hspace{1.5mm} 11$\times$3$\times$9 (P2) & 297 & 396 (99.0\%) & 3176 (99.3\%) & 0 & 159 (81.5\%) & 5382.27 (+19.5\%) & 44.13 & 121.96 (+18.3\%) \\
5. \hspace{1.5mm} 12$\times$4$\times$6 (P1) & 288 & 360 (90.0\%) & 2934 (91.7\%) & 16 & 144 (73.8\%) & 5031.19 (+11.7\%) & 40.68 & 123.68 (+20.0\%) \\
6. \hspace{1.5mm} 12$\times$3$\times$8 (P2) & 288 & 384 (96.0\%) & 3092 (96.6\%) & 0 & 156 (80.0\%) & 5225.05 (+16.0\%) & 42.28 & 123.58 (+19.9\%) \\
\hline
\hline
CHARM & 384 & 384 (96.0\%) & 3086 (96.4\%) & 0 & 80 (41.0\%) & 4504.46 (+0\%) & 43.69 & 103.10 (+0\%) \\
\Xhline{2.5\arrayrulewidth}
\end{tabular}
}
\label{tb:MaxEVA_results_fp32}
\end{table*}

\begin{table*}[t]
 \centering
 \renewcommand{\arraystretch}{1.2}
\caption{Evaluation of several MaxEVA configurations for int8 (results for CHARM obtained from \cite{dac23automm}).}
\resizebox{1.00\textwidth}{!}{
\begin{tabular}{c|c|c|c|c|c|c|c|c}
\Xhline{2.5\arrayrulewidth}
\textbf{MaxEVA Cfg.}  & \textbf{GEMM}  & \textbf{Total}  & \textbf{Memory} & \textbf{DMA}  & \textbf{PLIOs}  & \textbf{Throughput} & \textbf{Power} & \textbf{Energy Eff.}\\
\textbf{X$\times$Y$\times$Z (pat.)} & \textbf{kernels} & \textbf{AIE cores} & \textbf{banks} & \textbf{banks} & \textbf{} & \textbf{(TOPS)} & \textbf{(W)} & \textbf{(TOPS/W)}\\
\hline
\hline
1. \hspace{1.5mm} 13$\times$4$\times$6 (P1) & 312 & 390 (97.5\%) &  3112 (97.3\%) & 18 & 154 (79.0\%) & \textbf{77.01 (2.19$\times$)}& 66.83 & 1.152 \\
2. \hspace{0.2mm} 10$\times$3$\times$10 (P2) & 300 & 400 (100\%) &  3194 (99.8\%) & 0 & 160 (82.1\%) & 76.08 (2.16$\times$) & 65.52 & \textbf{1.161} \\
3. \hspace{1.5mm} 11$\times$4$\times$7 (P1) & 308 & 385 (96.3\%) &  3096 (96.8\%) & 18 & 149 (76.4\%) & 75.67 (2.15$\times$) & 66.79 & 1.133 \\
4. \hspace{1.5mm} 11$\times$3$\times$9 (P2) & 297 & 396 (99.0\%) &  3178 (99.3\%) & 0 & 159 (81.5\%) & 74.66 (2.12$\times$) & 65.83 & 1.134 \\
5. \hspace{1.5mm} 12$\times$4$\times$6 (P1) & 288 & 360 (90.0\%) &  2918 (91.2\%) & 16 & 144 (73.8\%) & 71.25 (2.02$\times$) & 62.13 & 1.147 \\
6. \hspace{1.5mm} 12$\times$3$\times$8 (P2) & 288 & 384 (96.0\%) &  3080 (96.3\%) & 0 & 156 (80.0\%) & 72.93 (2.07$\times$) & 63.24 & 1.153 \\
\hline
\hline
CHARM & 192 & 192 (48.0\%) & -- & -- & -- & 35.19 (1$\times$) & -- & -- \\
\Xhline{2.5\arrayrulewidth}
\end{tabular}
}
\label{tb:MaxEVA_results_int8}
\end{table*}

When further comparing the 13$\times$4$\times$6 MaxEVA solution to CHARM, we can observe from Table \ref{tb:MaxEVA_results_fp32} (row 1) that our method utilizes the AIE array slightly more (390 \textit{vs.} 384 AIE cores).
However, we use considerably fewer cores for GEMM kernels (312 \textit{vs.} 384), while
the remaining (390-312=78) AIE cores are used to run Add kernels.
Also notice that CHARM has only GEMM kernels.
Therefore, our solution is also able to achieve less AIE core power consumption (25.62 W \textit{vs.} 26.95 W), because the cores that run the fp32 Add kernels remain idle most of the time.
However, our implementation uses more memory banks than CHARM (3138 \textit{vs.} 3086 out of 3200 available), which leads to higher data memory power consumption (18.21 W \textit{vs.} 16.74 W).
When computing the total AIE power as the summation of AIE core power and data memory power \cite{AIE_XPE}, we observe that our 13$\times$4$\times$6 design exhibits slightly higher power than CHARM (43.83 W \textit{vs.} 43.69 W).
Hence, our highest throughput solution presents also 20.4\% higher energy efficiency compared to CHARM.
We note here that our method of input broadcasting and output adder tree reduction, utilizes efficiently the available PLIOs (79\% for 13$\times$4$\times$6).
On the contrary, CHARM severely under-utilizes the device's PLIOs (only 41\%), which acts as a performance bottleneck for their design.
Finally, we observe a very small DMA usage of 18 banks due to the \emph{``T"--like} shapes of pattern $P1$ (see Fig. \ref{fig:placement_patterns}), as expected.

\subsubsection{MaxEVA \textit{vs.} state-of-the-art CHARM for int8 precision}
Since int8 CHARM implementation is not open-sourced, we perform a qualitative comparison of performance.
In \cite{dac23automm} the authors report GEMM throughput of 28.15 TOPS for int8 CHARM design, when operating at 1 GHz frequency.
To fairly compare with our results, we scale the aforementioned value to 1.25 GHz (our frequency), thus becoming  35.19 TOPS.
In contrast, MaxEVA presents int8 maximum throughput of 77.01 TOPS, which is 2.19$\times$ higher than CHARM (Table \ref{tb:MaxEVA_results_int8}).
To get more confidence, we do a similar qualitative comparison for fp32 results. When scaling for fp32, we get a CHARM performance of 4342.33 GOPS at 1.25GHz.
But our experimental results in Table \ref{tb:MaxEVA_results_fp32} show a performance of 4504.46 GOPS for fp32 CHARM implementation.
This small performance difference of 3.73\% is expected because the authors in \cite{dac23automm} measure the end-to-end performance on the VCK190.
Thus, they experience sources of performance degradation, including the required zero padding \cite{dac23automm}.
However, this small difference indicates that our experiments are accurate, and also validates our $\sim$2.19$\times$ performance gain for int8 over CHARM.
This substantial performance gain is because CHARM utilizes only 192 AIE cores (48\%) for int8, due to routing congestion issues \cite{dac23automm}.
On the contrary, MaxEVA utilizes efficiently the entire AIE array, by mapping 390 cores (97.5\%) for the highest throughput design (row 1 in Table \ref{tb:MaxEVA_results_int8}).
Finally, we note that due to the absence of open-source code, power for int8 CHARM cannot be calculated through the XPE tool, thus we are not able to present energy efficiency comparison.

\subsubsection{Placement Patterns Comparison}
To provide a comprehensive evaluation of the proposed placement patterns, we show the two top-ranked solutions for each pattern in Tables \ref{tb:MaxEVA_results_fp32}, \ref{tb:MaxEVA_results_int8} (rows 1--4).
Based on the results, in general, we observe that pattern $P2$ has higher total AIE core and memory usage compared to $P1$, because it uses more Add kernels.
However, higher AIE core usage does not necessarily lead to higher core power consumption.
For instance, pattern $P2$ 10$\times$3$\times$10  design utilizes the entire AIE array (400 cores, feasible routing for pattern $P2$ since no DMA is used), but exhibits lower AIE core power than $P1$  13$\times$4$\times$6 design (25.54 W \textit{vs.} 25.62 W and 47.44 W \textit{vs.} 48.65 W for fp32 and int8, respectively).
This is attributed to the fact that $P2$ has fewer GEMM kernels than $P1$ (300 \textit{vs.} 312), and more cores that run Add kernels which remain mostly idle (100 \textit{vs.} 78).
However, when also including the memory power, the total power consumption depends on the number of memory banks used as well.
In particular, when comparing 10$\times$3$\times$10 with 13$\times$4$\times$6, the former shows slightly higher total power for fp32 (44.66 W \textit{vs.} 43.83 W), while for int8 its power is lower (65.52 W \textit{vs.} 66.83 W) than the latter.

In general, we observe from Tables \ref{tb:MaxEVA_results_fp32}, \ref{tb:MaxEVA_results_int8} that the higher the number of GEMM kernels, the higher the throughput.
However, this does not always hold true.
For instance, for int8 precision, the 10$\times$3$\times$10 design presents slightly higher throughput than 11$\times$4$\times$7 (76.08 \textit{vs.} 75.67 TOPS), despite the fact that it has fewer GEMM kernels (300 \textit{vs.} 308).
This very small performance difference ($<$1\%) is due to memory conflicts (leading to a few stalls), caused by
dissimilarities in buffer optimizations from the AIE PnR tool \cite{AI_Engine_programming_environment}.


To quantify the effect of DMA usage on GEMM performance, we also implement the highest \textit{common} solution (same number of GEMM kernels) between our two placement patterns (Tables \ref{tb:MaxEVA_results_fp32}, \ref{tb:MaxEVA_results_int8}, rows 5--6).
In particular, when comparing 12$\times$4$\times$6 ($P1$) with 12$\times$3$\times$8 ($P2$), which both have 288 GEMM kernels, we notice that throughput is higher in $P2$ for both precisions.
For instance, for int8, $P1$ attains 71.25 TOPS, while $P2$ achieves 72.93 TOPS.
This is attributed to the DMA resources used in $P1$, which increase latency compared to $P2$, where no DMA is used.
However, from Tables \ref{tb:MaxEVA_results_fp32}, \ref{tb:MaxEVA_results_int8} we observe that $P2$ has higher energy efficiency for int8 (1.153 \textit{vs.} 1.147 TOPS/W), while for fp32 the opposite occurs (123.58 \textit{vs.} 123.68 GOPS/W).
This arises from the fact that cores running Add kernels remain idle for significantly fewer cycles for int8 compared to fp32 (Table \ref{tb:single_AI_engine_results}).
To this end, we observe a higher percentage difference of total power for fp32 when comparing the aforementioned $P1$ and $P2$ solutions (40.68 W \textit{vs.} 42.28 W for fp32, and 
62.13 W \textit{vs.} 63.24 W for int8).
We notice that although in most cases $P2$ and $P1$ present higher energy efficiency for int8 and fp32, respectively, this relationship is complicated and depends on the number of GEMM kernels, the total cores used, as well as the memory banks and switch routing (as optimized by the AIE PnR tool).

Throughout all design points, 13$\times$4$\times$6 ($P1$) exhibits both highest throughput (5442.11 GOPS, 20.8\% over CHARM) and energy efficiency (124.16 GOPS/W, 20.4\% higher than CHARM), for fp32 precision.
However, for int8, 13$\times$4$\times$6 ($P1$) has the highest throughput (77.01 TOPS, 2.19$\times$ higher than CHARM), while 10$\times$3$\times$10 ($P2$) exhibits the greatest energy efficiency (1.161 TOPS/W).
Finally, all of our optimized designs present very high resource utilization, using up to 100\% AIE cores, 99.8\% AIE memory and 82.1\% PLIOs.




\begin{figure}[t]
\centering
\subfloat[fp32]{\includegraphics[width=0.48\linewidth]{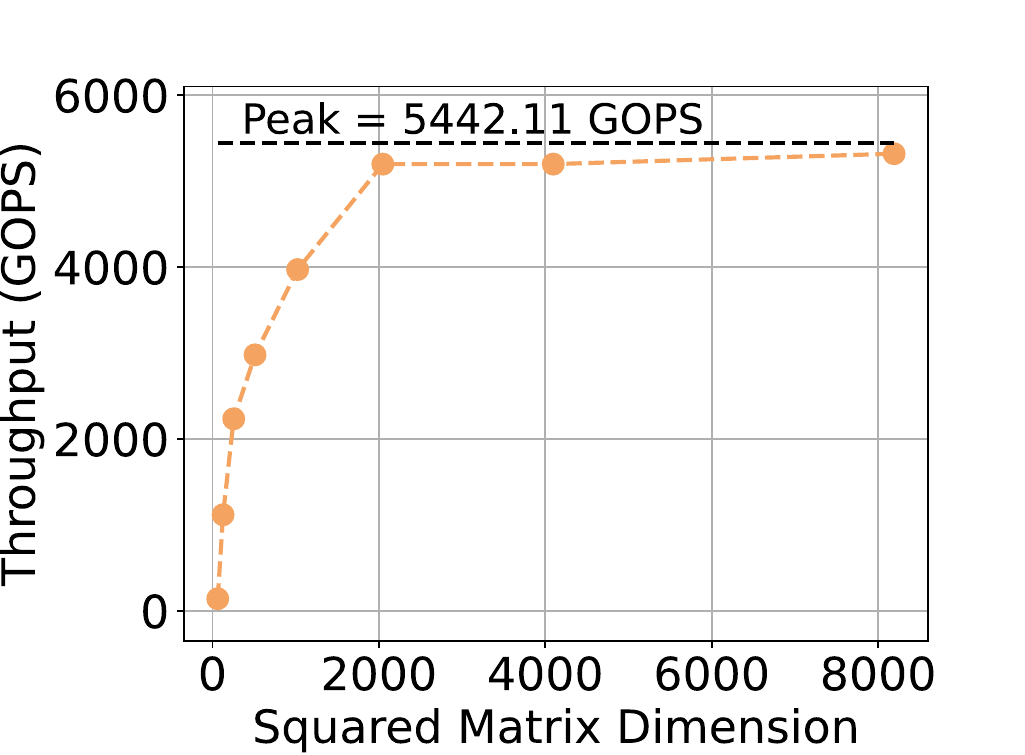}
\label{fig:throughput_mat_size_fp32}}
 \hfill 
\subfloat[int8]{\includegraphics[width=0.46\linewidth]{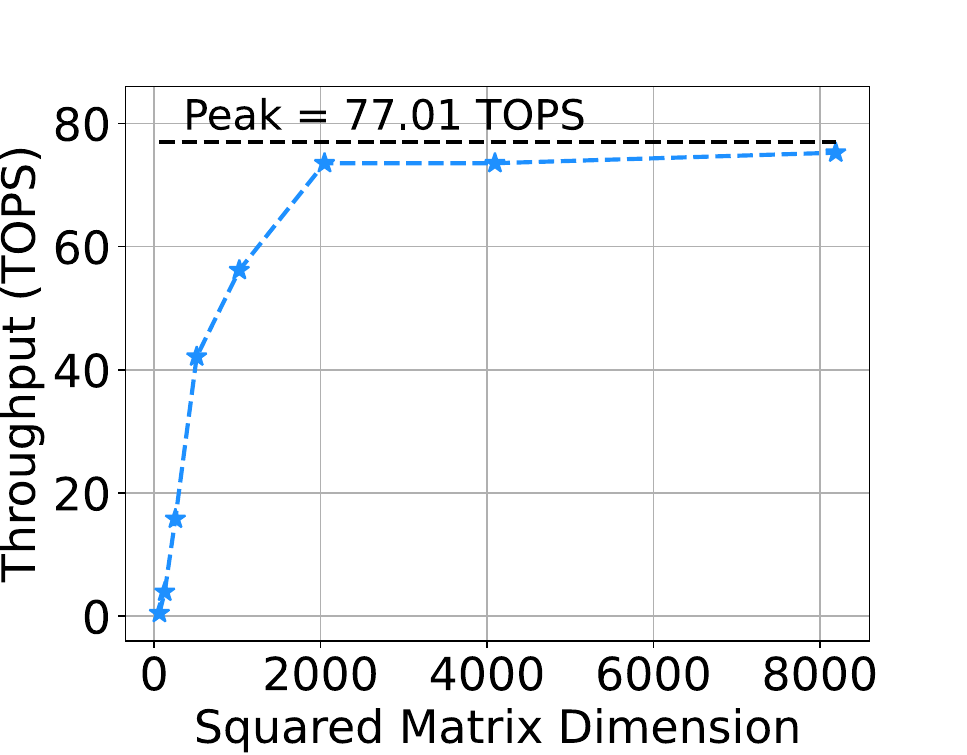}
\label{fig:throughput_mat_size_int8}}
\caption{Variation of throughput for different square matrix sizes for the 13$\times$4$\times$6 design.}
\label{fig:throughput_per_matrix_size}
\end{figure}

\subsubsection{Variation of Performance under Different Matrix Sizes}
We also explore the performance variation when altering the input matrix sizes (as powers of two) of the highest throughput design (Fig. \ref{fig:throughput_per_matrix_size}).
The throughput is estimated by supposing that tiling is performed in PL for large matrix sizes, and also the PL does not cause any stalls (commonly attained in practice \cite{dac23automm}).
As expected, we observe that as the matrix size increases, the throughput also increases, and for large enough matrices it converges to its peak value.
This is ascribed to zero padding in matrices such that they fit the native GEMM size of the 13$\times$4$\times$6 design.
In particular, the 13$\times$4$\times$6 design is able to perform a GEMM of 416$\times$128$\times$192 and 416$\times$512$\times$192 size for fp32 and int8, respectively.
To this end, we notice that for square matrices larger than $\sim$2K$\times$2K$\times$2K, less padding is needed throughout tiling, leading to almost peak performance.

Going a step further, we estimate the performance of full DNN inference, under the same assumptions as above.
More specifically, when considering the MLP used in \cite{charm2023fpga}, MaxEVA achieves a throughput of 4735.94 GOPS.
In contrast, when scaling the reported results from \cite{charm2023fpga} to 1.25 GHz, we get  3670.88 GOPS, showcasing a higher MaxEVA performance of 29\% over CHARM.

\section{Chapter Summary \& Impact}
\label{sec:MaxEVA:Sumamry}
The Versal AIE architecture introduces a new paradigm in reconfigurable computing, posing several unique design challenges. To address these challenges, we propose MaxEVA, a framework that maximizes GEMM efficiency on Versal AIE by effectively leveraging the AIE's architectural characteristics and resolving performance bottlenecks from multiple perspectives. Our experimental results demonstrate performance gains over the state-of-the-art design of up to 2.19$\times$ higher throughput and 20.4\% higher energy efficiency. MaxEVA is broadly applicable across Versal AIE platforms and GEMM-based DL workloads, providing a foundation for the extensions presented in the following chapters

\chapter{Efficient Approaches for GEMM Acceleration on
Leading AI-Optimized FPGAs}
\label{ch:Versal_vs_Stratix}

In this chapter, we present the frameworks developed for GEMM optimization on two leading AI-optimized FPGAs: the AMD Versal ACAP and the Intel Stratix 10 NX \cite{Versal_vs_Stratix_FCCM_2024}\footnote{This chapter is based on joint work published in \cite{Versal_vs_Stratix_FCCM_2024}. The dissertation author led the project and the publication; the Stratix 10 NX technical implementation was developed primarily by Dimitrios Gourounas.}. We extend the MaxEVA framework, detailed in Chapter~\ref{ch:MAXEVA}, to support arbitrary matrix sizes, and we introduce a new framework for the Stratix 10 NX device that efficiently exploits its in-fabric AI TBs. 
This work provides insights and guidelines for optimizing GEMM-based
applications on both platforms, while also delving into
their programmability trade-offs and associated challenges.

\section{Versal ACAP \& Stratix 10 NX Characteristics}
\label{sec:Versal_vs_Stratix:Hardware_Charact}

Table \ref{tb:table_hardware} demonstrates the characteristics of the two exemplar devices, where we showcase our proposed methods; the Versal VC1902 and Stratix 10 NX 2100.
Both are large chips with roughly equal number of logic elements/cells, and similar on-chip memory capacity. 
Additionally, both devices have nearly equal \textit{theoretical peak} throughput (int8) capabilities, under similar power envelopes (135 \emph{vs.} 143 TOPS, and 165 \emph{vs.} 125 W, for Versal~\cite{VCK_50000_installation} and Stratix~\cite{langhammer2021stratix}, respectively).

Besides their distinct architectures, the two devices also present differences in their DRAM technology and the manufacturing nodes (Table \ref{tb:table_hardware}).
Versal has 5$\times$ lower bandwidth (BW) than Stratix. 
Moreover, Versal is manufactured in a 7\textit{nm} TSMC process, while Stratix uses 14\textit{nm} Intel.
The main focus of this paper is to provide a comprehensive evaluation of various aspects in GEMM optimization,
emphasizing architecture-specific methodologies, which are largely agnostic to DRAM and manufacturing technology.
Thus, we perform experiments on designs that operate within on-chip memory.
However, to enable a complete and thorough analysis we extensively examine off-chip memory considerations and requirements for both devices.

\begin{table}[t]
\centering
\caption{Hardware platform characteristics.}
\resizebox{0.77\linewidth}{!}{
\begin{threeparttable}
 \centering
\setlength\tabcolsep{4.0pt}
\renewcommand{\arraystretch}{1.05}
\begin{tabular}{c|c c c}
\Xhline{2.5\arrayrulewidth}
\multicolumn{2}{c}{\textbf{Device}} & \textbf{Versal VC1902} & \textbf{Stratix 10 NX 2100}\\
\hline
\hline
\multirow{4}{*}{\rotatebox[]{90}{\textbf{FPGA}}} 
& \textbf{Logic Elements/Cells}  & 1968K & 2073K \\
& \textbf{On-chip Memory}  & 20.5 MB & 16.75 MB\\
& \textbf{DSP Slices}  & 1968 & --\\
& \textbf{Tensor Blocks}  & -- & 3960\\
\hline
\multicolumn{2}{c}{\textbf{AIE Cores}} & 400 & -- \\
\multicolumn{2}{c}{\textbf{AIE Memory}} & 12.5 MB & -- \\
\hline
\multicolumn{2}{c}{\textbf{Processing System}} & ARM A72 + R5F & -- \\
\hline
\multicolumn{2}{c}{\textbf{DRAM Technology}} & DDR4 & HBM2\\
\multicolumn{2}{c}{\textbf{Peak DRAM BW}} & 102.4 GB/s & 512 GB/s \\
\hline
\multicolumn{2}{c}{\textbf{Theor. Peak TOPS (int8)*}} & 135 TOPS & 143 TOPS\\
\multicolumn{2}{c}{\textbf{Peak Power}} & 165 W & 125 W \\
\multicolumn{2}{c}{\textbf{Process}} & 7nm TSMC & 14nm Intel \\

\Xhline{2.5\arrayrulewidth}
\end{tabular}

\begin{tablenotes}
      \small
      \item* 
      Versal's throughput is primarily attributed to the AIE, providing a peak of 128 TOPS at 1.25 GHz.
      DSPs present only 7 TOPS at 600 MHz, hence are not considered in this work.
      Stratix 10 NX peak throughput is reported in~\cite{IntelStratix10NX_WhitePaper}, for operation at 600MHz.
\end{tablenotes}


\end{threeparttable}
}
\label{tb:table_hardware}
\end{table}

\section{GEMM Design \& Optimization}
\label{sec:Versal_vs_Stratix:GEMM_Design}
In this section, we describe the design, mapping, and optimization of GEMM-based workloads on both Versal ACAP and Stratix 10 NX FPGAs.


\subsection{GEMM Implementation on Versal ACAP}
\label{subsec:GEMM_Architecture_Versal}


We leverage the MaxEVA open-source code \cite{maxeva_github}, and extend it to include on-chip buffers in the PL, tiling logic, as well as Load/Store units to communicate with DDR.
The PL is designed using Vitis HLS, as extensively used in prior works on Versal to enhance productivity
\cite{charm2023fpga, H_GCN_FPL2022, Stensil_AIE_FPGA23, CNN_acc_Versal_FPL22}.
Optimization of the PL design is attained through analytical modeling for maximization of on-chip data reuse (to reduce DDR BW requirements, as VC1902 has limited BW, Table \ref{tb:table_hardware}).
Moreover, this method effectively resolves severe memory over-utilization issues caused by Vitis HLS.
In the following sections, we provide a brief overview of the MaxEVA AIE design, and we elaborate on the PL design and optimization.

\subsubsection{GEMM Multi-Level Tiling Scheme}
\label{subsubsec:GEMM_Tiling_Versal}

\begin{figure}[tbp]
\centering
\vspace{-0.4cm}
\includegraphics[width=0.70\linewidth]{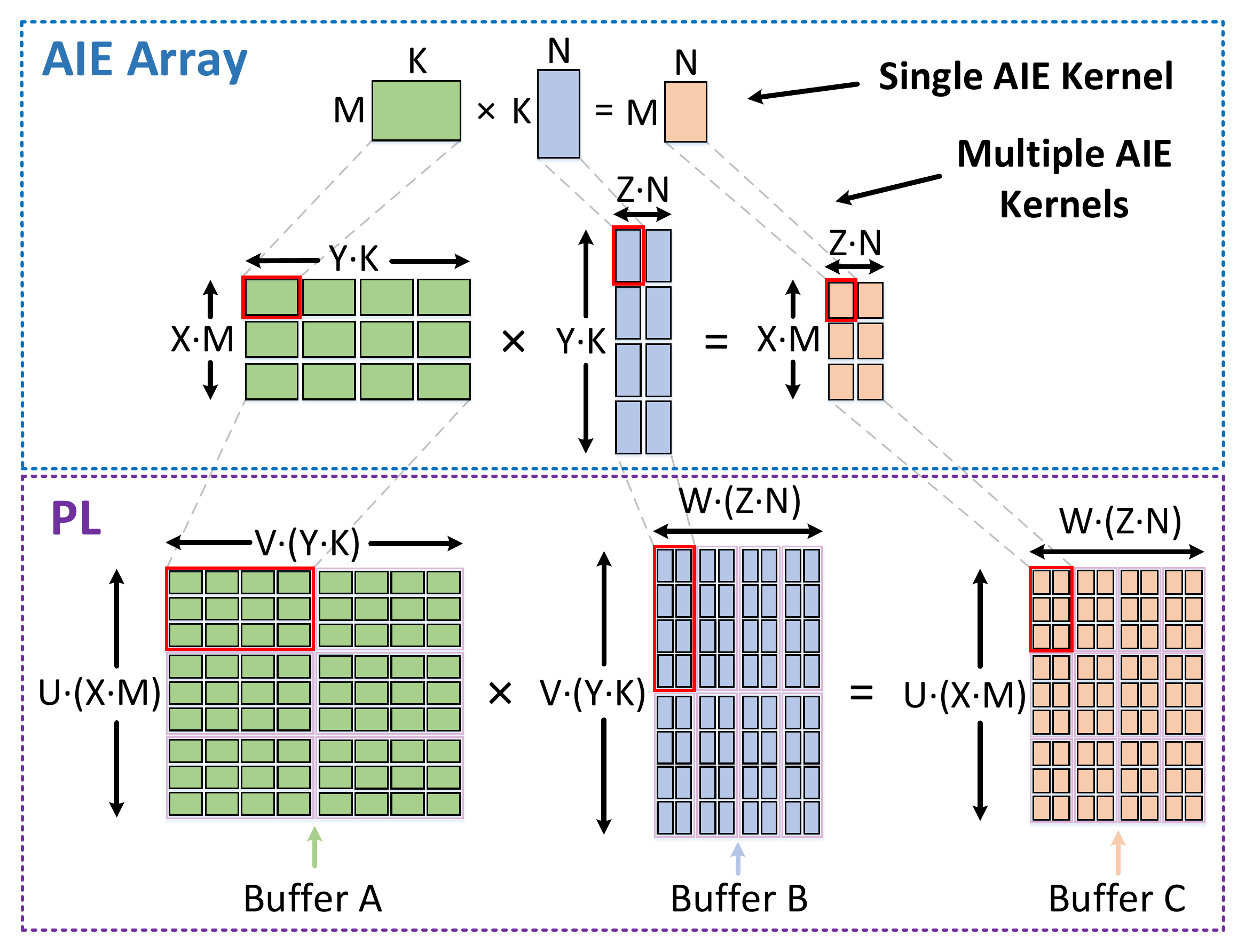}
\caption{Multi-level tiling scheme for GEMM on Versal ACAP.}
\label{fig:GEMM_tiling_Versal}
\end{figure}

Fig. \ref{fig:GEMM_tiling_Versal} illustrates the tiling scheme used in MaxEVA for the AIE, as well as our tiling method on PL.
Each AIE core executes a matrix multiplication (MatMul) kernel of $M$$\times$$K$$\times$$N$ size (\textit{first tiling level}).
The parameters $X, Y, Z$ determine the multiple MatMul kernels running on the entire AIE array, as discussed in the next section (\textit{second tiling level}).
We incorporate an additional level of buffering in the PL, introducing three new parameters: $U, V, W$ (\textit{third tiling level}).
The AIE-specific parameters ($X, Y, Z, M, K, N$) are optimized by utilizing the MaxEVA framework, while the PL-specific parameters ($U, V, W$) are optimized using our proposed PL optimization procedure (introduced in Sec. \ref{subsubsec:PL_Optimization}).
All the aforementioned parameters determine the
$A$, $B$ and $C$ matrix sizes supported out of on-chip memory in the PL, as depicted in Fig. \ref{fig:GEMM_tiling_Versal}.
Therefore, we define two GEMM sizes.
First, the \textit{compute} GEMM size, \emph{i.e.}, $(X \cdot M) \times (Y \cdot K) \times (Z \cdot N)$, running on the AIE array.
Second, the \textit{native} buffer size, \emph{i.e.}, $(U \cdot X \cdot M) \times (V \cdot Y \cdot K) \times (W \cdot Z \cdot N)$, of the data stored in the on-chip buffers inside the PL.


\subsubsection{GEMM Mapping on AIE Array}
\label{subsubsec:GEMM_Mapping_AIE}

The limited number of AIE-PL tiles on Versal devices (notably VC1902 has only 39 AIE-PL tiles out of 50 columns in the AIE \cite{Versal_AI_DC_AC_switching}), is one of the main challenges in GEMM design.
To overcome the limited PLIO bottleneck, MaxEVA utilizes the following two techniques \cite{Taka2023MaxEVA}.
First, the number of input PLIO ports is reduced by broadcasting input data to multiple AIEs.
Second, the output PLIO ports are decreased by performing adder tree reduction (via Add kernels) on the AIE.
This approach exploits only the most efficient circuit-switching AIE mechanism, as opposed to packet-switching used in \cite{charm2023fpga, charm_DAC23}.
Readers unfamiliar with MaxEVA are encouraged to read Chapter~\ref{ch:MAXEVA} first, which describes the framework in detail. In the following sections, we describe our PL implementation and optimization procedure.





\subsubsection{PL Implementation}
\label{subsubsec:PL_Implementation}
\begin{figure}[tbp]
\centering
\includegraphics[width=0.78\linewidth]{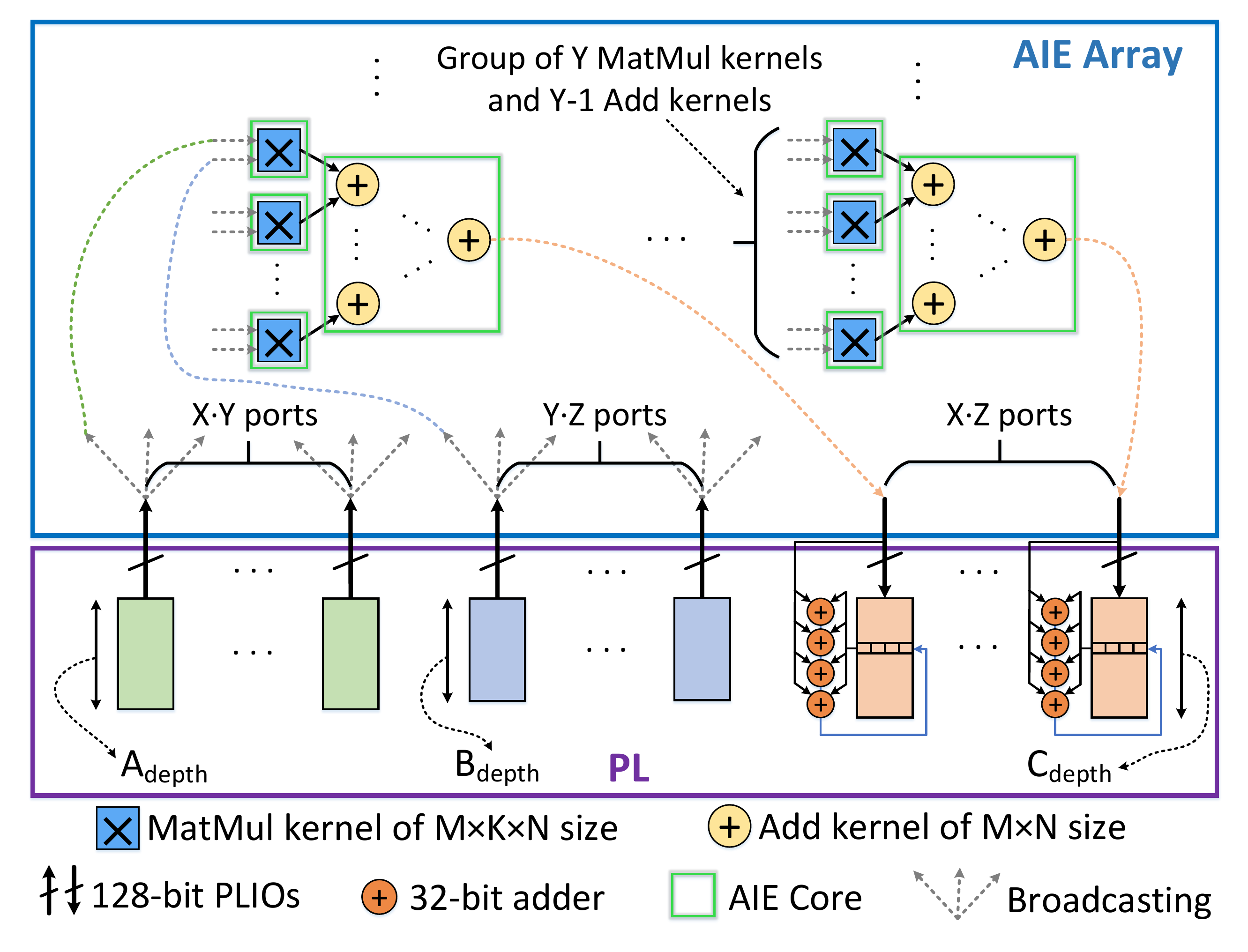}
\caption{GEMM accelerator design on Versal AIE and PL.}
\label{fig:GEMM_design_AIE_PL}
\end{figure}

The lower part of Fig. \ref{fig:GEMM_design_AIE_PL} (PL) shows a high-level block diagram of the PL design.
The 
$A$, $B$ and $C$ matrices are stored in on-chip buffers, exploiting the PL BRAM and URAM resources.
To provide sufficient bandwidth to/from the AIE, each buffer is partitioned (utilizing the $HLS$ $pragma$ $array\_partition$), to exactly match the corresponding AIE PLIO ports.
We note that although not shown in Fig. \ref{fig:GEMM_design_AIE_PL} for simplicity, we employ  double-buffering to effectively overlap GEMM computation with external off-chip DRAM communication.

We set the PLIO width to 128-bits to ensure rate matching between AIE and PL without performance loss \cite{Taka2023MaxEVA, AI_Engine_programming_guide}.
In addition, while the data type of the $A$ and $B$ buffers is int8, all accumulations are performed in 32-bits.
Thus, when sending data from the $A$ and $B$ buffers to the AIE, we concatenate 16 8-bit values for each input PLIO to form a 128-bit vector.
In contrast, when receiving data from the output PLIOs of the AIE, we pack 4 32-bit values and store them in the $C$ buffer.
The \emph{logical} size of the PL buffers is shown in Fig. \ref{fig:GEMM_tiling_Versal}.
However, in the \emph{physical} implementation, the buffers have a width of 128-bits to match the PLIO width, and are partitioned into smaller buffers
(Fig. \ref{fig:GEMM_design_AIE_PL}).
The partition factors ($\{A,B,C\}_{part}$) and depths ($\{A,B,C\}_{depth}$) of the buffers are expressed as:
\begin{align}
&A_{part} = 2 \cdot X \cdot Y, \quad A_{depth} = U \cdot V \cdot M \cdot K /16 \label{eq:A_buffer_part_depth}\\
&B_{part} = 2 \cdot Y \cdot Z, \quad B_{depth} = V \cdot W \cdot K \cdot N /16 \label{eq:B_buffer_part_depth}\\
&C_{part} = 2 \cdot X \cdot Z, \quad C_{depth} = U \cdot W \cdot M \cdot N /4 \label{eq:C_buffer_part_depth}
\end{align}
We multiply the partition factor by 2 for double-buffering. We also divide the depth by 16 and 4 to match the 128-bit packing.



Partial results from the AIE are accumulated in the PL to handle the reduction across tiles in the 
$V \cdot Y \cdot K$ dimension in Fig. \ref{fig:GEMM_tiling_Versal}, or for larger matrices if needed.
As depicted in Fig. \ref{fig:GEMM_design_AIE_PL}, we implement  4 32-bit adders in soft logic for each partitioned $C$ buffer ($4 \cdot X \cdot Z$ adders in total, all executing in parallel).
Each new partial result from every PLIO port (AXI4-Stream interface) is accumulated to its corresponding $C$ buffer address every clock cycle.
This PL logic is 
pipelined (using the $pragma$ $pipeline$) 
with an initiation interval (II) 
of 1, such that in every clock cycle a new partial result can be accumulated.
To this end, one load (for the current PLIO values) and one store operation (for the previous PLIO values) is required for every partitioned $C$ buffer.
Thus, we configure the $C$ buffers in \emph{simple} dual-port mode \cite{Versal_acap_memory_resources_manual, Vitis_HLS_guide} ($pragma$ $bind\_storage$).
This ensures a stall-free PL implementation that does not introduce any throughput degradation in the entire design.

In contrast, the input buffers $A$, $B$ are configured in single-port mode, since either only a load (send data to AIE) or store operation (receive data from DRAM) is required in each cycle for double-buffering.
Moreover, Load/Store units are implemented in the PL (not shown in Fig. \ref{fig:GEMM_design_AIE_PL}) to communicate with DRAM.
Finally, it is important to note that our implementation is symmetric in terms of the first and last dimensions in both the \textit{compute} GEMM size and the \textit{native} buffer size (refer to Sec. \ref{subsubsec:GEMM_Tiling_Versal} for definitions and Sec. \ref{subsubsec:GEMM_performance_versal} for evaluation).

\subsubsection{Memory Optimization Strategy}
\label{subsubsec:PL_Optimization}

To maximize the data reuse of the on-chip buffers, we propose an optimization methodology based on analytical modeling.
Our model utilizes the multiple configurations of BRAMs and URAMs to identify the optimal values of the $U, V, W$ parameters, as well as the buffer mapping to BRAMs and URAMs.
Although Vitis HLS includes the capability to automatically map buffers to memory resources ($impl$=$AUTO$ in $pragma$ $bind\_storage$), we found that it fails to find an operational mapping in several cases (Sec. \ref{subsubsec:PL_model_estimation}).
This automatic mapping generated by Vitis HLS leads to severe over-utilization of memory resources, and thus, failure to place and route (PnR). 
To the best of our knowledge, this HLS limitation has not been identified in any prior work. 
Several works have focused on optimizing the logical-to-physical memory mapping during the synthesis/PnR phases \cite{memory_mapping_slr, memory_mapping_old, memory_mapping_power}, while others \cite{memory_mapping_chow} propose tools to assist designers at the user-level 
with automatic memory mapping targeting BRAMs, but not URAMs.
Our approach focuses on overcoming this HLS limitation by identifying an optimal mapping to both BRAMs and URAMs, and subsequently guiding the HLS tool according to this mapping (through $impl$=$\{BRAM/URAM\}$ in $pragma$ $bind\_storage$).

The inputs of our model include a MaxEVA solution ($X, Y, Z, M, K, N$ parameters), and PL-specific parameters, \emph{i.e.}, BRAM, URAM configurations supported in Versal devices \cite{Versal_acap_memory_resources_manual} and the available on-chip memory resources.
The model produces as outputs the optimal $U, V, W$ parameters, as well as the $A$, $B$ and $C$ buffer mapping to BRAMs/URAMs.





\begin{figure}[t]
\centering
\subfloat[1K$\times$36 blocks]
{\includegraphics[width=0.48\linewidth]{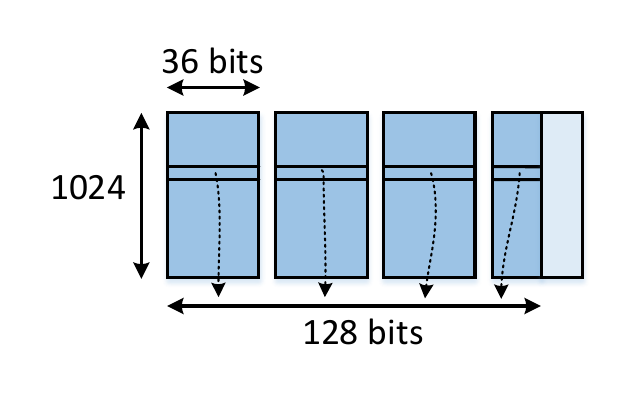}
\label{fig:1K_36}}
\subfloat[2 independent 1K$\times$18 blocks]{\includegraphics[width=0.44\linewidth]{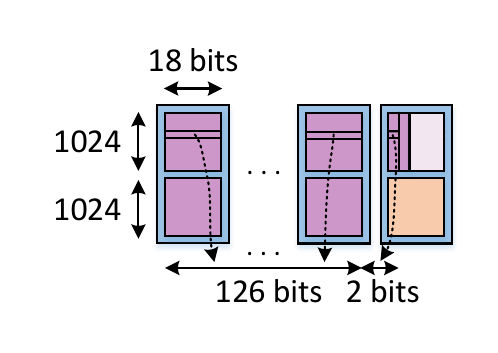}
\label{fig:2K_18}} 
\caption{BRAM configurations example and proposed modeling.} 
\label{fig:BRAM_config_examples}
\end{figure}

Besides the optimal parameter finding and mapping to memory resources, our model also estimates the BRAM/URAM utilization with 100\% accuracy in all cases (Sec. \ref{subsubsec:PL_model_estimation}).
First, we model the BRAM/URAM utilization based on the \textit{depth} of each partitioned buffer (all buffers have 128-bits width and are highly partitioned).
Fig. \ref{fig:1K_36} shows an example of our modeling when $512 < depth \le 1K$.
In this case, BRAMs are configured as 1K$\times$36 and 4 36K BRAMs are needed to construct the 128-bitwidth buffer.
Observe that a portion of the rightmost BRAM in Fig. \ref{fig:1K_36} is not utilized in this situation.
To this end, we define BRAM/URAM \textit{efficiency} as the fraction of the \textit{logical} buffer size to the total size of memory blocks used.
For instance, when assuming that all 1K entries are used in Fig. \ref{fig:1K_36}, the BRAM \textit{efficiency} is determined by the utilized width, \emph{i.e.}, 128/(36$\cdot$4) = 88.89\%.

In Fig. \ref{fig:2K_18}, we show another example when $1K < depth \le 2K$.
In this case, BRAMs are configured as 2 \textit{independent} 1K$\times$18 blocks.
With 7 2K$\times$18 blocks, 126-bits can be mapped.
The remaining 2-bits can be efficiently mapped by packing 2K$\times$2-bits on a single 1K$\times$18 block.
This is possible since 2K$\times$2-bits can be \textit{logically} viewed as 1K$\times$4-bits, where additional multiplexing logic is needed to determine the corresponding 2-bits, based on the address.
Thus, 7.5 36K BRAMs (or 15 18K blocks) are 
required in total. 
The remaining 18K BRAM (outlined in orange in Fig. \ref{fig:2K_18}), can be used for other purposes, since it is physically independent.

Similarly, we also model the cases when  $depth \le 512$ (2 36K BRAMs), and $2K < depth \le 4K$ (15 36K BRAMs) \cite{Versal_acap_memory_resources_manual}.
In addition, since URAMs support one configuration (4K$\times$72), 2 URAMs are required for $depth \le 4K$.
In eq. \ref{eq:BRAMs_per_depth_function} and \ref{eq:URAMs_per_depth_function}, we summarize the number of BRAMs ($f_{B}$) and URAMs ($f_{U}$), respectively, as functions of the buffer depth.
\begin{align}
f_{B}(depth) & =
\begin{cases}
    2, & \text{if $depth \le 512$}\\
    4, & \text{if $ 512 < depth \le 1K$}\\
    7.5, & \text{if $ 1K < depth \le 2K$}\\
    15, & \text{if $ 2K < depth \le 4K$}\\   
\end{cases}
\label{eq:BRAMs_per_depth_function}
\\[5pt]
f_{U}(depth) & = 2, \quad  \text{if $depth \le 4K$}
\label{eq:URAMs_per_depth_function}
\end{align}
Afterwards, we establish a constraint that limits each buffer's depth to 4K (eq. \ref{eq:depth_constraints}), since all buffers are highly partitioned with a relatively small depth.
This constraint allows a very high BRAM, URAM \textit{efficiency}, as shown in Sec. \ref{subsubsec:GEMM_performance_versal}. 
\begin{align}
\{A_{depth}, \ B_{depth}, \ C_{depth}\} \le 4K
\label{eq:depth_constraints}
\end{align}
Finally, we impose constraints to ensure that the total number of utilized BRAMs and URAMs does not exceed the device's available resources ($B_{36K}$ and $U_{288K}$).
Eq. \ref{eq:BRAMs_constraints_ex} and \ref{eq:URAMs_constraints_ex} show an example of the constraints for BRAMs and URAMs, respectively. 
In this example, buffers $A$, $B$ have been mapped to BRAMs, while buffer $C$ is mapped to URAMs.
\begin{gather}
A_{part} \cdot f_{B}(A_{depth}) +  B_{part} \cdot f_{B}(B_{depth}) \le B_{36K} \label{eq:BRAMs_constraints_ex}\\
C_{part} \cdot f_{U}(C_{depth}) \le U_{288K}
\label{eq:URAMs_constraints_ex}
\end{gather}
\noindent Constraints similar to  \ref{eq:BRAMs_constraints_ex}, \ref{eq:URAMs_constraints_ex} are applied for \textit{all} permutations of mapping buffers $A$, $B$ and $C$ to BRAMs and URAMs.

The solution of $U, V, W$ and buffer mapping to BRAMs, URAMs can be formulated as an integer programming (IP) optimization problem with the aforementioned constraints.
We solve the IP exhaustively by setting the maximization of on-chip data reuse as the objective.
The data reuse of all buffers is encapsulated in the product of $U$$\cdot$$V$$\cdot$$W$.
Notice from Fig. \ref{fig:GEMM_tiling_Versal} that buffers $A$ and $B$ are reused $W$ and $U$ times in GEMM, respectively, while $C$ is reused and updated (during accumulation) $V$ times.
In addition, note that maximizing data reuse also leads to maximization of the PL buffer sizes, under the resource constraints.
We report, implement and explore the trade-offs of multiple top-ranked solutions in Sec. \ref{subsec:Versal_results}.



\subsection{GEMM Implementation on Stratix 10 NX}
\label{subsec:GEMM_Architecture_Stratix}

We implement a configurable MatMul accelerator consisting of a control logic and a spatial 2D layout on the Stratix 10 NX architecture. 
The accelerator's TBs operate out of local, on-chip memory consisting of $A$, $B$ and $C$ buffers (Fig.~\ref{fig_tensor_block_diagram}) of size $M^\prime$$\cdot$$K^\prime$, $K^\prime$$\cdot$$N^\prime$ and $M^\prime$$\cdot$$N^\prime$, respectively.
The 
$M^\prime$$\times$$K^\prime$$\times$$N^\prime$ MatMul size is defined as the \textit{native} buffer size, similar to Versal. 
We conduct a DSE to optimize for accelerator throughput and employ analytical modeling to optimize for on-chip data reuse.
Below we delineate the TB layout, the dataflow, the memory architecture, optimization strategies and the automatic code generation tool we developed.

\subsubsection{TB Layout}
\label{subsubsec:Stratix_TB_layout}
\begin{figure}[!t]
\centering
\vspace{-0.20cm}
\hspace{-0.40cm}
\includegraphics[width=0.80\linewidth]{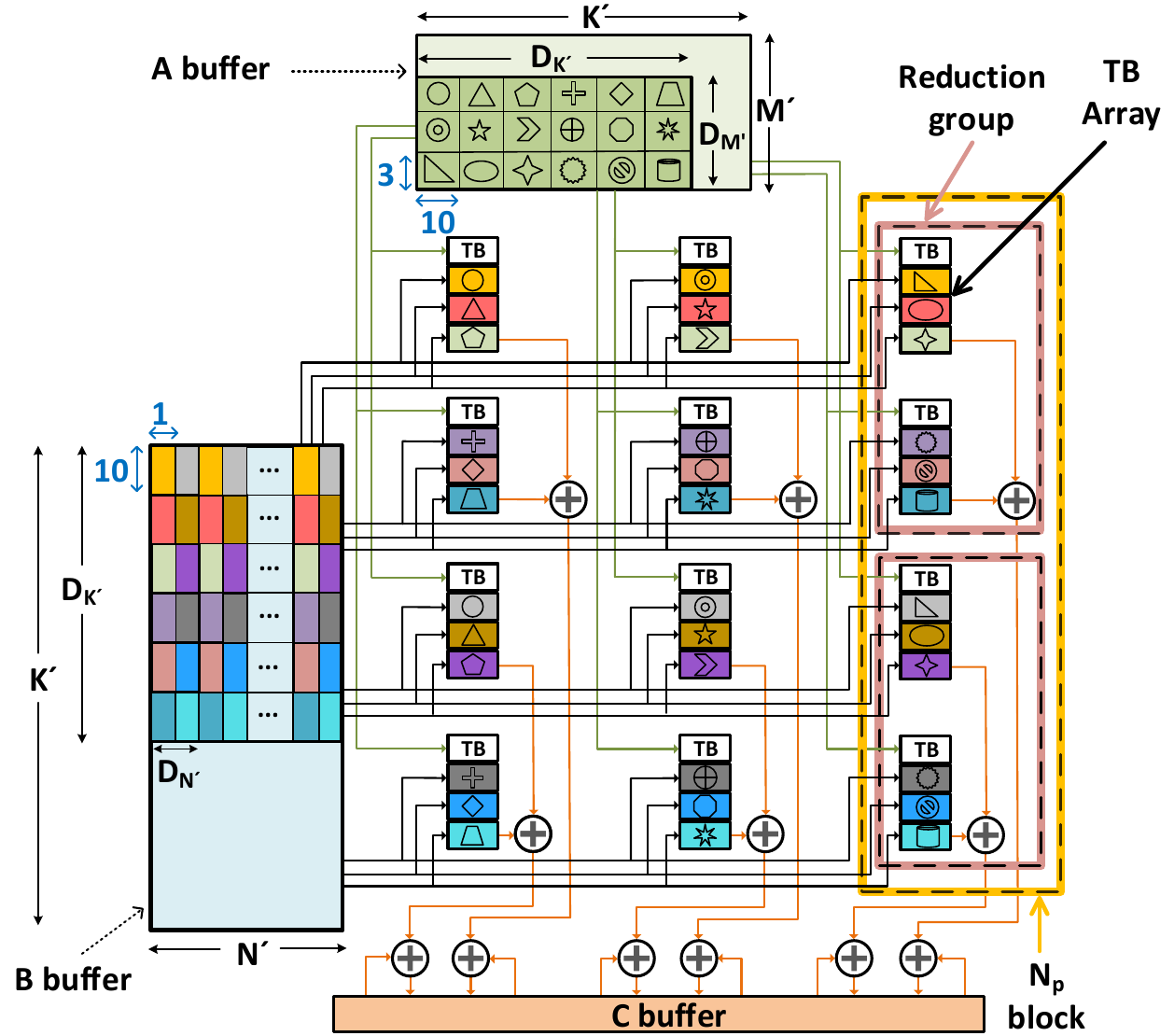}
\caption{2D TB layout and GEMM dataflow on Stratix 10 NX.}
\label{fig_tb_layout}
\end{figure}

We utilize groups of TB \textit{arrays} of configurable length that run in parallel and operate in the cascade loading mode.
An \textit{array's} first TB (colored white in Fig.~\ref{fig_tb_layout}) serves as its point of entry for $A$ data (Fig.~\ref{fig_tb_layout}), performing no computation. 
Subsequent TBs in the \textit{array} receive their $A$ data (and store them in their registers) via their $casc\_data\_in$ port and their $B$ data through their $data\_in$ port (Fig.~\ref{fig_tensor_block_diagram}). The latency of loading $A$ blocks is three cycles per TB, but can be hidden when overlapped with dot-product operations. While $bank$ $0$ (Fig.~\ref{fig_tensor_block_diagram}) is being multiplied with a set of $B$ blocks, $bank$ $1$ can be loaded and vice-versa. An \textit{array's} accumulated outputs are exposed through the $data\_out$ port of its final TB.

\noindent We define four architecture parameters for the TB layout:

\vspace{-5pt}

\begin{itemize}
\item \textbf{Parameter \arraylength}: Length of a TB \textit{array}, equal to four in Fig.~\ref{fig_tb_layout}. 
This length can be less than or equal to 36.

\item \textbf{Parameter \Kexpansion}: Set of \textit{arrays} that work in parallel across the $K^\prime$ dimension, which we refer to as a \textit{reduction group} (\Kexpansion[] equal to two in Fig.~\ref{fig_tb_layout}). The $data\_out$ outputs of all the \textit{arrays} in a \textit{reduction group} are fed into its corresponding adder tree, the output of which is accumulated at the $C$ buffer (Fig.~\ref{fig_tb_layout}). These adders are implemented in soft logic. 

\item \textbf{Parameter \Nexpansion}: Set of \textit{reduction groups} that contain the same $A$ blocks, but get multiplied with different $B$ blocks (equal to two in Fig.~\ref{fig_tb_layout}). 
\Nexpansion[] allows exploiting parallelism across $N^\prime$. We refer to this set of \textit{reduction groups} as an \Nexpansion[] \textit{block}.

\item \textbf{Parameter \Mexpansion}: Number of \Nexpansion[] \textit{blocks} that allow parallelism across $M^\prime$, equal to three in Fig. \ref{fig_tb_layout}. Each \Nexpansion[] \textit{block} uses different $A$ blocks, but the same $B$ blocks.
\end{itemize}

\subsubsection{Dataflow}
\label{subsubsec:Stratix_TB_dataflow}
Each TB holds a 3$\times$10 block of $A$ in its registers, represented by a shape in Fig.~\ref{fig_tb_layout}, and it multiplies it with a 10$\times$1 block of $B$ in each clock cycle. This $B$ block belongs to a set of $N^\prime/\Nexpansion$ 10$\times$1 blocks, represented by a color in the $B$ Buffer (Fig.~\ref{fig_tb_layout}). The TB will sequentially multiply its $A$ block with all $B$ blocks in the colored set. Given enough $B$ blocks in this set (\emph{i.e.}, a large enough $N^\prime$), the TB's register loading latency can be fully hidden.
Different TBs of the same \textit{array} hold separate $A$ blocks (shapes) and process different sets of $B$ blocks (colors). Partial dot-products are propagated along an \textit{array} through cascade connections and added to the next dot-products on-the-fly. This process takes two clock cycles per TB (Sec.~\ref{subsec:Stratix_10_Nx_architecture}). Therefore, a TB starts processing data two cycles after its previous TB. Finally, an \textit{array's} accumulated dot-products are exposed at its last TB.

Different \textit{arrays} in the same \textit{reduction group} hold separate blocks of $A$ (shapes) and process separate sets of $B$ blocks (colors) in parallel, each generating a new output every clock cycle. 
\textit{Reduction groups} of the same \Nexpansion[] \textit{block} hold the same $A$ blocks (shapes), but get multiplied with different sets of $B$ blocks (colors) in parallel, allowing parallelism across $N^\prime$. Finally, different \Nexpansion[] \textit{blocks} contain different sets of $A$ blocks (shapes), but they all get multiplied with the same $B$ blocks (colors) in parallel, allowing parallelism across $M^\prime$.

The total number of utilized TBs is equal to 
$\arraylength \cdot \Kexpansion \cdot \Nexpansion \cdot \Mexpansion$.
We define the \textit{compute} GEMM size as $(\Mexpansion\cdot3)$$\times$$([\arraylength-1]\cdot\Kexpansion\cdot10)$$\times$$(\Nexpansion)$. 
This is the size of the blocks the accelerator processes at the lowest memory hierarchy level. 
We refer to this GEMM size as $D_{M^\prime}$$\times$$D_{K^\prime}$$\times$$D_{N^\prime}$.
The multiplication by 3 and 10 accounts for the size of the $A$, $B$ blocks. We subtract by 1 for the wasted TB of each \textit{array} and we multiply by \Kexpansion[] for all \textit{arrays} in a \textit{reduction group}. 
Due to tiling over the \textit{compute} GEMM size, $M^\prime$, $K^\prime$ and $N^\prime$ need to be multiples of $D_{M^\prime}$, $D_{K^\prime}$ and $D_{N^\prime}$, respectively. 
Moreover, $N^\prime$ must be sufficiently large to allow hiding the TB register loading latency.

\subsubsection{Architectural Considerations}
\label{subsubsec:Stratix10_Architectural_considerations}
Below we elaborate on architectural considerations related to 
the
four architecture parameters, along with the constraints they introduce.

\textbf{Parameter \arraylength[]}: As mentioned, the TB \textit{chain} granularity is equal to 36. 
Hence, as multiple \textit{arrays} may fit in a chain, the \textit{array} length \arraylength[] must be a factor 
of 36 to avoid fragmentation during placement of a design with a large number
of TBs. 
Sound values for \arraylength[] are 36, 18, 12 and 9. Low \arraylength[] leads to more wasted TBs (first TB in every \textit{array}).
Meanwhile, this increases flexibility in executing different matrix sizes, as they do not require a high 
reduction dimension ($K^\prime$). Large \arraylength[] increases latency for loading $A$, but also results in higher peak throughput.  
A \arraylength[] of 36 has a loading latency of 36$\cdot$3 = 108 cycles. However, because each $A$ block is multiplied with a set of $B$ blocks, if this set contains at least 108 blocks, depending on $N^\prime$, 
this latency can be hidden. 
    
\textbf{Parameter \Kexpansion}: A TB \textit{array} processes $(\arraylength$$-$$1)$$\cdot$$10$ elements across the reduction dimension ($K^\prime$) each cycle. \Kexpansion[] increases the parallelism across $K^\prime$ (\emph{i.e.}, $D_{K^\prime}$), while keeping the signal fan-out low. 
However, if $D_{K^\prime}$ is large, but $K^\prime$ is small, some TBs in a \textit{reduction group} will be under-utilized.
    
\textbf{Parameter \Nexpansion}: 
In order to utilize more TBs, parallelism across $N^\prime$ and $M^\prime$ is also essential. \Nexpansion[] requires broadcasting $A$ blocks across all \textit{reduction groups} in an \Nexpansion[] \textit{block}, while \Mexpansion[] requires broadcasting $B$ blocks across all \Nexpansion[] \textit{blocks} (Fig.~\ref{fig_tb_layout}). However, each \textit{array} has only one input for $A$ data (at its initial TB), but a considerably larger amount of inputs for $B$ data (equal to $\arraylength$$-$$1$). As a result, \Mexpansion[] increases the fanout of a lot more signals than \Nexpansion[] does, which increases FPGA routing congestion.
Finally, a higher \Nexpansion[] leads to a larger minimum required value of $N^\prime$ to hide the TB loading latency.
    
\textbf{Parameter \Mexpansion}: 
\Mexpansion[] enables parallelism across $M^\prime$, but also increases the fan-out of $B$ blocks, which, as mentioned, can lead to significant routing congestion in the FPGA, adversely affecting the attainable clock frequency and performance.

\subsubsection{Memory Architecture}
\label{subsubsec:Stratix_Memory}
Evidently,
when a large number of TBs are used, high BRAM bandwidth is needed to feed them blocks of $A$ and $B$. 
We employ a memory architecture that is designed to (i) enable high compute throughput and (ii) to maximize 
data reuse, which will amortize 
off-chip BW requirements. 
In a similar manner to Versal, we maintain separate buffers for $A$, $B$ and $C$ and we utilize double-buffering. 
The Load/Store units perform read-only operations on $C$ buffers and write-only operations on $A$ and $B$ buffers. 
The compute logic performs read and write operations on $C$ and read-only operations on $A$ and $B$. 
M20K blocks configured in \textit{simple} dual-port mode are used to implement all buffers. For $A$ and $B$, the compute and Load/Store units have their own port for reading and writing, respectively. For double-buffering, we double the depth of $A$ and $B$, and split $C$ to two equal-sized buffers.
A more detailed description of the architecture of buffers $A$, $B$ and $C$ is provided below.


\textbf{Buffers $A$, $B$}: As illustrated in Fig.~\ref{fig_tb_layout}, a separate $A$ block must be fed to each \textit{array} of a \textit{reduction group} and across different \Nexpansion[] \textit{blocks}. Additionally, each TB in an \textit{array} (except for $TB_0$), receives separate $B$ blocks every cycle. Similarly, all \textit{arrays} in all \textit{reduction groups} of an \Nexpansion[] \textit{block}  require different blocks of $B$ each clock cycle. Also, the bitwidth of the $A$ and $B$ ports is 80-bits. Therefore, the partition factors $A_{part}$, $B_{part}$ of $A$, $B$ and the depth $A_{depth}$, $B_{depth}$ of each smaller partitioned buffer of $A$, $B$ are:
\begin{gather} 
B_{part}=(\arraylength-1) \cdot \Kexpansion \cdot \Nexpansion \label{eq:NX_Bpart}\\
B_{depth} = 2\cdot K^\prime\cdot N^\prime/(B_{part}\cdot10)
\label{eq:NX_Bdepth}\\
A_{part}= \Mexpansion \cdot \Kexpansion, \ \ A_{depth} = 2\cdot M^\prime\cdot K^\prime/(A_{part}\cdot10)
\label{eq:NX_Apart_Adepth}
\end{gather}
We multiply by 2 for double-buffering and we also multiply by 10, because 80-bits is 10 bytes. 
The 80-bit width of these buffers is implemented using M20Ks operating in the 512$\times$40, 1024$\times$20 or 2048$\times$10 configurations~\cite{Intel_M20K_guide}. We model the number of M20Ks required to implement each buffer, $f_{M_{80}}$, as a function of the depth. For the $A$ buffer, since $A_{depth}$ is sufficiently large, all M20K configurations lead to equivalent, accurate solutions in our model. However, $B_{depth}$ was always less than 1024. If $depth \leq 512$, two M20Ks in 512$\times$40 mode can be used. If $512 < depth \leq 1024$, both 512$\times$40 and 1024$\times$20 configurations lead to a usage of four M20Ks. As a result, in all our designs $f_{M_{80}}$ is estimated as:
\begin{equation} 
f_{M_{80}} (depth) = 2 \cdot \lceil depth/512 \rceil 
\label{eq:NX_fM40}
\end{equation}
\noindent The symbol $\lceil \ \rceil$ denotes rounding up to the next integer.

\textbf{Buffer $C$}: A TB has three $data\_out$ ports, so \textit{reduction groups} generate three distinct $C$ values each.
The partition factor $C_{part}$ and the depth $C_{depth}$ of each smaller buffer is:
\begin{equation}
    C_{part}= \Mexpansion \cdot \Nexpansion \cdot 3 \cdot 2, \ \ C_{depth} = M^\prime\cdot N^\prime\cdot2/C_{part}
    \label{eq:NX_Cpart_Cdepth}
\end{equation}
\noindent We multiply by 2 for double-buffering. 
The valid M20K configurations for the 32-bitwidth of the $C$ buffers are 2048$\times$8, 1024$\times$16 or 512$\times$32.
The number of M20Ks, $f_{M_{32}}$, required for $C$ buffers is also modeled as a function of the depth. Since the depth of the $C$ buffers is sufficiently large, all above M20K configurations led to equivalent, accurate solutions in our model, estimated by:
\begin{equation} 
f_{M_{32}} (depth) = \lceil depth/512 \rceil 
\label{eq:NX_fM32}\\
\end{equation}
\subsubsection{Optimization Strategies}
\label{subsubsec:Stratix_Optimization}

Similar to Versal, we formulate the selection of $M^\prime$, $K^\prime$, $N^\prime$ for a given TB configuration as an IP problem and solve it exhaustively to optimize for data reuse, by maximizing the product $M^\prime$$\cdot$$K^\prime$$\cdot$$N^\prime$.
We impose two constraints on the IP. First, the utilized M20Ks must not exceed the device's available M20K blocks ($B_{M20K}$). Second, $N^\prime$ is set to be sufficiently large to hide TB register loading latency:
\begin{equation}
    A_{part} \cdot f_{M_{80}}(A_{depth}) + B_{part} \cdot f_{M_{80}}(B_{depth}) + C_{part} \cdot f_{M_{32}}(C_{depth}) \leq B_{M20K}
\label{eq:NX_M20K_constraint}
\end{equation}
\begin{equation}
    N^\prime \geq \arraylength \cdot 3 \cdot \Nexpansion[]
    \label{eq:Nprime_constraint} 
\end{equation}
Since throughput is directly related to operating frequency, we implement several optimizations to shorten critical paths. These include replication of control logic and insertion of a configurable number of pipeline stages along the data and address datapaths.
Additionally, we conduct an extensive exploration on various TB architecture parameters, to find configurations that maximize frequency, and thus, throughput (see Sec.~\ref{subsec:Stratix_results}).

In an effort to reduce signal fan-out, we also tried inserting registers between \Nexpansion[]{} \textit{blocks} in a systolic fashion in order to propagate $B$ blocks, thus minimizing broadcasting. However, our experiments showed no frequency improvements, while greatly increasing ALM usage (up to $\sim$40\%, as opposed to a maximum of $\sim$19\% with the proposed design, see Table~\ref{tab_NX_results}). Such a large number of soft-logic pipeline registers is undesirable, as it can reduce energy efficiency and also render logic resources unavailable, as mentioned in~\cite{langhammer2021stratix}. Thus, the systolic distribution approach was not considered in our final designs.



\subsubsection{Automatic RTL Code Generation}
\label{subsubsec:Stratix_Design_Generation}
We develop a Python-based tool that automatically generates the aforementioned architecture's RTL code, including the control logic, TB layout and memory. The input configurations of the tool are the four architecture parameters, the 
$M^\prime$, $K^\prime$, $N^\prime$  dimensions, 
and the number of pipeline stages for both address and data.

\section{Evaluation}
\label{sec:Versal_vs_Stratix:Evaluation}
In this section, we discuss and compare various aspects of GEMM design and optimization on both Versal and Stratix devices.

\subsection{Versal DSE Evaluation}
\label{subsec:Versal_results}

For the Versal AIE, we obtain the two most efficient solutions from MaxEVA.
As found in \cite{Taka2023MaxEVA}, solution $P1$ 13$\times$4$\times$6 ($X$$\times$$Y$$\times$$Z$) shows the highest throughput, while the $P2$ 10$\times$3$\times$10 presents the highest energy efficiency.
Both use a 32$\times$128$\times$32 ($M$$\times$$K$$\times$$N$) single AIE MatMul kernel with 95\% throughput efficiency.
For the PL, we utilize our model (Sec. \ref{subsubsec:PL_Optimization}), which takes as input a MaxEVA solution ($X, Y, Z, M, K, N$ parameters) and produces the optimal values of $U, V, W$.
We perform DSE on the 5 top-ranked $U$$\times$$V$$\times$$W$ solutions, for each of the two MaxEVA solutions (10 designs in total).

We use the AMD/Xilinx Vitis 2022.1 version to implement and compile our designs.
The PL part is designed using Vitis HLS, while AIE-PL linking is achieved via the V++ compiler.
Throughout all experiments, the AIE frequency is set to its maximum value of 1.25 GHz, while the PL frequency ranges from 275--300 MHz, depending on the PL configuration.
To calculate the throughput of our designs, we use hardware emulation in Vitis, while
power is estimated through the post-implementation Vivado Power Analysis Tool \cite{Vivado_power_analysis}.

\subsubsection{Model Estimation}
\label{subsubsec:PL_model_estimation}

In Table \ref{tb:HLS_results}, we present 4 top-ranked solutions of our PL optimization procedure (parameters $U$$\times$$V$$\times$$W$ and buffer mapping to BRAM/URAM resources).
First, we observe that our model estimates the BRAM/URAM utilization with 100\% accuracy in all cases.
For example, for solution 4$\times$2$\times$4 ($P1$), the model suggests (``Model Est." column) that buffers $\{A, B, C\}$ should be mapped to $\{\mathcal{B, U, U}\}$, where $\mathcal{B}$, $\mathcal{U}$ denote BRAM, URAM, respectively.
When we guide the HLS tool to use this mapping, 
both our model and HLS synthesis report identical BRAM/URAM utilization, \emph{i.e.}, 780 (81\%) / 408 (88\%).
Letting the HLS tool instead to automatically map buffers (``HLS AUTO" column), results in a severe over-utilization of URAMs (616 or 133\%), without any BRAM usage.
Vivado PnR attempts to implement this HLS AUTO solution by mapping the surplus URAMs to BRAMs.
However, it generates an error reporting a 119.4\% BRAM and 99.8\% URAM utilization.
A similar result is observed for another solution, \emph{i.e.}, 4$\times$2$\times$4 ($P2$).
However, for the other two solutions shown in Table \ref{tb:HLS_results}, HLS AUTO is able to successfully find an efficient mapping.
In these cases, both our model and HLS AUTO produce exactly the same BRAM/URAM utilization, \emph{e.g.}, 416 (43\%) / 408 (88\%), for 2$\times$2$\times$8 ($P1$).
We note that HLS AUTO fails to implement 5 of the 10 top solutions (Table \ref{tb:Versal_top10_designs}), justifying the necessity of our approach.

\begin{table}[t]
 \centering
\caption{Optimization model estimation for various solutions and comparison with HLS AUTO mapping.}
\setlength\tabcolsep{2.6pt}
\renewcommand{\arraystretch}{1.00}
\resizebox{0.82\textwidth}{!}{
\begin{tabular}{c|ccc|cc}
\Xhline{2.5\arrayrulewidth}

\textbf{U$\times$V$\times$W}  &  
\multicolumn{3}{c|}{\textbf{Model Estimation}} & \multicolumn{2}{c}{\textbf{HLS AUTO}}\\

\cline{2-4} \cline{5-6}
\textbf{(MaxEVA P.)} & \textbf{\textbf{\{A, B, C\}}} & \textbf{BRAMs} & \textbf{URAMs} & \textbf{BRAMs} & \textbf{URAMs}\\
\hline
\hline
4$\times$2$\times$4 (P1) & $\{\mathcal{B, U, U}\}$ & 780 (81\%) & 408 (88\%) & \textbf{0 (0\%)} & \textbf{616 (133\%)}\\

4$\times$2$\times$4 (P2) & $\{\mathcal{B, B, U}\}$ & 900 (93\%) & 400 (86\%) & \textbf{0 (0\%)} & \textbf{640 (138\%)}\\

2$\times$2$\times$8 (P1) & $\{\mathcal{B, U, U}\}$ & 416 (43\%) & 408 (88\%) & 416 (43\%) & 408 (88\%)\\

2$\times$8$\times$2 (P2) & $\{\mathcal{U, U, B\}}$ & 800 (83\%) & 240 (52\%) & 800 (83\%) & 240 (52\%)\\

\Xhline{2.5\arrayrulewidth}

\end{tabular}
}

\label{tb:HLS_results}
\end{table}


\subsubsection{GEMM Performance}
\label{subsubsec:GEMM_performance_versal}
In Table \ref{tb:Versal_top10_designs}, we show various metrics for our 10 top solutions.
In all solutions, throughput is calculated on their \textit{native} buffer sizes 
(Sec. \ref{subsubsec:GEMM_Tiling_Versal})
Overall, we observe that all designs exhibit high throughput ranging from 75.4--77.01 TOPS.
However, to maintain such high throughput, several designs require higher DRAM BW compared to the VC1902's BW (102.4 GB/s).
We note that we calculate DRAM BW
as the worst-case of concurrent loads for buffers $A$, $B$ and stores for $C$ (all as 8-bits due to quantization in DL).
With more sophisticated data reuse techniques as in \cite{XVDPU_AIE_FPL22}, this requirement can be amortized. 
However, in this work, we consider the worst-case BW scenario for the sake of generality.
Therefore, we only examine designs that stay within the BW of the VC1902 device (highlighted in bold in Table \ref{tb:Versal_top10_designs}).

From these designs, 2$\times$2$\times$8 ($P1$) shows both the highest throughput, \emph{i.e.}, 76.93 TOPS, and the best energy efficiency, \emph{i.e.}, 0.938 TOPS/W.
For all valid solutions, throughput ranges from 75.40--76.93 TOPS, which is 58.9--60.1\% of the \textit{theoretical peak} throughput of VC1902, and the same as the state-of-the-art MaxEVA~\cite{Taka2023MaxEVA}.
Additionally, energy efficiency ranges from 0.911--0.938 TOPS/W.
In all cases, we notice a very high resource utilization, up to 94\% BRAMs, 88\% URAMs, and 100\% AIE cores, with a small LUT usage of up to 11\%.
Moreover, we observe high RAM \textit{efficiency} of 75.7--90.2\%, as a direct result of our modeling and optimization methodology.
Finally, although not shown in Table \ref{tb:Versal_top10_designs} for brevity, the swapping of the first and last GEMM dimensions results in equivalent solutions (symmetrical design, see Sec. \ref{subsubsec:PL_Implementation}).
For instance, solution 2$\times$2$\times$8 ($P1$) can be swapped to 8$\times$2$\times$2 ($P1$), resulting in a \textit{native} buffer size of 1536$\times$1024$\times$832 (\textit{compute} GEMM size becomes 192$\times$512$\times$416 in this case).

\begin{table*}[t]
\centering
\caption{Evaluation of 10 top-ranked GEMM designs on Versal VC1902.}
\renewcommand{\arraystretch}{1.00}

\resizebox{1.00\textwidth}{!}{
\begin{tabular}{c|c|c|c|c|c|c|c|c}
\Xhline{2.5\arrayrulewidth}

\textbf{U$\times$V$\times$W} &
\textbf{LUTs} &
\textbf{BRAMs} &
\textbf{URAMs} &
\textbf{AIE} &
\textbf{PL Fq.} &
\textbf{Thrpt.} &
\textbf{Power} &
\textbf{En. Eff.} \\

\textbf{(MaxEVA P.)} &
 &
 &
 &
\textbf{cores} &
\textbf{(MHz)} &
\textbf{(TOPS)} &
\textbf{(W)} &
\textbf{(TOPS/W)} \\

\hline
\hline

2$\times$8$\times$2 (P1) & 85K (9\%) & 630 (65\%) & 304 (66\%) & 390 (98\%) & 300 & 77.01 & 78.6 & 0.980 \\

\textbf{2$\times$2$\times$8 (P1)} & \textbf{91K (10\%)} & \textbf{422 (44\%)} & \textbf{408 (88\%)} & \textbf{390 (98\%)} & \textbf{290} & \textbf{76.93} & \textbf{82.0} & \textbf{0.938} \\

\textbf{3$\times$2$\times$5 (P1)} & \textbf{94K (10\%)} & \textbf{792 (82\%)} & \textbf{408 (88\%)} & \textbf{390 (98\%)} & \textbf{278} & \textbf{76.72} & \textbf{82.7} & \textbf{0.932} \\

\textbf{4$\times$2$\times$4 (P1)} & \textbf{90K (10\%)} & \textbf{792 (82\%)} & \textbf{408 (88\%)} & \textbf{390 (98\%)} & \textbf{278} & \textbf{76.72} & \textbf{82.3} & \textbf{0.928} \\ 

2$\times$4$\times$4 (P1) & 97K (11\%) & 792 (82\%) & 408 (88\%) & 390 (98\%) & 278 & 76.72 & 82.8 & 0.927 \\

\hline

2$\times$8$\times$2 (P2) & 92K (10\%) & 806 (83\%) & 240 (52\%) & 400 (100\%) & 300 & 76.08 & 78.3 & 0.971 \\

2$\times$7$\times$2 (P2) & 92K (10\%) & 806 (83\%) & 240 (52\%) & 400 (100\%) & 300 & 76.08 & 77.8 & 0.977 \\

2$\times$6$\times$2 (P2) & 91K (10\%) & 806 (83\%) & 240 (52\%) & 400 (100\%) & 300 & 76.08 & 77.5 & 0.982 \\

\textbf{4$\times$2$\times$4 (P2)} & \textbf{100K (11\%)} & \textbf{912 (94\%)} & \textbf{400 (86\%)} & \textbf{400 (100\%)} & \textbf{275} & \textbf{75.40} & \textbf{82.8} & \textbf{0.911} \\

4$\times$2$\times$3 (P2) & 100K (11\%) & 912 (94\%) & 400 (86\%) & 400 (100\%) & 275 & 75.40 & 82.0 & 0.919 \\

\Xhline{2.5\arrayrulewidth}

\end{tabular}
}

\label{tb:Versal_top10_designs}
\end{table*}

\subsection{Stratix DSE Evaluation}
\label{subsec:Stratix_results}


We conducted a DSE on 100 designs with different TB architecture parameters 
to optimize for performance and energy efficiency. The TB utilization of all explored designs is 85\%--91\%. The \textit{native} buffer sizes (Sec. \ref{subsec:GEMM_Architecture_Stratix}) were set based on our IP solver. 
The RTL (Verilog) code 
was automatically generated using our Python tool, and afterwards implemented on Intel Quartus 2021.1 (with a patch for Stratix 10 NX support from Intel). We use 
ModelSim~\cite{ModelSim_user_guide} to calculate throughput and the Quartus Power Analyzer to estimate power~\cite{Quartus_power_analysis}.

\begin{table*}[t]
\centering
\renewcommand{\arraystretch}{0.9}

\caption{Evaluation of 10 top-ranked GEMM designs on Stratix 10 NX.}

\resizebox{1.00\textwidth}{!}{
\begin{tabular}{c|c|c|c|c|c|c|c}
\Xhline{2.5\arrayrulewidth}

\multirow{2}{*}{\textbf{TB config.}} &
\multirow{2}{*}{\textbf{ALMs}} &
\multirow{2}{*}{\textbf{BRAMs}} &
\multirow{2}{*}{\textbf{TBs}} &
\textbf{Freq.} &
\textbf{Thrpt.} &
\textbf{Power} &
\textbf{En. Eff.} \\

 &
 &
 &
 &
\textbf{(MHz)} &
\textbf{(TOPS)} &
\textbf{(W)} &
\textbf{(TOPS/W)} \\

\hline
\hline

18$\times$16$\times$4$\times$3 & 124K (18\%) & 6304 (92\%) & 3456 (87\%) & 349 & 68.00 & 51.1 & 1.331 \\

18$\times$8$\times$8$\times$3 & 123K (17\%) & 6064 (89\%) & 3456 (87\%) & 345 & 67.21 & 50.2 & 1.340 \\

\textbf{9$\times$16$\times$5$\times$5} & \textbf{127K (18\%)} & \textbf{5840 (85\%)} & \textbf{3600 (91\%)} & \textbf{350} & \textbf{66.94} & \textbf{52.5} & \textbf{1.275} \\

12$\times$8$\times$6$\times$6 & 100K (14\%) & 6144 (90\%) & 3456 (87\%) & 338 & 64.00 & 48.6 & 1.317 \\

18$\times$16$\times$3$\times$4 & 108K (15\%) & 6272 (92\%) & 3456 (87\%) & 327 & 63.71 & 47.3 & 1.347 \\

9$\times$16$\times$6$\times$4 & 131K (19\%) & 6464 (94\%) & 3456 (87\%) & 342 & 62.88 & 50.7 & 1.241 \\

18$\times$8$\times$3$\times$8 & 81K (12\%) & 6064 (89\%) & 3456 (87\%) & 321 & 62.40 & 46.5 & 1.342 \\

\textbf{9$\times$8$\times$10$\times$5} & \textbf{124K (18\%)} & \textbf{5840 (85\%)} & \textbf{3600 (91\%)} & \textbf{320} & \textbf{61.21} & \textbf{48.7} & \textbf{1.257} \\

18$\times$8$\times$5$\times$5 & 101K (14\%) & 6150 (90\%) & 3600 (91\%) & 301 & 61.08 & 45.4 & 1.346 \\

18$\times$4$\times$8$\times$6 & 91K (13\%) & 6080 (89\%) & 3456 (87\%) & 312 & 60.69 & 46.2 & 1.315 \\

\Xhline{2.5\arrayrulewidth}

\end{tabular}
}

\label{tab_NX_results}
\end{table*}

Table \ref{tab_NX_results} shows the evaluation of the top 10 designs.
Configurations are a combination of the 4 TB architecture parameters ($\arraylength\times\Kexpansion\times \Nexpansion\times\Mexpansion$). Designs are ranked based on their throughput when running GEMM on their \textit{native} buffer sizes.
Our solutions achieve high throughput, up to 68 TOPS, which is 47.6\% of the \textit{theoretical peak} of the NX 2100 device. 
We note that our throughput is directly related to our achieved frequencies, which are similar to prior work~\cite{hpipex_NX,boutros2020beyond,langhammer2021stratix}. The maximum achieved energy efficiency is 1.347 TOPS/W. All designs present a small ALM  (12-19\%), but very high BRAM (85--94\%) and TB (87--91\%) utilization.
Moreover, they present very high RAM \textit{efficiency} (81.2--90\%) and low BW requirements (79.3--92.6 GB/s).

Overall, we notice that the top designs have a \arraylength[] of 18, 12 or 9. Designs with a \arraylength[] of 36 achieve lower frequency and require a longer time for PnR, as they have less flexibility during placement.
Note that design 9$\times$16$\times$5$\times$5 has a lower throughput than 18$\times$16$\times$4$\times$3, despite having higher TB utilization and operating at a higher frequency.
This is attributed to a smaller \arraylength[], and therefore more wasted TBs operating in parallel load mode. Notice also how designs 18$\times$16$\times$4$\times$3 and 18$\times$8$\times$8$\times$3 perform better than designs 18$\times$16$\times$3$\times$4 and 18$\times$8$\times$3$\times$8, respectively. 
They achieve a higher frequency, due to a lower \Mexpansion[], which leads to lower overall signal fan-out and FPGA routing congestion. 


\section{Insights \& Discussion}
\label{sec:Versal_vs_Stratix:Insights}

Next, we discuss and compare various aspects of GEMM design and optimization on both Versal and Stratix devices. 

\begin{figure}[t]
\centering
\subfloat[]{\includegraphics[width=0.45\linewidth]{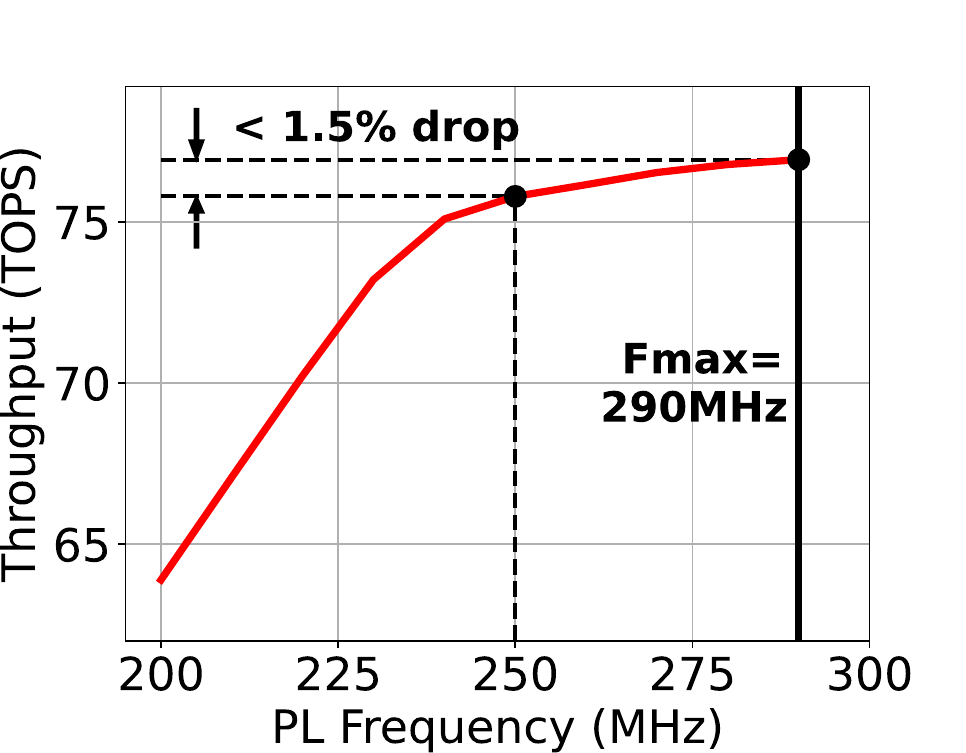}
\label{fig:throughput_PL_freq}}
\hfill
\subfloat[]{\includegraphics[width=0.45\linewidth]{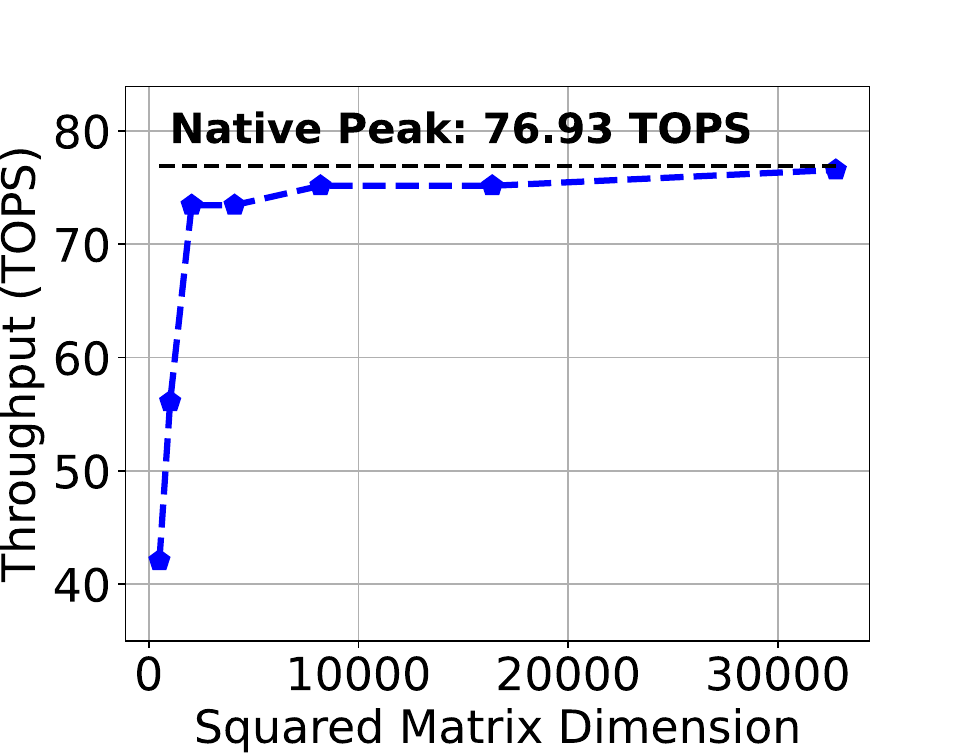}
\label{fig:throughput_mat_size_2x2x8_P1}}
\caption{Performance \emph{vs.} PL frequency (a), and square matrix dimensions (b) of 2$\times$2$\times$8 ($P1$) on Versal VC1902.}
\label{fig:throughput_per_matrix_size_PL_freq}
\end{figure}

\textbf{\textit{1) FPGA Frequency:}} We examine the reliance of our solutions on the attainable FPGA frequency. While Versal's AIE array operates at a fixed frequency (1.25 GHz), the PL (FPGA) frequency varies from 275-300 MHz across all top solutions (Table \ref{tb:Versal_top10_designs}).
Despite the PL frequency variation, we observe a 
stable performance ($\sim$2\%) for Versal.
We further depict the impact of PL frequency on throughput in Fig. \ref{fig:throughput_PL_freq} for design 2$\times$2$\times$8 ($P1$).
We notice a negligible throughput decrease ($<$1.5\%)  for a wide frequency range (290 to 250 MHz).
This is attributed to higher AIE computation time compared to communication with PL, illustrating the \textit{weak} dependence of performance on a \textit{wide} PL frequency range. 
However, for frequencies lower than 250 MHz performance degrades more severely ($\sim$16\% from 250 to 200 MHz).

Conversely, for Stratix, we observed a higher performance range ($\sim$12\%) across all solutions in Table \ref{tab_NX_results}.
This is attributed to the direct relationship of performance to frequency, illustrating the \textit{strong} dependence of the NX on PL frequency.




\textbf{\textit{2) GEMM Scalability}}: Furthermore, we explore the scalability of our solutions when altering the matrix dimensions.
For Versal, in Fig. \ref{fig:throughput_mat_size_2x2x8_P1}, we present the throughput of the best overall design 2$\times$2$\times$8 ($P1$), for square matrices (powers-of-two) ranging from 512 to 32K.
Zero-padding is applied to align to the \textit{compute} GEMM size (Table \ref{tb:Versal_top10_designs}).
We notice that our design scales effectively, leading to almost its \textit{native} peak 
throughput for dimensions $\sim$2K and higher.
We note here that \textit{native} peak refers to the \textit{achieved} throughput of our designs when running GEMM on their \textit{native} buffer sizes (Tables \ref{tb:Versal_top10_designs}, \ref{tab_NX_results}).

\begin{figure}[t]
\centering
\subfloat[]{\includegraphics[width=0.45\linewidth]{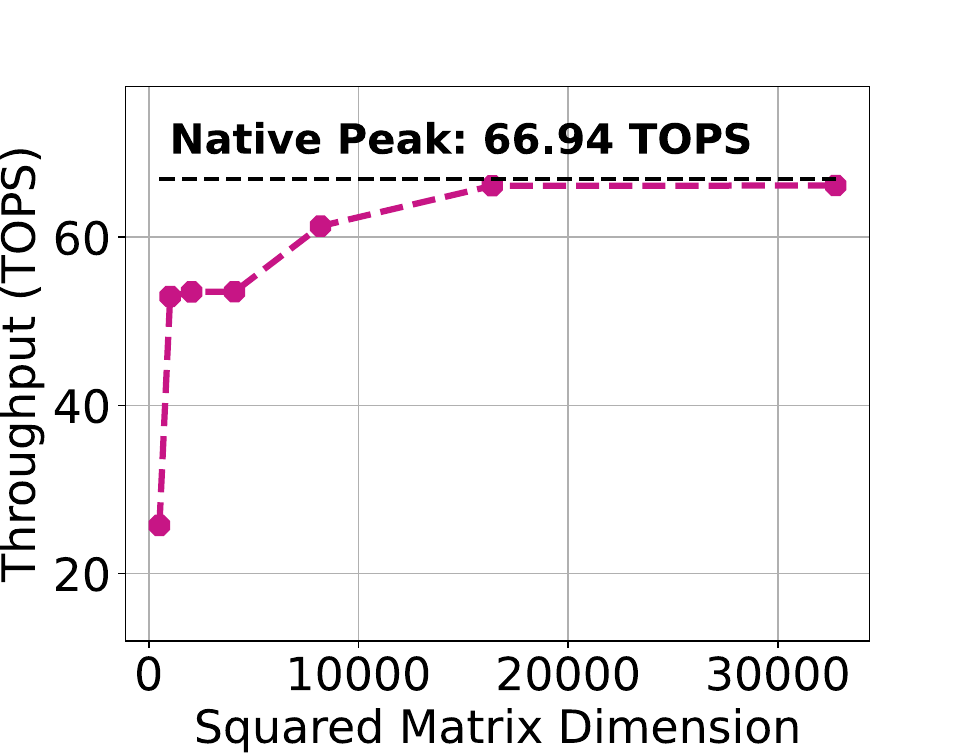}
\label{fig:high_K_design_NX}}
\hfill
\subfloat[]{\includegraphics[width=0.45\linewidth]{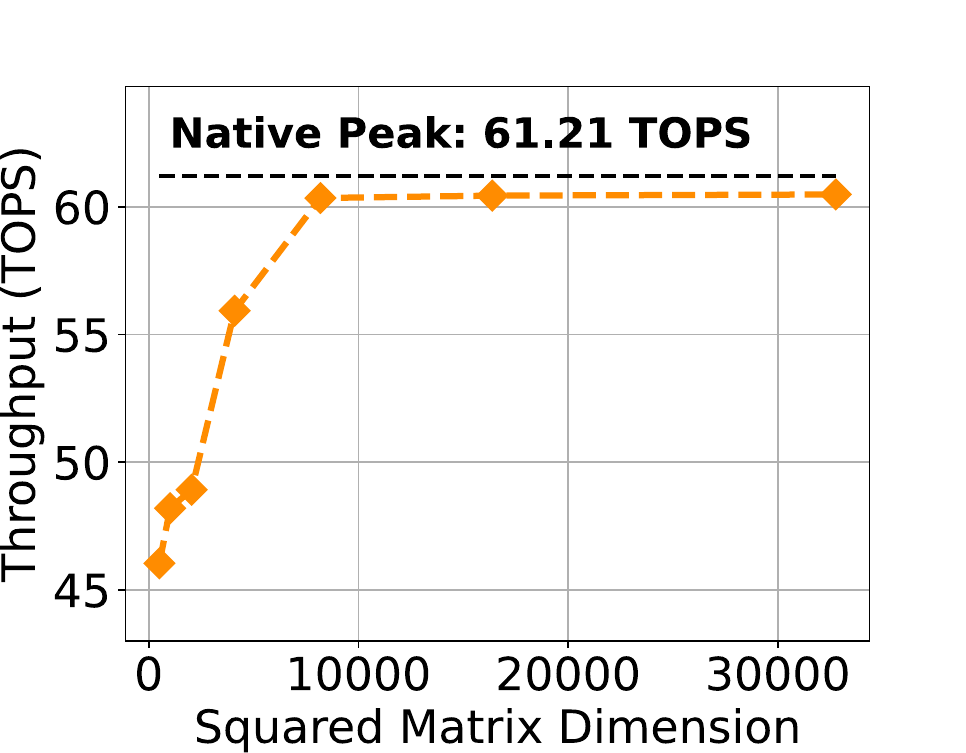}
\label{fig:low_K_design_NX}}
\caption{Performance \emph{vs.} square matrix dimensions of designs  9$\times$16$\times$5$\times$5 (a) and 9$\times$8$\times$10$\times$5 (b) on Stratix 10 NX.}
\label{fig:stratix10_NX_results}
\end{figure}

Similarly, for Stratix, we show the scalability of the two bolded designs in Table~\ref{tab_NX_results}.
The first (Fig. \ref{fig:high_K_design_NX}) has a high $D_{K^\prime}$, and presents one of the highest throughput among all top solutions.
The second (Fig. \ref{fig:low_K_design_NX}) has the lowest $D_{K^\prime}$. 
As $D_{K^\prime}$ is much higher than $D_{M^\prime}$ and $D_{N^\prime}$ (Table~\ref{tab_NX_results}), designs with flexible (small) $D_{K^\prime}$ require less zero-padding along the reduction dimension ($K^\prime$) and scale better.
We observe that, while the design in Fig.~\ref{fig:high_K_design_NX} has higher \textit{native} peak throughput, the design in Fig.~\ref{fig:low_K_design_NX} scales better, due to lower $D_{K^\prime}$.
Overall, GEMM scalability on both devices directly depends on the \textit{compute} GEMM size. 
By properly adjusting this size, our methods can be exploited in straightforward fashion to target specific matrix dimensions \emph{e.g.}, long and narrow matrices. 
However, for the sake of generality, we demonstrate GEMM scalability on square matrices, without targeting specific sizes.





\textbf{\textit{3) Achieved Performance}}:
As mentioned, Versal achieves $\sim$60\% of its \textit{theoretical peak} throughput. 
This can be ascribed to multiple reasons.
First, the non-ideal (95\%) efficiency~\cite{Taka2023MaxEVA, charm2023fpga} of the AIE MatMul kernels.
Second, the necessity to introduce Add kernels on the AIE, which do not contribute to throughput, but occupy some of the cores (Sec. \ref{subsubsec:GEMM_Mapping_AIE}).
Third, the inevitable stalls caused by memory conflicts on the AIE array \cite{AI_Engine_programming_environment}.
In contrast, Stratix achieves $\sim$47\% of its \textit{theoretical peak} throughput. This is because the \textit{theoretical peak} assumes 100\% TB utilization with no wasted TBs in cascade mode, and a very high operating frequency of 600MHz~\cite{langhammer2021stratix}, which are infeasible to attain in practice.

Our results in Tables \ref{tb:Versal_top10_designs}, \ref{tab_NX_results} indicate that, on average, Versal attains 19.8\% higher throughput, while Stratix presents 41.6\% higher energy efficiency on GEMM workloads.
However, we note that the focus of this study strays from a competitive comparison among the two AI-optimized FPGA architectures.
Instead, we emphasize the distinct design
methods required for each device, due to their considerably different architecture styles (out-of-fabric \emph{vs.} in-fabric).
A broader set of full AI-workloads would be needed to explore complicated trade-offs between the two architectures, which we leave as future work.

\textbf{\textit{4) Programmability Trade-Offs}}: The two technologies require completely different programming methods.
In particular, Versal utilizes high-level C/C++ software for programming the AIE array. 
Efficient, high-level programming constructs can enable high performance of the AIE array's SIMD processors on vectorizable workloads (with 95\% AIE kernel efficiency).
For productivity purposes, the PL part is typically programmed in HLS. 
However, HLS inefficiencies can lead to performance degradation, which can be 
mitigated by exploiting sophisticated methods (Sec. \ref{subsubsec:PL_Optimization}).
Finally, AMD/Xilinx provides automation tools, such as Vitis V++, to reduce the complexity of integrating the entire system.


On the contrary, no high-level tools are available for programming Stratix 10 NX.
In particular, Intel HLS~\cite{Intel_HLS_guide} has no support for TBs, thus requiring exclusive coding in RTL for design, verification, and full-system integration.
This resulted in $\sim$40$\times$ higher number of lines of code in Stratix compared to Versal.
Nonetheless, our Python-based generator significantly reduced this complexity.
Moreover, because the performance of Stratix is directly analogous to frequency, unlike Versal, more programming effort is required to meet timing goals, and an extensive DSE is needed to identify high-throughput designs. In our case, this led to $\sim$5$\times$ higher total tool compilation time for Stratix compared to Versal.
While each individual design required 3--6 hours to compile on both devices, the extensive exploration on Stratix greatly increased the design space. Finally, based on our estimations, it might be 1.5--2$\times$ more productive to program Versal due to the 
increased programming
effort required by Stratix. However, we note that design productivity is highly dependent on designer skills.





\textbf{\textit{5) GEMM Optimization Insights}}:
Both devices exhibit design challenges due to their high complexity.
To address complexity, systematic methodologies, \emph{e.g.}, analytical modeling, are essential for optimization of GEMM-based applications.

Versal AIE is a novel architecture that introduces several new design challenges.
A comprehensive exploitation of AIE architectural attributes, \emph{e.g.}, local memory sharing and circuit switching, is crucial to optimize performance in GEMM \cite{Taka2023MaxEVA}.
Moreover, challenges, such as the reduced number of AIE-PL tiles and AIE routing congestion, require refined techniques to prevent performance degradation (Sec. \ref{subsubsec:GEMM_Mapping_AIE}).
Finally, the limited DRAM BW of Versal devices introduces additional challenges in maintaining the high throughput of the AIEs.



In contrast, Stratix 10 NX preserves a traditional FPGA architecture, while introducing in-fabric TBs embedded in the PL.
The inherent complexity of TBs necessitates the
consideration of multiple operating modes to effectively map GEMM workloads (Sec.~\ref{subsec:GEMM_Architecture_Stratix}).
Moreover, as throughput is directly related to frequency, performance depends on both low-level RTL optimizations and the intricacies of the FPGA architecture. RTL techniques, \emph{e.g.}, replication of control logic and insertion of pipeline stages, are essential in shortening critical paths. 

\section{Chapter Summary \& Impact}
\label{sec:Versal_vs_Stratix:Sumamry}

In this chapter, we propose frameworks to optimize GEMM-based applications targeting the two leading AI-optimized FPGA architectures. We present a thorough evaluation of several aspects in GEMM design by efficiently leveraging the unique and substantially different architectural attributes of the AMD Versal ACAP and Altera/Intel Stratix 10 NX devices. Our experimental results show that our frameworks achieve up to 77 and 68 TOPS (int8) throughput, with up to 0.94 and 1.35 TOPS/W energy efficiency, for Versal and Stratix, respectively. This study provides fundamental insights regarding the architectural characteristics, programmability trade-offs, challenges, and limitations inherent in both Versal and Stratix AI-optimized FPGAs, motivating the hardware-level sparsity support explored in the next chapter.




\chapter{Systolic Sparse Tensor Slices: FPGA Building Blocks for Sparse
and Dense AI Acceleration}
\label{ch:SST_Slices}

In this chapter, we present the architectural details of our proposed SSTs \cite{SST_slices_FPGA_2025}, their integration into SST-based GEMM implementations, and the resulting inference speedups on sparse DL workloads compared to dense in-fabric FPGA acceleration. 
Moreover, we show that sparse accelerators exploiting SSTs offer substantial performance, scalability, and area advantages over traditional FPGAs.

\section{Architecture \& Design Overview}
\label{sec:SST_Slices:Architecture_Design}

\subsection{Fine-Grained Structured Sparsity}
\label{subsec:Fine_grained_structured_sparsity}
\begin{figure}[t]
\centering
\subfloat[]
{\includegraphics[width=0.32\linewidth]{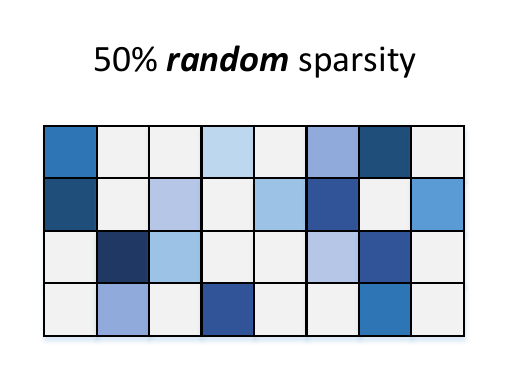}
\label{fig:random_sparsity}}
\subfloat[]{\includegraphics[width=0.685\linewidth]{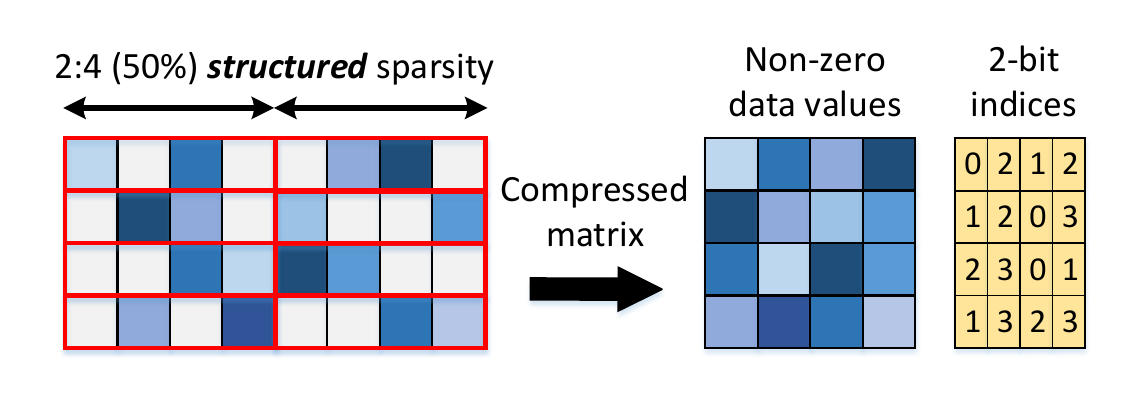}
\label{fig:structured_sparsity}} 


\caption{50\% unstructured sparse matrix (a) and 2:4 (50\%) structured sparse matrix along with compressed format (b).} 
\label{fig:sparsity_struct_unstruct}
\end{figure}

In this work, we leverage the regular patterns of fine-grained structured sparsity to architect SST slices with low area overhead. 
Fig. \ref{fig:random_sparsity} depicts a 50\% unstructured sparse matrix, where the non-zero data are distributed \textit{randomly}, \emph{i.e.,} there is no specific pattern of their locations.
In contrast, in Fig. \ref{fig:structured_sparsity}, the 2:4 structured sparsity pattern is illustrated, which has the same sparsity level (50\%), but in every group of four \textit{consecutive} elements
there are two non-zero values. 
Notice that the location of the two non-zero values can vary significantly within the four-element group, offering \textit{fine-grained} sparsity flexibility.  
These types of constraints enable low area hardware enhancements, allowing for efficient  exploitation of sparsity.

A 2:4 sparse matrix can be efficiently stored in  compressed format by saving only the non-zero values.
The location of each non-zero data is encoded using 2-bit indices, as shown in Fig. \ref{fig:structured_sparsity}.
Similar to 2:4, for 1:4 (75\%) sparsity one every four consecutive elements is non-zero, while for 1:3 (66.7\%) sparsity there is one non-zero every three consecutive elements.  
For all aforementioned patterns, 2-bit indices are required to encode the location of each non-zero element.
This is a very efficient compressed format, as we show in Sec. \ref{subsec:AIE_ML_comparison}, where comparison among other formats is performed.

\subsection{Systolic Sparse Tensor Slices Architecture}
\label{subsec:Sparse_Tensor_slices_architecture}

In this section, we present an
overview of the proposed SST slices. 
The core compute unit of the SSTs is a 4$\times$4 systolic array (SA) \cite{Kung_SA_1982}, as depicted in Fig. \ref{fig:ST_architecture}.
A 2D SA consists of homogeneous processing elements (PEs), where each PE performs a multiply--accumulate (MAC) operation and forwards the input operands to the neighboring PEs. 
This architecture allows maximization of data reuse
in GEMM, 
while also delivering high performance due to its regular and highly scalable design.
To this end, SAs have become a prime architecture in many DNN accelerators \cite{TPUV2_v3_2021, TPUv42021, S2TA_HPCA_2022, Vegeta_HPCA_2023, SA_CNN_FPGA_2017, SA_attention_FPGA_TECS_2023, Scale_sim_2020}.
In this work, we show that incorporating SST slices in FPGAs leads to high performance and scalable dense and sparse GEMM accelerators.

\subsubsection{SST Operation} 
\begin{figure}[tbp]
\centering
\includegraphics[width=0.75\linewidth]{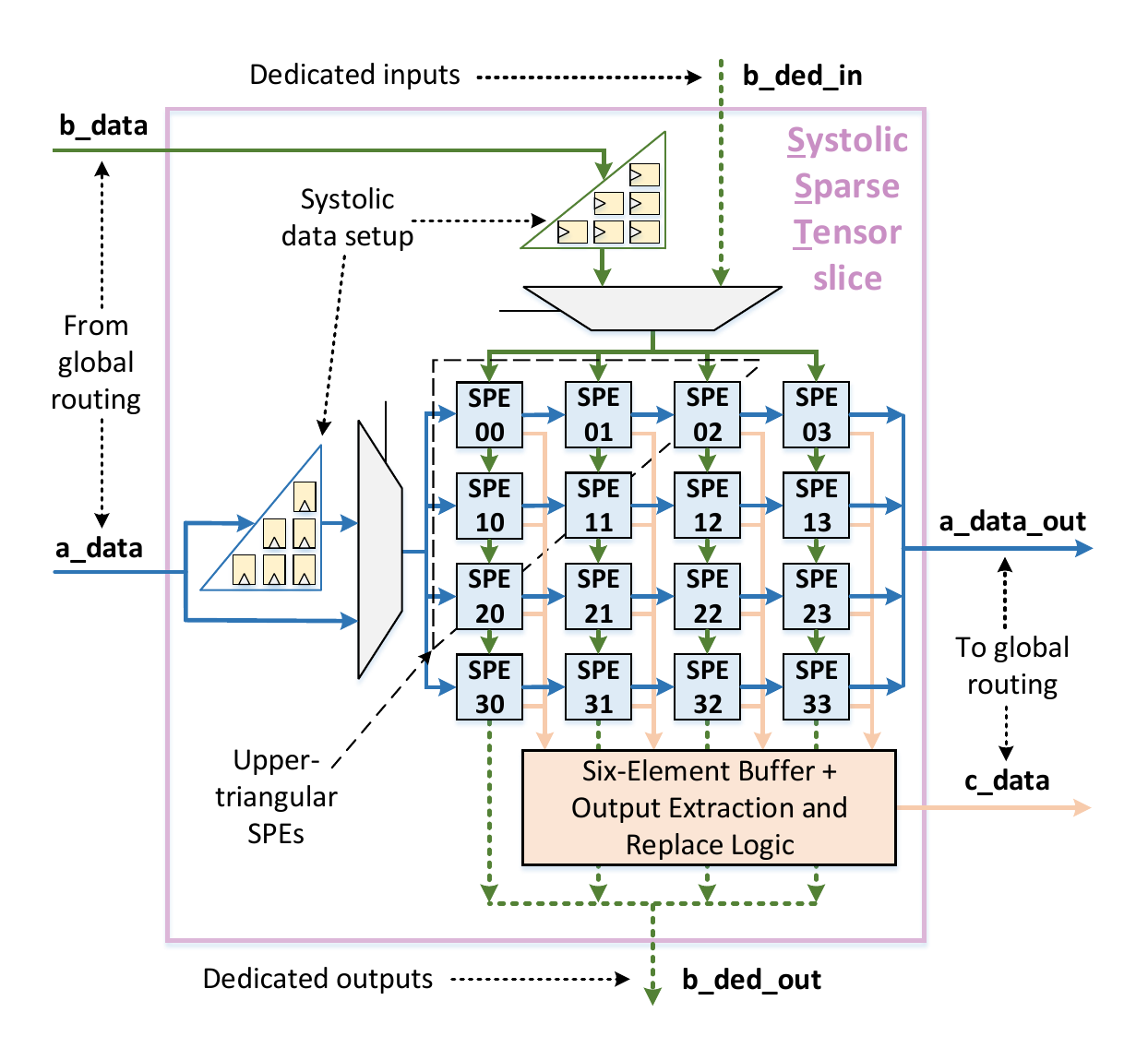}


\caption{Systolic Sparse Tensor slice architecture.}
\label{fig:ST_architecture}


\end{figure}

We enhance the systolic PEs with sparse features by introducing sparse processing elements (SPEs), while maintaining the properties of the SAs discussed above. 
Our SSTs utilize an output stationary SA, consisting of 16 SPEs.
The accumulations remain stationary in the SPEs, while input operands are propagated to their neighbors every clock cycle.
The $a\_data$ of an input matrix $A$ are propagated and reused across SPEs horizontally, while the $b\_data$ of an input matrix $B$ are propagated and reused vertically (Fig. \ref{fig:ST_architecture}).
The matrix $A$ can be either sparse or dense (typically to map \textit{weights}), while $B$ is dense (typically to map input \textit{activations}).
The architecture of the SPEs is delineated in Sec. \ref{subsec:Sparse_Processing_Element}.

Besides the 4$\times$4 SPE grid, we also implement pipeline registers to delay the input operands for systolic data setup \cite{TPU_paper2017} (arranged in triangular manner in Fig. \ref{fig:ST_architecture}).
Systolic setup is needed at the interface for loading input matrices (typically from on-chip memory), when chaining multiple SSTs to construct larger SA grids (Sec. \ref{subsec:Matrix_multiplication_mapping}).
Multiplexers are used to select either the systolic setup
or directly the input data, and are configured \textit{statically} (during bitstream loading).

The SSTs support both int8 and bf16 precisions, while accumulations are realized int32 and fp32, respectively, similar to Nvidia GPUs \cite{Nvidia_accelerate_sparse_2021} and Google TPUs \cite{TPUV2_v3_2021}.
When the SA operation is completed, the output matrix $C$ is extracted via the $c\_data$ output ports.
We note that SPEs finish their operation in a \textit{diagonal} fashion cycle after cycle.
In the first cycle, $SPE00$ finishes, in the second cycle both $SPE01$ and $SPE10$ finish, etc. (Fig. \ref{fig:ST_architecture}). 
However, to maintain \textit{regularity}, thus simplifying the downstream logic (typically in CLBs), we extract the output values in a column-wise manner (four values per cycle).
To achieve that, we introduce a six-element buffer (implemented as a register array), to store the data before getting extracted column-wise.
This is particularly important in sustaining 100\% SPE utilization (16 MACs per cycle) at the steady-state (matrices processed one after the other).

Since SPEs complete their operation diagonally, the six upper-triangular SPE outputs, shown in Fig. \ref{fig:ST_architecture}, need to be stored in the buffer.
Suppose a cycle $T$ where the outputs of the biggest diagonal, \emph{i.e.,} $SPEs$ $\{03, 12, 21, 30\}$ are generated.
In cycle $T$, the values of the first column, \emph{i.e.,} $SPEs$ $\{00, 10, 20, 30\}$ can be extracted, where $SPEs$ $\{00, 10, 20\}$ are loaded from the buffer, while only $SPE30$ is directly extracted.
In cycle $T + 1$, the values of $SPEs$ $\{03, 12, 21\}$ replace the position of $SPEs$ $\{00, 10, 20\}$ in the buffer (since they have been extracted).
Therefore, the locations of the six-element buffer are being effectively reused, and in cycle $T+1$, the second column can be extracted, \emph{i.e.,} $SPEs$ $\{01, 11, 21, 31\}$.
In a similar fashion, the rest two columns are extracted in cycles $T+2$ and $T+3$, respectively, while the buffer locations are efficiently reused due to replacement.

\subsubsection{Global Routing Interface \& Dedicated Wires}

As illustrated in Fig. \ref{fig:ST_architecture}, the $a\_data$ and $b\_data$ input ports as well as $c\_data$ output ports are connected to the global FPGA routing resources.
When chaining multiple SSTs, $a\_data$ and $b\_data$ are forwarded to their next in the chain SSTs, horizontally and vertically, respectively.
Horizontally, the data are propagated via the $a\_data\_out$ using the FPGA routing resources.
However, vertically, we utilize \textit{dedicated} wires to propagate the data (via the $b\_ded\_out$ ports) and connect them to the next SST in the \textit{same} FPGA column (via the $b\_ded\_in$ ports).
These vertical dedicated wires provide efficient connections without the usage of the global routing resources, matching the columnar nature of the modern FPGA fabric \cite{FPGA_architecture_2021, FPGA_for_DL_2024}.
This approach significantly reduces routing resources, as opposed to  \cite{TS_Aman_FPGA_2021, Aman_TS_TRETS_2022}, where all inputs/outputs (I/Os) of the in-fabric tensor slices are connected to global routing (see detailed comparison in Sec. \ref{subsec:Dedicated_wires_benefits}).

Besides the I/Os shown in Fig. \ref{fig:ST_architecture}, the SSTs include also several dynamic control signals.
In particular, an input $enable$ signal is used to control the operation of the SSTs.
This signal should be deasserted when the operation of the SST needs to be stalled.
Moreover, an $accumulate$ signal is utilized to control the accumulation in the SPEs.
This signal remains asserted when accumulation is desirable in the SPEs, \emph{e.g.,} during tiling to process larger matrices.
However, for every new matrix operation the $accumulate$ signal should be deasserted for one cycle.
Another input signal is the  $d\_type$, which  dynamically selects the precision, \emph{i.e.,} int8 or bf16.
Finally, a 2-bit $sparsity\_level$ signal is employed to dynamically select the sparsity level of the matrix $A$, \emph{i.e.,} dense, 2:4, 1:3, 1:4.
This signal is utilized internally in the SSTs to control the operation of the supported sparsity modes, as discussed in the next section.

Regarding the outputs of the SST, an $accumulate\_out$ signal is used for systolic distribution of the $accumulate$ signal when chaining multiple SSTs.
This systolic distribution makes the control logic significantly simpler, since the $accumulate$ is set only for the first SST in the 2D array layout, while being distributed to the remaining SSTs, as explained in Sec. \ref{subsec:Matrix_multiplication_mapping}.
Another output signal is the $valid\_out$, which is asserted when the output $c\_data$ are extracted column-wise (Fig. \ref{fig:ST_architecture}). 
This signal simplifies the downstream logic, since only $valid\_out$ needs to be checked for valid output data.

\subsection{Sparse Processing Element}
\label{subsec:Sparse_Processing_Element}

\begin{figure}[tbp]
\centering
\subfloat[Dense]
{\includegraphics[width=0.50\linewidth]{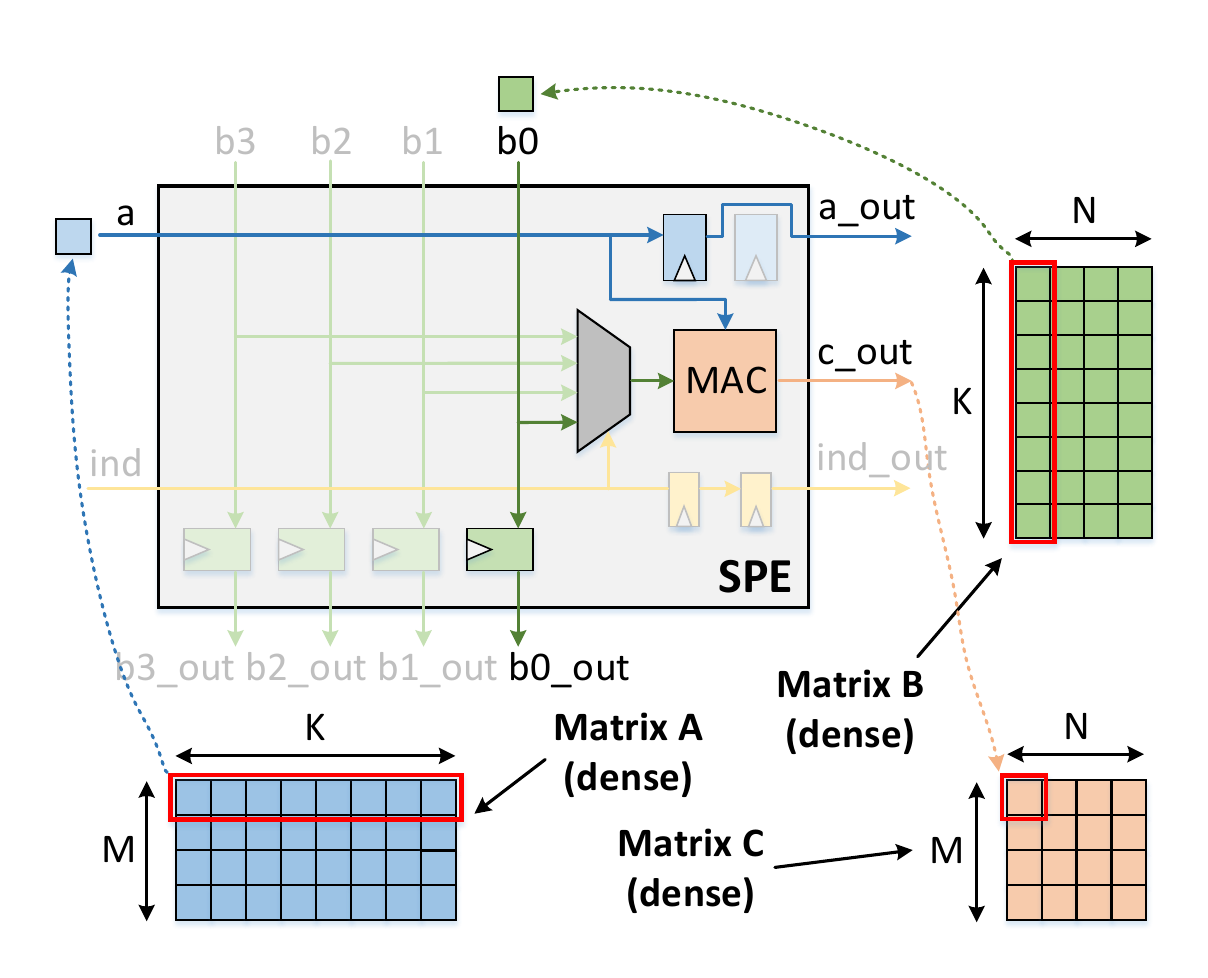}
\label{fig:dense_mode}}
\subfloat[2:4 (50\%)]{\includegraphics[width=0.50\linewidth]{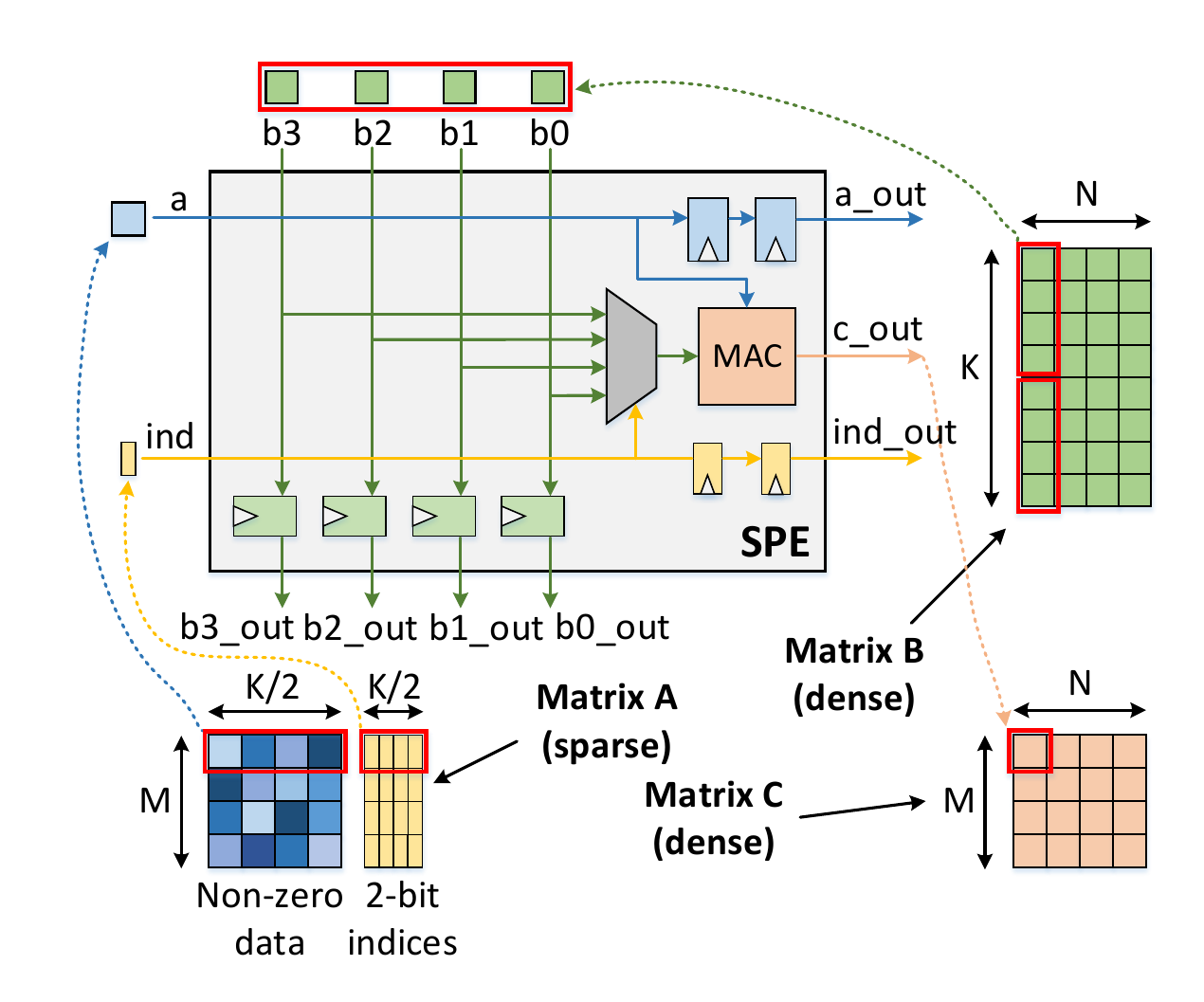}
\label{fig:2_4_mode}} 


\subfloat[1:4 (75\%)]{\includegraphics[width=0.50\linewidth]{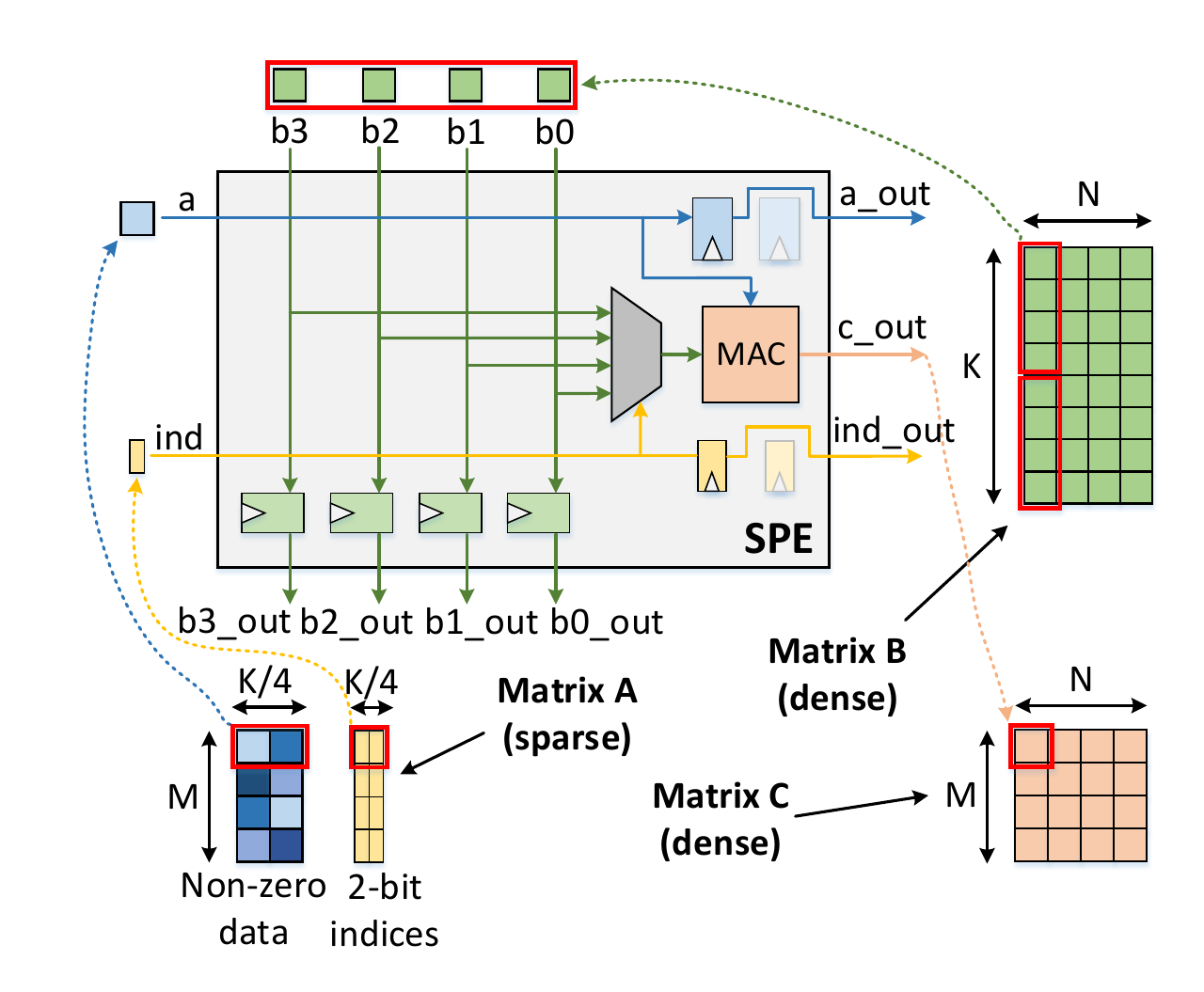}
\label{fig:1_4_mode}}
\subfloat[1:3 (66.7\%)]{\includegraphics[width=0.50\linewidth]{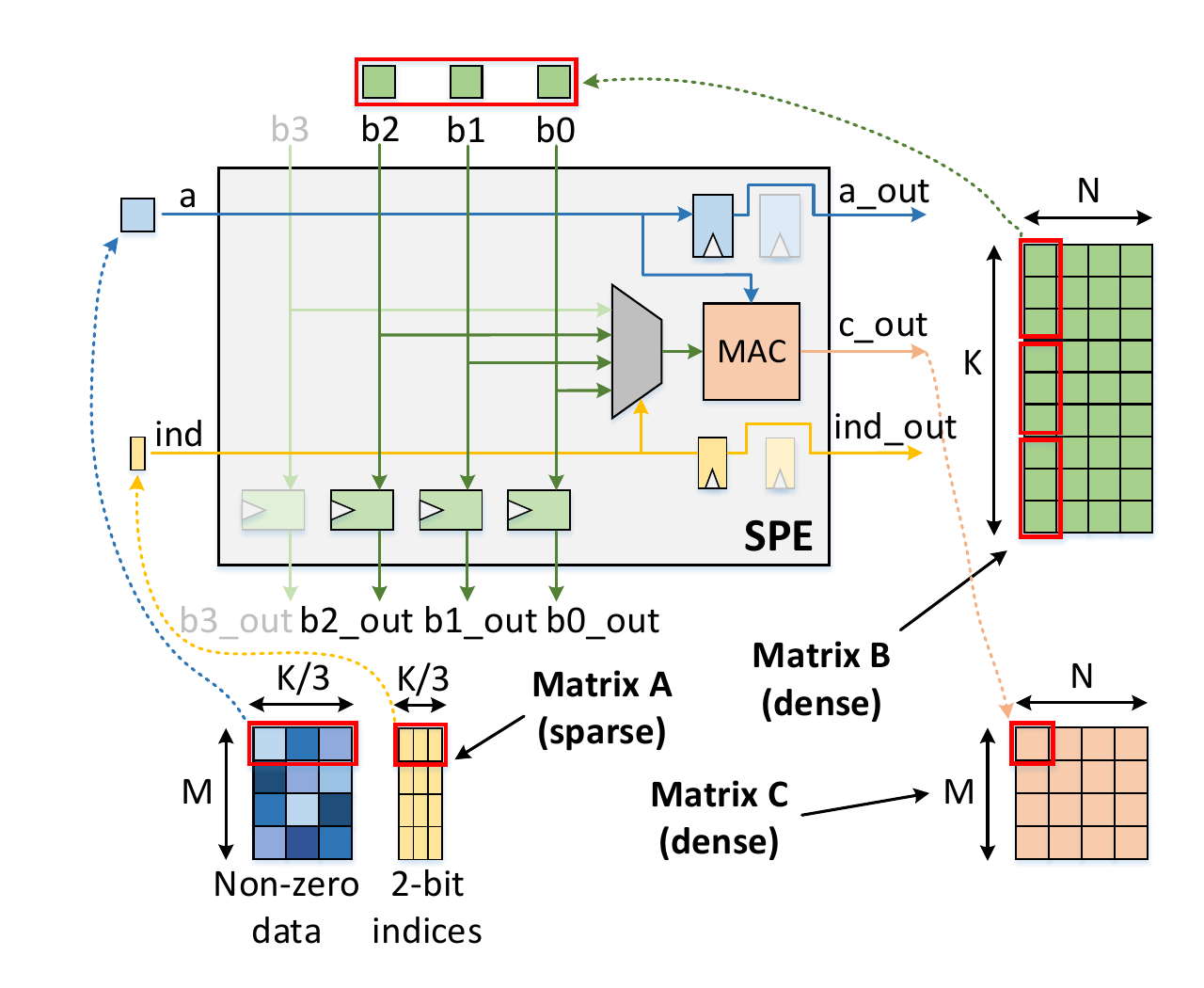}
\label{fig:1_3_mode}} 

\caption{Sparsity modes in systolic sparse elements of the SST slices (multiplexing logic omitted for clarity).}
\label{fig:SPE_sparse_modes}
\end{figure}

The SPE architecture in all supported sparsity modes is illustrated in Fig. \ref{fig:SPE_sparse_modes}.
When the SPE is configured in dense mode (Fig. \ref{fig:SPE_sparse_modes}a), it operates as a regular dense PE.
In particular, one MAC operation is performed every cycle, while the elements of matrices $A$ and $B$ are forwarded to their neighboring SPEs (via pipeline registers) horizontally and vertically, respectively.
The output value of matrix $C$ in each SPE is calculated after $K$ cycles, when considering the $M$$\times$$K$$\times$$N$ matrix dimensions depicted in Fig. \ref{fig:SPE_sparse_modes}a.

In 2:4 mode (50\% sparsity), matrix $A$ is stored in a compressed format of $M$$\times$$K/2$ size for both values and the 2-bit indices, as shown in Fig. \ref{fig:SPE_sparse_modes}b.
In order to achieve speedup over the dense case, we increase the number of ports of each SPE in the vertical dimension, to load four elements of the matrix $B$ in parallel.
Furthermore, the 2-bit index is also loaded in the SPE to select the corresponding $B$ values (via a 4:1 multiplexer), which need to get multiplied with each $A$ value.
However, in 2:4 sparsity, for every four values of the matrix $B$, two MAC operations need to be performed.
Hence, the four $B$ values remain in the SPE registers for two clock cycles (the four green registers in Fig. \ref{fig:SPE_sparse_modes}b are loaded every two cycles).
To correctly synchronize the systolic operation, two pipeline stages are required for both the $A$ values and their indices (blue and yellow registers in Fig. \ref{fig:SPE_sparse_modes}b), which are loaded every cycle. 
In this manner, one MAC operation is performed every cycle, ensuring 100\% utilization.
Moreover, the output value of matrix $C$ in each SPE is calculated in $K/2$ cycles, achieving 2$\times$ acceleration over dense operation.

\begin{table}[t]
\centering
\caption{Summary of supported sparsity levels in SST slices.}

\setlength\tabcolsep{9pt}
\renewcommand{\arraystretch}{1.0}
\resizebox{0.70\linewidth}{!}{
\begin{tabular}{c|cc|c|c}
\Xhline{2.5\arrayrulewidth}

\textbf{Sparsity}  &  
\multicolumn{2}{c|}{\textbf{Compres. ratio}} & \textbf{Speedup} & \textbf{SPE} \\

\cline{2-3}

\textbf{level}  & \textbf{int8} & \textbf{bf16} 
 & \textbf{over dense} & \textbf{util.} \\

\hline
\hline

\textbf{Dense (0\%)} & 1$\times$ & 1$\times$ & 1$\times$ & 100\% \\ 

\textbf{2:4 (50\%)} & 1.6$\times$ & 1.78$\times$ & 2$\times$ & 100\% \\ 

\textbf{1:3 (66.7\%)} & 2.4$\times$ & 2.67$\times$ & 3$\times$ & 100\% \\ 

\textbf{1:4 (75\%)} & 3.2$\times$ & 3.56$\times$ & 4$\times$ & 100\% \\

\Xhline{2.5\arrayrulewidth}

\end{tabular}
}

\label{tb:sparsity_summary_benefits}


\end{table}

Similar to 2:4, for 1:4 (75\%) sparsity, four $B$ values are loaded in the SPE.
However, in this case the $B$ values are loaded every cycle, while only one pipeline stage is required for the $A$ value and its index, as shown in Fig. \ref{fig:SPE_sparse_modes}c.
The output value needs $K/4$ cycles to be calculated, offering 4$\times$ speedup over the dense case.
To fill the sparsity gap between 50\% and 75\%, we leverage the same hardware enhancements for 2:4 and 1:4 sparsity to also support the 1:3 (66.7\%) pattern.
In particular, as illustrated in Fig. \ref{fig:SPE_sparse_modes}d, the 1:3 operation is similar to 1:4, with the main difference being that three $B$ values are loaded every cycle instead of four.
The indices ensure that the fourth $B$ value is never selected, thus no additional hardware is required.
In this case, $K/3$ cycles are needed for every SPE output value, providing 3$\times$ acceleration.
Finally, we note that similar to 1:3, 2:3 sparsity can also be supported by directly utilizing the 2:4 mode, leading to 33.3\% sparsity.
However, a matrix is typically considered sparse when it has sparsity of 50\% or higher \cite{AMD_AIE_ML_kernel_guide}.
Hence, in this work, we do not consider the 2:3 pattern.

In Table \ref{tb:sparsity_summary_benefits}, we present a summary of the supported sparsity levels.
First, we show the \textit{compression ratio}, \emph{i.e.,} the memory reduction over dense storage due to compressed format, for both int8 and bf16 precisions.
For all sparsity levels, bf16 offers a higher compression ratio over int8, \emph{e.g.,} 1.78$\times$ \emph{vs.} 1.6$\times$ for 2:4 sparsity. 
This is because 2-bit indices are required for both 8-bit and 16-bit data types, resulting in relatively lower overhead for the compressed representation in bf16 compared to int8.
This compressed format substantially reduces both on-chip and off-chip memory requirements, achieving up to 3.56$\times$ reduction (Table \ref{tb:sparsity_summary_benefits}).

Second, we notice that every sparsity level is
translated to its corresponding speedup, \emph{e.g.,} 4$\times$ for 1:4 (75\%) sparsity, achieving 100\% SPE utilization in all cases (Table \ref{tb:sparsity_summary_benefits}).
For all sparsity levels, data reuse is maximized since both indices and values are propagated and reused horizontally and vertically, similar to dense operation.

\subsection{GEMM Design Utilizing Multiple SST Slices}
\label{subsec:Matrix_multiplication_mapping}

In this section, we describe parametric GEMM implementations for both dense and sparse configurations, utilizing multiple SST slices.
Both implementations are highly regular and scale effectively on the FPGA fabric, attaining high frequencies as shown in Sec. \ref{subsec:Sparse_GEMM_implementation}.

\subsubsection{Dense Implementation}

\begin{figure}[tbp]
\centering
\includegraphics[width=0.76\linewidth]{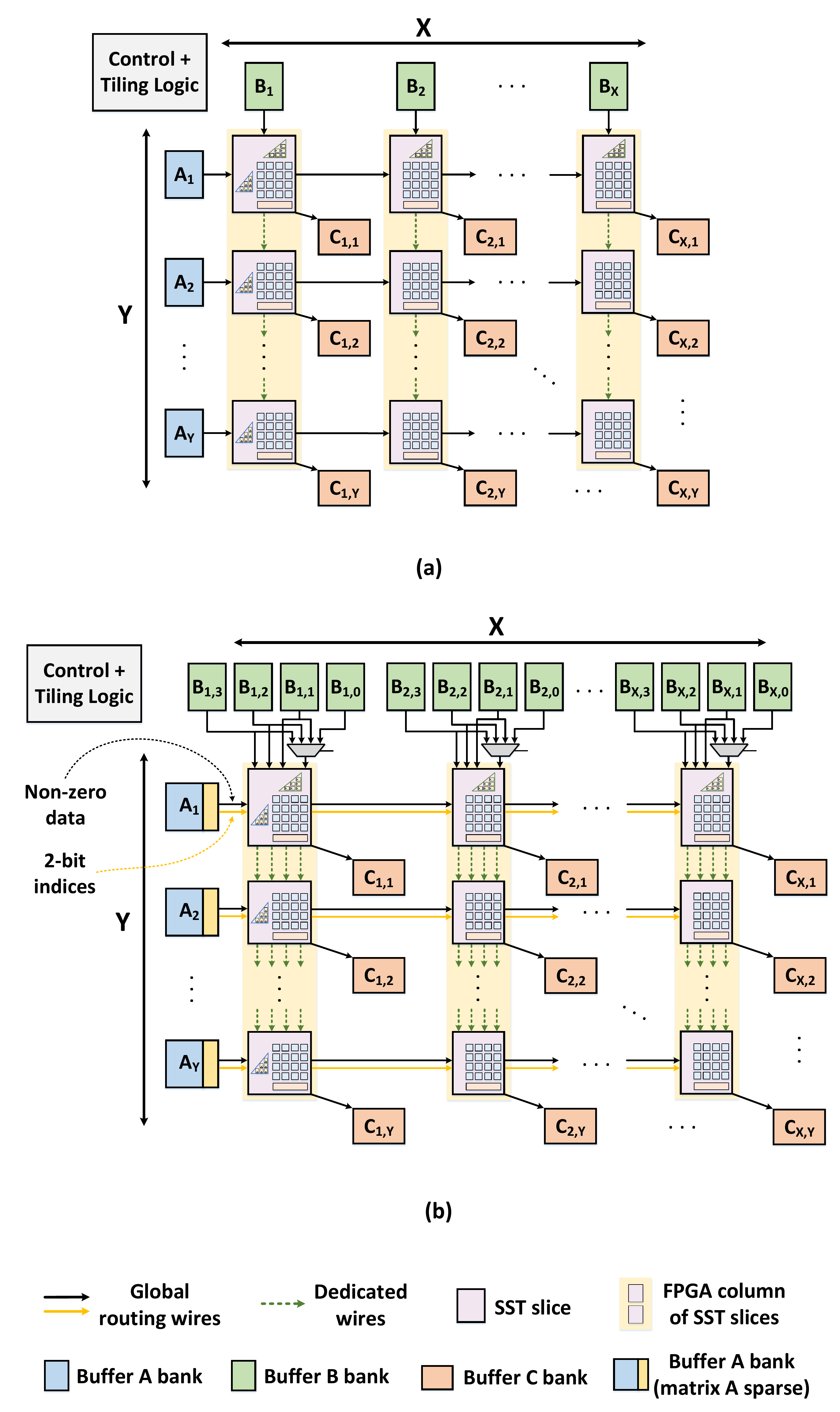}
\caption{2D systolic GEMM design: dense implementation (a) and dynamic configuration of all supported sparsity modes (b).}
\label{fig:GEMM_2D_array_SSTs}
\end{figure}

Fig. \ref{fig:GEMM_2D_array_SSTs}a depicts a parametric GEMM accelerator comprising a 2D array of SST slices, which are configured in \textit{dense} mode (Sec. \ref{subsec:Sparse_Processing_Element}). 
The 2D array consists of $Y \cdot X$ SST slices, implementing a total SA size of $(Y \cdot 4) \times (X \cdot 4)$.
This size is denoted as the \textit{native} size of the GEMM accelerator. 
On-chip memory buffers are implemented to store the input matrices $A$, $B$ and the output matrix $C$.
The input buffers $A$, $B$ are located in the left and top edges of the 2D array, respectively, and are partitioned into banks, providing sufficient bandwidth to feed the SSTs.
In particular, for buffer $A$, $Y$ banks are required, while for buffer $B$, $X$ banks are needed.
Since each SST slice includes a 4$\times$4 SA, each bank needs to provide a bandwidth of 32-bits per cycle when SSTs are configured for int8 precision, while for bf16, 64-bits per cycle are required.
Regarding the output buffer $C$, $X \cdot Y$ banks are needed due to the output stationary architecture of the SSTs, each receiving an output of 128-bits per cycle (four 32-bit values as explained in Sec. \ref{subsec:Sparse_Tensor_slices_architecture}).

The data from the buffers $A$, $B$ are propagated in a systolic fashion between the SST slices, as illustrated in Fig. \ref{fig:GEMM_2D_array_SSTs}a.
Horizontally, the $A$ data are propagated via the global routing resources of the FPGA fabric.
Vertically, the $B$ data are inserted via global routing wires in the first SST at each vertical chain ($Y$ SSTs in Fig. \ref{fig:GEMM_2D_array_SSTs}a), while being forwarded to the next SSTs via dedicated wires.
It is important to note here that the SSTs comprising each vertical chain are \textit{physically} contiguous in the FPGA.
Nevertheless, in the horizontal dimension, the $Y$ chains might not follow the \textit{logical} arrangement shown in Fig. \ref{fig:GEMM_2D_array_SSTs}a inside the FPGA fabric, due to the routing flexibility of FPGAs.
This depends on decisions made by the FPGA place and route (PnR) algorithm.
Finally, notice the (static) systolic data setup configuration specifically for the SST slices that interface with the buffers $A$, $B$ (left and top edge of the 2D array).

Control logic, mapped to the CLB resources of the FPGA, is utilized to orchestrate the entire operation of the GEMM design.
We also implement tiling logic (in CLBs) to exploit data reuse in GEMM, as well as to support arbitrary GEMM sizes based on the available on-chip memory resources.
When mapping an arbitrary GEMM of $M^\prime$$\times$$K^\prime$$\times$$N^\prime$ dimensions, $M^\prime$ must be a multiple of $(Y \cdot 4)$, while $N^\prime$ must be a multiple of $(X \cdot 4)$, since the \textit{native} size of the accelerator is $(Y \cdot 4) \times (X \cdot 4)$.
Note that there is no constraint on the reduction $K^\prime$ dimension.
Finally, although not shown in Fig. \ref{fig:GEMM_2D_array_SSTs}a, $accumulate$ signals utilized during tiling (Sec. \ref{subsec:Sparse_Tensor_slices_architecture}) are propagated in a systolic fashion among the 2D array of SSTs, similar to the $A$, $B$ data.
The control logic sets the $accumulate$ signal only for the first SST in both vertical and horizontal dimensions (\emph{i.e.,} the SST fed by buffers $A_{1}$ and $B_{1}$), which significantly simplifies the overall logic.

\subsubsection{Dynamic Sparse Configuration}

Fig. \ref{fig:GEMM_2D_array_SSTs}b illustrates a parametric GEMM design that is \textit{dynamically} configured to support all the sparsity modes in the SSTs, \emph{i.e.,} dense, 2:4, 1:3 and 1:4.
This dynamic configuration is particularly important for layer-wise sparsity exploitation in DNNs, since each layer might require different sparsity level for optimal trade-off between DNN accuracy and speedup (Sec. \ref{subsec:Performance_estimation_DNNs}).
We note that the implementation is similar to the dense design (Fig. \ref{fig:GEMM_2D_array_SSTs}a), with main differences lying in the design of buffers $A$ and $B$, as well as in the vertical and horizontal propagation of the data.
Horizontally, the $A$ data are kept in the buffer banks in compressed format for the sparse modes (2:4, 1:3 and 1:4).
In this case, both non-zero data and indices are loaded in the SSTs and are propagated horizontally (see Sec. \ref{subsec:Sparse_Processing_Element}).
More specifically, for int8, each buffer $A$ bank needs to provide a bandwidth of 40-bits per cycle, due to the additional 8-bits indices for the four vertical SPEs at the interface of each SST.
Similarly, for bf16, 72-bits per cycle are required.

Vertically, four $B$ banks are needed to feed the SSTs due to the 4$\times$ increase in ports for 2:4 and 1:4 sparsity acceleration (Sec. \ref{subsec:Sparse_Processing_Element}).
Each $B$ bank provides the same bandwidth as the dense design (Fig. \ref{fig:GEMM_2D_array_SSTs}a), \emph{i.e.,} 32-bits and 64-bits per cycle for int8 and bf16, respectively.
Similar to the dense design, the $B$ data are inserted in the first SST at each vertical chain and dedicated wires are used to propagate them vertically (Fig. \ref{fig:GEMM_2D_array_SSTs}b).
These dedicated wires are particularly important for the sparse design, since otherwise 4$\times$ more vertical wires would be required to use global routing compared to the dense design (in the case where \textit{only} non-dedicated wires are employed).

Besides sparse operation, the design in Fig. \ref{fig:GEMM_2D_array_SSTs}b supports also dense computation.
This is because dense computation might still be needed, even if all weights in DNNs are sparse.
For instance, in Transformer-based DNNs \cite{Attention_all_you_need_2017, BERT_2019, ViT_2020}, the QKV (Query, Key, Value) self-attention GEMM computations do not involve weights, and are typically computed as dense.
For dense operation, only one $B$ bank is sufficient at each vertical chain, \emph{e.g.,} $B_{x,0}$ (Fig. \ref{fig:GEMM_2D_array_SSTs}b).
However, multiplexing logic can be employed to 
utilize the remaining $B_{x,1}$, $B_{x,2}$, $B_{x,3}$ banks.
This is particularly important to ensure efficient utilization of on-chip memory resources, which leads to maximized data reuse and thus optimized energy efficiency \cite{Versal_vs_Stratix_FCCM_2024, Taka2023MaxEVA}.
Finally, for 1:3 sparsity, only banks $B_{x,0}$, $B_{x,1}$, $B_{x,2}$ are required, leaving bank $B_{x,3}$ unused. 
However, similar to the dense operation, additional multiplexing logic can be employed to reuse this bank when it comprises multiple BRAMs, which we do not explore in this work (see Sec. \ref{subsec:Sparse_GEMM_implementation} for implementation details).

The dynamic configuration among all sparsity levels 
is implemented in the control logic (using CLBs).
However, FPGA accelerators can be designed in a custom fashion depending on the sparsity of each DNN.
For instance, a specific DNN might require only 1:3 sparsity and dense computation across all of its layers.
The control and tiling logic for sparsity is similar to the dense design (Fig. \ref{fig:GEMM_2D_array_SSTs}a), showcasing a marginal increase in CLB resources (see Sec. \ref{subsec:Sparse_GEMM_implementation}).

\section{Evaluation}
\label{sec:SST_Slices:Evaluation}

\subsection{FPGA Architecture}
\label{subsec:FPGA_architecture}

The traditional FPGA architecture comprises 
CLBs, BRAMs, DSP slices and routing resources, \emph{i.e.,} connection and switch boxes (CBs and SBs).
We enrich the FPGA architecture with our proposed in-fabric SST slices.
The SST columns repeat every 15 FPGAs columns, occupying only 14\% of the total FPGA area, which is sufficient for DNN applications as also found in \cite{Aman_TS_TRETS_2022} (refer to Fig. \ref{fig:GEMM_SST_3x3_int8} for the layout of the proposed FPGA architecture).
We exploit the automated transistor sizing tool, COFFE \cite{3D_FPGAs_COFFE_Boutros_FPT23, COFFE2_TRETS_2019}, to model the delays and areas of the FPGA components. 
These delays and areas are subsequently used to describe the FPGA architecture in the Verilog-to-Routing (VTR) tool flow \cite{VTR_8_2020}, facilitating FPGA architecture exploration.
We utilize the 7nm FinFET ASAP7 predictive process design kit (PDK) \cite{ASAP7_2016}, for the SPICE simulations conducted by COFFE.
Specifically, we exploit the typical corner (\emph{TT}) ASAP7 model \cite{ASAP7_2016, ASAP7_invited_2017}. 
Finally, COFFE was run with cost function of $area \times delay$ for four transistor sizing iterations (2--4 iterations are typically sufficient 
\cite{Original_COFFE_2013}).

\begin{table}[t]
\centering
\caption{FPGA logic tile and routing parameters in COFFE.}

\resizebox{0.84\linewidth}{!}{
\begin{tabular}{cc|cc}
\Xhline{2.5\arrayrulewidth} 

\textbf{Parameter} & \textbf{Value} & \textbf{Parameter} & \textbf{Value} \\

\hline
\hline

FLEs per CLB ($N$) & 10 & Frac. LUT size ($K$) & 6\\
FLE independent inputs & 2 & Adders per FLE & 2\\
Channel width ($W$) & 300 & Wire length ($L$) & 4/16\\
CLB inputs ($I$) & 60 & CLB outputs ($O$) & 40\\
FLE outputs to routing ($O_{r}$) & 4 & Feedback FLE outputs ($O_{fb}$) & 2\\
SB flexibility ($F_{s}$) & 3 & Input connection flex. ($F_{cin}$) & 0.15\\
Output connection flex. ($F_{cout}$) & 0.1 & Input crossbar flex. ($F_{clocal}$) & 0.5\\

\Xhline{2.5\arrayrulewidth}

\end{tabular}
}

\label{tb:FPGA_tile_architectural_parameters}


\end{table}

We model a modern Agilex-like FPGA similar to \cite{TS_Aman_FPGA_2021, Aman_TS_TRETS_2022}.
Table \ref{tb:FPGA_tile_architectural_parameters} shows the FPGA logic (CLB) and routing (CBs and SBs) architectural parameters used in COFFE.
Each CLB contains 10 fracturable logic elements (FLEs), which consist of 6-input lookup tables (LUTs), registers and two adder bits.
Since COFFE does not model 7nm optimized memory circuits, we obtain the model of the 20Kbit BRAMs used in the recent work \cite{3D_FPGAs_COFFE_Boutros_FPT23} for 7nm FPGAs.
The delay and area values of these BRAMs are conservatively estimated using the 14nm Stratix 10 architecture values, as described in \cite{3D_FPGAs_COFFE_Boutros_FPT23}. 
Finally, we leverage the DSP used in \cite{Aman_TS_TRETS_2022}, which supports multiple precisions and modes, closely matching the commercial Agilex DSP \cite{Intel_Agilex_DSP_2020}.
However, since this DSP is modeled in 22nm, we scale down its area and delay values to 7nm, exploiting the scaling factors from \cite{Scale_VLSI_2017}.

\subsection{SST Slices Implementation}
\label{subsec:SST_slices_implementation}

We implement the SST slices utilizing the  ASAP7 7.5--track standard cell library \cite{ASAP7_invited_2017, ASAP7_github}.
Specifically, we use the regular threshold voltage (RVT) cells of the typical (\emph{TT}) corner of the library.
COFFE's hybrid flow was exploited, where the core of the SSTs is implemented using the standard cell flow, while the interface to the programmable routing (local crossbar, drivers for dedicated wires, etc.) is implemented in the full custom flow using SPICE simulations.
The standard cell flow uses Synopsys Design Compiler for synthesis, Cadence Innovus for PnR and Synopsys Primetime for timing analysis.
The full custom flow uses the \emph{TT} ASAP7 SPICE model and COFFE was run with the same configurations as in the previous section.

\begin{table}[t]
\centering
\caption{Area and frequency of SST \emph{vs.} SDT\_GIO slices (post PnR).}

\renewcommand{\arraystretch}{1.0}
\resizebox{0.74\linewidth}{!}{
\begin{tabular}{c|ccc}
\Xhline{2.5\arrayrulewidth}

\multicolumn{2}{c}{\textbf{In-fabric slice}} & \textbf{SST} & \textbf{SDT\_GIO}\\
\hline
\hline
& Standard-cell core & 4530.8 (+22.9\%) & 3687.8 \\

& Input crossbar & 1511.4 & 985.6\\

\textbf{Area} & Dedicated output routing & 40.2 & 0\\

\textbf{($\boldsymbol{\mu m^2}$)} & Switch Box (SB) & 1374.3 & 1603.4\\

& Connection Box (CB) & 734.9 & 571.6\\

& \textbf{Total tile} & \textbf{8191.6 (+19.6\%)} & \textbf{6848.4}\\

\hline

\textbf{Freq.} & 
int8 & 928.6 & 935.3 \\

\textbf{(MHz)} & bf16 & 838.1 & 850.5\\

\Xhline{2.5\arrayrulewidth}

\end{tabular}
}

\label{tb:SST_delay_area_vs_SDT_GIO}


\end{table}

To accurately quantify the area overhead of the sparsity enhancements described in Sec. \ref{subsec:Sparse_Processing_Element}, we remove the sparsity features of the SST slices, to implement a 2D systolic dense tensor slice.
We retain all the design features delineated in Sec. \ref{subsec:Sparse_Tensor_slices_architecture}, except the dedicated wires.
For this dense tensor slice, we utilize the approach proposed in \cite{TS_Aman_FPGA_2021, Aman_TS_TRETS_2022}, where all I/O ports are connected to global routing.
We refer to this tensor slice as SDT\_GIO, which is similar to \cite{TS_Aman_FPGA_2021, Aman_TS_TRETS_2022}.
The dense SDT\_GIO slices are used as the baseline comparison with our proposed sparse SSTs, while also allowing for quantification of the routing savings due to dedicated wires in SSTs 
(Sec. \ref{subsec:Dedicated_wires_benefits}).

Table \ref{tb:SST_delay_area_vs_SDT_GIO} shows the area and frequency results for both the SST and SDT\_GIO slices obtained from COFFE. 
At the standard-cell core level, we observe a low area overhead of 22.9\% for the sparsity enhancements of the SST slices.
For both SST and SDT\_GIO slices, we implement a 50\% populated input crossbar to enhance their routability inside the FPGA fabric \cite{COFFE2_TRETS_2019}. 
Each SST slice has 333 global routing inputs and 202 outputs, along with 256 dedicated inputs/outputs.
In contrast, the SDT\_GIO slice has 259 global routing inputs and 258 outputs, without dedicated I/Os.
Given the aforementioned global routing I/Os, the SST slices need to access 6 SBs, thus spanning 6  CLB (logic) tiles.
However, since more outputs are needed for the SDT\_GIOs (primarily due to absence of dedicated wires), they need to span 7 CLB tiles.
Therefore, when calculating the total tile area for both slices (standard-cell core and routing interface), we obtain a low area increase of 19.6\% for 
SSTs \emph{vs.} SDT\_GIOs (Table \ref{tb:SST_delay_area_vs_SDT_GIO}).
Finally, we observe that both slices can operate at high frequencies for the supported precisions ($>$ 838 MHz in all cases), similar to commercial 7nm FPGA hard blocks \cite{Versal_DSP_frequencies, Agilex_5_frequencies}.



\subsection{Wirelength Gains Utilizing Dedicated Wires}
\label{subsec:Dedicated_wires_benefits}

To quantify the benefits of the proposed dedicated wires in SSTs, we implement \textit{dense} GEMMs utilizing our SSTs slices (Fig. \ref{fig:GEMM_2D_array_SSTs}a) and compare them with \textit{dense} GEMMs using SDT\_GIOs slices, which support only global routing.
To this end, similar to the enhanced FPGA architecture with SST slices, we create an architecture that has SDT\_GIOs instead of SSTs. 
To facilitate the exploration, we implement a parametric Python script that generates the Verilog code for arbitrary GEMM sizes.
The Verilog designs are synthesized and implemented on the two aforementioned FPGA architectures (one with SST and one with SDT\_GIO slices), exploiting the VTR flow.
In all cases, we perform a 10 seed-sweep in VTR, and report the design that attains the maximum FPGA frequency.



\begin{figure}[tbp]
\centering
\subfloat[int8]{\includegraphics[width=0.48\linewidth]{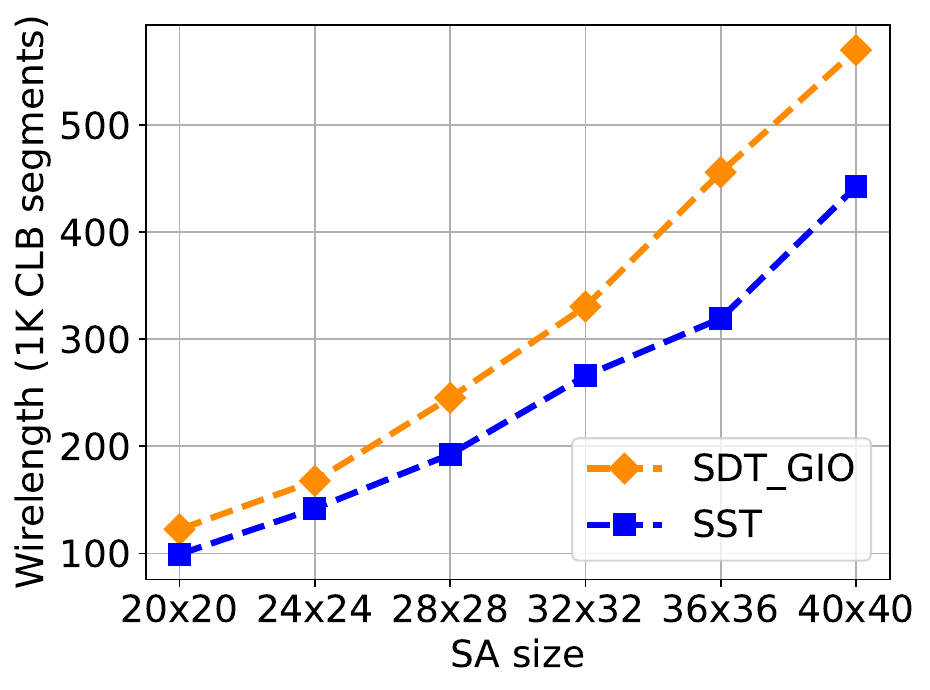}
\label{fig:wirelength_SST_vs_GIO_int8}}
\hfill
\subfloat[bf16]{\includegraphics[width=0.48\linewidth]{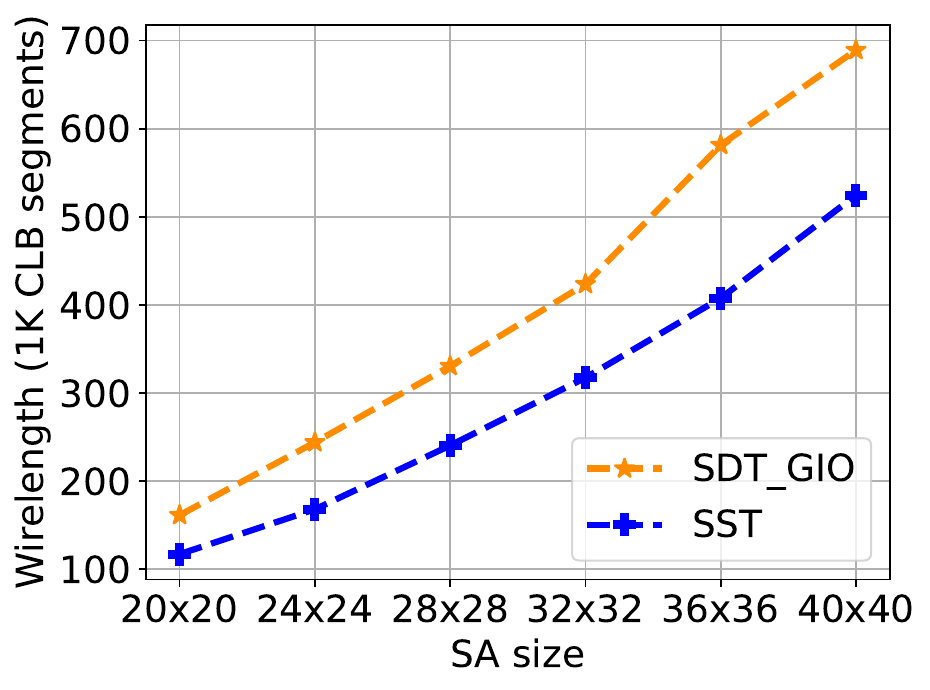}
\label{fig:wirelength_SST_vs_GIO_bf16}}

\caption{Routing wirelength comparison of \textit{dense} GEMM mapped to SST and SDT\_GIO slices, for various SA sizes.}
\label{fig:wirelength_SST_vs_GIO}
\end{figure}

Fig. \ref{fig:wirelength_SST_vs_GIO} illustrates the total routing wirelength of the two dense GEMM accelerators, for various SA sizes at both int8 and bf16 precisions.
These designs use 512-element deep banks for buffers $A$, $B$ and $C$ (Sec. \ref{subsec:Matrix_multiplication_mapping}), exploiting the 512x40-bit BRAM configuration mode (also found in recent Intel FPGAs \cite{Intel_Agilex_5_M20Ks, Intel_Agilex_7_M20Ks, Intel_Stratix_10_M20Ks}).
For the SA sizes shown in Fig. \ref{fig:wirelength_SST_vs_GIO}, we observe a significant wirelength reduction due to dedicated wires in the SSTs, ranging from 15.5--29.9\% for int8 precision, compared to the SDT\_GIO-based GEMM designs.
Similarly, for bf16, the wirelength reduction ranges from 23.9--31.2\%, showcasing the importance of vertical dedicated wires for 2D GEMM implementations on FPGA architectures.

Regarding the maximum attainable FPGA frequency, both dense GEMM implementations 
achieve high frequencies ranging from 668--731 MHz.
In particular, the smallest designs, \emph{i.e.,} 20x20 SA size, attain the highest frequency of $\sim$731 MHz for both SSTs and SDT\_GIOs at int8 and bf16 precisions.
Similarly, the largest designs, \emph{i.e.,} 40x40 SA size, 
both attain $\sim$668 MHz. 
We notice that dense GEMMs mapped to SST and SDT\_GIO slices attain almost similar FPGA frequencies ($<$1\% difference for all tested SA sizes depicted in Fig. \ref{fig:wirelength_SST_vs_GIO}).
This can be explained by the fact that the critical path is typically in the control logic implemented in CLBs, which is the same for both dense GEMMs mapped to SSTs and SDT\_GIOs. 


\subsection{Sparse SST-based GEMM Implementation}
\label{subsec:Sparse_GEMM_implementation}

In this section, we present the \textit{sparse} GEMM implementation shown in Fig. \ref{fig:GEMM_2D_array_SSTs}b, which allows dynamic configuration for all supported sparsity modes in SSTs, \emph{i.e.,} dense, 2:4, 1:3 and 1:4.
We use two baselines for our comparison.
First, we compare with a GEMM design mapped to SDT\_GIOs using a FPGA architecture that replaces SSTs with SDT\_GIOs. 
Since SDT\_GIOs have a 2D dense systolic dataflow that does not support sparse computation (Sec. \ref{subsec:SST_slices_implementation}), 
we retain the zeros of the sparse matrices, therefore operating similarly to dense. 
Hence, this implementation is similar to that shown in the previous section (Sec. \ref{subsec:Dedicated_wires_benefits}), but it has the same amount of on-chip memory as the sparse SST-based design. 
In particular, we implement four memory banks in the vertical dimension to feed each SDT\_GIO (similar to Fig. \ref{fig:GEMM_2D_array_SSTs}b).
This allows for the same data reuse opportunities as in the SST-based design, thus enabling a fair comparison.
Second, we implement the same sparse SST-based design, mapped to CLBs and DSPs of a traditional FPGA architecture, \emph{i.e.,} without SSTs and SDT\_GIOs.
Finally, we note that all tested designs utilize 512-element deep banks and are based on a 10 seed-sweep maximization of FPGA frequency, similar to Sec. \ref{subsec:Dedicated_wires_benefits}.

\subsubsection{FPGA Frequency}
\begin{figure}[tbp]
\centering
\subfloat[int8]{\includegraphics[width=0.48\linewidth]{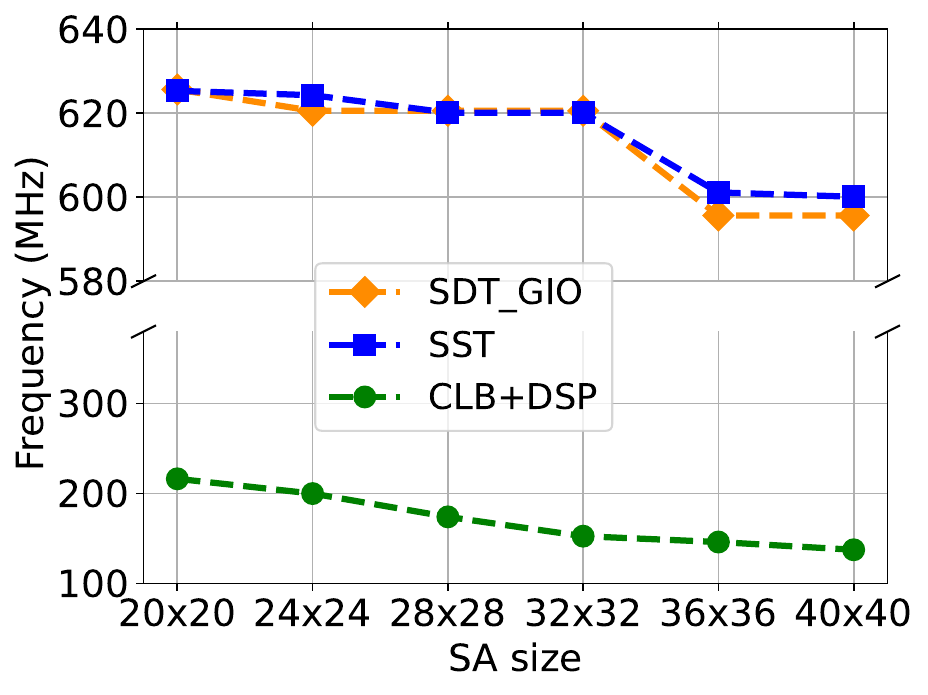}
\label{fig:freq_SST_GIO_DSP_int8}}
\hfill
\subfloat[bf16]{\includegraphics[width=0.48\linewidth]{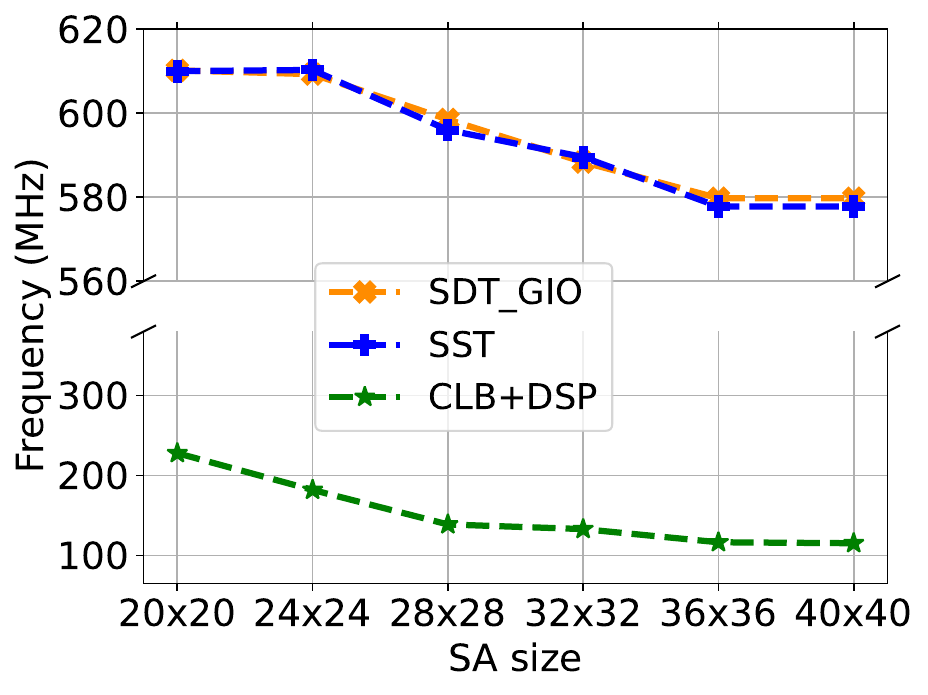}}
\label{fig:freq_SST_GIO_DSP_bf16}


\caption{FPGA frequency comparison of \textit{sparse} GEMM using SSTs \emph{vs.} SDT\_GIOs \emph{vs.} CLBs+DSPs, for various SA sizes.}
\label{fig:freq_SST_vs_GIO_vs_DSP}
\end{figure}

Fig. \ref{fig:freq_SST_vs_GIO_vs_DSP} depicts the FPGA frequency of the sparse GEMMs mapped to SST slices, SDT\_GIO slices as well as CLBs and DSPs.
First, we observe the high FPGA frequency of GEMMs mapped to SST and SDT\_GIO slices, ranging from 596--625 MHz for int8 and from 578--610 MHz for bf16. 
We note that the slightly lower frequencies of the SDT\_GIO-based GEMMs compared to the previous section are attributed to the higher BRAM usage.
For all tested SA sizes, we notice a negligible difference in FPGA frequency ($<$1\%) between the SST-based and the SDT\_GIO-based GEMMs. 
However, when comparing the SST-based \emph{vs.} the CLB+DSP GEMMs, we observe considerably higher frequencies, ranging from 2.8--4.4$\times$ and from 2.7--5$\times$ for int8 and bf16, respectively.

Second, note that the sparse GEMMs utilizing SSTs scale effectively when increasing the SA size.
In particular, we observe a minor decrease in FPGA frequency (4.6\% and 5.2\% for int8 and bf16, respectively), when considering the 40x40 \emph{vs.} the 20x20 SA size.
In contrast, the sparse GEMM mapped to CLBs and DSPs, does not scale as effectively, where we notice a substantial frequency drop of 36.6\% and 49.6\% for int8 and bf16, respectively.
These results emphasize the significant performance advantages and scalability of in-fabric sparse hard blocks compared to traditional FPGAs.





\subsubsection{GEMM Area}

\begin{figure}[tbp]
\centering
\subfloat[int8]{\includegraphics[width=0.48\linewidth]{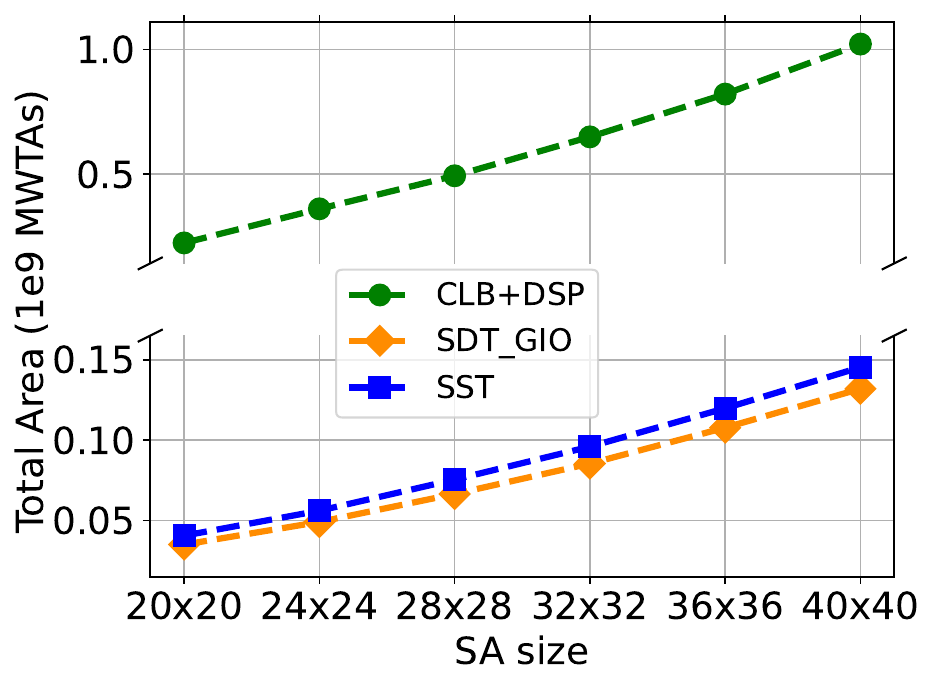}}
\label{fig:area_SST_GIO_DSP_int8}
\hfill
\subfloat[bf16]{\includegraphics[width=0.48\linewidth]{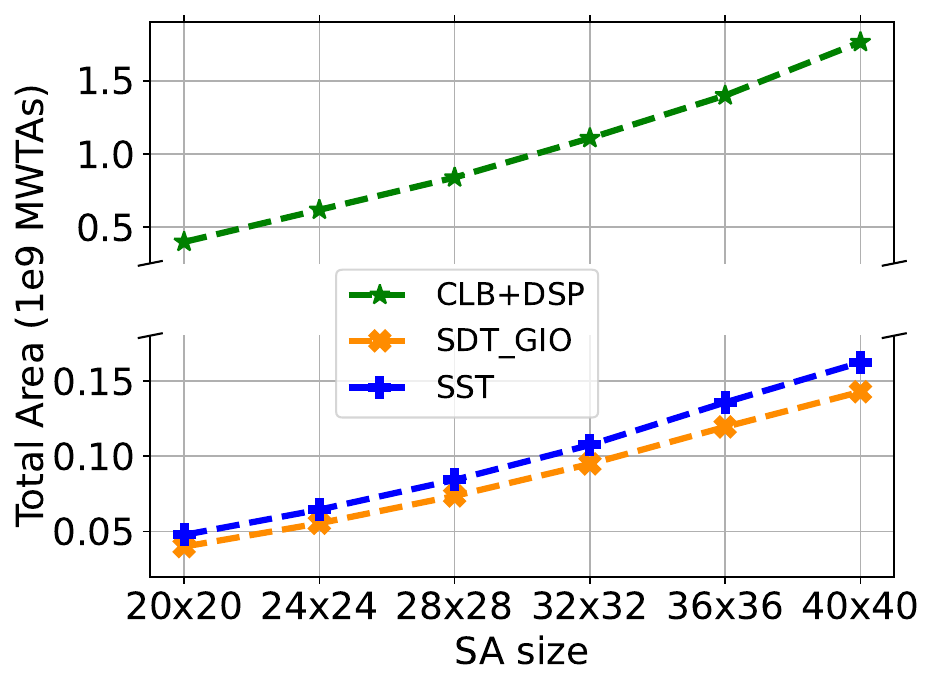}}
\label{fig:area_SST_GIO_DSP_bf16}

\caption{Total area comparison of various \textit{sparse} GEMM implementations utilizing SSTs \emph{vs.} SDT\_GIOs \emph{vs.} CLBs+DSPs.}
\label{fig:area_SST_vs_GIO_vs_DSP}

\end{figure}

We calculate the total area of our GEMM designs as the summation of the utilized logic area and the FPGA routing area.
While the utilized logic area is reported in VTR, the tool  does not report the used routing area.
Thus, we estimate the routing area by summing the area of all utilized multiplexers in both the SB and CB resources of our designs.
Fig. \ref{fig:area_SST_vs_GIO_vs_DSP} shows the total area of our sparse GEMM designs in minimum width transistor area (MWTA) units \cite{VTR_8_2020}.
As shown in Table \ref{tb:SST_delay_area_vs_SDT_GIO}, we observed a 19.6\% area overhead of the SSTs \emph{vs.} the SDT\_GIOs at the \textit{tile level}.
However, here we calculate this area overhead at the \textit{GEMM implementation level}, as this provides more accurate estimations for DNN accelerators.
When comparing the sparse GEMM mapped to SSTs \emph{vs.} the GEMM using SDT\_GIOs, we observe a small area increase ranging from 10.2--15.9\% for int8, and from 13.3--19.4\% for bf16, across all SA sizes.
We note that the lowest area increase, \emph{i.e.,} 10.2\% and 13.3\% for int8 and bf16, respectively, occurs for the highest, 40x40 SA size.
This is mainly attributed to the dedicated wires in SSTs, where the increase of the SA size (thus the total slices), leads to more routing area in the GEMM design using SDT\_GIOs. 

As shown in Fig. \ref{fig:area_SST_vs_GIO_vs_DSP}, we notice a substantial area increase of the sparse GEMM mapped to CLBs and DSPs \emph{vs.} using SSTs, of up to 7$\times$ for int8, and 10.9$\times$ for bf16.
Moreover, we observe that this area difference increases with the SA size.
In particular, for 20x20 SA size, we obtain a difference of 5.5$\times$ and 8.3$\times$ for int8 and bf16, respectively.
However, for 40x40 size, the difference increases to 7$\times$ and 10.9$\times$ for int8 and bf16, respectively.
Hence, our results exhibit the significant area efficiency and scalability of the proposed sparse GEMM designs using in-fabric SSTs, 
over traditional FPGAs.








\subsubsection{Resource Usage \& Efficiency of Sparse SST-based GEMM}

Table \ref{tb:GEMM_SST_vs_SDT_GIO_vs_DSP_40x40_SA} presents the resource usage and several evaluation metrics of the sparse GEMM designs, for SA size of 40x40.
First, we notice a small CLB increase for the SST \emph{vs.} the SDT\_GIO GEMM, of 26.2\% and 29.1\% for int8 and bf16, respectively.
This is ascribed to the additional control logic of including all supported sparsity levels, compared to only control logic for dense (since for SDT\_GIOs the zeros of the sparse matrices are retained, effectively operating similar to dense).
Second, note that the BRAM usage is the same in all GEMM designs, despite the additional bandwidth requirement due to the indices of the compressed sparse format (Sec. \ref{subsec:Matrix_multiplication_mapping}).
This is due to the use of the 512x40-bit BRAM mode (Sec. \ref{subsec:Dedicated_wires_benefits}). 
For example, for int8, buffer $A$ banks in the dense design of Fig. \ref{fig:GEMM_2D_array_SSTs}a, need a bitwidth of 4$\cdot$8=32-bits, which is smaller than the 40-bits width of the BRAMs.
However, for the sparse operation of Fig. \ref{fig:GEMM_2D_array_SSTs}b, they need additionally 8-bits for the indices (40-bits total), which is exactly the same as the BRAMs bitwidth.
Similar conclusions can be drawn for bf16.
Hence, no additional BRAMs are needed for the higher bandwidth requirements of sparse matrices in compressed format.

\begin{table*}[t]
 \centering
\caption{Evaluation and comparison of sparse GEMM designs using SSTs \emph{vs.} SDT\_GIOs \emph{vs.} CLBs+DSPs (SA size: 40x40).}
\vspace{-0.10cm}
\setlength\tabcolsep{2.8pt}
\renewcommand{\arraystretch}{1.15}
\resizebox{1.00\textwidth}{!}{
\begin{tabular}{c|c|c|c|c|c|c|c|c|c|c|c}
\Xhline{2.5\arrayrulewidth}


\multirow{2}{*}{\textbf{Pr.}} & \textbf{GEMM} & \multirow{2}{*}{\textbf{CLBs}} & \multirow{2}{*}{\textbf{BRAMs}} & \textbf{DSP} & \textbf{SST} & \textbf{SDT\_GIO} & \textbf{Freq.} & \textbf{Area} & \textbf{Wirelength} & \textbf{Eff. Thrpt.} & \textbf{Area Eff.}\\

& \textbf{Cfg.} &  &  & \textbf{slices} & \textbf{slices} & \textbf{slices} & \textbf{(MHz)} & \textbf{(GMWTA)} & \textbf{(CLB seg. units)} & \textbf{(TOPS)} & \textbf{(TOPS/GMWTA)}\\

\hline
\hline

\multirow{3}{*}{\rotatebox[]{90}{\textbf{int8}}} & SST  & 665 & 450 & 0 & 100 & -- & 601 &  0.145 & 493K & \textbf{7.69} & \textbf{53.03}\\

& SDT\_GIO & 527 & 450 & 0 & -- & 100 & 596 & 0.132 & 578K & 1.91 & 14.47\\

& CLB+DSP & 22650 & 450 & 400 & -- & -- & 138 & 1.022 & 7373K & 1.77 & 1.732 \\

\hline

\multirow{3}{*}{\rotatebox[]{90}{\textbf{bf16}}} & SST & 719 & 500 & 0 & 100 & -- & 578 &  0.162 & 684K & \textbf{7.40} & \textbf{45.68}\\

& SDT\_GIO & 557 & 500 & 0 & -- & 100 & 580 & 0.143 & 725K & 1.86 & 13.01\\

& CLB+DSP & 27065 & 500 & 2400 & -- & -- & 116 & 1.764 & 15212K & 1.48 & 0.839 \\

\Xhline{2.5\arrayrulewidth}

\end{tabular}
}
\label{tb:GEMM_SST_vs_SDT_GIO_vs_DSP_40x40_SA}


\end{table*}

Table \ref{tb:GEMM_SST_vs_SDT_GIO_vs_DSP_40x40_SA} shows the \textit{effective} throughput and area efficiency of our GEMM designs.
The \textit{effective} throughput (TOPS) is calculated on the GEMM \textit{native} size (40x40 in this case, see Sec. \ref{subsec:Matrix_multiplication_mapping}), and similar to \cite{Nvidia_accelerate_sparse_2021, NVIDIA_A100, AMD_CDNA_3} when acceleration can be attained from sparsity.
In particular, since 1:4 sparsity is included in our SST-based and CLB+DSP designs, their \textit{effective} throughput is increased by a factor of four.
When comparing the SST-based GEMM \emph{vs.} the SDT\_GIO-based design, we observe a throughput gain of 4.03$\times$ and 3.98$\times$ for int8 and bf16, respectively.
Moreover, we observe 3.66$\times$ (int8) and 3.51$\times$ (bf16) higher area efficiency (TOPS/GMWTA, area in Giga MWTAs).
In addition, when comparing with the GEMM using CLBs and DSPs, we obtain 4.34$\times$ (int8) and 5$\times$ (bf16) higher throughput for the SST-based GEMM.
Regarding the area efficiency, an immense 30.62$\times$ (int8) and 54.45$\times$ (bf16) difference is observed, primarily due to the increased usage
of the traditional FPGA resources.
To this end, notice the very high CLB and wirelength usage for the CLB+DSP \emph{vs.} the SST-based GEMM. 
For instance, for int8, there is a 34.1$\times$ higher CLB and a 15$\times$ higher wirelength usage.

\begin{figure}[tbp]
\centering
\includegraphics[width=0.75\linewidth]{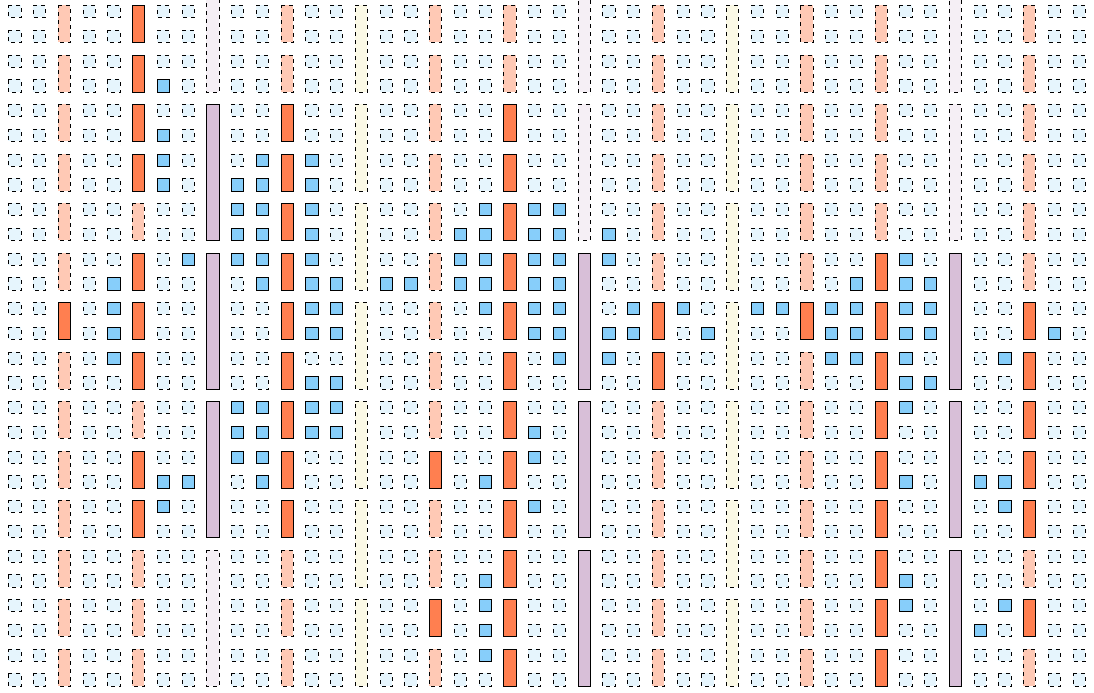}
\caption{SST-based GEMM implemented in VTR (SA size: 12x12). Blue: CLB, Orange: BRAM, Yellow: DSP, Purple: SST.}
\label{fig:GEMM_SST_3x3_int8}
\end{figure}

Finally, in Fig. \ref{fig:GEMM_SST_3x3_int8} we show a screenshot of a sparse SST-based design implemented in VTR.
Notice the physically contiguous vertical SST chains due to dedicated wires.
Also, observe the 2D \textit{physical} layout after VTR PnR, compared to the \textit{logical} 2D array of Fig. \ref{fig:GEMM_2D_array_SSTs}b.

\subsection{DNN Speedup Estimation Using SST Slices}
\label{subsec:Performance_estimation_DNNs}

We provide a speedup estimation for actual DNNs when leveraging our proposed SST slices.
In particular, we utilize the DeiT \cite{DeiT_2021} and ConvNeXt \cite{Convnext_2022} models, which attain state-of-the-art accuracy for vision tasks.
We apply two types of sparsity to determine the optimal sparsity configuration for each aforementioned models. 
First, we utilize \textit{uniform} sparsity, \emph{i.e.,} all layers have the same sparsity, by pruning weights based on their magnitudes \cite{Nvidia_accelerate_sparse_2021}.
Second, we exploit \textit{layer-wise} sparsity by adopting the neural architecture search (NAS) methodology proposed in \cite{Huang_2024_CVPR} to identify the optimal sparsity level of each layer, aiming to minimize accuracy degradation. 
Accuracy is evaluated on the ImageNet-1K dataset \cite{deng2009imagenet}, with int8 quantization applied using the post-training method proposed in \cite{PTQ4ViT_arixv2022}.

For speedup estimation, we develop an analytical model based on our results in Sec. \ref{subsec:Sparse_GEMM_implementation}, comparing the sparse SST-based GEMM to the SDT\_GIO-based GEMM, with the latter operating on dense matrices.
Our model exploits the matrix sizes for each layer of the DeiT and ConvNeXt models, and applies zero padding to align with the \textit{native} size of the GEMM accelerator (Sec. \ref{subsec:Matrix_multiplication_mapping}).
We utilize our largest implemented GEMM design, \emph{i.e.,} 40x40 SA size, and assume 100 GB/s bandwidth for DRAM modeling, which is typical in modern FPGAs \cite{Speedster_product_brief, VCK_5000}.
We note here that utilizing GEMM computations provides a reliable model for estimating speedup, since GEMM is the core computation in contemporary DNNs, accounting for more than 90\% of the total execution time \cite{adolf2016fathom, wang_gemm_breakdown}.
Other operations \emph{e.g.,} softmax, layernorm, can be effectively overlapped by DNN accelerators.
Moreover, GEMM-based estimation (convolutions implemented as GEMM for ConvNeXt) is also performed in multiple works targeting speedup calculations due to sparsity exploitation \cite{HighLight_MIT_2023, Vegeta_HPCA_2023, Sparse_tensor_GPUs_2019, STA_arxiv_2020, SDP_sparse_2023}.

 
Table \ref{tb:DNN_sparsity_results} presents the speedup estimation and accuracy for our two ViT models, \emph{i.e.,} DeiT-S and DeiT-B, and the ConvNeXt-S model. 
When applying uniform 2:4 sparsity, we observe a negligible accuracy decrease in DeiT-B for bf16 (0.07\%), or even slightly higher accuracy for int8 (by +0.05\%).
However, we notice a larger decrease for DeiT-S, \emph{e.g.,} 0.48\% for int8.
This is due to the fact that smaller models (DeiT-S with 4.7 GOPS) exhibit less redundancy compared to larger models (DeiT-B with 17.6 GOPS), making them more resilient to sparsity.
Note that even with uniform 2:4 sparsity across all layers, dense computation is still needed (Table \ref{tb:DNN_sparsity_results}), due to QKV computation in ViTs (Sec. \ref{subsec:Matrix_multiplication_mapping}).
In addition, for ConvNeXt-S, we retain its depth-wise convolution layers as dense, since we observed a severe accuracy degradation when attempting to sparsify them.
For instance, when applying 2:4 sparsity to all other layers except the (36) depth-wise layers in ConvNeXt-S, we obtain an accuracy degradation of only 0.47\%.
However, uniform 2:4 sparsity can result in significant speedup, \emph{e.g.,} 1.93$\times$ in ConvNeXt-S, 
compared to dense operation.
We note that both int8 and bf16 cases deliver nearly indistinguishable speedups over dense computation, since the frequency difference between the SST-based and SDT\_GIO-based GEMM accelerators  is less than 1\% (Sec. \ref{subsec:Sparse_GEMM_implementation}).


When applying layer-wise sparsity (all supported levels), we observe higher speedup in all cases, \emph{e.g.,} 2.35$\times$ in DeiT-B.
In this case, for DeiT-B, accuracy degradation is still negligible, \emph{e.g.,} 0.12\%. 
However, we observe higher degradation for DeiT-S (1.61\% for int8) and ConvNeXt-S (0.69\% for int8), since they are both small models. 
When applying higher sparsity (uniform 1:3 and 1:4), accuracy degrades even further, but higher speedup is attained, showcasing the accuracy-speedup trade-off.
If we constraint the accuracy degradation to be $\sim$1\% (acceptable in the vast majority of applications \cite{Accuracy_degradation_1_percent_NEURIPS2020}), the optimal speedup is attained for different sparsity configurations in DNNs (bolded in Table \ref{tb:DNN_sparsity_results}).
Specifically, for DeiT-S, uniform 2:4 is optimal (1.88$\times$ speedup), while for DeiT-B, uniform 1:4 shows best results (3.52$\times$ speedup).
For ConvNeXt-S, layer-wise sparsity (all supported levels) exhibits optimal results (2.07$\times$ speedup).
However, when allowing higher accuracy degradation (for instance, within 2\%) higher speedup can be attained, \emph{e.g.,} 3.63$\times$ for uniform 1:4 in ConvNeXt-S.
Finally, notice the higher weight memory reduction as sparsity increases (Table \ref{tb:DNN_sparsity_results}).
Our results demonstrate the importance of supporting multiple sparsity levels in the SST slices.


\begin{table*}[t]


\centering
\caption{Speedup estimation, accuracy and weight memory reduction of various DNNs when mapped to SSTs \emph{vs.} SDT\_GIOs.}
\setlength\tabcolsep{2.5pt}
\renewcommand{\arraystretch}{1.0}
\resizebox{1.00\textwidth}{!}{
\begin{tabular}{c|c|c|c|c|c|c|c|c|c||c|c}
\Xhline{2.5\arrayrulewidth}


\textbf{DNN} & \textbf{Sparsity} & \multicolumn{4}{c|}{\textbf{Number of layers}} & \textbf{In-fabric} & \multirow{2}{*}{\textbf{Speedup}} & \multicolumn{2}{c||}{\textbf{Top-1 accuracy (\%)}} & \multicolumn{2}{c}{\textbf{Weight red.}} \\

\cline{3-6} 
\cline{9-12} 

\textbf{model} & \textbf{levels} & \textbf{Dense} & \textbf{2:4} & \textbf{1:3} & \textbf{1:4} & \textbf{slice} & & \textbf{int8} & \textbf{bf16} & \textbf{int8} & \textbf{bf16} \\

\hline
\hline
 
 & dense & 48 & -- & -- & -- & SDT\_GIO & 1$\times$ & 79.56 (-0.00\%) & 79.79 (-0.00\%) & 1$\times$ & 1$\times$\\

\textbf{DeiT-S} & \textbf{[dense, 2:4]} & \textbf{--} & \textbf{48} & \textbf{--} & \textbf{--} & \textbf{SST} & \textbf{1.88$\times$} & \textbf{79.08 (-0.48\%)} & \textbf{79.20 (-0.59\%)} & \textbf{1.58$\times$} & \textbf{1.76$\times$} \\

\textbf{(4.7 GOPS)} & [dense, 2:4, 1:3, 1:4] & 3 & 26 & 16 & 3 & SST & 2.14$\times$ & 77.95 (-1.61\%) & 78.33 (-1.46\%) & 1.81$\times$ & 2.02$\times$ \\

& [dense, 1:3] & -- & -- & 48 & -- & SST & 2.61$\times$ & 73.36 (-6.20\%) & 73.95 (-5.84\%) & 2.32$\times$ & 2.61$\times$ \\

\hline

 & dense & 48 & -- & -- & -- &  SDT\_GIO & 1$\times$ & 81.40 (-0.00\%) & 81.80 (-0.00\%) & 1$\times$ & 1$\times$\\

\textbf{DeiT-B} & [dense, 2:4] & -- & 48 & -- & -- & SST & 1.91$\times$ & 81.45 (+0.05\%) & 81.73 (-0.07\%) & 1.59$\times$ & 1.77$\times$ \\

\textbf{(17.6 GOPS)} & [dense, 2:4, 1:3, 1:4] & 6 & 15 & 13 & 14 & SST & 2.35$\times$ & 81.28 (-0.12\%) & 81.60 (-0.20\%) & 2.05$\times$ & 2.26$\times$ \\

& [dense, 1:3] & -- & -- & 48 & -- & SST & 2.75$\times$ & 81.10 (-0.30\%) & 81.25 (-0.55\%) & 2.36$\times$ & 2.64$\times$ \\

& \textbf{[dense, 1:4]} & \textbf{--} & \textbf{--} & \textbf{--} & \textbf{48} & \textbf{SST} & \textbf{3.52$\times$} & \textbf{80.37 (-1.03\%)} & \textbf{80.69 (-1.11\%)} & \textbf{3.12$\times$} & \textbf{3.50$\times$} \\

\hline

& dense & 108 & -- & -- & -- & SDT\_GIO & 1$\times$ & 82.99 (-0.00\%) & 83.09 (-0.00\%) & 1$\times$ & 1$\times$\\

\textbf{ConvNeXt-S}& [dense, 2:4] & 36 & 72 & -- & -- & SST & 1.93$\times$ & 82.52 (-0.47\%) & 82.62 (-0.47\%) & 1.46$\times$ & 1.66$\times$ \\

\textbf{(8.7 GOPS)}& \textbf{[dense, 2:4, 1:3, 1:4]} & \textbf{36} & \textbf{59} & \textbf{7} & \textbf{6} & \textbf{SST} & \textbf{2.07$\times$} & \textbf{82.30 (-0.69\%)} & \textbf{82.39 (-0.70\%)} & \textbf{1.52$\times$} & \textbf{1.74$\times$} \\

& [dense, 1:3] & 36 & -- & 72 & -- & SST &  2.81$\times$ & 81.44 (-1.55\%) & 81.53 (-1.56\%) & 1.95$\times$ & 2.32$\times$ \\

& [dense, 1:4] & 36 & -- & -- & 72 & SST & 3.63$\times$ & 81.05 (-1.94\%) & 81.16 (-1.93\%) & 2.34$\times$ & 2.89$\times$ \\

\Xhline{2.5\arrayrulewidth}

\end{tabular}
}
\label{tb:DNN_sparsity_results}


\end{table*}


\subsection{Versal AIE-ML Sparsity Comparison}
\label{subsec:AIE_ML_comparison}

The new Versal AIE-ML \cite{AMD_AIE_ML_architecture_manual, AMD_AIE_ML_kernel_guide} includes out-of-fabric 
processors that support 2:4 sparsity.
Therefore, we aim to compare the attainable acceleration 
due to sparsity, as well as the efficiency of the compressed sparse format between our SSTs and the AIE-ML.
We exploit AMD's \textit{optimized} codes 
for dense and sparse GEMM kernels for the AIE-ML \cite{AMD_dense_GEMM_int8_github, AMD_AIE_user_guide_2023.2}, using 16-bits for the outputs as this leads to substantially higher compute utilization \cite{AMD_dense_GEMM_int8_github}.
We compile and simulate the AIE-ML designs using Vitis 2023.2 on the VEK280 platform \cite{AMD_VEK_280}.

\begin{figure}[tbp]

\centering
\subfloat[]{\includegraphics[width=0.48\linewidth]{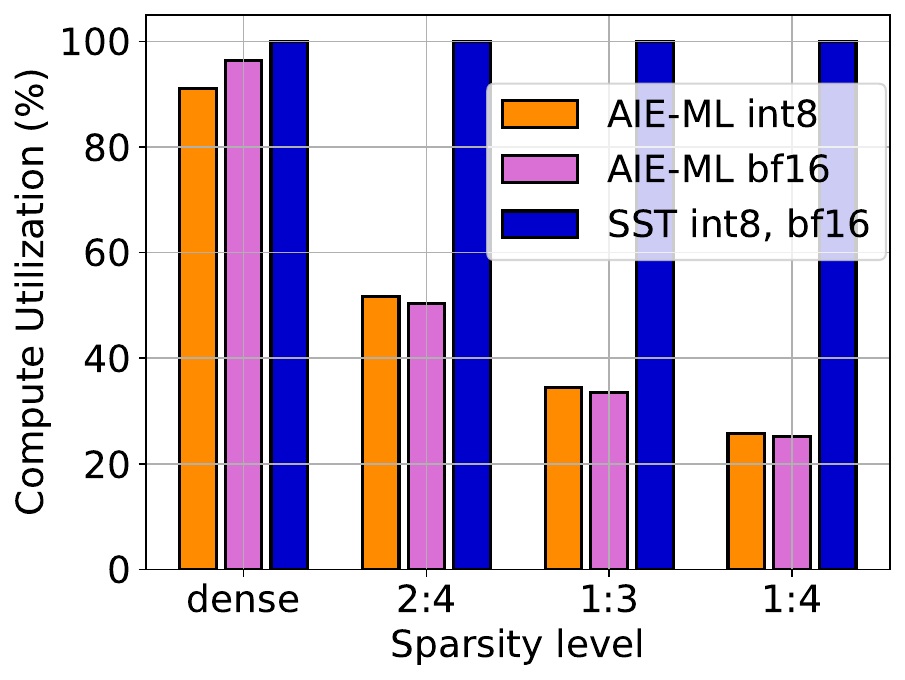}
\label{fig:compute_utilization_SST_vs_AIE_ML}}
\hfill
\subfloat[]{\includegraphics[width=0.455\linewidth]{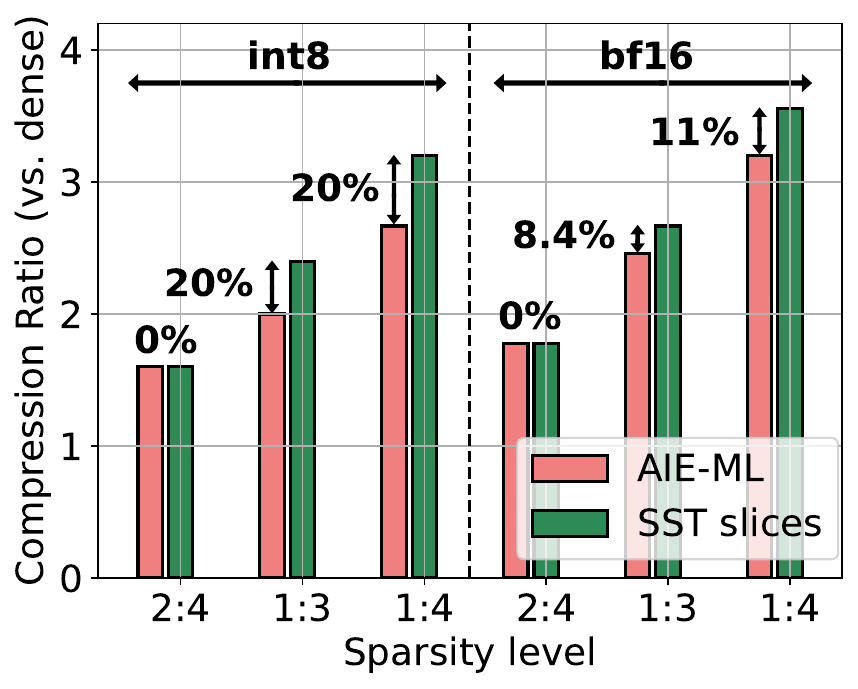}
\label{fig:compression_ratio_SST_vs_AIE_ML}}


\caption{Compute utilization (a) and compression ratio (b) comparison of a 64$\times$64$\times$64 GEMM using SSTs \emph{vs.} AIE-ML.}
\label{fig:utilization_compres_ratio_SST_vs_AIE_ML}
\end{figure}

Fig. \ref{fig:compute_utilization_SST_vs_AIE_ML} shows the compute utilization of a 64$\times$64$\times$64 GEMM, mapped to both SST slices and the AIE-ML.
We notice that the AIE-ML achieves higher than 90\% utilization in dense GEMM for both precisions. 
However, for 2:4 sparsity, we observe a substantially lower utilization, \emph{i.e.,} 51.6\% for int8 and 50.4\% for bf16.
We note that the compute utilization is calculated as the \textit{effective} utilization of only non-zero computation in the sparse case.
Thus, we obtain a marginal speedup of 13\% (int8) and 4.5\% (bf16), for the 2:4 sparse GEMM over the dense case for the AIE-ML.
Moreover, for higher sparsity, \emph{i.e.,} 1:3 and 1:4, the compute utilization in AIE-ML drops more significantly (Fig. \ref{fig:compute_utilization_SST_vs_AIE_ML}).
While higher than 2:4 sparsity can be stored in a compressed format in the AIE-ML, only 2:4 sparsity is inherently supported.
Consequently, no further acceleration can be achieved, leading to significantly decreased \textit{effective} utilization.

This marginal speedup due to sparsity in the AIE-ML is primarily attributed to limited vector load/store bandwidth, causing the sparse GEMM to become I/O bound
\cite{AMD_dense_GEMM_int8_github, AMD_AIE_ML_architecture_manual}.
In contrast, as depicted in Fig. \ref{fig:compute_utilization_SST_vs_AIE_ML}, our SST-based GEMM achieves 100\% utilization for all supported sparsity levels.
This is because we architect our SST slices in a fashion that guarantees the corresponding speedup of each sparsity level, \emph{e.g.,} 4$\times$ for 1:4 sparsity (refer to Sec. \ref{subsec:Sparse_Processing_Element} for design details).

Fig. \ref{fig:compression_ratio_SST_vs_AIE_ML} shows the compression ratios achieved by the SSTs \emph{vs.} the AIE-ML.
The AIE-ML uses a bitmap-based compressed format
\cite{AMD_AIE_ML_kernel_guide}, which is also used in various works accelerating structured sparsity \cite{S2TA_HPCA_2022, STA_arxiv_2020, N_M_sparse_transformers_FPGA_VLSI_2022, Fine_grained_Neural_ODE_FPGA_2023, LAMPS_FCCM_2024}.
Instead, we utilize an index-based format (Sec. \ref{subsec:Fine_grained_structured_sparsity}).
For 2:4 sparsity, both approaches present the same compression ratio (Fig. \ref{fig:compression_ratio_SST_vs_AIE_ML}).
However, for 1:3 and 1:4, our approach leads to higher compression ratio, up to 20\% for int8 and 11\% for bf16, showcasing the storage efficiency of the index-based format.
Moreover, as shown in \cite{N_M_sparse_transformers_FPGA_VLSI_2022}, bitmap achieves higher compression ratio over traditional sparse formats, \emph{e.g.,} CSR, CSC, COO \cite{Sparsity_Hoefler_2021}, for sparsity levels of 50--87.5\%.
This highlights that the supported index-based sparse format provides superior compression over other formats.

\subsection{Impact on Non-AI Benchmarks}
\label{subsec:Non_AI_benchmarks}

We demonstrate the flexibility of our enhanced FPGA architecture with SST slices, by 
exploring the maximum attainable frequency of non-AI benchmarks.
We select non-AI benchmarks from the VTR benchmark suite \cite{VTR_8_2020}, which target various application domains, \emph{i.e.,} $sha$, $mmcl$, $arm\_core$, $stereovision2$, $LU32PEEng$, $bgm$, $raygentop$, and $blob\_merge$  (see \cite{VTR_benchmarks} for domains).
When comparing our enhanced FPGA with a traditional FPGA on the aforementioned benchmarks, we observe, on average, a negligible decrease (\textbf{$<$1\%}) in maximum frequency. 
Hence, our proposed FPGAs maintain the performance and flexibility of traditional FPGAs, while offering highly efficient sparse and dense DNN acceleration (Sec. \ref{subsec:Dedicated_wires_benefits}--\ref{subsec:Performance_estimation_DNNs}).

\section{Insights \& Discussion}
\label{sec:SST_Slices:Insights}

\subsection{Very High Sparsity in DNN Layers}
\label{subsec:high_sparsity_DNN}

In the previous sections, we exhibit the importance and efficiency of our proposed in-fabric SST slices for accelerating sparse DNNs.
However, our SSTs support up to 75\% (1:4) structured sparsity.
Prior work \cite{EIE_ISCA_2016, Song_Han_learning_2015} has shown that a few DNN layers can have very high \textit{random} sparsity, \emph{e.g.,} $>$90\%. 
In this case, \textit{unstructured} sparse FPGA accelerators mapped to traditional CLB and DSP resources, \emph{e.g.,}
\cite{Fowers_sparse_FCCM_2014, Abhishek_sparsity_FPL_2020, unstructured_sparsity_CNN_FPGA_FCCM_2019}, can be exploited to efficiently compute these few high-sparsity layers.
All other DNN layers with less sparsity can be mapped to our SSTs (Sec. \ref{subsec:Sparse_GEMM_implementation} \& \ref{subsec:Performance_estimation_DNNs}), enabling both \textit{structured} and \textit{unstructured} sparse FPGA accelerators to operate synergistically, for maximized efficiency.
Moreover, for very high sparsity ($>$90\%), the CSR, CSC, COO formats utilized in unstructured sparse accelerators, offer the most efficient compression \cite{Sparsity_Hoefler_2021}.
Instead, for lower sparsity, the index-based format utilized in SSTs is the most efficient (Sec. \ref{subsec:AIE_ML_comparison}).
This emphasizes the custom flexibility of the enhanced FPGAs with in-fabric SST slices in supporting highly efficient sparse DNN accelerators.

\subsection{Efficient In-Fabric Slices}
\label{subsec:in_fabric_slices}

In-fabric slices have attained commercial success in 
AI-optimized FPGAs, \emph{e.g.,} the Intel tensor blocks \cite{Stratix_10_NX_FPGA_2021} and the new Intel AI-enhanced DSPs \cite{Sergey_Intel_TB_Agilex_5_FCCM_2024, Intel_Agilex_5_tensor_blocks}.
Moreover, as we showed in Chapter \ref{ch:Versal_vs_Stratix}, the comparison of the leading in-fabric (Intel Stratix 10 NX) to out-of-fabric (AMD Versal ACAP) AI-optimized FPGAs, resulted in $\sim$44\% higher energy efficiency in GEMM, favoring the in-fabric solutions. 
However, these efficient in-fabric slices primarily target dense AI acceleration, rendering them insufficient for most DNNs, which are sparse.
Instead, our proposed flexible SST slices support both dense and sparse computation (multiple sparsity levels), showcasing substantial advantages 
over dense in-fabric blocks and traditional FPGAs (Sec. \ref{subsec:Sparse_GEMM_implementation}), for actual sparse DNN workloads (Sec. \ref{subsec:Performance_estimation_DNNs}).
Note that although we are inspired by prior academic research that utilize dense 2D systolic in-fabric blocks \cite{TS_Aman_FPGA_2021, Aman_TS_TRETS_2022}, the methodology described in Sec. \ref{sec:SST_Slices:Architecture_Design} for introducing sparsity support can be generalized in straightforward fashion to other in-fabric FPGA blocks.

\section{Chapter Summary \& Impact}
\label{sec:SST_Slices:Sumamry}
Structured sparsity support has been incorporated in state-of-the-art GPUs, \emph{e.g.,} Nvidia H100 and AMD MI300, as well as the new Versal AIE-ML processors.
Although FPGA architectures have been enriched with in-fabric blocks for DNN acceleration, these blocks are primarily designed for dense operation.
This leads to insufficient computation for most contemporary DNN models, which display varying levels of sparsity.
To effectively address this challenge, we propose flexible in-fabric blocks, named SST slices, that support multiple levels of structured sparsity. 
We show that our sparse GEMM accelerators exploiting the SST slices offer substantial performance, scalability, and area advantages over traditional FPGAs. 
Demonstration on various state-of-the-art sparse DNN models utilizing our SST slices, exhibits up to 3.52$\times$ speedup with marginal accuracy degradation compared to dense in-fabric acceleration.

\chapter{Striking the Balance: GEMM Performance Optimization Across
Generations of Ryzen AI NPUs}
\label{ch:RyzenAI_NPUs}

In this chapter, we present the methodology we propose to optimize GEMM workloads across the two current NPU generations, XDNA and XDNA2 \cite{GEMM_NPU_Taka_FPGA_Poster,GEMM_NPU_Taka_arxiv}. 
As an evolution of the Versal AI Engine, these NPUs introduce new architectural constraints that require rethinking several of the optimization strategies developed in earlier chapters. Our methodology addresses key performance bottlenecks at the system level and provides significant insights for various aspects in GEMM optimization.

\section{GEMM Design \& Optimization}
\label{sec:RyzenAI_NPUs:GEMM_Design}
Our approach consists of a parametrized GEMM mapping that exploits the unique features of the NPU architecture, combined with an exhaustive parameter search based on an integer programming formulation of the performance characteristics of our chosen mapping.
In this section, we describe the details of both.

\subsection{Multi-Level Tiling Method}
\label{subsec:GEMM_tiling}

\begin{figure}[tbp]
\centering
\includegraphics[width=0.70\linewidth]{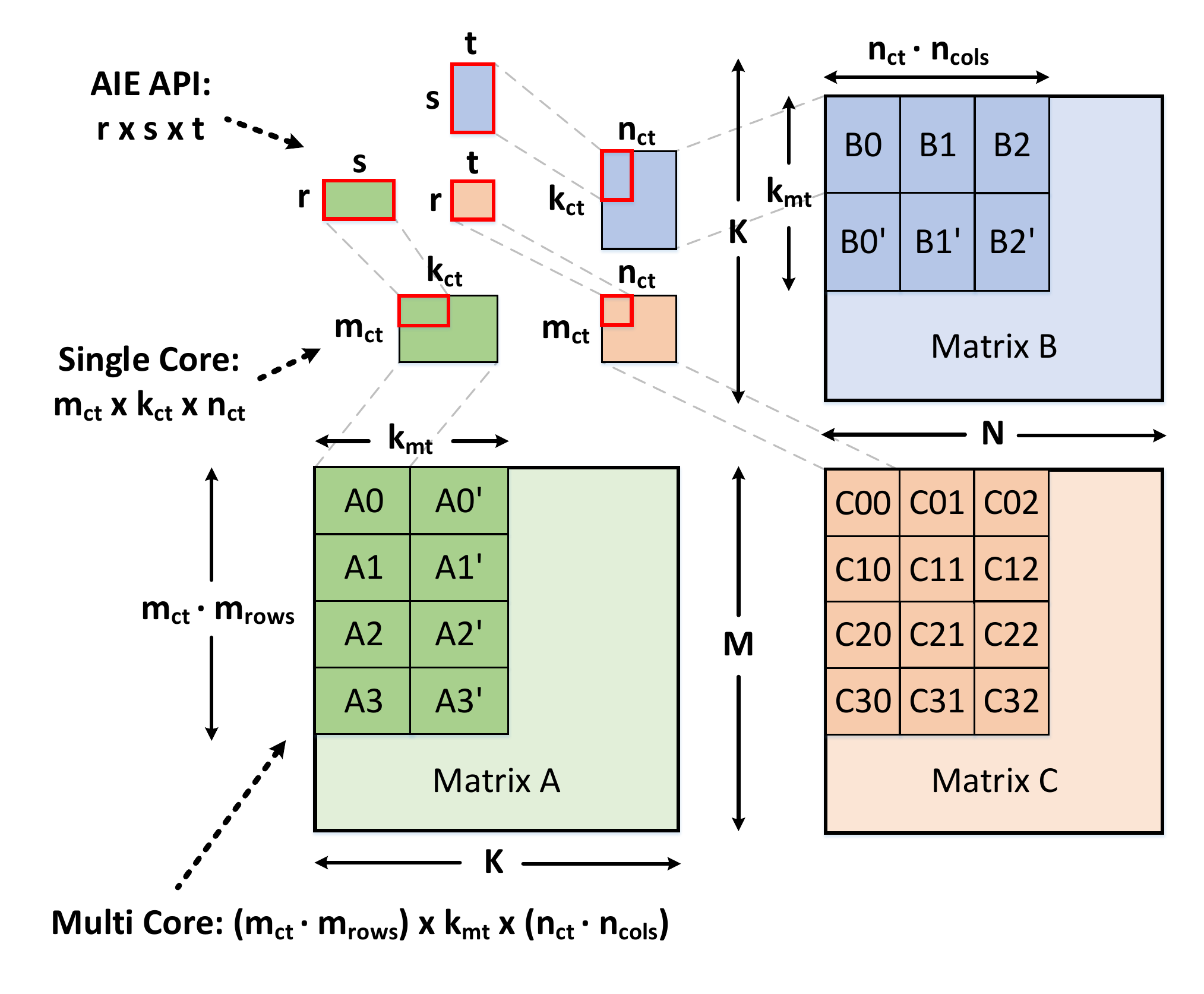}
\caption{Proposed GEMM multi-level tiling scheme.}
\label{fig:GEMM_tiling}
\end{figure}

Fig. \ref{fig:GEMM_tiling} illustrates the multi-level tiling scheme we use to partition the input and output matrices.
The inner-most (first) tiling level is defined by the supported shapes of the AIE API library for the single-core GEMM kernel \cite{AMD_AIE_API_user_guide_2025.1}, and is expressed via the parameters $r$ $\times$ $s$ $\times$ $t$.
The AIE API provides multiple optimized modes for each supported precision and enables portability across NPU generations.
The second tiling level is determined by the GEMM kernel that is supported out of local L1 memory, denoted as $m_\text{ct}$ $\times$ $k_\text{ct}$ $\times$ $n_\text{ct}$.
Since the NPU is a multi-core architecture, we partition GEMM across multiple cores to exploit spatial-level parallelism and maximize performance.
The third tiling level achieves this partitioning and corresponds to the GEMM size operating on the entire NPU array (parameters and mapping delineated in the next section).
Finally, the outer-most (fourth) tiling level is dictated by the final GEMM sizes of input matrices $A$ (typically \textit{activations}) and $B$ (typically \textit{weights}), and output matrix $C$, expressed as $M$ $\times$ $K$ $\times$ $N$.

\subsection{GEMM Mapping Strategy on NPUs}
\label{subsec:GEMM_mapping_strategy}


\subsubsection{\textbf{NPU Array Mapping \& Core Design}} 
\label{subsubsec:NPU_array_mapping}

\begin{figure}[tbp]
\centering
\includegraphics[width=1.00\linewidth]{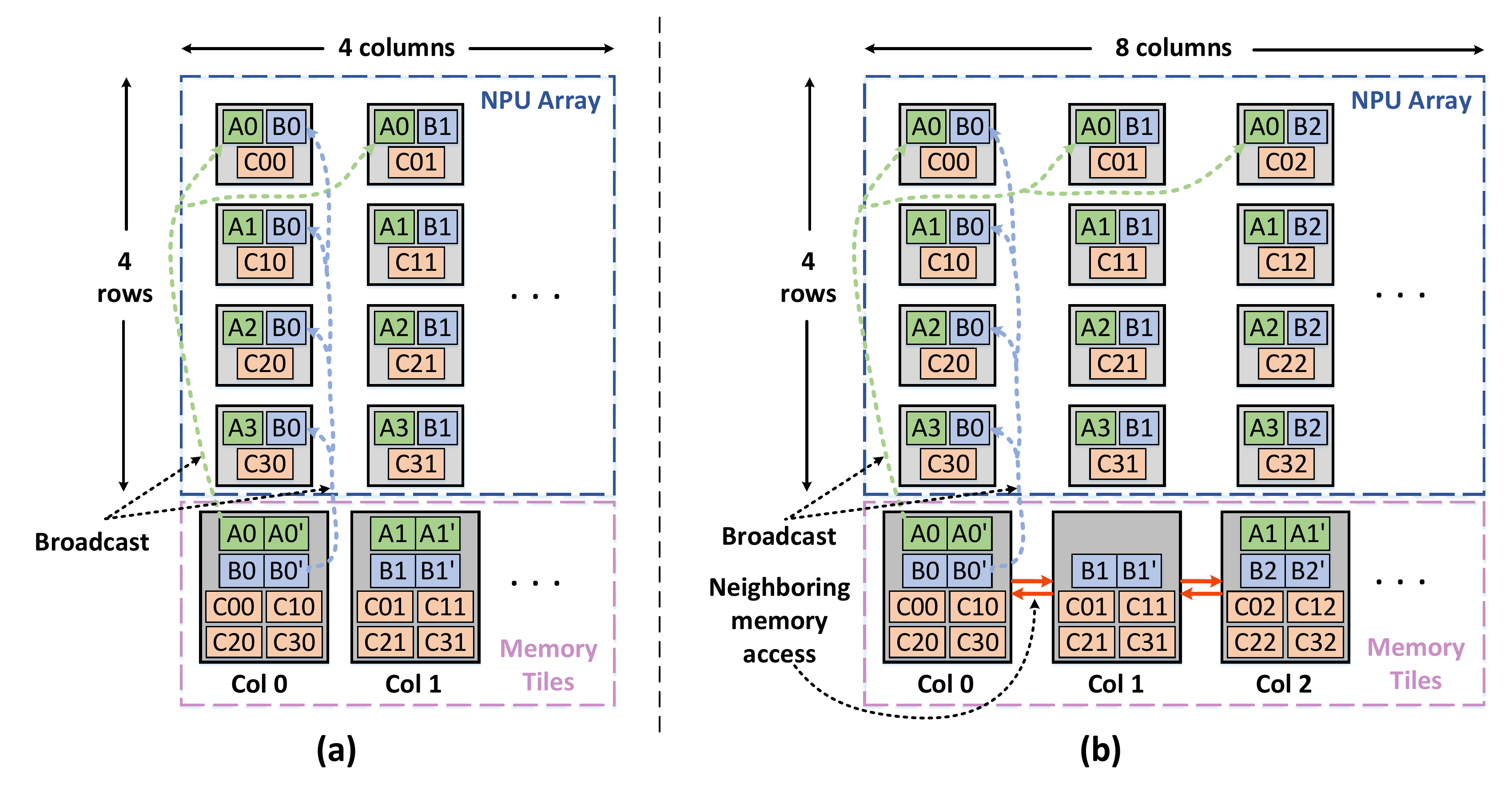}
\caption{GEMM mapping strategy on XDNA (a) and XDNA2 (b).}
\label{fig:GEMM_mapping}
\end{figure}

Our mapping strategy is to parallelize GEMM in \textit{space} across the $M$ and $N$ dimensions, while reduction across the $K$ dimension is performed in \textit{time}.
In this manner, all NPU cores perform the \textit{same} GEMM computation on different data and operate \textit{independently}.
This leads to maximized performance, since no data communication occurs between the cores, as opposed to Versal devices, where the reduction dimension is partitioned across multiple cores, \emph{e.g.,} \cite{Taka2023MaxEVA, GAMA_FPL_2025}.
By using the broadcast feature of the NPU architecture, we can attain spatial parallelization and exploit the inherent data reuse of the GEMM algorithm.
This parallelization is expressed via the number of single-core tiles in the $M$ and $N$ dimensions, and is defined by the parameters $m_\text{rows}$ and $n_\text{cols}$, respectively (Fig \ref{fig:GEMM_tiling}). 
In this work, we map the parameters $m_\text{rows}$ and $n_\text{cols}$ in a straightforward fashion to the number of rows and columns of the two NPU architectures, respectively.
In particular, we broadcast each input $A$ tile across one row in the NPU array, \emph{e.g.,} $A0$ in Fig. \ref{fig:GEMM_mapping}a and Fig. \ref{fig:GEMM_mapping}b for XDNA and XDNA2, respectively.
Similarly, each $B$ tile is broadcast across one column of NPU cores. 
Note that due to the absence of a ShimTile in the last column of XDNA, we choose to map GEMM across 4 rows and 4 columns for this architecture, similar to \cite{GEMM_example_mlir_aie, roesti2025unlocking, fang2025dato_paper}.
This enables a \textit{symmetric} 4 $\times$ 4 ($m_\text{rows}$ $\times$ $n_\text{cols}$) GEMM implementation, as depicted in Fig. \ref{fig:GEMM_mapping}a.
For XDNA2, we utilize the entire 4 $\times$ 8 array, which results in an \textit{asymmetric} mapping (Fig. \ref{fig:GEMM_mapping}b).


Each core performs a GEMM of $m_\text{ct}$ $\times$ $k_\text{ct}$ $\times$ $n_\text{ct}$ size.
To handle the reduction across $K$, the single-core kernel also loads previous partial results of $m_\text{ct}$ $\times$ $n_\text{ct}$ size, performs accumulation and stores the updated results back to the output tiles. 
For example, the upper-left core in Fig. \ref{fig:GEMM_mapping}b, sequentially loads tiles $A0$ and $B0$, $A0'$ and $B0'$, etc., in order to perform reduction.
In this fashion, the output $C$ tiles remain stationary in the L1 memory of each core (output stationary mapping).
When all tiles complete their accumulation ($K$/$k_\text{ct}$ tiles in total), the output tile is transferred to an L2 MemTile, and eventually to DRAM. The MemTile design is detailed in Sec. \ref{subsubsec:MemTiles_design}.
Subsequently, a fast vectorized kernel initializes the output $C$ tile to zero, preparing it for the next accumulation in GEMM tiling.

To overlap GEMM computation with DMA transfers of matrix tiles, we employ double-buffering on both L1 and L2 for the input tiles $A$ and $B$.
However, we retain the output $C$ tiles as a single buffer on each core.
Since the output tiles are transferred only once for each complete reduction, when $K$/$k_\text{ct}$ is sufficiently large, the infrequent added latency of a single-buffered output transfer becomes negligible.
At the same time, this design choice frees up valuable L1 memory, thereby enabling higher flexibility in tiling parameter optimization.
The larger tile sizes enabled by this additional memory ultimately lead to higher end-to-end GEMM performance in the general case (refer to Sec. \ref{subsubsec:single_output_buffer}).



\subsubsection{\textbf{Memory Tile Design}}
\label{subsubsec:MemTiles_design}

The input tiles are temporarily stored in L2 MemTiles before being broadcast to each NPU core.
Note that we load multiple input tiles across the reduction dimension $K$ into MemTiles, \emph{e.g.,} $A0$ and $A0'$ in Fig. \ref{fig:GEMM_mapping}a and \ref{fig:GEMM_mapping}b.
This is an important aspect of the mapping; 
loading multiple tiles allows us to access a larger amount of \textit{contiguous} data from DRAM, where matrices are stored in regular order (\textit{row-} and \textit{column-major}).
Such long contiguous reads result in increased DRAM BW utilization, which is essential for maximizing 
GEMM performance at the system level.
In this work, we retain matrices $A$ and $C$ in \textit{row-major} order, while matrix $B$ is either in \textit{row-} or \textit{column-major} order.
To this end, we introduce another parameter, $k_\text{mt}$, which specifies the size of the tiles loaded into L2.
In particular, for matrix $A$, each L2 MemTile loads a tile of $m_\text{ct}$ $\times$ $k_\text{mt}$ size.
Similarly, when $B$ is in \textit{column-major} order, each L2 MemTile loads a tile of $k_\text{mt}$ $\times$ $n_\text{ct}$ size.
Storing matrix $A$ in \textit{row-major} and $B$ in \textit{column-major} provides sufficient contiguous DRAM access for both matrices, which leads to higher GEMM performance (Sec. \ref{subsubsec:GEMM_sweeps}).
However, when $B$ is in \textit{row-major}, MemTiles load the same tile as CompTiles, \emph{i.e.,} $k_\text{ct}$ $\times$ $n_\text{ct}$, since contiguous 
data are accessible across the $n_\text{ct}$ dimension. 


Due to the 4 $\times$ 4 symmetric design of XDNA, each MemTile holds the same amount of input tiles, as illustrated in Fig. \ref{fig:GEMM_mapping}a.
Specifically, each MemTile broadcasts $B$ tiles within its own column.
The $A$ tiles are broadcast across the four rows; we map them in a regular fashion to the four MemTiles: the MemTile in column 0 holds $A0$ (which will be broadcast across row 0), the MemTile in column 1 holds $A1$ (which will be broadcast across row 1), etc.
In contrast, for XDNA2, we map the four $A$ tiles to eight MemTiles in an alternating pattern across even columns due to its 4 $\times$ 8 asymmetric design.
The mapping for $B$ remains the same.
In this design, even MemTiles hold more tiles than their odd counterparts, as depicted in the \textit{logical} view of Fig. \ref{fig:GEMM_mapping}b.
This particular mapping aids the IRON tool to leverage the NPU architectural feature of directly accessing the memory of the neighboring MemTile \cite{AMD_AIE_ML_architecture_manual}.
Therefore, when buffer sizes exceed the capacity of a specific MemTile, IRON \textit{physically} allocates buffers to a neighboring MemTile.

Observe that four output $C$ tiles are aggregated across each column into a MemTile (Fig. \ref{fig:GEMM_mapping}a and \ref{fig:GEMM_mapping}b).
This is because four $C$ tiles need to be transferred concurrently, while ShimTiles provide only two S2MM channels.
Hence, we exploit the six S2MM channels available in each MemTile \cite{AMD_AIE_ML_architecture_manual} to temporarily store the four output tiles, before they are transferred to DRAM via ShimTiles.

We define the \textit{native} GEMM size, as ($m_\text{ct}$ $\cdot$ $m_\text{rows}$) $\times$ $k_\text{mt}$ $\times$ ($n_\text{ct}$ $\cdot$ $n_\text{cols}$).
This corresponds to the GEMM size that operates natively on the entire NPU array, while also ensuring high performance.
Moreover, note that although we arbitrarily map matrix $A$ across rows and matrix $B$ across columns, the reverse mapping is equally feasible, yielding symmetrical solutions across the $M$ and $N$ dimensions.
Finally, although an input stationary mapping can also be employed, it would not be adequate to efficiently support arbitrary GEMM dimensions. 
Specifically, partial results would need to be temporarily stored in MemTiles, and subsequently reloaded to CompTiles for exploitation of data reuse in GEMM.
This would require three input channels for efficient GEMM computation, while CompTiles provide two inputs channels.




\subsection{On-The-Fly Tensor Transformations}
\label{subsec:DMA_transformations}

\begin{figure}[tbp]
\centering
\includegraphics[width=0.60\linewidth]{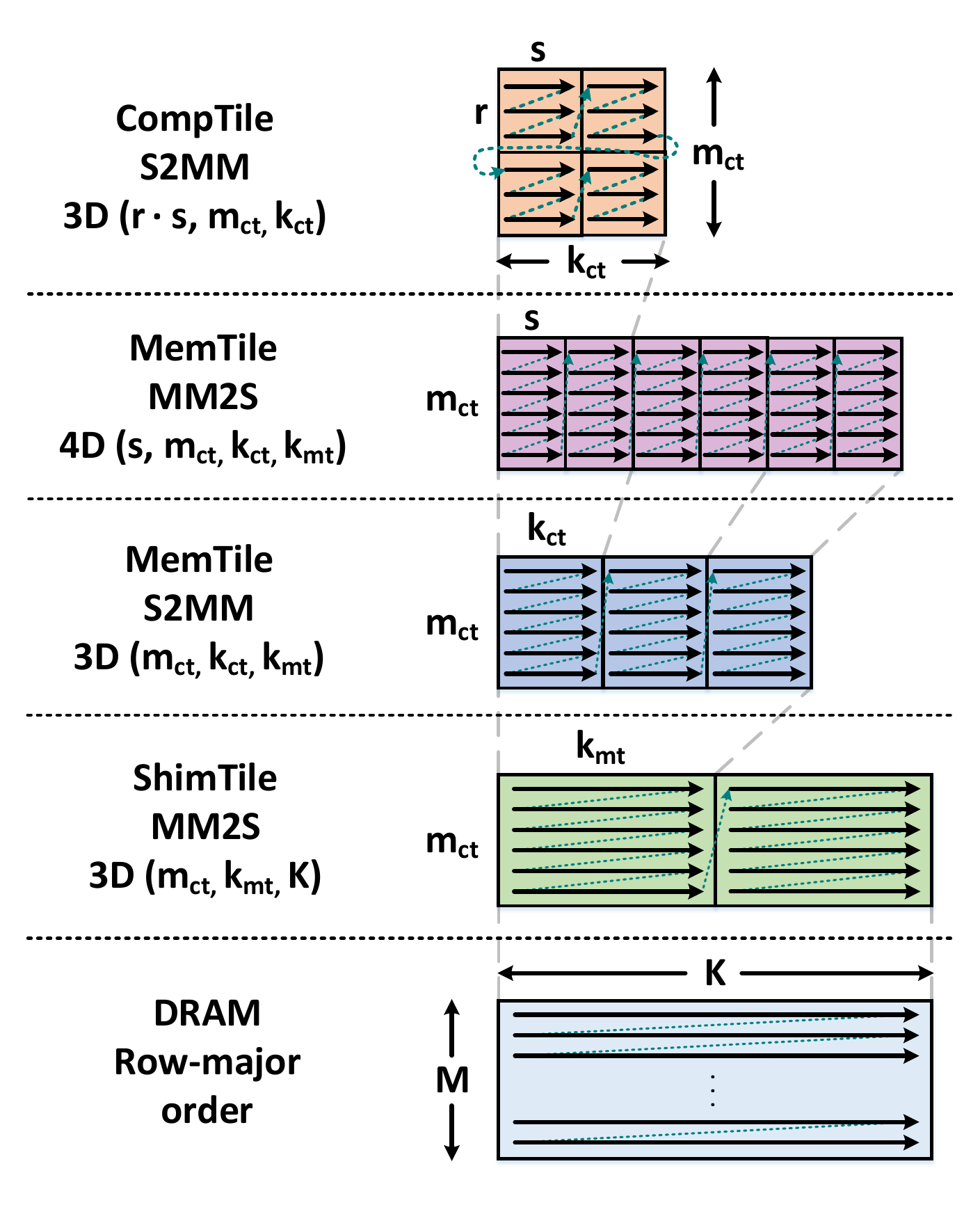}
\caption{On-the-fly DMA transformations for matrix A.}
\label{fig:DMA_transformations_mat_A}
\end{figure}

We extensively exploit the multi-dimensional addressing feature of DMAs 
to reorganize data into tiled layouts, as needed by the NPU cores.
This enables matrices to be stored in standard order in DRAM (\textit{row-} and \textit{column-major}).
The single-core GEMM kernels assume that matrices are pre-tiled \cite{AMD_AIE_API_user_guide_2025.1}.
In particular, the kernels running on each core expect $r\times s \times t$-sized tiles and both data \textit{within} tiles as well as the tiles \textit{themselves} to be in \textit{row-major} order, as illustrated in the upper part of Fig. \ref{fig:DMA_transformations_mat_A}, for the case of matrix $A$.
The aforementioned distribution of tiles across multiple cores, along with the requirement for contiguous DRAM accesses necessitates multiple data layout transformations, as explained below.

As matrices are transferred from DRAM to NPU, DMAs apply a series of transformations depending on the addressing capabilities of each NPU tile. 
Fig. \ref{fig:DMA_transformations_mat_A} shows the transformations of each DMA channel associated with the transfer of matrix $A$ tiles.
Initially, a ShimTile MM2S channel reads a tile of $m_\text{ct}$ $\times$ $K$ size from DRAM. 
Since ShimTiles support 3D addressing, the \textit{row-major} $m_\text{ct}$ $\times$ $K$ tile is transformed into multiple smaller $m_\text{ct}$ $\times$ $k_\text{mt}$  tiles via the MM2S channel (parameters: $m_\text{ct}$, $k_\text{mt}$, $K$).
Before each $m_\text{ct}$ $\times$ $k_\text{mt}$ tile is stored in the MemTile, another 3D transformation occurs at the S2MM MemTile channel, partitioning it into several $m_\text{ct}$ $\times$ $k_\text{ct}$ tiles (parameters: $m_\text{ct}$, $k_\text{ct}$, $k_\text{mt}$).

Each MemTile holds a $m_\text{ct}$ $\times$ $k_\text{mt}$ tile, which it sequentially transmits to the corresponding CompTiles as a series of smaller $m_\text{ct}$ $\times$ $k_\text{ct}$ tiles.
Since the CompTiles expect pre-tiled data, the data layout of each of the smaller tiles requires transformation.
Describing this transfer requires five parameters, namely $r$, $s$, $m_\text{ct}$, $k_\text{ct}$, $k_\text{mt}$.
However, MemTiles only support 4D addressing. 
To circumvent this, we decompose the transformation into two separate transformations, by utilizing the hardware's data layout transformation features in both the MM2S MemTile (output) channel and the S2MM CompTile (input) channel.
First, the MM2S MemTile channel partitions data into several $m_\text{ct}$ $\times$ $s$ tiles, as depicted in Fig. \ref{fig:DMA_transformations_mat_A} (parameters: $s$, $m_\text{ct}$, $k_\text{ct}$, $k_\text{mt}$).
This enables address \textit{linearization} within the $r$ $\times$ $s$ tile, thereby allowing a subsequent 3D transformation in the S2MM CompTile channel to reorganize data into the required layout
(\textit{effective} parameters: $r \cdot s$, $m_\text{ct}$, $k_\text{ct}$).

Address generation in the NPU DMAs occurs at 32-bit granularity \cite{AMD_AIE_ML_architecture_manual, AMD_AIE_ML_kernel_guide}.
DMAs alone cannot perform layout transformations at smaller-precision data types, \textit{e.g.,} individual elements of int8 or bf16 data types.
However, shuffle instructions running on the AIE cores can be utilized to support this operation and enable the fine-grained data swizzling.
In our application, this becomes relevant in use cases where an int8 or bf16 matrix $B$ is stored in \textit{column-major} order in DRAM.
To this end, we modify the GEMM kernel to utilize shuffling instructions, by using the AIE API transpose function \cite{transpose_AIE_API}, such that both data \textit{within} tiles and the tiles themselves are in \textit{column-major} order. 
Subsequently, we apply similar transformations to matrix $B$ as demonstrated for matrix $A$, across the NPU hierarchy (see Fig. \ref{fig:DMA_transformations_mat_A}).

Furthermore, when matrix $B$ is in \textit{row-major}, only one 4D transformation is required in the MemTiles (parameters: $s$, $t$, $k_\text{ct}$, $n_\text{ct}$), since each MemTile holds a $k_\text{ct}$ $\times$ $n_\text{ct}$ tile (Sec. \ref{subsec:GEMM_mapping_strategy}).
In a similar fashion, matrix $C$ tiles (\textit{row-major}) entail a single 4D transformation in the MemTiles (parameters: $r$, $t$, $m_\text{ct}$, $n_\text{ct}$).

\begin{figure}[tbp]
\centering
\includegraphics[width=0.70\linewidth]{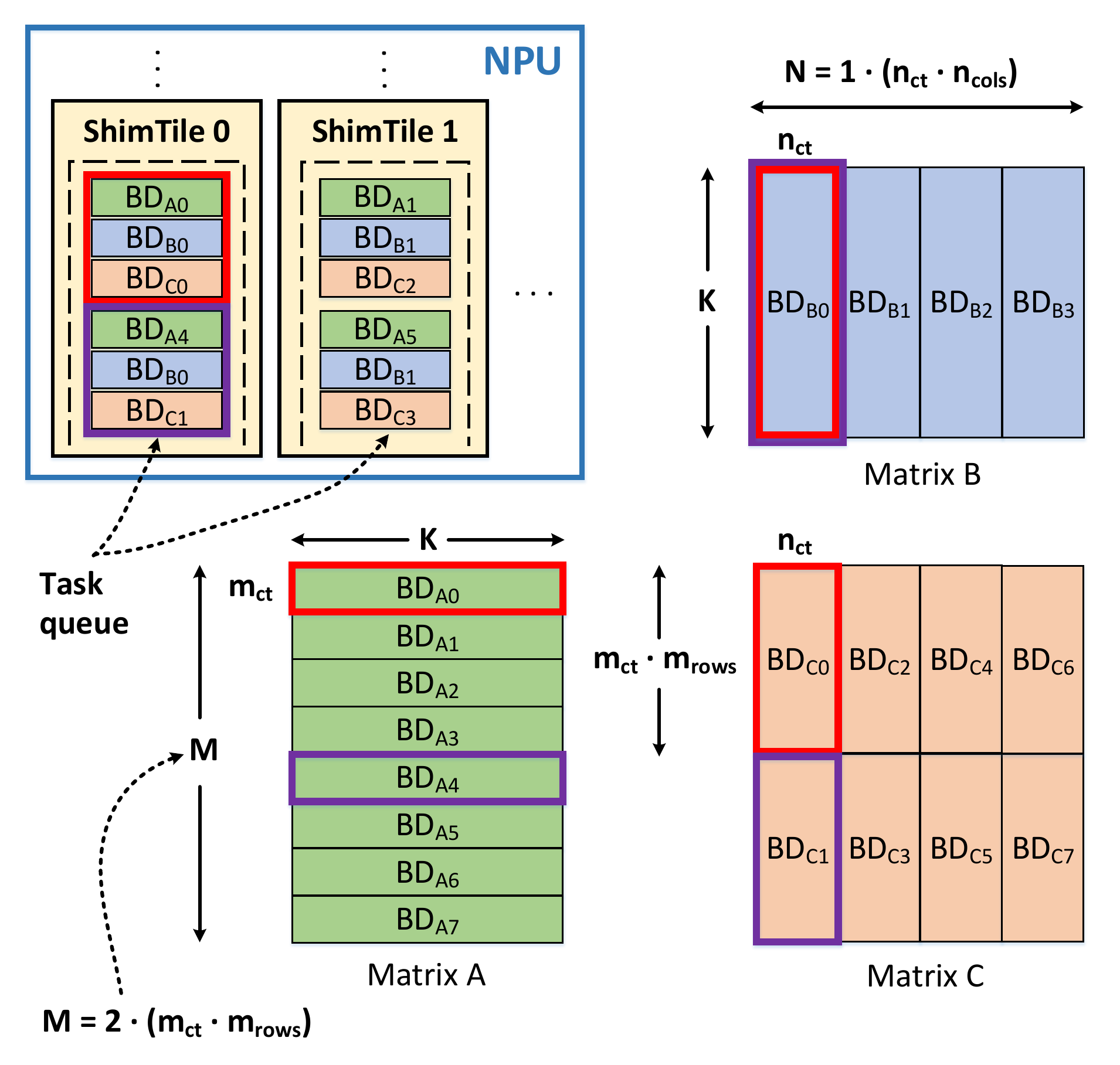}
\caption{Simplified view of outer-most (fourth) GEMM tiling level, determined by NPU--DRAM transfers ($m_\text{rows}$, $n_\text{cols}=4$).}
\label{fig:DRAM_NPU_data_movement}
\end{figure}

\subsection{Data Movement Between NPU \& DRAM}
\label{subsec:npu_main_memory_data_movement}

The outer-most (fourth) level of GEMM tiling encompasses data movement between the NPU and DRAM. 
The NPU interfaces with DRAM via ShimTiles and the control of data movement is orchestrated by the on-chip command processor.
ShimTiles are equipped with an input task queue to facilitate DMA transfers, where tasks are submitted sequentially.
When a task terminates, it issues a task-completion token, which the command processor uses to synchronize between multiple DRAM transfers.
Each ShimTile has access to 16 BDs \cite{AMD_AIE_ML_architecture_manual}, which are used to specify DMA transfers.
When a complex data movement pattern requires more than 16 BDs on a ShimTile, we can reuse (reconfigure) BDs. 
Before reconfiguring a BD, it is necessary to properly synchronize and ensure that the previous transfer associated with the BD has completed \cite{runtime_data_movement_mlir_aie}.

Fig. \ref{fig:DRAM_NPU_data_movement} illustrates a simplified view of the data movement between the NPU and DRAM.
Each BD is utilized to describe data movements in a fine-grained manner.
In particular, for matrix $A$, each BD defines one ShimTile DMA transfer of $m_\text{ct}$ $\times$ $K$ size.
Similarly, for matrix $B$, each BD defines a transfer of $K$ $\times$ $n_\text{ct}$. 
For matrix $C$, each BD describes a transfer of ($m_\text{ct}$ $\cdot$ $m_\text{rows}$) $\times$ $n_\text{ct}$, since $m_\text{rows}$ output tiles are aggregated per column.
The command processor program in our implementation maps and inserts BDs into ShimTile task queues as determined by the GEMM mapping strategy, \emph{e.g.,} $BD_{A0}$,  $BD_{B0}$, and $BD_{C0}$ to ShimTile 0, $BD_{A1}$, $BD_{B1}$, and $BD_{C2}$ to ShimTile 1, etc.
Moreover, $BD_{A4}$, $BD_{B0}$, and $BD_{C1}$ are also pushed into the queue of ShimTile 0, as dictated by GEMM tiling.
Similarly, our program enqueues BDs describing the GEMM transfers into the input task queues of each ShimTile in a sequential fashion.
We note that the simplified example in Fig. \ref{fig:DRAM_NPU_data_movement} is directly applicable to XDNA ($n_\text{cols}=4$, see Sec. \ref{subsec:GEMM_mapping_strategy}), while the BD mapping for XDNA2 can be determined in a straightforward fashion.

Depending on the target GEMM dimensions, \emph{i.e.,} $M$, $K$, $N$, more than the maximum of 16 BDs in each ShimTile might be required.
BD reconfiguration is needed in this situation, which might result in performance degradation (Sec. \ref{subsubsec:BD_reconfiguration}).
To address this challenge, we propose the following procedure, which enables DMA data transfers to overlap with BD reconfiguration.
Initially, we submit to the task queue five BDs for each of the three $A$, $B$, and $C$ DMA transfers.
This efficiently utilizes 15 out of the 16 BDs available in each ShimTile.
Each DMA transfer begins immediately once its associated BD reaches the front of the queue.
For instance, $BD_{A0}$ and $BD_{B0}$ transfers start first in Fig \ref{fig:DRAM_NPU_data_movement}.
Subsequently, the command processor waits for a task-completion token for each output transfer in a sequential manner, \emph{e.g.,} first for $BD_{C0}$, then for $BD_{C1}$, etc.
Notice that the command processor only needs to wait for completion of each BD associated with the output matrix, \emph{e.g.,} $BD_{C0}$, since once it completes, the corresponding input BDs, \emph{e.g.,} $BD_{A0}$ and $BD_{B0}$, have also finished.
Therefore, once each output BD completes, the three retired BDs can be safely reconfigured, and the next three BDs are inserted into the queue (if available).
This ensures that 15 BDs are in the queue in the steady-state operation, allowing DMA data movement to overlap efficiently with BD reconfiguration.
This process is repeated iteratively until GEMM tiling is completed.

The fine-grained BD description of NPU--DRAM data movement discussed above allows supporting very large GEMM dimensions.
GEMM dimensionality is limited by the NPU registers bitwidth utilized in multi-dimensional tensor addressing \cite{AMD_AIE_ML_kernel_guide}.
For instance, when storing $B$ in \textit{column-major}, the GEMM programming example in \cite{GEMM_example_mlir_aie} allows only a reduction $K$ dimension of up to $\sim$4K for bf16 on XDNA2, while our approach allows sizes $>$64K in all dimensions.



\subsection{Analytical Modeling Optimization}
\label{subsec:analytical_modeling_optimization}

To maximize GEMM performance, we propose an optimization methodology based on analytical modeling.
First, we focus on single-core performance to gain insights for the on-chip compute part, 
and afterwards we extend our methodology to optimize performance at the system level by incorporating off-chip DRAM BW constraints.

\subsubsection{\textbf{Single-Core GEMM Optimization}} 
\label{subsubsec:singe_core_optimization}

Our model utilizes architectural parameters, such as the peak compute throughput of the cores ($peak\_MACs$ in MACs/cycle) and the DMA BW ($DMA\_BW$ in Bytes/cycle), in order to identify the optimal $m_\text{ct}$, $k_\text{ct}$, $n_\text{ct}$ parameters that maximize performance.
We define the efficiency ($eff$) as the fraction of the attained compute throughput to the peak throughput of the core.
Eq. \ref{eq:compute_cycles} expresses the compute cycles of the GEMM kernel ($C_{comp}$), while Eq. \ref{eq:comm_A_cycles} and \ref{eq:comm_B_cycles} formulate the number of cycles needed for the DMA transfers of $A$ ($CA_{comm}$) and $B$ ($CB_{comm}$) tiles, respectively.
Here, $ty(\cdot)$ indicates the data type size (in Bytes) of the matrices.
\begin{gather}
C_{comp} = m_\text{ct} \cdot k_\text{ct} \cdot n_\text{ct}/ (eff \cdot peak\_MACs) \label{eq:compute_cycles}\\
CA_{comm} = m_\text{ct} \cdot k_\text{ct} \cdot ty(A)/ DMA\_BW \label{eq:comm_A_cycles}\\
CB_{comm} = k_\text{ct} \cdot n_\text{ct} \cdot ty(B)/ DMA\_BW
\label{eq:comm_B_cycles}
\end{gather}
Next, we define the constraint in Eq. \ref{eq:compute_bound_constr} to ensure that the single-core GEMM remains compute bound (not bounded by the DMA BW for $A$ and $B$ tiles).
Furthermore, Eq. \ref{eq:memory_constr_NPU} restricts the GEMM buffers to fit within the 64KB L1 memory capacity (with 1KB reserved for stack).
Finally, we also impose the straightforward constraint that $m_\text{ct}$, $k_\text{ct}$, $n_\text{ct}$ need to be multiples of $r$, $s$, $t$, respectively (not shown).
\begin{gather}
C_{comp} \geq \{CA_{comm},\ CB_{comm}\} \label{eq:compute_bound_constr} \\
\begin{split}
\{2 \cdot m_\text{ct} \cdot k_\text{ct} \cdot ty(A) + 2 \cdot k_\text{ct} \cdot n_\text{ct} \cdot ty(B) + m_\text{ct} \cdot n_\text{ct} \cdot ty(C)\} \leq 63\,\text{KB} \label{eq:memory_constr_NPU}
\end{split}
\end{gather}
The solution of $m_\text{ct}$, $k_\text{ct}$, $n_\text{ct}$ can be formulated as an integer programming (IP) optimization problem utilizing the aforementioned constraints.
The IP is solved exhaustively by setting the maximization of the number of MACs ($m_\text{ct}$ $\cdot$ $k_\text{ct}$ $\cdot$ $n_\text{ct}$) as the main objective.
This increases data reuse in GEMM, thereby maximizing the overall efficiency.
Furthermore, due to the output stationary GEMM mapping, we impose a second objective to minimize the output $C$ tile, \emph{i.e.,} the product $m_\text{ct}$ $\cdot$ $n_\text{ct}$.
This is essential in reducing the number of loads/stores for accumulations 
and decreasing memory stalls caused by bank conflicts.
Evidently, the two optimization objectives lead to increased $k_\text{ct}$ and reduced $m_\text{ct}$, $n_\text{ct}$ (sufficiently large to not become DMA BW bound).
Moreover, notice that we do not impose DMA constraints on the output $C$ buffer, as opposed to $A$ and $B$ (Eq. \ref{eq:comm_A_cycles}, \ref{eq:comm_B_cycles}), while also retaining $C$ as a single buffer (Eq. \ref{eq:memory_constr_NPU}). 
This significantly increases the search space, thus increasing performance in the \textit{general} GEMM case (Sec. \ref{subsubsec:single_output_buffer}), which is an essential aspect of the system-level optimization discussed below.


\subsubsection{\textbf{System-Level NPU Array Optimization}} 
\label{subsubsec:system_level_optimization}

First, we analytically express the DRAM accesses for each matrix in GEMM. 
Eq. \ref{eq:A_DRAM_footprint} captures the DRAM reads needed for matrix $A$ ($A_{mem}$).
The first term, $m_\text{ct} \cdot m_{\text{rows}} \cdot K \cdot ty(A)$, represents the DRAM reads (in Bytes) during GEMM tiling along the $K$ dimension.
This is due to the output stationary mapping and because  $A$ is broadcast across rows (Sec. \ref{subsec:GEMM_mapping_strategy}).
The second term, $N/(n_\text{ct} \cdot n_\text{cols})$, describes the repeat factor of the aforementioned read accesses due to tiling along the $N$ dimension. 
In a similar manner, the third term, $M/(m_\text{ct} \cdot m_\text{rows})$, captures the repeat across the $M$ dimension.
\begin{gather}
A_{mem} = 
\left(m_\text{ct} \cdot m_{\text{rows}} \cdot K \cdot ty(A)\right)
\left(\frac{N}{n_\text{ct} \cdot n_\text{cols}}\right)
\left(\frac{M}{m_\text{ct} \cdot m_\text{rows}}\right) \nonumber \\
\Rightarrow \
A_{mem} = M \cdot K \cdot N \cdot ty(A)/(n_\text{ct} \cdot n_\text{cols})
\label{eq:A_DRAM_footprint}
\end{gather}
Similarly, Eq. \ref{eq:B_DRAM_footprint} represents the DRAM reads for matrix $B$ ($B_{mem}$), while Eq. \ref{eq:C_DRAM_footprint} shows the DRAM writes for the output matrix $C$ ($C_{mem}$).
We note here that these equations provide an elegant and compact representation of data reuse and tiling scheme in GEMM.
\begin{gather}
B_{mem} = M \cdot K \cdot N \cdot ty(B)/(m_\text{ct} \cdot m_\text{rows}) \label{eq:B_DRAM_footprint} \\
C_{mem} = M \cdot N \cdot ty(C)
\label{eq:C_DRAM_footprint}
\end{gather}
Furthermore, Eq. \ref{eq:NPU_time} models the GEMM compute time on the NPU ($T_{comp}$), while Eq. \ref{eq:DRAM_time} expresses the total DRAM access time ($T_{mem}$).
Here, $peak\_TOPS$ denotes the \textit{theoretical peak} throughput of the NPU array, calculated at the maximum operating frequency.
Furthermore, $DRAM\_BW$ is the \textit{effective} DRAM BW achieved during GEMM execution on the NPU.
Note that, since all NPU cores execute the same GEMM kernel \textit{independently}, the single-core efficiency $eff$, as defined in Sec. \ref{subsubsec:singe_core_optimization}, directly corresponds to the entire NPU array efficiency in Eq. \ref{eq:NPU_time}.
\begin{gather}
T_{comp} = 2 \cdot M \cdot K \cdot N /(eff \cdot peak\_TOPS) \label{eq:NPU_time} \\
T_{mem} = (A_{mem} + B_{mem} + C_{mem})/ DRAM\_BW
\label{eq:DRAM_time}
\end{gather}
As discussed previously (Sec. \ref{subsubsec:singe_core_optimization}), minimizing $m_\text{ct}$ and $n_\text{ct}$, as well as maximizing $k_\text{ct}$ is essential to attain high efficiency, and thus maximized GEMM performance (equivalent to minimized $T_{comp}$).
However, from Eq. \ref{eq:A_DRAM_footprint} and \ref{eq:B_DRAM_footprint}, we notice that $n_\text{ct}$ and $m_\text{ct}$ parameters appear on the denominator of DRAM accesses for $A$ and $B$, respectively.
Therefore, the DRAM access time, $T_{mem}$, increases as $m_\text{ct}$ and $n_\text{ct}$ decrease.
This highlights the \textit{inverse} relationship between compute and memory time: \textit{as one increases, the other decreases}.
Thus, the optimal GEMM performance is attained at the \textit{balanced} point where they intersect (\emph{i.e.,} $T_{comp}$ $\approx$ $T_{mem}$).



In order to find the aforementioned optimal \textit{balanced} point we empirically set starting values for $m_\text{ct}$, $k_\text{ct}$, $n_\text{ct}$, and $eff$ parameters, and perform the iterative procedure explained below (starting values reduce iterations to typically $<$5).
These starting values are set based on the results of the single-core kernel optimization (Sec. \ref{subsubsec:singe_core_optimization}), and the effective $DRAM\_BW$ during GEMM execution (measured via micro-benchmarking, refer to Sec. \ref{subsubsec:optimal_balanced_GEMM_kernel}).
First, we measure the actual GEMM performance on the NPU device as well as the single kernel efficiency for the starting values, verifying that GEMM is memory bound ($T_{comp}$ $<$ $T_{mem}$, due to low $m_\text{ct}$, $n_\text{ct}$, and high $k_\text{ct}$).
In each iteration, we decrease the parameter $k_\text{ct}$ (as a multiple of $s$), and solve exhaustively an IP similar to the previous Sec. (\ref{subsubsec:singe_core_optimization}).
However, for that specific iteration, we fix the $k_\text{ct}$ parameter and the objective is to maximize the product of $m_\text{ct}$ $\cdot$ $n_\text{ct}$.
This leads to maximized $m_\text{ct}$, $n_\text{ct}$ values, given the DMA BW and L1 memory constraints, ensuring a highly optimized GEMM kernel (maximized number of MACs).
This is essential in identifying the optimal \textit{balanced} point, because it results in the smallest possible increment of $T_{comp}$ in each iteration, while also ensuring maximized possible GEMM performance given the specific parameters of that iteration.
We note here that the efficiency ($eff$) for a specific iteration is estimated based on each previous point measurements of each iteration.
We iteratively measure GEMM performance on the NPU device of the top-ranked solution for each IP, verifying that in each step performance is higher compared to the previous.
While GEMM performance is increasing in each step, at some point we observe lower performance, and the iteration stops.
Evidently, at this specific point GEMM has become compute bound ($T_{comp}$ $>$ $T_{mem}$), while also performance is lower.
Therefore, the \textit{balanced} point where GEMM performance is maximized has been identified ($m_\text{ct}$, $k_\text{ct}$, $n_\text{ct}$ parameters of the previous iteration).


\section{Evaluation}
\label{sec:RyzenAI_NPUs:Evaluation}

For the experimental evaluation, we use two representative mini PCs, corresponding to two NPU generations.
The Minisforum UM790 Pro \cite{Phoenix_mini_PC}, equipped with Ryzen 9 7940HS processor (\textit{Phoenix Point}) \cite{Phoenix_processor}, is used for XDNA.
For XDNA2, we use the ASRock 4$\times$4 Box \cite{Krackan_mini_PC}, featuring the AMD Ryzen AI 7 350 processor (\textit{Krackan Point}) \cite{Krackan_processor}.
Both mini PCs have dual-channel DDR5-5600 MT/s DRAM.
The CPUs run Ubuntu 24.04 LTS and serve as the host.
In particular, the CPUs allocate 
buffers in DRAM using Xilinx Runtime (XRT) \cite{xrt_github_repo}, configure the NPU, invoke the NPU for GEMM execution, and process the completion notification.
Finally, throughout all experiments, NPUs are configured at their maximum performance level, \emph{i.e.,} \textit{turbo} mode \cite{xrt_smi_commands}, utilizing XDNA driver commands \cite{xdna_driver_github_repo}.

\subsection{Single-Core GEMM Performance}
\label{subsec:single_core_GEMM_performance}


We exploit the AIE API \cite{AMD_AIE_API_user_guide_2025.1} to design highly optimized single-core GEMM kernels.
The kernels are compiled using the \emph{xchesscc} tool, employing various compiler directives, \emph{e.g.,} software pipelining and loop unrolling/flattening, to attain high efficiency.
Performance is assessed via hardware profiling utilizing the NPU trace unit \cite{AMD_AIE_ML_architecture_manual, tracing_mlir_aie}, thereby enabling cycle accurate measurements.
For int8, besides full output precision (int32), we also perform precision reduction to 16- and 8-bits, which is a common technique to increase GEMM performance in AIE architectures \cite{Vitis_tutorials, Charm_v2_2024, AMA_FPL_2024, GAMA_FPL_2025}.
Moreover, for bf16, we retain the output precision to bf16.
Finally, for XDNA2 we emulate the bf16 precision utilizing the bfp16 hardware datapath, which leads to increased GEMM performance (emulation attained via a specific \emph{xchesscc} flag at compile time, as mentioned in \cite{AMD_AIE_API_user_guide_2025.1}).

\begin{table}[t]
\centering
\caption{Single-core GEMM results for XDNA \& XDNA2.}
\setlength\tabcolsep{5.8pt}
\renewcommand{\arraystretch}{1.00}
\resizebox{0.70\linewidth}{!}{
\begin{threeparttable}
\begin{tabular}{c|c|c|c|c}

\Xhline{2.5\arrayrulewidth}

\multirow{2}{*}{\rotatebox[]{90}{\textbf{Dev.}}} & \textbf{Precision}  &  \textbf{Kernel Size} & \textbf{Throughput}  & \textbf{L1 Core}\\

& \textbf{In-Out}  &  \textbf{$m_\text{ct}$ $\times$ $k_\text{ct}$ $\times$ $n_\text{ct}$} & \textbf{MACs/cycle} & \textbf{Mem. (KB)}\\

\hline
\hline

\multirow{4}{*}{\rotatebox[]{90}{\textbf{XDNA}}} & int8-int8 & 64 $\times$ 232 $\times$ 64 & 233.0 & 62.0 (97\%)\\ 

& int8-int16 & 64 $\times$ 216 $\times$ 64 & 217.6 & 62.0 (97\%)\\ 

& int8-int32 & 48 $\times$ 280 $\times$ 48 & 192.0 & 61.5 (96\%)\\ 

& bf16-bf16 & 64 $\times$ 104 $\times$ 64 & 112.6 & 60.0 (94\%)\\ 

\hline

\multirow{4}{*}{\rotatebox[]{90}{\textbf{XDNA2}}} & int8-int8 & 64 $\times$ 232 $\times$ 64 & 450.6 & 62.0 (97\%)\\ 

& int8-int16 & 64 $\times$ 216 $\times$ 64 & 419.8 & 62.0 (97\%)\\ 

& int8-int32 & 48 $\times$ 280 $\times$ 48 & 384.0 & 61.5 (96\%)\\ 

& bf16-bf16 & 48 $\times$ 152 $\times$ 48 & 158.1 &   61.5 (96\%)\\ 

\Xhline{2.5\arrayrulewidth}

\end{tabular}


\end{threeparttable}
}

\label{tb:single_core_GEMM_results}

\end{table}



In Table \ref{tb:single_core_GEMM_results}, we present the top-ranked solutions of the single-core optimization procedure described in Sec. \ref{subsubsec:singe_core_optimization}.
For int8 precision, we attain very high throughput, ranging from 192.0--233.0 MACs/cycle and from 384.0--450.6 MACs/cycle for XDNA and XDNA2, respectively. 
Similarly for bf16, we achieve 112.6 MACs/cycle for XDNA, and 158.1 MACs/cycle for XDNA2.
Observe that in both cases, XDNA2 attains higher throughput compared to XDNA, since it has higher peak compute throughput capabilities per core.
We note here that, since performance is measured via hardware tracing, the results in Table \ref{tb:single_core_GEMM_results} include the inevitable memory stalls due to bank conflicts, thus reflecting the \textit{actual} NPU performance (when not bounded by 
DRAM BW).
Finally, very high L1 memory usage is achieved across all solutions, ranging from 94--97\%.


\subsection{GEMM Performance on NPU Array}
\label{subsec:npu_array_GEMM_performance}

In this section, we present GEMM performance on the entire NPU array.
Performance is evaluated using \textit{wall-clock} time, thereby capturing the \textit{actual} performance observed by the users (includes OS overheads, NPU dispatch time, etc.)  \cite{wall_clock_time_mlir_aie}.
All reported results represent the average of 100 runs.

\subsubsection{\textbf{Optimal Balanced GEMM Kernel}} 
\label{subsubsec:optimal_balanced_GEMM_kernel}


As shown in the previous section, GEMM kernels can attain very high throughput.
However, when using the optimum compute kernel sizes of Table \ref{tb:single_core_GEMM_results}, we observe that GEMM performance on the NPU array remains low.
For instance, for int8-int16 precision, we obtain only 17.86 TOPS on XDNA2 at $\sim$4K square GEMM size. 
However, the peak compute capability of this kernel on the XDNA2 array is 48.36 TOPS, when calculated at maximum operating frequency (identified through XDNA driver commands \cite{xdna_driver_github_repo}, \emph{i.e.,} 1 GHz and 1.8 GHz for XDNA and XDNA2, respectively).
This implies that GEMM is memory bound at this specific kernel size, due to low values of $m_\text{ct}$ and $n_\text{ct}$ (inverse relationship of compute and memory). 
Therefore, we leverage the optimization methodology described in Sec. \ref{subsubsec:system_level_optimization} in order to identify the optimal \textit{balanced} kernel, \emph{i.e.,} where compute and memory become balanced.
First, we use micro-benchmarking to estimate the \textit{effective} DRAM BW that is available to the NPU when running GEMM workloads.
For micro-benchmarking, we imitate GEMM transfers from DRAM to NPU array and vice-versa, observing $\sim$15 GB/s and $\sim$50 GB/s for XDNA and XDNA2, respectively.
Afterwards, we exploit these values to select starting points for the iterative optimization procedure of Sec. \ref{subsubsec:system_level_optimization}.


Tables \ref{tb:top_ranked_solutions_GEMM_XDNA_array} and \ref{tb:top_ranked_solutions_GEMM_XDNA2_array} present the two top-ranked solutions, for XDNA and XDNA2, respectively.
First, notice that kernel sizes have lower $k_\text{ct}$ and higher $m_\text{ct}$, $n_\text{ct}$ compared to kernels in Table \ref{tb:single_core_GEMM_results}.
This decreases their compute throughput; for example, 96$\times$112$\times$96 for int8-int16 (Table \ref{tb:top_ranked_solutions_GEMM_XDNA_array}) achieves 192.0 MACs/cycle, compared to 217.6 MACs/cycle of 64$\times$216$\times$64 (Table \ref{tb:single_core_GEMM_results}).
However, GEMM performance on the NPU array increases, since compute and memory become balanced.
For example, when using the 96$\times$112$\times$96 kernel, GEMM performance on XDNA for $\sim$4K GEMM size is 5.85 TOPS (Table \ref{tb:top_ranked_solutions_GEMM_XDNA_array})  compared to 4.03 TOPS of 64$\times$216$\times$64 (not shown in Table \ref{tb:single_core_GEMM_results}). 


Second, observe that when further increasing the product of $m_\text{ct} \cdot n_\text{ct}$ (thus decreasing $k_\text{ct}$ to fit in L1), GEMM performance on NPU array drops.
For instance, for int8-int8, kernel 160$\times$64$\times$144 has higher $m_\text{ct} \cdot n_\text{ct}$ product compared to 144$\times$72$\times$144 (22.5K \emph{vs.} 20.3K), but performance is lower: 36.13 \emph{vs.} 37.35 TOPS (Table \ref{tb:top_ranked_solutions_GEMM_XDNA2_array}).
This is because at this point GEMM has become compute bound, but compute has also become lower (throughput drops to 322.6 from 343.0 MACs/cycle).
The peak compute throughput on the entire XDNA2 array when attaining single-core throughput of 322.6 MACs/cycle is 37.16 TOPS.
However, the actual GEMM performance is 36.13 TOPS for $\sim$4K GEMM size, as shown in Table \ref{tb:top_ranked_solutions_GEMM_XDNA2_array}.
This $\sim$3\% difference is mainly attributed to DMA transfers of $C$ tiles (single output buffer), occurring at the end of each complete reduction across $K$ (every $K$/$k_\text{ct}$ tiles).
Another smaller contribution  comes from the vectorized zeroing kernel executing every $K$/$k_\text{ct}$ tiles (Sec. \ref{subsec:GEMM_mapping_strategy}), which is typically $<$10\% of GEMM kernel time (thus $<$0.15\% for the $\sim$4K GEMM of this particular example).
Both output DMA transfers and zeroing kernel cycles have been verified using NPU tracing.
Note that as $K$ becomes higher, these effects are further amortized.
Other contributions include the DRAM transfer time for initial $A$ and $B$ tiles, along with final output $C$ tiles, as well as overheads due to \textit{wall-clock} time measurements, \emph{e.g.,} NPU dispatch time \cite{wall_clock_time_mlir_aie}.
In summary, the bolded solutions across all precisions in Tables \ref{tb:top_ranked_solutions_GEMM_XDNA_array} and \ref{tb:top_ranked_solutions_GEMM_XDNA2_array} represent the optimal \textit{balanced} point between compute and memory, where end-to-end GEMM performance is maximized.

\begin{table}[t]
 \centering
\caption{Evaluation of two top-ranked solutions for XDNA across various data types (matrix $B$ in \textit{column-major}).}
\setlength\tabcolsep{3.0pt}
\renewcommand{\arraystretch}{1.2}
\resizebox{1.00\textwidth}{!}{
\begin{tabular}{c|c|c|c|c|c|c||c|c}
\Xhline{2.5\arrayrulewidth}

\textbf{Precision} & \textbf{Kernel Size} & \textbf{Product} & \textbf{Thrghpt.} & \textbf{L1 Core} & \textbf{L2 Total} & \textbf{Peak} & \textbf{GEMM Size} & \textbf{Actual} \\

\textbf{In-Out} & \textbf{\textbf{$m_\text{ct}$ $\times$ $k_\text{ct}$ $\times$ $n_\text{ct}$}} & \textbf{$m_\text{ct}$ $\cdot$ $n_\text{ct}$} & \textbf{MACs/cyc} & \textbf{Mem. (KB)} & \textbf{Mem. (KB)} & \textbf{TOPS} & \textbf{$M$ $\times$ $K$ $\times$ $N$} & \textbf{TOPS} \\

\hline
\hline

\multirow{2}{*}{\textbf{int8-int8}} & \textbf{112 $\times$ 112 $\times$ 112} & \textbf{12.3K} & \textbf{212.5}  & \textbf{61.3 (96\%)} & \textbf{980 (48\%)} & \textbf{6.80} & \textbf{4032 $\times$ 4032 $\times$ 4032} & \textbf{6.52}\\

 & 112 $\times$ 104 $\times$ 128 & 14.0K & 207.4 & 62.8 (98\%) & 1004 (49\%) & 6.63 & 4032 $\times$ 4160 $\times$ 4096 & 6.48\\

\hline

\multirow{2}{*}{\textbf{int8-int16}} & \textbf{96 $\times$ 112 $\times$ 96} & \textbf{9.0K} & \textbf{192.0} & \textbf{60.0 (94\%)} & \textbf{960 (47\%)} & \textbf{6.14} &  \textbf{4224 $\times$ 4032 $\times$ 4224} & \textbf{5.85}\\

 & 80 $\times$ 104 $\times$ 128 & 10.0K & 186.9 & 62.3 (97\%) & 996 (49\%) & 5.98 & 4160 $\times$ 4160 $\times$ 4096 & 5.75 \\

\hline

\multirow{2}{*}{\textbf{int8-int32}} & \textbf{80 $\times$ 88 $\times$ 96} & \textbf{7.5K} & \textbf{146.0} & \textbf{60.3 (94\%)} & \textbf{964 (47\%)} & \textbf{4.67} &  \textbf{4160 $\times$ 4224 $\times$ 4224} & \textbf{4.42}\\

 & 64 $\times$ 80 $\times$ 128 & 8.0K & 133.1 & 62.0 (97\%) & 992 (48\%) & 4.26 & 4096 $\times$ 4160 $\times$ 4096 & 4.09 \\

\hline

\multirow{2}{*}{\textbf{bf16-bf16}} & \textbf{96 $\times$ 56 $\times$ 96} & \textbf{9.0K} & \textbf{99.8}  & \textbf{60.0 (94\%)} & \textbf{960 (47\%)} & \textbf{3.19} &  \textbf{4224 $\times$ 4032 $\times$ 4224} & \textbf{3.12}\\

 & 96 $\times$ 48 $\times$ 112 & 10.5K & 97.3 & 60.0 (94\%) & 960 (47\%) & 3.11 & 4224 $\times$ 4032 $\times$ 4032 & 3.02 \\

\Xhline{2.5\arrayrulewidth}

\end{tabular}
}
\label{tb:top_ranked_solutions_GEMM_XDNA_array}


\end{table}

\begin{table}[t]
 \centering
\caption{Evaluation of two top-ranked solutions for XDNA2 across various data types (matrix $B$ in \textit{column-major}).}
\setlength\tabcolsep{3.0pt}
\renewcommand{\arraystretch}{1.2}
\resizebox{1.00\textwidth}{!}{
\begin{threeparttable}
\begin{tabular}{c|c|c|c|c|c|c||c|c}
\Xhline{2.5\arrayrulewidth}

\textbf{Precision} & \textbf{Kernel Size} & \textbf{Product} & \textbf{Thrghpt.} & \textbf{L1 Core} & \textbf{L2 Total} & \textbf{Peak} & \textbf{GEMM Size} & \textbf{Actual} \\

\textbf{In-Out} & \textbf{\textbf{$m_\text{ct}$ $\times$ $k_\text{ct}$ $\times$ $n_\text{ct}$}} & \textbf{$m_\text{ct}$ $\cdot$ $n_\text{ct}$} & \textbf{MACs/cyc} & \textbf{Mem. (KB)} & \textbf{Mem. (KB)} & \textbf{TOPS} & \textbf{$M$ $\times$ $K$ $\times$ $N$} & \textbf{TOPS} \\

\hline
\hline

\multirow{2}{*}{\textbf{int8-int8}} & \textbf{144 $\times$ 72 $\times$ 144} & \textbf{20.3K} & \textbf{343.0} & \textbf{60.8 (95\%)} & \textbf{2106 (51\%)} & \textbf{39.52} & \textbf{4032 $\times$ 4320 $\times$ 4608} & \textbf{37.35}\\

 & 160 $\times$ 64 $\times$ 144 & 22.5K & 322.6 & 60.5 (95\%) & 2064 (50\%) & 37.16 & 4480 $\times$ 4224 $\times$ 4608 & 36.13\\
\hline

\multirow{2}{*}{\textbf{int8-int16}} & \textbf{128 $\times$ 72 $\times$ 112} & \textbf{14.0K} & \textbf{307.2} & \textbf{61.8 (97\%)} & \textbf{2084 (51\%)} & \textbf{35.39} &  \textbf{4096 $\times$ 4320 $\times$ 4480} & \textbf{30.77}\\

 & 160 $\times$ 64 $\times$ 96 & 15.0K & 271.4 & 62.0 (97\%) & 2016 (49\%) & 31.26 & 4480 $\times$ 4224 $\times$ 4608 & 29.59 \\
\hline

\multirow{2}{*}{\textbf{int8-int32}} & \textbf{96 $\times$ 64 $\times$ 96} & \textbf{9.0K} & \textbf{256.0} & \textbf{60.0 (94\%)} & \textbf{2016 (49\%)} & \textbf{29.49} &  \textbf{4224 $\times$ 4224 $\times$ 4608} & \textbf{24.74}\\

 & 128 $\times$ 56 $\times$ 80 & 10.0K & 209.9 & 62.3 (97\%) & 2036 (50\%) & 24.18 & 4096 $\times$ 4032 $\times$ 4480 & 21.67 \\

\hline

\multirow{2}{*}{\textbf{bf16-bf16}} & \textbf{112 $\times$ 48 $\times$ 96} & \textbf{10.5K} & \textbf{137.2} & \textbf{60.0 (94\%)} & \textbf{2496 (61\%)} & \textbf{15.81} &  \textbf{4032 $\times$ 4224 $\times$ 4608} & \textbf{14.52}\\

 & 160 $\times$ 40 $\times$ 80 & 12.5K & 124.1 & 62.5 (98\%) & 2400 (59\%) & 14.30 & 4480 $\times$ 4160 $\times$ 4480 & 13.67 \\

\Xhline{2.5\arrayrulewidth}

\end{tabular}


\end{threeparttable}
}
\label{tb:top_ranked_solutions_GEMM_XDNA2_array}


\end{table}


We note that results in Tables \ref{tb:top_ranked_solutions_GEMM_XDNA_array}, \ref{tb:top_ranked_solutions_GEMM_XDNA2_array} correspond to matrix $B$ in \textit{column-major} (results for $B$ in \textit{row-major} are presented in Sec. \ref{subsubsec:GEMM_sweeps}). 
Across all results, the contiguous parameter $k_\text{mt}$, which is utilized to increase the effective DRAM BW  (and hence GEMM performance), is configured as shown below (Sec. \ref{subsubsec:contiguous_kmt_parameter}).
Finally, the optimization procedure requires less than 30 minutes to identify the optimum solution for each data type (evaluated on the corresponding mini PCs).
The overall execution time is dominated by the compilation time of \emph{xchesscc} and IRON (up to 5 minutes per iteration), while the exhaustive search takes less than 1 s in all cases. 



\subsubsection{\textbf{Contiguous $k_\text{mt}$ Parameter}} 
\label{subsubsec:contiguous_kmt_parameter}

\begin{figure}[t]
\centering
\subfloat[XDNA]
{\includegraphics[width=0.42\linewidth]{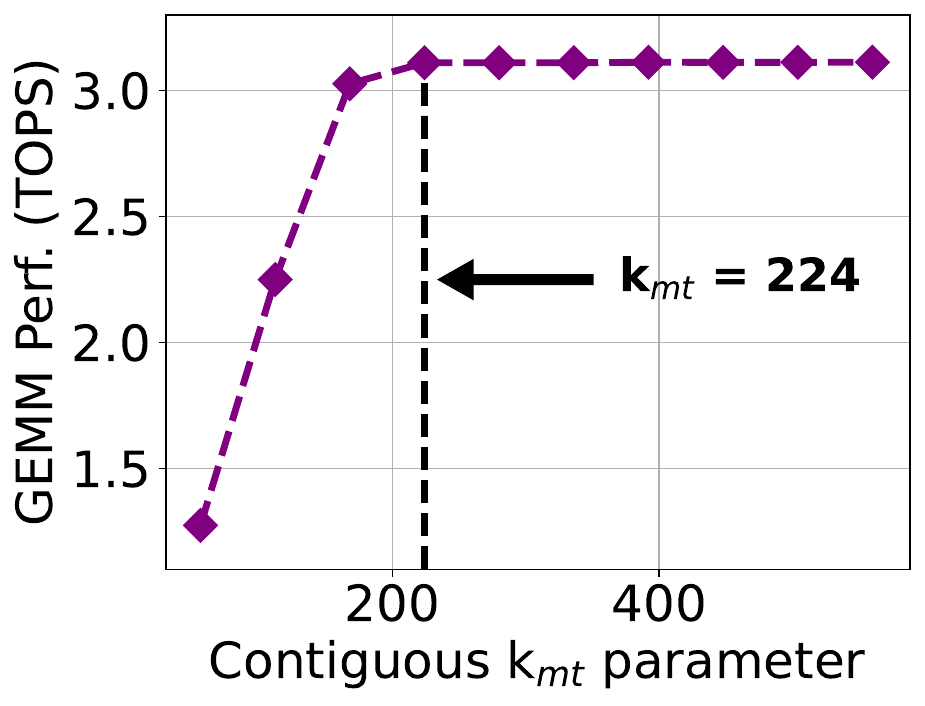}
\label{fig:contiguous_bf16_bf16_XDNA}}
\hfill
\subfloat[XDNA2]{\includegraphics[width=0.42\linewidth]{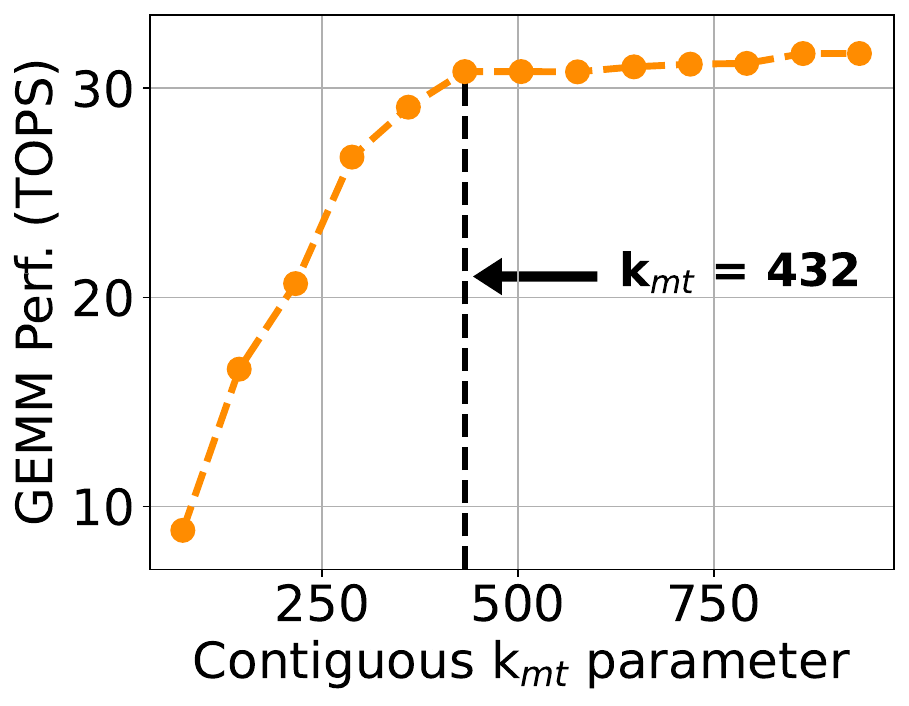}
\label{fig:contiguous_int8_int16_XDNA2}} 

\caption{GEMM performance while varying parameter $k_\text{mt}$ for bf16-bf16 96$\times$56$\times$96 (a) and int8-int16 128$\times$72$\times$112 (b).} 
\label{fig:GEMM_performance_vs_contiguous_param_k_mt}
\end{figure}

In this section, we present the impact of the contiguous $k_\text{mt}$ parameter on GEMM performance.
Fig. \ref{fig:GEMM_performance_vs_contiguous_param_k_mt} depicts the GEMM performance when varying the parameter $k_\text{mt}$ and utilizing the optimal \textit{balanced} kernels for two arbitrary data types ($\sim$4K GEMM size and $B$ in \textit{column-major}).
For instance, for XDNA, when $k_\text{mt}$ is equal to $k_\text{ct}$ ($=$ 56 in Fig. \ref{fig:contiguous_bf16_bf16_XDNA}), GEMM performance is very low, \emph{i.e.,} 1.27 TOPS.
As $k_\text{mt}$ increases (in multiples of $k_\text{ct}$) performance improves.
This is because it enables higher \textit{effective} DRAM BW, by traversing more contiguous elements for both $A$ (\textit{row-major}) and $B$ (\textit{column-major}) matrices. 
However, at some point, this enhancement becomes saturated, \emph{i.e.,} higher $k_\text{mt}$ provides marginal improvement.
Therefore, we \textit{empirically} select the smaller value where GEMM performance becomes saturated for each data type, \emph{i.e.,} $k_\text{mt}$=224 in Fig. \ref{fig:contiguous_bf16_bf16_XDNA}.
Note that selecting the smallest value while also maintaining high GEMM performance is essential, since it reduces any potential zero padding needed to align the final GEMM dimensions with the \textit{native} GEMM size (Sec. \ref{subsec:GEMM_mapping_strategy}).
For example, for the bf16-bf16 case, the \textit{native} GEMM size operating natively on the entire 4$\times$4 XDNA array is 384$\times$224$\times$384.
We note here that this results in reduced L2 memory utilization, ranging from 47--61\% across all data types (Tables \ref{tb:top_ranked_solutions_GEMM_XDNA_array} \& \ref{tb:top_ranked_solutions_GEMM_XDNA2_array}).
While maximized L2 memory usage can be attained for higher $k_\text{mt}$ values (\emph{e.g.,} 96\% for $k_\text{mt}$=560 in Fig. \ref{fig:contiguous_bf16_bf16_XDNA}), this only leads to marginal performance improvement in GEMM ($<$1\%).

Similarly, for XDNA2, for int8-int16, we set $k_\text{mt}$=432 (Fig. \ref{fig:contiguous_int8_int16_XDNA2}).
In this case, the \textit{native} GEMM size on the XDNA2 array becomes 512$\times$432$\times$896.
We note here that the three latest points of Fig. \ref{fig:contiguous_int8_int16_XDNA2} are enabled via the neighboring memory sharing in MemTiles, as a direct result of the GEMM mapping strategy on XDNA2 (Sec. \ref{subsubsec:MemTiles_design}).
For all other data types, we set the parameter $k_\text{mt}$ in a similar fashion. 
Specifically, for XDNA, we set $k_\text{mt}$ equal to 448 for int8-int8 and int8-int16, while for int8-int32 we set it to 352.
On XDNA2, we use 432 for int8-int8, and 384 for both int8-int32 and bf16-bf16.

It is important to mention here that the non-optimized GEMM example in \cite{GEMM_example_mlir_aie}, cannot support sufficient contiguous elements.
For a fair comparison, we utilize our optimized \textit{balanced} kernels and modify their implementation to support single output $C$ buffers (allowing it to fit in L1).
For instance, for the data types shown in Fig. \ref{fig:GEMM_performance_vs_contiguous_param_k_mt}, we attain 2.4$\times$ and 3.6$\times$ higher performance for XDNA and XDNA2, respectively.
These results highlight the importance of accessing sufficient contiguous elements in GEMM performance.





\subsubsection{\textbf{Roofline GEMM Performance Sweeps}} 
\label{subsubsec:GEMM_sweeps}



\begin{figure}[tbp]
\centering
\subfloat[int8-int8 XDNA]
{\includegraphics[width=0.45\linewidth]{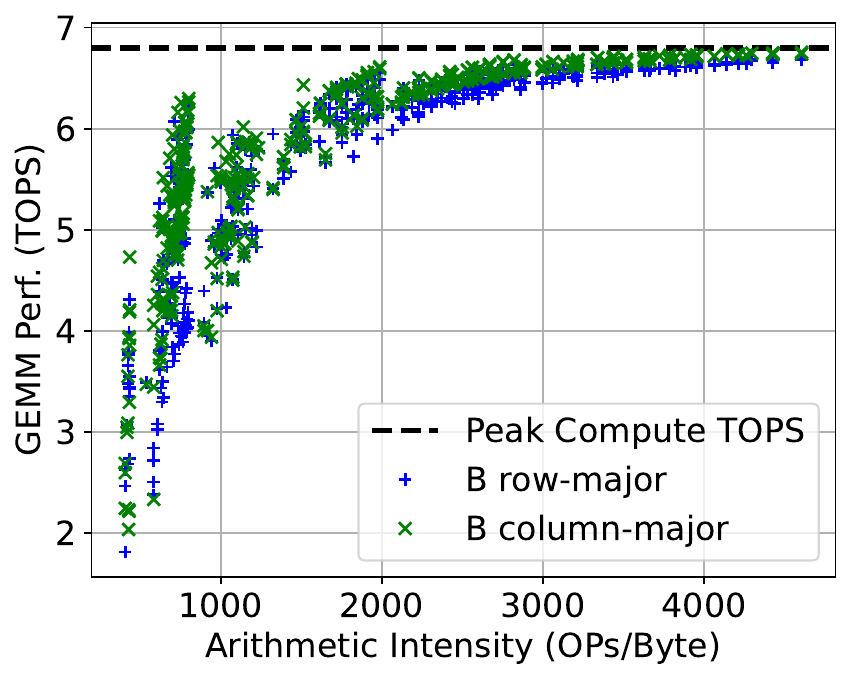}
\label{fig:GEMM_sweep_int8_int8_phoenix}}
\hfill
\subfloat[int8-int16 XDNA]{\includegraphics[width=0.45\linewidth]{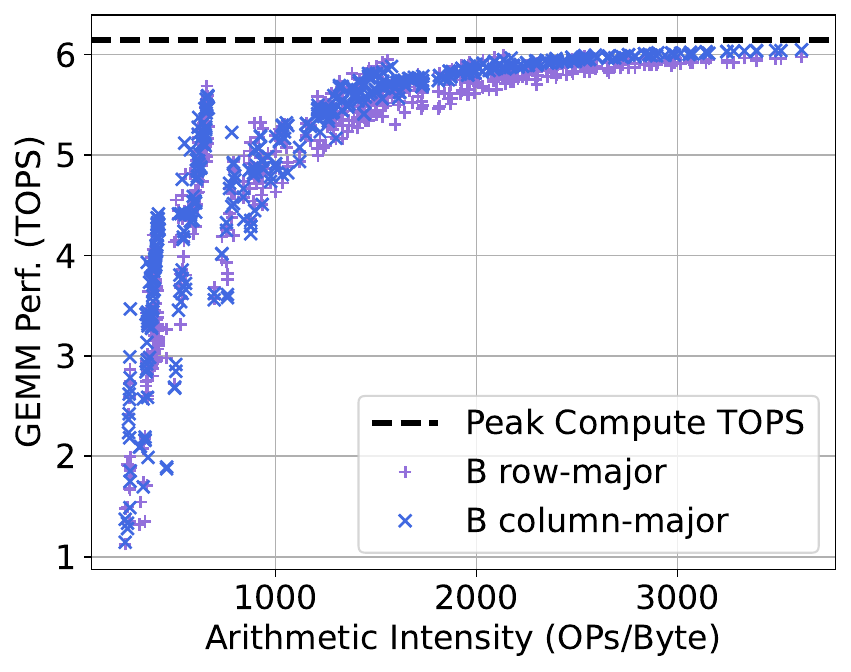}
\label{fig:GEMM_sweep_int8_int16_phoenix}} 
\hfill
\\
\subfloat[bf16-bf16 XDNA]{\includegraphics[width=0.46\linewidth]{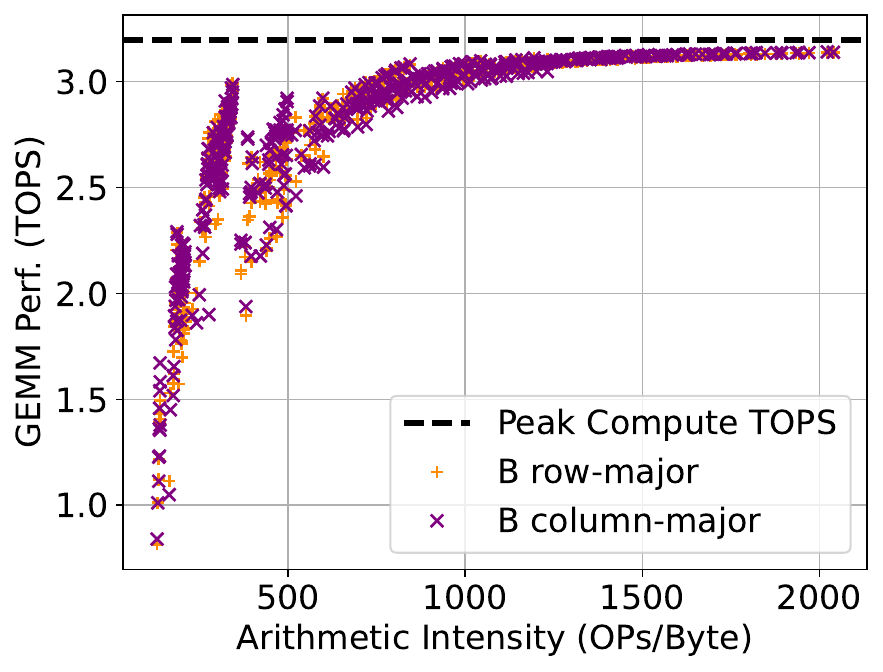}
\label{fig:GEMM_sweep_bf16_bf16_phoenix}} 
\hfill
\subfloat[int8-int8 XDNA2]{\includegraphics[width=0.45\linewidth]{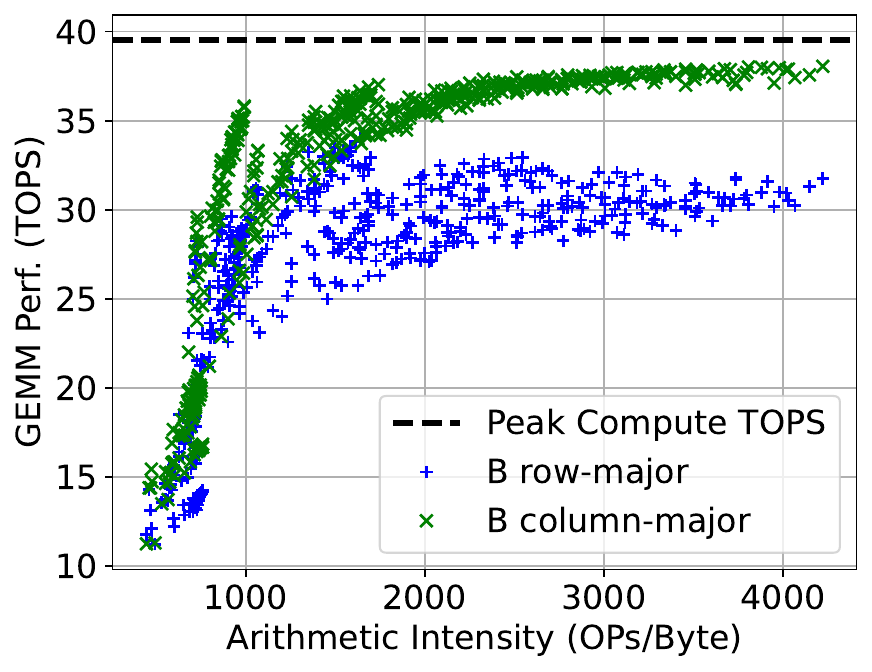}
\label{fig:GEMM_sweep_int8_int8_kracken}}
\\
\subfloat[int8-int16 XDNA2]{\includegraphics[width=0.45\linewidth]{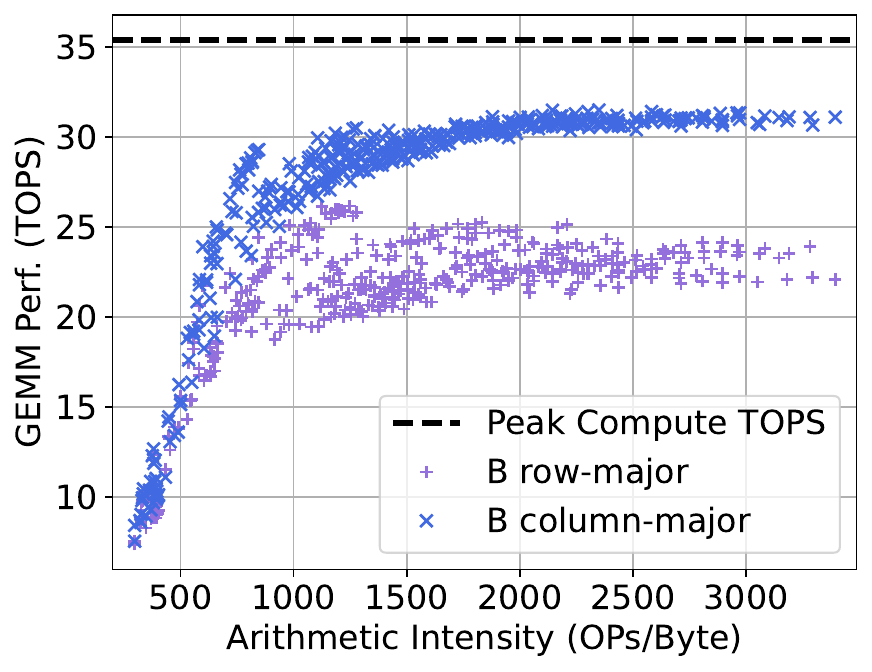}
\label{fig:GEMM_sweep_int8_int16_kracken}} 
\hfill
\subfloat[bf16-bf16 XDNA2]{\includegraphics[width=0.45\linewidth]{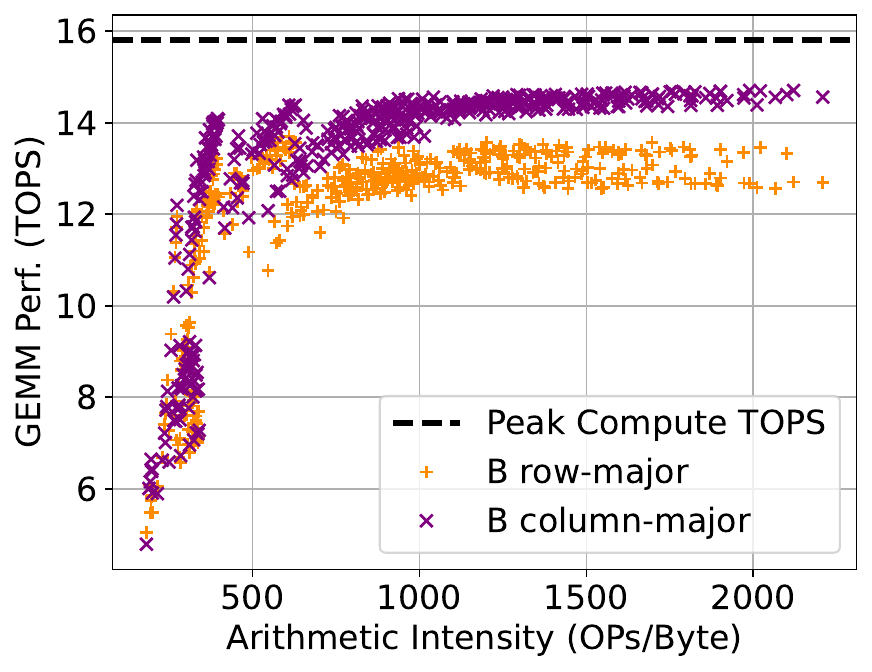}
\label{fig:GEMM_sweep_bf16_bf16_kracken}} 
\caption{Roofline GEMM performance sweeps for various matrix sizes on XDNA \& XDNA2.} 
\label{fig:GEMM_sweeps}
\end{figure}

In Fig. \ref{fig:GEMM_sweeps} we show roofline GEMM performance sweeps (in linear scale) for various data types. 
Each point represents a matrix size that is a multiple of the \textit{native} GEMM size (using the optimal \textit{balanced} kernel).
We select more than 400 points for each case (separately for $B$ in \textit{column-} and \textit{row-major}), up to 8K-sized matrices, \textit{without} favoring any particular $M$, $K$, $N$ dimension.
First, notice that when arithmetic intensity (ARI) is low, \emph{i.e.,} small matrix sizes, performance is limited.
In this case, GEMM is severely memory bound.
However, as ARI increases, GEMM performance improves, and becomes more stabilized after a specific value. 
In all cases, storing matrix $B$ in \textit{column-major} provides, on average, higher performance compared to \textit{row-major}.
This is because of accessing sufficient contiguous data for both $A$ and $B$ matrices (determined by $k_\text{mt}$ parameter).
However, for $B$ in \textit{row-major}, the contiguous access is limited to the $n_\text{ct}$ parameter, while only for $A$ (\textit{row-major}), $k_\text{mt}$ contiguous data are traversed.
To this end, for XDNA, we observe, on average, 4.8\%, 4.4\%, and 0.57\% higher performance, for int8-int8, int8-int16, and bf16-bf16, respectively.

Moreover, we notice that for XDNA2 the difference between \textit{column-} and \textit{row-major} is higher.
In particular, we observe, on average, 19.1\%, 25.2\%, and 8.7\% higher performance, for int8-int8, int8-int16, and bf16-bf16, respectively.
This difference between XDNA and XDNA2 is presumably attributed to complex interaction between the NPU NoC, the SoC-level fabric and DRAM \cite{NPU_arch_MICRO_2024, NPU_HOT_CHIPS_2023}, which affects the \textit{effective} DRAM BW that NPU perceives (although both mini PC devices are equipped with the same DRAM technology).
Also, we note that for both XDNA and XDNA2, the difference between \textit{column-} and \textit{row-major} for bf16 is lower compared to int8.
This is because of accessing a larger number of bytes for bf16 across the $n_\text{ct}$ dimension (when $B$ is in \textit{row-major}), which increases GEMM performance in this case, thereby lowering their difference.


For XDNA, we note that performance gets stabilized  after a specific ARI value for $B$ in both \textit{row-} and \textit{column-major} formats (almost resembling a line in Fig. \ref{fig:GEMM_sweep_int8_int8_phoenix}, \ref{fig:GEMM_sweep_int8_int16_phoenix}, \ref{fig:GEMM_sweep_bf16_bf16_phoenix}).
Moreover, for XDNA2, we observe that for $B$ in \textit{column-major}, GEMM performance also resembles a stable line.
However, for $B$ in \textit{row-major} it displays a more scattered distribution (Fig. \ref{fig:GEMM_sweep_int8_int8_kracken}, \ref{fig:GEMM_sweep_int8_int16_kracken}, \ref{fig:GEMM_sweep_bf16_bf16_kracken}).
This is because XDNA2 has higher reliance on the  \textit{effective} DRAM BW (due to attaining significantly higher absolute TOPS values), which is substantially increased and stabilized when accessing sufficient contiguous data across both $A$ and $B$ matrices.
For example, for int8-int16 on XDNA2, we measure a variability of only 5\% for $B$ in \textit{column-major}, while for \textit{row-major} the variability is 19\% (ARI $>$ 1600).
Finally, across all points in GEMM sweeps, XDNA attains up to 6.76 (int8-int8), 6.05 (int8-int16), 4.57 (int8-int32), and 3.14 (bf16-bf16) TOPS.
Similarly, XDNA2 achieves up to 38.05 (int8-int8), 31.52 (int8-int16), 25.31 (int8-int32), and 14.71 (bf16-bf16) TOPS (int8-int32 sweep omitted for brevity).

\section{Insights \& Discussion}
\label{sec:RyzenAI_NPUs:Insights}

\subsection{Performance Across Multiple GEMM Sizes in DL Workloads} 
\label{subsubsec:GEMM_perf_multiple_sizes}


Modern DL workloads perform GEMMs with a wide range of  sizes across their layers. 
Our employed output stationary mapping allows \textit{arbitrary} GEMM dimensions to be supported, by applying zero-padding to align with the \textit{native} GEMM size (Sec. \ref{subsec:GEMM_mapping_strategy}).
Zero-padding can be applied efficiently by utilizing the NPU's architectural support for on-the-fly zero-padding in MemTile channels \cite{AMD_AIE_ML_architecture_manual}; leveraging this feature is left for future work.

Switching between different GEMM sizes can incur critical performance inefficiencies.
To this end, one approach could be to reconfigure the NPU array with a dedicated GEMM design for each size.
When quantifying the reconfiguration latency of the entire GEMM design, we measure a delay of 3.4 ms and 4.9 ms on XDNA and XDNA2, respectively.
However, this reconfiguration latency is comparable to the GEMM execution time (\emph{e.g.,} a $\sim$4K square GEMM for int8-int16 on XDNA2 takes 5.2 ms).
This underscores that reconfiguring the entire design can impose substantial overheads.

When retaining the \textit{same} GEMM design on the NPU, only two parameters require reconfiguration across different problem sizes ($M, K, N$): (i) the total number of output tiles $M\cdot N/(m_\text{ct} \cdot n_\text{ct})$, and (ii) the number of tiles across the reduction dimension $K/k_\text{ct}$ \cite{roesti2025unlocking}.
According to our measurements, this negligible reconfiguration does not incur any noticeable performance overhead at the system level.
Hence, identifying the optimal parameters, \emph{i.e.,} $m_\text{ct}$, $k_\text{ct}$, $n_\text{ct}$, $k_\text{mt}$, Sec. \ref{subsec:npu_array_GEMM_performance}, and reusing them across different GEMM sizes is essential for high-performance DL deployment.
Finally, we note that the system-level GEMM results presented are specific to the mini PCs used in this evaluation.
However, our optimization methodology is generalizable to any NPU device in the current two generations.


\subsection{Single Output Buffer} 
\label{subsubsec:single_output_buffer}

Due to output stationary mapping, we retain the output $C$ tiles as single buffers and apply double-buffering only for inputs $A$ and $B$ (Sec. \ref{subsec:GEMM_mapping_strategy}).
This is crucial in maximizing performance in the \textit{general} GEMM case, because it enables significantly higher flexibility in single-core parameter optimization.
In particular, it allows increased values for $m_\text{ct}$ $\times$ $k_\text{ct}$ $\times$ $n_\text{ct}$ parameters, compared to utilizing double-buffering (constrained by L1 memory size).
This results in identifying a \textit{balanced} kernel that has higher performance, thereby enabling higher GEMM performance at the system level.
For example, when using the optimization methodology described in Sec. \ref{subsubsec:system_level_optimization} and apply double-buffering for $C$, we identify the 112$\times$48$\times$96 size as the optimal \textit{balanced} kernel for int8-int16 data type on XDNA2.
For $\sim$4K square GEMM, this kernel provides 26.1 TOPS.
However, the 128$\times$72$\times$112 kernel of Table \ref{tb:top_ranked_solutions_GEMM_XDNA2_array} (single $C$ buffer), provides 30.77 TOPS, representing an 18\%  performance improvement.
Similarly, on XDNA, double-buffering on $C$ provides 2.76 TOPS (80$\times$40$\times$96 kernel), while single buffer offers 3.12 TOPS (96$\times$56$\times$96 kernel on Table \ref{tb:top_ranked_solutions_GEMM_XDNA_array}, 13\% higher).
The transfer of the output $C$ tiles in the single  buffer case gets amortized as the reduction $K$ dimension becomes sufficiently high (typically $<$5\% degradation in GEMM performance when $K$/$k_\text{ct}$$>$20).





\subsection{NPU--DRAM Data Movement \& BD Reconfiguration} 
\label{subsubsec:BD_reconfiguration}

The procedure delineated in Sec. \ref{subsec:npu_main_memory_data_movement} enables efficient overlapping of NPU--DRAM data movement with BD reconfiguration.
To quantify its impact on performance, we modify our design to synchronize and reconfigure BDs sequentially (without overlap).
In this case, for int8-int16 on XDNA2, we notice only 22.21 TOPS at $\sim$4K square GEMM, while the overlapped design of Table \ref{tb:top_ranked_solutions_GEMM_XDNA2_array} exhibits 30.77 TOPs (28\% decrease for non-overlapped design).
Similarly, for XDNA, for int8-int16 (Table \ref{tb:top_ranked_solutions_GEMM_XDNA_array}), we observe a 27\% degradation in performance.
This highlights the critical role of overlapping DMA transfers with BD reconfiguration in attaining high GEMM performance.




\section{Chapter Summary \& Impact}
\label{sec:RyzenAI_NPUs:Sumamry}
In this chapter, we present a methodology to maximize GEMM performance on Ryzen AI NPUs. We observe an \textit{inverse} relationship between compute and off-chip memory utilization and identify the optimal balance point at which performance is maximized, using a combination of analytical modeling and hardware profiling. Our methodology attains state-of-the-art GEMM performance and generalizes across both current AMD Ryzen AI NPU generations, achieving throughput of up to 6.76 TOPS (XDNA) and 38.05 TOPS (XDNA2) for int8 precision. Similarly, for bf16 precision, our GEMM implementations attain up to 3.14 TOPS (XDNA) and 14.71 TOPS (XDNA2). This study provides key insights into the performance aspects that govern GEMM optimization on reconfigurable NPU architectures.

\chapter{Conclusion and Future Work}
\label{ch:Summary}

Modern DL workloads are predominantly deployed on GPUs and specialized ASIC accelerators. However, their rapid evolution demands platforms that combine high efficiency with architectural flexibility. Reconfigurable architectures are well positioned to offer this combination, yet their large-scale adoption for DL acceleration has yet to be fully realized.
This dissertation addresses this gap through a comprehensive evaluation of three prominent AI-optimized reconfigurable platforms: the Versal ACAP, Intel Stratix 10 NX, and AMD Ryzen AI NPUs. Using GEMM as the primary workload, this evaluation yields insights into their performance and efficiency characteristics.
Building on these insights, this dissertation proposes architectural enhancements that enable efficient acceleration of both dense and sparse DL workloads. The proposed in-fabric sparsity support addresses a key gap in the literature on structured sparse acceleration.
Together, these contributions advance the case for reconfigurable architectures as an efficient and adaptable platform for DL acceleration.

\section{Summary of Contributions}
\label{sec:Summary:DissertationSummary}

First, we target the AMD Versal AIE, whose array of programmable vector processors introduces a new paradigm for GEMM acceleration on reconfigurable hardware. We propose MaxEVA, a framework to map and optimize GEMM workloads onto the AIE by exploiting its architectural features and resolving performance bottlenecks from multiple angles. On the VC1902 device, MaxEVA reaches up to 5.44 TOPS (fp32) and 77.01 TOPS (int8) throughput, outperforming the prior state-of-the-art by up to 2.19$\times$ in throughput and 20.4\% in energy efficiency, and establishes the foundation the remaining work builds on.

Second, we extend this foundation in two ways: we add a memory hierarchy level to MaxEVA to support arbitrary matrix sizes beyond what fits in on-chip memory resources of the Versal FPGA, and we introduce a parallel framework for the Intel Stratix 10 NX, whose in-fabric tensor blocks take a fundamentally different architectural approach from the AIE. The two frameworks achieve up to 77 and 68 TOPS (int8) throughput, with up to 0.94 and 1.35 TOPS/W energy efficiency, for Versal and Stratix respectively.
Notably, Stratix 10 NX delivers $\sim$44\% higher energy efficiency, suggesting that in-fabric hardened compute blocks are a promising path to efficient FPGA acceleration for DL.
Direct comparison also yields practical insights and guidelines on GEMM optimization, programmability, architectural attributes, and limitations across both platforms.

Third, building on the efficiency advantage of in-fabric hardened AI blocks, we address a key limitation of existing architectures: these blocks are optimized primarily for dense computation and cannot efficiently exploit the structured sparsity present in most modern DL workloads.
To this end, we architect the SSTs, in-fabric hard blocks that support 2:4 and 1:4 structured sparsity, as well as a new 1:3 sparsity mode, in addition to dense operation.
Applied to GEMM accelerators, SST-based designs achieve up to 5$\times$ higher FPGA frequency and 10.9$\times$ lower area than traditional FPGA logic, with up to 3.52$\times$ speedup on sparse ViT and CNN models at minimal area cost of up to 13.3\%. Where the first two contributions optimize dense GEMM on existing devices, this work proposes a new hardware architecture to close this important sparsity gap.

Finally, we extend this work beyond FPGAs to AMD's Ryzen AI NPUs, a direct architectural evolution of the Versal AIE. We present a common methodology for optimizing GEMM across both NPU generations, XDNA and XDNA2, achieving up to 6.76 and 38.05 TOPS (int8), and 3.14 and 14.71 TOPS (bf16), respectively. While many optimization principles from earlier work carry over to this architecture, the NPU's distinct execution model and system-level constraints require a new methodology tailored specifically to XDNA and XDNA2, underscoring that generalizing across reconfigurable AI hardware demands adaptation rather than direct reuse.

\section{Future Work}
\label{sec:Summary:FutureWork}

First, for MaxEVA, the proposed mapping methodology extends directly to other special cases of GEMM, such as matrix-vector multiplication (GEMV), which we leave as future work.
Moreover, for the Versal \textit{vs.} Stratix comparison, future work could broaden the evaluation to include actual end-to-end DL workloads rather than GEMM alone, as well as additional devices, to provide a more holistic view of the advantages and disadvantages of each platform.

Second, for the SST slices, our current design supports static sparsity targeting DNN weights; future work could extend this support to dynamic sparsity in activations. Additional modes could further increase flexibility, such as element-wise computation within SSTs, similar to those explored in prior tensor-slice designs \cite{TS_Aman_FPGA_2021, Aman_TS_TRETS_2022}. 
GEMV is already supported directly by the SSTs at full utilization with a batch size of four. Improving utilization at a batch size of one is possible by adding multiplexing logic, as shown in \cite{Aman_TS_TRETS_2022}, and is left for future work.
Future work could explore multiple input crossbars to further reduce the SSTs’ area overhead, as well as a broader architectural study of area allocation and the trade-offs between area and FPGA routability introduced by the SST routing circuitry. 
Finally, future work could pursue an in-depth exploration of the architectural placement and organization of FPGA building blocks, \emph{e.g.}, logic and memory blocks, SST slices, etc., to optimize the architecture for DL workloads without significantly degrading the performance of non-DL workloads.

Third, for the GEMM optimization methodology on Ryzen AI NPUs, XDNA2 adds hardware support for bfp16 precision, in which multiple values share a common exponent. This introduces additional challenges for data layout transformations that exploit the NPU’s multi-dimensional DMA addressing, which we leave for future work.
Additionally, similar to the other works in this dissertation, the proposed optimization methodology extends directly to GEMV, which we also leave as future work.

Throughout this dissertation, we deliberately focus on GEMM as the primary workload. 
As the dominant computational kernel in most modern DL models, it provides a natural and representative basis for studying and optimizing reconfigurable architectures for DL.
A key direction for future work is to build on the architectural insights and optimization techniques developed here to support full end-to-end DL workload deployment on these devices, extending beyond  GEMM optimization to complete model execution.

\appendices

\sloppy
\bibliographystyle{unsrtnat}
\bibliography{references}        



\end{document}